\documentclass[a4paper,notitlepage, 10pt]{report}

\usepackage{wrapfig}
\usepackage{epigraph}
\usepackage{gensymb}
\usepackage{slashed}
\usepackage{float}
\usepackage{paralist}

\usepackage[english]{babel}

\usepackage[utf8x]{inputenc} 

\usepackage[T1]{fontenc} 

\usepackage[a4paper,top=3cm,bottom=2cm,left=3cm,right=3cm,marginparwidth=1.75cm]{geometry} 

\usepackage{titling} 

\usepackage{extarrows}

\usepackage{verbatim}

\usepackage{wasysym}

\usepackage{arydshln}

\usepackage{multirow}

\usepackage{booktabs}

\usepackage{amsmath}

\usepackage{bbm}

\usepackage{cite}

\usepackage{mathrsfs}

\usepackage{braket}

\usepackage{physics}

\usepackage{mathtools}

\usepackage[makeroom]{cancel}

\usepackage{xfrac}

\usepackage{amssymb}

\usepackage{mathrsfs}

\usepackage[colorinlistoftodos]{todonotes}

\usepackage{color}

\usepackage[colorlinks=true, allcolors=a]{hyperref}

\usepackage{mathalfa}

\usepackage[myheadings]{fullpage}

\usepackage{fancyhdr}

\usepackage{lastpage}

\usepackage{graphicx, setspace}

\usepackage[font=small, labelfont=bf]{caption}

\usepackage{fourier}

\usepackage[protrusion=true, expansion=true]{microtype}

\usepackage{sectsty}

\usepackage[symbol]{footmisc}

\usepackage{float}

\usepackage{subcaption}

\usepackage{appendix}

\usepackage{afterpage}

\usepackage{babel}

\usepackage[acronym, toc]{glossaries}

\numberwithin{equation}{section}
\definecolor{a}{rgb}{0.9,0.2,0.5}

\providecommand{\sin}{} \renewcommand{\sin}{\hspace{2pt}\textrm{sen}}
 
\begin{document}
\newpage{}
\noindent\begin{minipage}[t][1\totalheight][c]{1\columnwidth}%
\begin{onehalfspace}
\begin{center}
\textsc{Federal University of Rio Grande do Sul}
\par\end{center}
\begin{center}
\textsc{Institute of Physics}
\par\end{center}
\end{onehalfspace}
\end{minipage}

\vspace{3cm}

\noindent
\begin{minipage}[c]{1\columnwidth}%
    \begin{center}
        \textbf{\LARGE Investigating Notions of Entropy and Its Production in High Energy Particle Collisions}
    \end{center}%
\end{minipage}

\vspace{2cm} 
\noindent
\begin{minipage}[c]{1\columnwidth}%
    \begin{center}
        \normalsize{Student: Gabriel Silveira Ramos} \\
        \normalsize{Advisor: Magno V. T. Machado}
    \end{center}%
\end{minipage}

\vspace{1.5cm} 
\begin{flushright}
    \begin{minipage}[c]{0.45\columnwidth}
        Thesis presented to the Institute of Physics at the Federal University of Rio Grande do Sul as a requirement for obtaining the title of Doctor of Philosophy in Physics through the Graduate Program in Physics.

        \vspace{0.5cm}
        
        Field of study: High Energy Particle Physics.
    \end{minipage}
\end{flushright}

\vspace{3cm}

\noindent\begin{minipage}[c]{1\columnwidth}%
\begin{onehalfspace}
\begin{center}
Porto Alegre - RS
\par\end{center}
\begin{center}
2024
\par\end{center}
\end{onehalfspace}

\end{minipage}

\newpage
\vspace*{\fill}
{ \raggedleft

In memory of Dorival Machado.

~
}

\newpage
\vspace*{\fill}
{ \raggedleft

\textit{Chaos was the law of nature; Order was the dream of man. \\
Henry Adams}

~
}

\chapter*{Acknowledgments}

\hspace{0.5cm}To my parents, for standing by my side once again, serving as a steadfast support.

To my advisor, {\it Magno V. T. Machado}. Just like a father, who patiently teaches us our first steps and is always available, the professor follows the same path. The difference is that, in the professor's case, the first steps involve learning which tools to use, which research direction to follow, and the best way to explain a concept.

{\it Anita Barros Vanzin}, whom I was fortunate to meet during these studies. Your presence made life lighter.

To my friends, {\it Lars Sanhudo de Souza}, {\it Vinicius Sanhudo de Souza}, {\it Diego Pierozan}, {\it Franco Tonietto}, {\it Felipe Costa}, {\it João Otávio May Buogo}, {\it Guilherme Petrini}, {\it Juliana Harmatiuk}, {\it Guilherme Ramos Broliato}, {\it Felipe da Tomazoni}, and {\it Kelvin Sayao}.

\chapter*{Abstract}

This work analyzes different notions of entropy and its production in $ep$ and heavy-ion collisions, focusing on the early stages of the collision. To this end, the importance of phenomena such as quantum entanglement and decoherence in entropy generation was studied. Using state-of-the-art methods for calculating entanglement entropy in the high-energy limit and synthesizing various approaches for its computation, phenomenological results were obtained from analytical expressions for the number of gluons. Results were also presented for the entanglement entropy in two-body elastic scattering based on the hadronic structure given by the model-independent Lévy method, at energy values typical of the RHIC, Tevatron, and LHC. Phenomenological models of unintegrated gluon distributions were used to calculate dynamic entropy in the QCD of dense gluonic states in $pp$ and $pA$ collisions at high energies, comparing it with decoherence entropy. The obtained results were contrasted with other notions of entropy in the literature, and theoretical uncertainties were discussed.\newline {\bf Keywords:} High-energy physics, entropy, entanglement.\footnote{This work was originally written in Portuguese and translated into English by the author.}

\newpage

\sectionfont{\scshape}

{\hypersetup{linkcolor=black}

\tableofcontents
}

\pagestyle{fancy}

\chapter{Introduction}
\label{chapter 1: Introduction}

The notion of entropy was proposed with the development of {\it Thermodynamic Theory} in the context of the first thermal machines. Its name was coined by {\it Rudolf Clausius}\footnote{{\it Clausius} wanted the new concept to sound like {\it energy}\cite{clausius1865}, where, in his own words and in free translation, he stated: "But as I consider it better for significant magnitudes to have terms from ancient languages that can be adopted unchanged in all modern languages, I propose calling the magnitude $S$ the entropy of a body, from the Greek word $\epsilon \nu \tau \rho o \pi \iota \alpha$, {\it transformation}. I intentionally formed the word entropy to be as similar as possible to the word energy; for the two magnitudes to be indicated by these words are so nearly allied in their physical meanings that a certain similarity in designation seems desirable."}\cite{HistFis}, after he had come into contact with the work of {\it Nicolas Sadi Carnot}\cite{carnot}, where the first developments on the impossibility of fully transforming energy into work began to become clearer. Thus, the natural saturation of the energy available in a transformation process aimed at producing work was discovered. The importance of this observable can be verified in the postulation of the {\it second law of thermodynamics}.

Thermodynamics is a phenomenological theory of great success, having been capable, as early as the 19th century, of providing a set of simple laws that would be verified in the thermal processes compiled in the technological developments of its era. However, the questions surrounding the behavior of matter at that time were still quite obscure; even the atomic theory was not well established, and a fundamental explanation of the behavior of matter observed in the laws of thermodynamics was only possible with the advent of {\it Statistical Mechanics}. This theory has, as its main founders, {\it James Clerk Maxwell} and {\it Ludwig Eduard Boltzmann}, the latter being responsible for a deeper understanding of entropy.

The {\it modus operandi} of Statistical Mechanics consists in characterizing a group of microscopic entities through probability theories, with the aim of explaining their macroscopic behavior. At the time of its development, the scientific community had not yet discovered {\it Quantum Mechanics} or {\it Theory of Relativity}, and the predominant theory for describing the dynamics of particles and bodies was {\it Classical Mechanics}, which includes {\it Newton's Laws} or, equivalently, the Lagrangian and Hamiltonian formulations.

Considering only particles, Classical Mechanics is capable of obtaining consistent results from initial conditions, such as the particle's potential and kinetic energy, its constraints and motion restrictions, as well as its position and velocity. However, when considering, for instance, water ($H_2O$), in which just one gram contains approximately $3.34\times 10^{22}$ molecules, the use of Classical Mechanics tools to solve Newton's Second Law requires the characterization of six times the number of molecules to resolve it, that is, $6 \times 3.34\times 10^{22}$ differential equations. Even if this calculation were feasible, a computer printing one coordinate per microsecond would take about 10 billion years to complete the process — nearly the age of the universe\cite{callen1960}. Moreover, even if it were possible to reduce this time using a supercomputer capable of finding the solutions in a few days, the human mind would not be able to evaluate the result of such a large number of variables.

The idea of {\it Boltzmann} and {\it Maxwell} was to describe this information using probability and statistical theories, thereby obtaining average values. After all, for macroscopic beings, such as humans, what really matters most of the time is knowing whether there will be oxygen available when breathing; whether the air temperature is not so high as to burn the lungs; and whether the air pressure is not so high as to rupture the eardrums\cite{greene2021ate}. Thus, the observables that truly matter are those referred to, in the language of thermodynamics, as intensive variables, such as pressure and temperature, or extensive variables, such as energy, volume, and entropy.

Entropy is also associated with the {\it irreversibility of natural processes}. Until then, dynamic theories did not seem to distinguish between past and future. For example, in a game of pool, when someone takes the opening shot, striking the cue ball to scatter the others that are arranged in a triangular formation, these balls eventually move until they come to rest. Using dynamic laws, it is possible to imagine the scene reversed: the scattered balls moving back to recompose the original triangular formation and, finally, the cue ball returning to its initial position\footnote{However, it is now known that in complex systems, this possibility is rendered unfeasible by classical chaos theory, which shows that small uncertainties in the initial conditions can grow exponentially, making the system unpredictable and, in many cases, irreversible.}. However, this scenario changes if an observer watches the scene with night vision goggles, perceiving the heat released in the collisions between the billiard balls\footnote{This excellent example is taken from Julian Barbour's The Janus Point: A New Theory of Time\cite{barbour2020janus}.}. This heat cannot return to do work on the balls, just as one does not expect the waters of a deep lake to stir and expel a stone from its interior, or for the shards of a broken glass on the floor to spontaneously reassemble. There is an order to the progression of natural phenomena, and this order is the one that increases entropy, in the so-called {\it arrow of time}.

From this perspective, every moment in the universe up to the present state is unique and exclusive, progressing in a determined direction: the one that increases entropy. From this, there are two major pieces of information that neither {\it Boltzmann}, {\it Maxwell}, nor any other 19th-century scientist knew: the Universe is expanding at an accelerated rate, and it had a beginning, called the {\it Big Bang}. The discovery of the universe's expansion is attributed to astronomy\footnote{The discovery of the universe's expansion is credited to the astronomer Edwin Hubble\cite{hubble1929relation}, who in 1929 observed that distant galaxies are moving away from Earth and, crucially, that the velocity at which they recede increases with distance—a relationship known as Hubble's Law.}; Statistical Mechanics unveiled the arrow of time, but nowadays, to understand the universe's earliest moments, particle physics is required, particularly in the study of the {\it Quark-Gluon Plasma} (QGP). This plasma was discovered in 2005\cite{adcox2005} from collisions involving heavy gold ions at the Relativistic Heavy Ion Collider (RHIC), and since then, new experiments have been conducted at both this accelerator and the Large Hadron Collider (LHC), including the A Large Ion Collider Experiment (ALICE), A Toroidal LHC Apparatus (ATLAS), and Compact Muon Solenoid (CMS).

Plasma is a state of matter that can be considered exotic under Normal Temperature and Pressure Conditions. Although it shares some properties with gases, the high temperature causes the kinetic energy of its constituents to globally exceed their binding energy, resulting in ionization. In the case of atoms and molecules, all matter dissociates into a highly ionized high-temperature gas. In this regime, the main theory at play is Electromagnetic Theory, and the force involved is electromagnetic in nature. Furthermore, as energy increases, regimes emerge where the very constituents of atomic nuclei (nucleons) and mesons are no longer fundamental but are instead formed by quarks. In this case, the relevant interaction is no longer the electromagnetic force but a much stronger and extremely short-range force: the strong force, mediated by gluons.

If "ordinary" plasma is rare under usual temperature and pressure conditions, a quark-gluon plasma is even rarer: its critical temperature\footnote{The critical temperature, $T_C$, is the temperature above which a substance can only exist in the gaseous phase, regardless of the applied pressure.} is approximately\footnote{It was recently discovered that some mesons composed of heavy quarks, such as the top quark or the charm quark, do not dissolve until the temperature reaches about 350~$MeV$ ($4.3\times10^{12}~K$)\cite{young}.} $175~$MeV ($2.0\times10^{12}~$K)\cite{rafelski,majumder}. These temperature values are associated with extremely high energies and matter densities. For comparison, the temperature of the Sun's surface is approximately $6000~$K. These extreme conditions are not the only distinctive feature of this state: quarks have fractional electric charge (relative to the elementary charge of the electron, $e$) and move at high velocities within the QGP, generating both electric and magnetic fields. Calculations involving electrodynamic theory yield magnetic field values on the order of $10^{18}~$G\cite{tuchin} for collisions involving gold ions, which can be up to thirty times higher for lead-ion collisions. For comparison, a {\it magnetar} (a neutron star characterized by an extremely strong magnetic field) has a magnetic field in the range of $10^{15}~$G\cite{kouveliotou}.

Thus, with temperatures and magnetic fields that can match or even surpass cosmological phenomena, the QGP only occurs in reactions involving extremely high energies. Furthermore, it is speculated that this state of matter was present in the early stages of the universe, according to the {\it Big Bang} theory\cite{muller}, more precisely, within the first 20 to 30 microseconds.

Thus, the intersections of the study between elementary particle physics and entropy hold a prominent position: the object of study is the elementary entities and their interactions, meaning that the analysis of entropy will be conducted in the most fundamental regime of matter, its most elementary form. From this, the study of entropy can be fundamentally extended to its {\it creation}. Therefore, the main objective of this work is to study entropy creation, initially in $ep$ collisions, and subsequently to understand more complex cases, such as heavy-ion collisions in ultrarelativistic regimes. This extension is a direct continuation of the work conducted in \cite{dissertacaogabriel}, expanding it to the regime of heavy nucleons in an attempt to better understand fundamental properties from the simplest hadronic collisions to the investigation of the early stages of the QGP.

Thus, since entropy will be investigated in collisions involving the strong force, the appropriate Quantum Field Theory is {\it Quantum Chromodynamics} (QCD). In this research program, much of the information about the behavior and properties of matter is derived from measurements of the particles produced and their spectra in the final states of heavy-ion collisions. These properties can be interpreted in terms of concepts from thermodynamics and relativistic hydrodynamics. At the energies available in current colliders, one of the most relevant quantities is the azimuthal quadrupole anisotropy of the collective flow, commonly referred to as the elliptic flow $v_2$. Excellent agreement for this quantity is achieved through hydrodynamic calculations when compared to the measured anisotropy of the flow of matter produced in nuclear collisions. Theoretically, the hypothesis of a rapid thermal equilibrium of matter on a timescale of approximately $1$ fm/c is required for experimental results to be accurately described. Such a short timescale is interpreted as problematic, as field theories fail to adequately explain how matter thermalizes so quickly.

In principle, we can distinguish five different stages in entropy production, namely: 1) decoherence of the initial nuclear wave functions, 2) thermalization of the glasma, 3) dissipation due to shear viscosity during hydrodynamic expansion, 4) hadronization accompanied by high viscosity of the collective system generated (known as bulk viscosity), and 5) viscous hadronic freeze-out. These different stages raise the question of how entropy is created in reactions, as there is currently only a partial understanding of the contribution of each stage to the final entropy \cite{ar2}.

In this context, one of the best-known quantities in heavy-ion collisions is the final entropy per unit of rapidity\cite{ar3}\footnote{A relativistic physical quantity, also known as the {\it hyperbolic parameter}, defined as $\frac{1}{2}\ln \left(\frac{1+\beta}{1-\beta}\right)$.}, $dS/dy$, which can be determined from the hadron spectrum in the final state combined with information about the source, extracted from correlations between identical particles using a method known as {\it Hanbury-Brown-Twiss} (HBT) interferometry. Alternatively, this quantity can also be obtained from the abundance of produced hadrons combined with the entropy per particle for a hadron gas in chemical equilibrium at a temperature of $T_C\approx$ 160 MeV, $S/N\approx$ 7.25. Significant discrepancies exist when comparing results obtained by different methods, which supports the thesis of entropy production during hadronic freeze-out and reflects the entropy production resulting from the decay of excited states of hadronic resonances, as well as the significant contribution of shear viscosity in a thermal hadron gas\cite{ar6,ar7}.

Thus, in this thesis, the production of entropy in the early stages of $ep$ and heavy-ion collisions will be investigated. To this end, different models will be addressed with the aim of characterizing entropy in these early stages of collisions in high-energy physics, generally restricting the analysis to simpler cases, such as $ep$ and $pp$ collisions, and later adapting them to $pA$ collisions. Specifically, three models were investigated to characterize entropy in the early stages of collisions: (I) obtaining the entropy density per unit of rapidity, $dS/dy$, using the so-called QCD Dynamical Entropy generated by the dense states of the QCD medium\cite{pescha}; (II) characterizing quantum entanglement between different QCD-specific observables\cite{karkar, peschanski2016, peschanski2019, cgcvenu} and calculating an entanglement entropy; and (III) entropy production due to the decoherence of hadronic wave functions in the early stages of the collision\cite{ar2}.

In studies on entanglement entropy, some authors argue that it contributes significantly to entropy generation in the high-energy regime\cite{karkar}. One way to evaluate this claim is through the analysis of alternative models of entropic characterization in this regime. Thus, some recently developed entropy models were included in this study, allowing for a comparison of their behaviors. Among them is the semi-classical entropy of {\it Wehrl} in the context of QCD\cite{hatahata}. In this model, quasi-statistical distributions are used to simulate a phase space. Another model addressed is the dynamical entropy\cite{pescha}, which allows for the definition of a calculable entropy based on unintegrated gluon distributions (UGD), which can be defined using the $k$-factorization theorem at high energies. This formulation of entropy at the microscopic level is established through analogies with {\it nonequilibrium Statistical Mechanics}. Finally, the last approach considers the effects of quantum decoherence on a specific notion of entropy\cite{sdeco}, associating the high entropy production in the early stages of collisions with the rapid decoherence involved in the process.

Thus, this thesis is organized as follows: Chapter 2 introduces the basic notions of entropy, covering fundamental concepts of thermodynamics and statistical mechanics, such as Shannon and von Neumann entropies, as well as their relationships, which serve as the foundation for the subsequent study. Chapters 3, 4, and 5 examine each of the phenomena associated with entropy production, namely, the dense states of the QCD medium, quantum entanglement, and decoherence, respectively. Chapter 6 compiles the main results obtained, including the analysis of entropy production per unit of rapidity and the comparison between the different models studied. Finally, Chapter 7 summarizes the conclusions of this work, highlighting the contributions to the understanding of entropy production in QCD systems within high-energy physics.

\chapter{Basic Entropy Notions}
\label{chapter 2: Basic Entropy Notions}

The study of entropy generation and characterization in high-energy physics encounters various formulations of this observable. Historically, as previously discussed, this concept emerged in Thermodynamic Theory, incorporating the second law. Later, with the advent of Quantum Mechanics and the confirmation of the atomic hypothesis, new approaches were developed, such as von Neumann entropy. A more modern perspective, developed by {\it Claude Shannon}, created an entirely new field of research: {\it Information Theory}.

These notions of entropy, although scattered across different theories, have close relationships, and this chapter aims to briefly introduce each of them, demonstrating cases where one notion overlaps with another. This distinction is important because the use of one or the other varies depending on the context and may cause confusion for scientists conducting the analysis.

\section{Entropy in Statistical Mechanics}
\label{chapter 2: Section 1 - Entropy in Statistical Mechanics}

\begin{figure}[ht]
    \centering
    \includegraphics[width=0.70\linewidth]{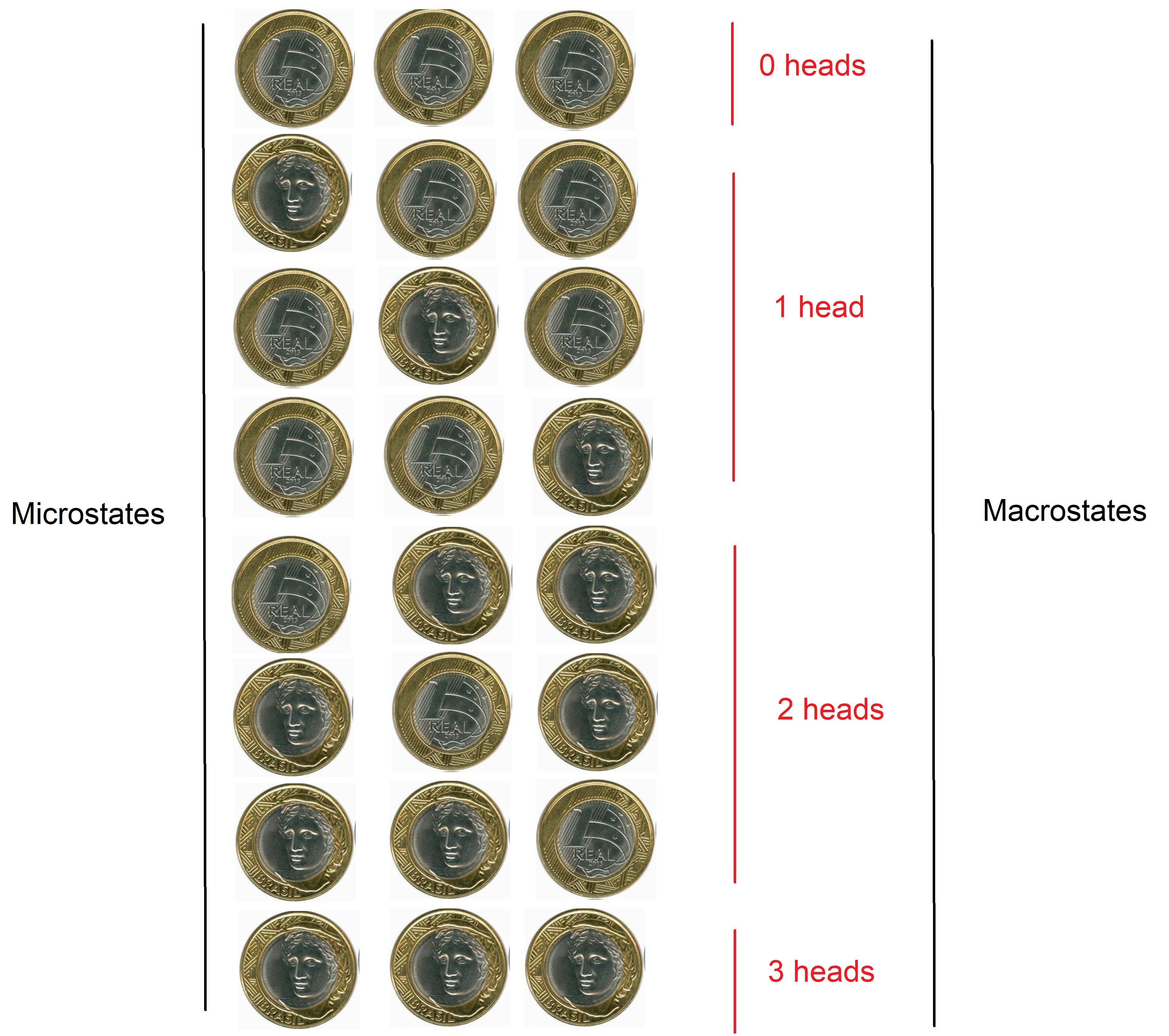}
    \caption{Possible microstates resulting from flipping three coins, organized into their respective macrostates.}
    \label{CK}
\end{figure}

Because of the wide range of observables referred to as "entropy", it is common to call the entropy calculated in Statistical Mechanics as {\it Thermodynamics Entropy}. The Statistical Mechanics modus operandi is based in initially exposes the microscopic characteristics of the elements that compose the system. Then, tools from probability and statistics are used to obtain the observables of interest, such as equations of state and intensive variables. A starting point is the fundamental postulate of statistical mechanics:

\begin{quote}
    {\it In a closed statistical system with fixed energy, all accessible microstates are equally probable}.
\end{quote}
To understand this postulate, it is necessary to define the concepts of microstate, macrostate, and multiplicity, as follows:

\begin{quote}
    {\bf Microstate:} Specific configuration of a system at a given instant.\\
    {\bf Macrostate:} Set of microstates that share the same observable global characteristic.\\
    {\bf Multiplicity:} Number of microstates corresponding to a given macrostate.
\end{quote}
For example, consider the case of three distinguishable and fair R\$ 1.00 coins (one brazilian real), meaning each has an equal probability of showing heads (H) or tails (T) when flipped. After flipping the coins and observing the results, configurations such as TTH, HHT, or HHH can be obtained. Each of these individual configurations is an example of a {\it microstate} of the system. Fig.~[\ref{CK}] illustrates all possible outcomes of the coin flips.

When analyzing the microstates resulting from the coin flips, an organization into groups is observed: 0 heads, 1 head, 2 heads, and 3 heads. These groups represent the {\it macrostates} of the system, where the global characteristic defining each group is the number of heads. For example, in the macrostate with {\it 1 head}, there are three possible configurations: HTT, THT, and TTH. In the macrostate with {\it 3 heads}, there is only one configuration, HHH. The number of microstates associated with a macrostate is called {\it multiplicity}, denoted here by $\Omega$, which depends on the number of heads (ranging from 0 to 3). Thus, when constructing a mathematical expression for the multiplicity as a function of the number of heads, for instance, for the macrostate with {\it 2 heads}, $\Omega_2 = 3$.

\begin{figure}[ht]
    \centering
    \includegraphics[width=0.50\linewidth]{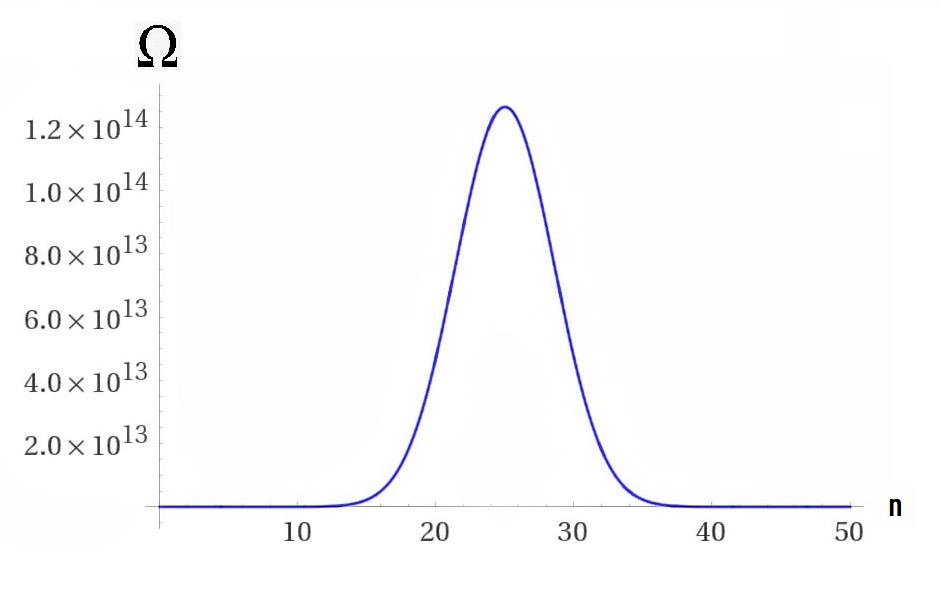}
    \caption{Multiplicity of microstates in the case of flipping 50 coins. The highest multiplicity is obtained for an equal number of 25 heads and 25 tails.}
    \label{50caras}
\end{figure}

Concluding this analysis, the difficulty in obtaining an expression for the multiplicity increases as the number of coins grows. Considering a statistical system composed of 50 coins, it is straightforward to determine that the number of microstates for 0 or 50 heads is only 1; however, the other values require a more elaborate analysis. To solve this mathematical problem, {\it combinatorial analysis} is used, where the result is given by a {\it simple combination}, expressed as:

\begin{equation}
    \Omega(N, n) = C^N_n=\frac{N!}{(N-n)!n!}.
    \label{permu}
\end{equation}
In this expression, $N$ is the number of coins, and $n$ is the number of heads. Defining $p$ as the number of tails, it is evident that Eq.~[\ref{permu}] is subject to the constraint $p+n=N$. Figure~[\ref{50caras}] shows the graph of the multiplicity for the flipping of 50 coins. It is notable that the situation with an equal number of heads and tails, 25 each, exhibits the highest multiplicity.

In a game where players must predict the outcome of coin flips, if they know the fundamental postulate of equilibrium statistical mechanics, where all microstates are equally probable, they have an advantage. This is because the multiplicity associated with $i$ heads, $\Omega_i$, and the total number of possible microstates, $\Omega_T$, is given by the sum:
\begin{equation}
    \Omega_T=\Omega_0+\Omega_1+...+\Omega_N=\sum_{i=0}^N\Omega_i.
\end{equation}
Thus, the probability of obtaining $i$ heads in a flip is given by:
\begin{equation}
    P(i)=\frac{\Omega_i}{\Omega_T}.
    \label{prop}
\end{equation}
Therefore, to maximize the chances of winning in the described game, one should bet on the state with the highest probability, which in this case is the macrostate with an equal number of heads and tails, as illustrated in Fig.~[\ref{50caras}].

Another important property in probability analysis lies in the characterization of the multiplicity of two or more sets that form a single system. To understand this, consider a set of six coins divided into two groups, \(A\) and \(B\), with three coins each. When the coins are flipped, the multiplicity of the composite system, denoted by \(\Omega_{AB}\), depends on the two sets \(A\) and \(B\). The mathematical expression for the composite multiplicity is given by the product of the individual multiplicities of \(A\) and \(B\):
\begin{equation}
    \Omega_{AB}=\Omega_A \Omega_B.
    \label{produto}
\end{equation}
For example, if after flipping, the first group of coins shows the macrostate of one head, with a multiplicity of $\Omega_1 = 3$, since this macrostate comprises the equally probable microstates HTT, THT, and TTH, the outcome of the second flip is entirely independent of the first. Therefore, considering all possible outcomes for the second group (0 to 3 heads, with a total multiplicity of $\Omega_B = 8$), the number of ways to obtain one head in the first group combined with any result in the second is $3 \times 8 = 24$. When all possible outcomes for \(A\) and \(B\) are considered, the total number of possible combinations is $64$. Although this example is simple, the method of counting composite multiplicity is a general property for composite systems, always described by Eq.~[\ref{produto}].

\begin{figure}[ht]
    \centering
    \includegraphics[width=0.4\linewidth]{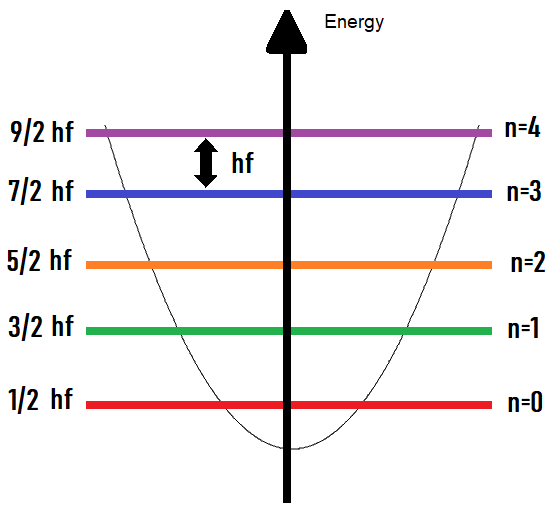}
    \caption{Energy levels in a quantum harmonic oscillator, $n=0$ to $n=4$.}
    \label{OHenergy}
\end{figure}

Now this formalism will be physically contextualized. In 1907, {\it A. Einstein} proposed a mathematical model to characterize the behavior of a solid \cite{einstein1907plancksche}. To understand the functioning of the model proposed by {\it Einstein}, it is worth revisiting the fundamental concepts associated with a {\it one-dimensional quantum harmonic oscillator}. Solving the time-independent {\it Schrödinger} equation for the oscillator's potential energy yields the quantized energy eigenvalues, $E_n$, given by:
\begin{equation}
    E_n=(n+1/2)hf.
    \label{OHE}
\end{equation}
In this equation, $h$ is {\it Planck's constant}, $f$ is the oscillation frequency, and $n$ is an integer. For $n=0$, the result is $E_0=hf$, the zero-point energy. The unit $hf$ is called the {\it quantum of energy}. Figure [\ref{OHenergy}] illustrates the allowed energy values in the oscillator.

The solid model proposed by {\it Einstein} consists of $N$ independent harmonic oscillators, where $N$ represents the number of atoms in the solid. Denoting the total energy as $U$, which results from the sum of the energy of all oscillators, we have:
\begin{equation}
    \begin{split}
        U &= E_1 + E_2 + \dots + E_N \\
          &= \left(n_1 + \frac{1}{2}\right)hf + \left(n_2 + \frac{1}{2}\right)hf + \dots + \left(n_N + \frac{1}{2}\right)hf \\
          &= \frac{N hf}{2} + \left(n_1 + n_2 + \dots + n_N\right)hf.
    \end{split}
\label{Total Internal Energy}
\end{equation}
Since the absolute value of energy has no physical significance, the sum of the zero-point energy, $Nhf/2$, can be set to zero. Additionally, defining the integer $r$ as the sum of the $n_i$'s:

\begin{equation}
    r=\sum_{i=0}^Nn_i.
\end{equation}
Thus, using Eq.~[\ref{Total Internal Energy}], $ U=rhf$.

\begin{figure}[ht]
    \centering
    \includegraphics[width=0.8\linewidth]{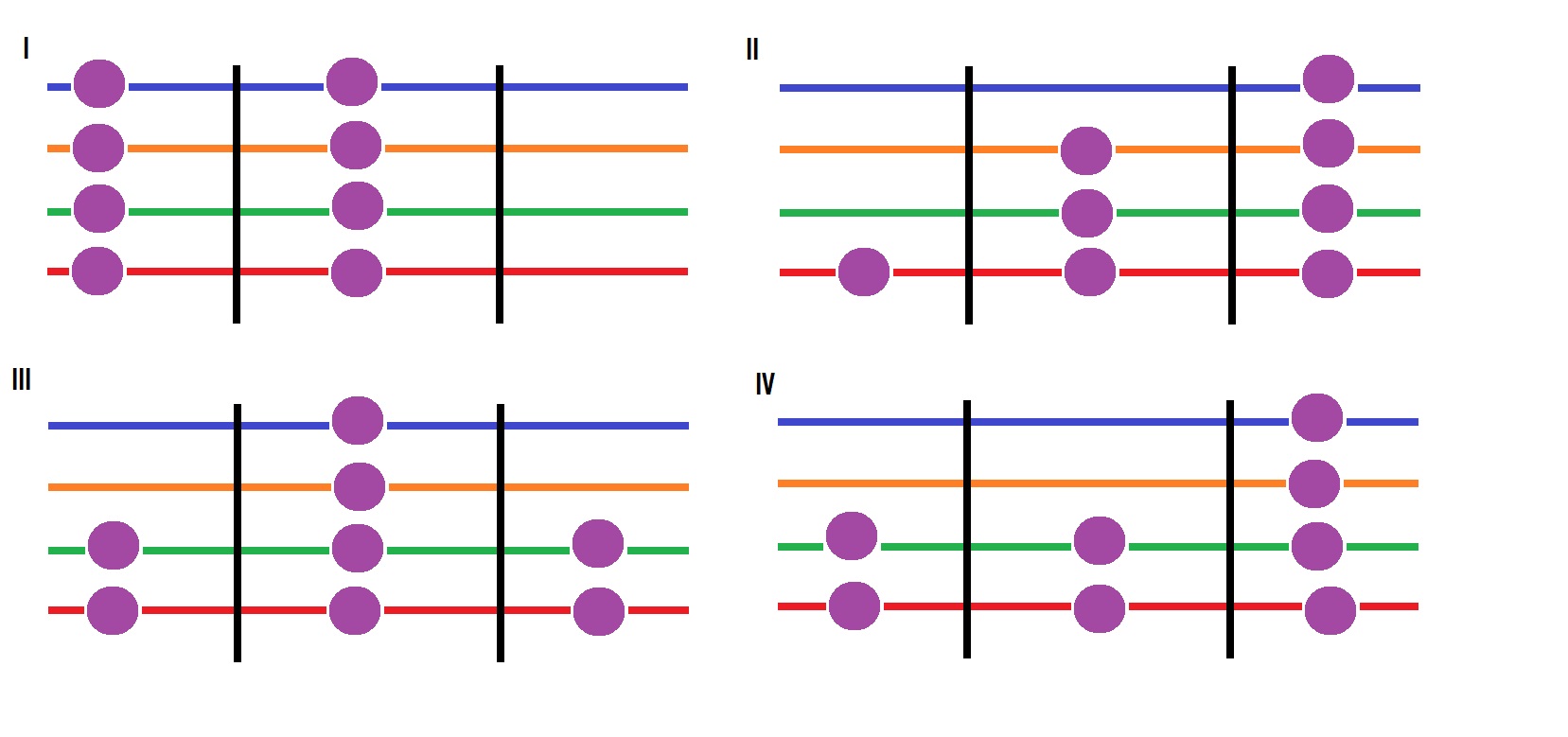}
    \caption{Some possible microstates for distributing 8 {\it quanta} among the 3 oscillators in the solid.}
    \label{caixas}
\end{figure}

To elucidate the energy distribution in {\it Einstein}'s solid, consider an example consisting of a solid made up of only three particles and eight {\it quanta} of energy. The objective is to determine the different possible ways of distributing the energy among the oscillators. Figure [\ref{caixas}] shows some possible distributions of energy in the solid. In $II$, for instance, the first oscillator has one {\it quantum}, the second has three, and the third has the remaining four.

\begin{figure}[h!]
    \centering
    \includegraphics[width=1.0\linewidth]{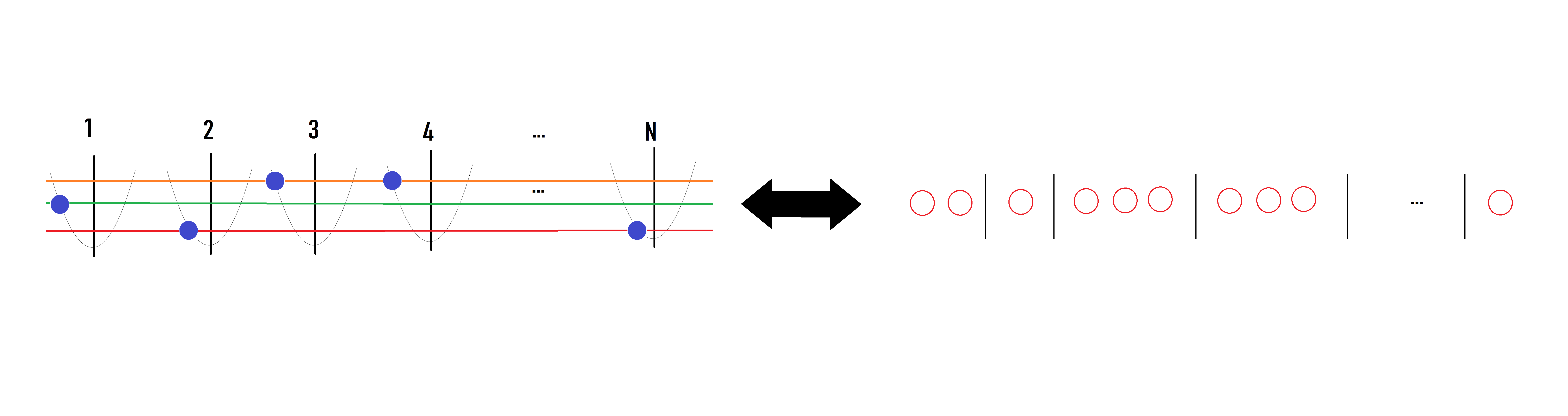}
    \caption{The problem of counting possible energy {\it quanta} distributions among oscillators in {\it Einstein}'s solid is analogous to distributing spheres among boxes.}
    \label{olhacaixa}
\end{figure}

This physical problem resembles a classic mathematical problem, where the {\it quanta} are replaced by spheres and the oscillators by boxes, as shown in Figure [\ref{olhacaixa}]. The question then becomes determining how many ways it is possible to distribute seven spheres among three boxes.

Fortunately, the problem becomes even simpler as it reduces to distributing lines among spheres: there will always be $N-1$ lines for $r$ spheres. For instance, observing case $II$ in Figure [\ref{caixas}], it is possible to conceive the representation: $\cdot | \cdot \cdot \cdot | \cdot\cdot\cdot\cdot$, where there are 10 elements, which are either bars or dots. The difference between the possible microstates lies in determining which of these elements are bars and which are dots. Thus, referring to Eq.~[\ref{permu}], which performed a similar counting by substituting $N \rightarrow N+r-1$ and $n \rightarrow r$, the multiplicity for distributing $r$ {\it quanta} of energy in {\it Einstein}'s solid composed of $N$ oscillators is obtained as:
\begin{equation}
\Omega(N, r) =\frac{(N+r-1)!}{(N-1)!r!}\approx\frac{(N+r)!}{N!r!}.
\label{Einstein Solid Multiplicity}
\end{equation}

Now, two identical solids, $A$ and $B$, with $N_A=N_B=N$, isolated from the universe, will be considered. These solids are free to exchange energy with each other over a long period of time. The total energy of the system, $E_T$, is the simple sum of the energy of system $A$, $E_A$, and the energy of system $B$, $E_B$, such that $E_T=E_A+E_B$. Therefore, the total energy is given by $E_T=(r_A+r_B)hf$. Furthermore, the interaction between the solids must respect energy conservation, as there are no external energy sources in the system isolated from the universe. Defining $K$ as a constant number, energy conservation imposes the constraint $r_A+r_B=K$. Thus, a system with 9 oscillators and 13 {\it quanta} of energy establishes the results shown in Table [\ref{(b)}].
\begin{table}[ht]
\centering
\caption{Multiplicity associated with a macrostate described by $r_A$, $r_B$ while respecting energy conservation.}
\label{(b)}
\begin{tabular}{|c|c|c|c|c|}
\hline
$r_A$       & $\Omega_A$ & $r_B$   & $\Omega_B$    & $\Omega_{AB}$    \\ \hline
0           & 1          & 13      & 203490        &     203490       \\
1           & 9          & 12      & 125970        &     1133730      \\
2           & 45         & 11      & 75582         &     $3.4 \times 10^6$   \\
3           & 165        & 10      & 43758         &     $7.22 \times 10^6$  \\
4           & 495        & 9       & 24310         &     $12.03 \times 10^6$ \\
5           & 1287       & 8       & 12870         &     $16.56 \times 10^6$ \\
6           & 3003       & 7       & 6435          &     $19.32 \times 10^6$ \\
7           & 6435       & 6       & 3003          &     $19.32 \times 10^6$ \\
8           & 12870      & 5       & 1287          &     $16.56 \times 10^6$ \\
9           & 24310      & 4       & 495           &     $12.03 \times 10^6$ \\
10          & 43758      & 3       & 165           &     $7.22 \times 10^6$  \\
11          & 75582      & 2       & 45            &     $3.4 \times 10^6$   \\
12          & 125970     & 1       & 9             &     1133730      \\
13          & 203490     & 0       & 1             &     203490       \\ \hline
\end{tabular}
\end{table}
For a more rigorous analysis, it would be necessary to perform calculations for $N$ on the order of $10^{23}$ particles to approach a more realistic level of the problem, as this number corresponds to the order of {\it Avogadro's number}, which represents the number of atoms in a mole of a solution or sample. Nevertheless, it is illustrative of the entropy characteristics that will be addressed in this thesis.

Finally, on the tombstone of Boltzmann in Vienna, the equation he authored, linking entropy to the number of multiplicities, is inscribed:
\begin{equation}
S=k_B\ln\Omega.
\label{boltzman entropy}
\end{equation}
where $k_B$ is Boltzmann's constant. Stirling's formula simplifies the calculation of the logarithm of factorials and is given by:
\begin{equation}
    \ln n! \approx n \ln n - n.
\end{equation}
Applying this approximation to the case of Einstein's solid, substituting the multiplicity [\ref{Einstein Solid Multiplicity}] into Boltzmann's entropy expression, and using the fact that $U=rhf=r\omega\hbar$, where $\omega$ is the angular frequency and $\hbar$ is the reduced Planck constant, we have:
\begin{equation}
     S(U, N) \approx k_B  \left[N\ln \left(1 + \frac{U}{N\hbar \omega}\right)+ \frac{U}{\hbar \omega} \ln\left( \frac{N\hbar \omega}{U}+1\right)\right].
\end{equation}
The temperature $T$ of the system is related to the entropy $S$ by the expression:
\begin{equation}
    \frac{1}{T} = \left(\frac{\partial S}{\partial U}\right).
\end{equation}
Thus:
\begin{equation}
    \frac{1}{T} \approx \frac{k_B}{\hbar \omega} \ln \left( \frac{N \hbar \omega}{U}+1 \right),\quad \therefore \quad U=\frac{N\omega\hbar}{e^{\omega\hbar/k_BT}-1}.
\end{equation}
Hence, the entropy of the one-dimensional Einstein solid, as a function of temperature and the number of particles (oscillators), $N$, is\footnote{For more details, see reference \cite{callen1960}.}:
\begin{equation}
S(N,T)=Nk_B\left[\frac{\omega\hbar}{k_BT}\frac{e^{-\frac{\omega\hbar}{k_BT}}}{1-e^{-\frac{\omega\hbar}{k_BT}}}-\ln\left(1-e^{-\frac{\omega\hbar}{k_BT}}\right)\right].
\label{Einstein Solid Entropy}
\end{equation}
This traditional example is sufficient to establish the general characteristics of entropy alongside the other notions that will be elaborated in this work.

\section{Chaos and the Maximum Entropy Principle}
\label{chapter 2: Section 2 - Chaos and the Maximum Entropy Principle}

Different characteristics of entropy related to the developments in the previous section will now be discussed. First, the concept of chaos is highlighted. In the case of flipping 50 coins, configurations where all coins show heads or all show tails constitute special cases. These states are very specific, just like the case where all 13 {\it quanta} of energy concentrate in a single oscillator in Einstein's solid. Although such configurations are possible, they are extremely unlikely, and in a set of particles or coins on the order of a mole, they become virtually impossible. Thus, configurations that deviate from equilibrium tend to be less probable.

\begin{wrapfigure}{r}{0.43\textwidth}
\begin{center}
\includegraphics[width=0.85\linewidth]{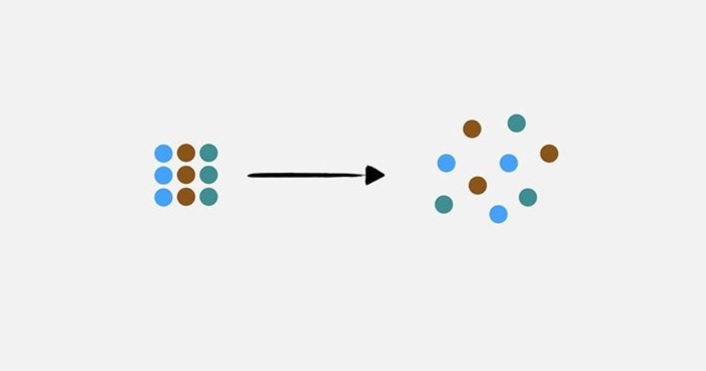}
\end{center}
\caption{Pictorial representation of increasing chaos, where an initially organized system evolves into a disorganized one. Reproduced from \cite{fermatslibrary}.}
\label{fermat}
\end{wrapfigure}

Brian Greene, in his popular science book {\it Until the End of Time}\cite{greene2021ate}, presents an excellent example to illustrate this tendency toward increasing chaos: one can imagine an organized room, with objects placed in a relatively orderly manner, no dust on the floor, and the bed neatly made. This is a specific and particular state, similar to flipping 50 coins and obtaining 49 heads. Heuristically, these states can be compared to states of low entropy. Over time, and due to various interactions, this state tends to transform into one of higher entropy: dust accumulates, clothes are scattered on the floor, and the bed requires fresh sheets. This initial low-entropy state is difficult to maintain, which is why it is necessary to clean and organize the room regularly. There is, therefore, a tendency toward \textquoteleft disorder', as systems naturally evolve toward states of higher entropy or chaos (Fig.~[\ref{fermat}]).

The dedicated engineer seeks to develop an engine capable of producing the maximum amount of work with the least amount of energy. However, one way to state the second law of thermodynamics is to assert that perfect heat engines do not exist; that is, in a natural process—such as the energy released during the combustion of gasoline in a car—besides the work that makes the wheels turn, heat is also released, which is useless for performing work. In thermodynamic theory, there are mathematical objects called potentials that describe the energy available to do work under different conditions. In this discussion, for isothermal processes, the Helmholtz potential, $F$, is particularly noteworthy:
\begin{equation}
F = U - TS,
\label{helmholtz potential}
\end{equation}
where $U$ is the internal energy (in the case of the car, the energy released by the combustion of gasoline), $T$ is the temperature, and $S$ is the entropy. The term $TS$, subtracted from the internal energy, indicates how much of the energy can be used for work, which is one of the reasons why the Helmholtz potential is also called free energy. Thus, the Helmholtz potential describes how, even in processes aiming for local organization, such as the example of the organized room, the energy available to perform work tends to decrease over time due to dissipation. In practical terms, this means that maintaining a low-entropy state requires a constant input of external work to compensate for inevitable losses, reinforcing the natural tendency toward states of higher entropy and greater disorder.

Finally, the analysis of Fig.~[\ref{50caras}], which shows that the maximum multiplicity for flipping 50 coins occurs with 25 heads and 25 tails, and Table~[\ref{(b)}], which indicates that a balanced division of the {\it quanta of energy} also presents the highest multiplicity, reveals that the equilibrium state is a state of {\it maximum entropy}. Analyzing the entropy expression proposed by {\it Boltzmann} in Eq.~[\ref{boltzman entropy}] and the examples discussed so far, it becomes evident that the most probable state is precisely the one that balances the system and increases the entropy, reaching the highest possible multiplicity $\Omega$, since, in this case, $S\propto \ln\Omega$. This property is called the {\it Maximum Entropy Principle}, and next, the entropies of Shannon and von Neumann will be introduced, which, like thermodynamic entropy, are also consistent with this principle.

\section{Shannon Entropy}
\label{chapter 2: Section 3 - Shannon Entropy}

To understand Shannon entropy, a brief introduction to the basic concepts of information theory is necessary. Initially, consider two events, $A$, with a probability $p$ of occurring, and $B$, with a probability $q$. In information theory, a quantity of information is associated with each event through a mathematical function called the information function $I(x)$, which measures the amount of information related to an event $X$ with probability $x$.

If the probability of event $A$ occurring is smaller than that of event $B$, then the information associated with event $A$, $I(p)$, must be greater than that of event $B$, $I(q)$ \cite{baez2024entropy}. Thus:
\begin{equation}
\text{If}\quad p<q, \quad \text{then}\quad I(p)>I(q).
\label{decrecent information}
\end{equation}
Information theory is based on the interpretation that less probable events carry more information. For example, a weather forecast stating “it will be sunny tomorrow during summer in Rio de Janeiro” represents a very likely event, as sunny days are common during summer in the city. This information does not cause much surprise and does not bring significant “novelty” — that is, it carries little information because the event was expected. Now, if the forecast predicts “snow in Rio de Janeiro tomorrow,” this would represent an extremely unlikely event, given the city's tropical climate. If this prediction were true, it would cause enormous surprise. This type of information would be much more “dense,” as the occurrence of such an unexpected event challenges common expectations.

Now, it is postulated that the joint occurrence of two independent events, $A$ and $B$, must result in additive information. Therefore:
\begin{equation}
I(pq) = I(p) + I(q).
\label{aditivity information}
\end{equation}
This is the requirement of information additivity. If Eqs.~[\ref{decrecent information}] and [\ref{aditivity information}] are satisfied, and if information is defined as a function with the domain $(0, 1]$, the only possible solution is:
\begin{equation}
I(x) = -\log_b(x),
\label{information}
\end{equation}
with $b > 1$. In physics, $b = e$, referred to as the natural base of information theory, or $b = 2$, referred to as the base of \textit{bits}, is commonly used.

For example, the flip of three fair coins resulting in HTT is an event with probability $1/2^3$, since each coin has a $1/2$ probability of landing on either side. The amount of information for this outcome is:
\begin{equation}
I(2^{-3})=- \log \left( \frac{1}{2^3} \right) = 3 \log 2.
\end{equation}
This is equivalent to 3 units called \textit{bits} (short for “binary digits”), as the information from each coin provides $\log 2$ of information. In other words, upon receiving information about an event that occurs with probability $1/8$, the observer gains “3 bits” of information.

Each unit of information measures the amount of “surprise” or “novelty” associated with an event, based on the general formula: $I(p) = -\log p$. While the most common unit is the bit, which represents the information associated with an event of probability $1/2$, there are many other units of information depending on the logarithmic base and the event's probability.

\begin{table}[ht]
\centering
\begin{tabular}{|c|c|c|}
\hline
\textbf{Event Probability} & \textbf{Unit of Information} & \textbf{Equivalent in Bits} \\
\hline
$1/2$    & 1 bit                & 1 bit               \\
$1/e$    & 1 nat                & \( \approx 1.44 \) bits \\
$1/3$    & 1 trit               & \( \approx 1.58 \) bits \\
$1/4$    & 1 crumb              & 2 bits              \\
$1/10$   & 1 hartley            & \( \approx 3.32 \) bits \\
$1/16$   & 1 nibble             & 4 bits              \\
$1/256$  & 1 byte               & 8 bits              \\
$1/8192$ & 1 kilobyte           & 8192 bits (or 1024 bytes) \\
\hline
\end{tabular}
\caption{Units of information and their equivalences in bits. Adapted from \cite{baez2024entropy}.}
\end{table}

\begin{figure}[ht]
\centering
\includegraphics[width=0.5\linewidth]{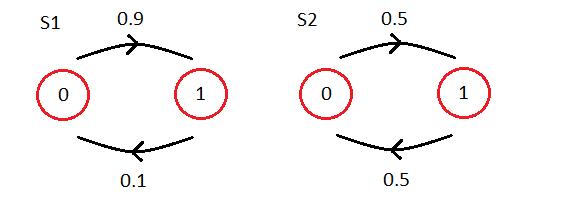}
\caption{Systems $S1$ (left) and $S2$ (right)}
\label{s1s2}
\end{figure}

Born in thermodynamics and verified as an emergent concept when relating microscopic to macroscopic treatments in statistical mechanics, entropy reappears as a fundamental concept in the works of {\it Claude Shannon} \cite{shannon}. This formulation associates entropy as an intrinsic characteristic of information, defined by Shannon entropy:
\begin{equation}
    H=-\sum_{i=1}^n p_i \log_b p_i=\expval{-\log_b p_i},
\end{equation}
In information theory, entropy is a measure of uncertainty. For example, consider two binary systems, $S1$ and $S2$ (Fig.~[\ref{s1s2}]), with states $0$ and $1$. In $S1$, the system has a probability $p_1=0.9$ that 0 becomes 1, and $p_2=0.1$ that 1 becomes 0. Thus:
\begin{equation}
    H_{S1}=-\left (\frac{9}{10} \log_2\frac{9}{10}+\frac{1}{10}\log_2\frac{1}{10}\right)\approx 0.47.
\end{equation}
On the other hand, in $S2$, there is only one possibility with a probability $p_1=0.5$ that 0 becomes 1, so $p_2=0.5$, and:
\begin{equation}
H_{S2}=-\left (\frac{1}{2} \log_2\frac{1}{2}+\frac{1}{2}\log_2\frac{1}{2}\right )=1.   
\end{equation}
This example reveals that $S1$ is less {\it uncertain} than $S2$, exhibiting a tendency in its construction.

With the current digital revolution, information theory and its central concept, Shannon entropy, are experiencing their apex, even though it was first published in 1948. At that time, the American mathematician was unsure about what to call his recent discovery, and in a conversation with {\it von Neumann}, he received convincing arguments\cite{conversa}.

\begin{quote}
    “My greatest concern was what to call it. I thought of calling it 'information,' but the word was overly used, so I decided to call it \textquoteleft uncertainty'. When I discussed it with John von Neumann, he had a better idea. Von Neumann told me, \textquoteleft You should call it entropy, for two reasons. In the first place your uncertainty function has been used in statistical mechanics under that name, so it already has a name. In the second place, and more important, no one really knows what entropy really is, so in a debate you will always have the advantage.”
\end{quote}

In the context of information theory, the notion of additivity plays a central role; however, other types of entropies have been proposed to capture behaviors, for example, non-additive ones, which are relevant in complex and interdependent systems, such as those observed in non-extensive statistical mechanics contexts. In this sense, the Tsallis and Rényi entropies are examples of entropic notions that arise as generalizations of Shannon entropy.

Initially, Tsallis entropy is defined as\cite{tsallis2009introduction}:
\begin{equation}
S_q=k_B \frac{1}{q - 1} \left( 1 - \sum_{i} p_i^q \right),
\label{tsallis_entropy}
\end{equation}
where $p_i$ are the probabilities of the events, and $q$ is a real parameter that controls the degree of non-additivity.

This entropic notion is related to Non-Extensive Statistical Mechanics. In an extensive system, macroscopic properties, such as entropy, are proportional to the system size. In other words, when two independent systems are combined, the total entropy is simply the sum of the individual entropies. However, for non-extensive systems (such as those found in complex networks, biological systems, and astrophysical systems), this additivity does not apply. These systems often exhibit long-range correlations, large-scale fluctuations, and a much more complex dynamics than those described by traditional statistical mechanics. Mathematically, extensivity can be defined as\cite{junior2013entropia}:
\begin{equation}
\lim_{N\to \infty}\frac{|S(N)|}{N}<\infty.
\end{equation}
Thus, a system classified as extensive has an asymptotic behavior with a number of subsystems $N$ such that there is a finite proportionality factor between $|S(N)|$ and $N$. Therefore, additivity with respect to a given composition law implies extensivity.

Using L'Hôpital's rule, when $q = 1$, Tsallis entropy recovers Shannon entropy:
\begin{equation}
\lim_{q \to 1} S_q = k_B\lim_{q \to 1} \frac{\frac{d}{dq}\left(1 - \sum_{i} p_i^q\right)}{\frac{d}{dq}\left(q - 1\right)}= -k_B\sum_{i} p_i^q \ln p_i \Big|_{q = 1} = -k_B\sum_{i} p_i \ln p_i = k_BH.
\end{equation}
However, for $q \neq 1$, Tsallis entropy satisfies a property called \textit{pseudo-additivity}:
\begin{equation}
\frac{S_q(A + B)}{k_B} = \frac{S_q(A)}{k_B} + \frac{S_q(B)}{k_B} + (1 - q) \frac{S_q(A)}{k_B} \frac{S_q(B)}{k_B},
\end{equation}
where $A$ and $B$ represent two independent systems. The additional term depends on the parameter $q$ and reflects the correlation between events, an important characteristic in complex and interdependent systems.

Rényi entropy\cite{renyi1961measures} is a generalization of Shannon entropy, developed to measure the diversity, uncertainty, and dispersion of probabilistic systems where it is necessary to give greater or lesser weight to events with very low or very high probabilities. It is commonly used in contexts involving rare or highly probable events. In practice, Rényi entropy allows adjusting the sensitivity of the uncertainty measure by varying the parameter $\alpha$, which controls the weighting of different probabilities.

Mathematically, Rényi entropy is given by:
\begin{equation}
S_{\alpha} = \frac{1}{1 - \alpha} \log \left( \sum_{i} p_i^{\alpha} \right),
\label{renyi_entropy}
\end{equation}
where $ \alpha $ is a real parameter that controls the "focus" of the entropy on events with high or low probabilities. When $ \alpha > 1 $, Rényi entropy is more sensitive to events with higher probabilities, while values of $ \alpha < 1 $ give more weight to events with lower probabilities. In the limit as $ \alpha \to 1 $, Rényi entropy recovers Shannon entropy, taking the form of a classical average uncertainty measure over the system:
\begin{equation}
\lim_{\alpha \to 1} S_{\alpha} = \lim_{\alpha \to 1} \frac{\frac{d}{d\alpha}\left[ \log \left( \sum_{i} p_i^{\alpha} \right) \right]}{\frac{d}{d\alpha}\left(1 - \alpha\right)} = -\sum_{i} p_i \ln p_i = H.
\end{equation}

Tsallis and Rényi entropies introduce a useful flexibility to the concept of entropy, allowing complex systems with interdependencies between events to be described by an alternative information measure to that proposed by Shannon.

\section{The Relation between the Shannon Entropy and Statistical Mechanics}
\label{chapter 2: Section 4 - The Relation between the Shannon Entropy and the Statistical Mechanics}

In this section, Shannon entropy is evaluated using the maximum entropy principle. To this end, the maximization follows the procedure of Lagrange multipliers. Initially, consider the case where the probability distribution must be normalized:
\begin{equation}
\sum_{n=1}^N p_n = 1.
\label{microcaninical bound}
\end{equation}
Thus, the functional to be maximized is given by:
\begin{equation}
\mathscr{L} = -\sum_{n=1}^N p_n \ln p_n - \lambda \left(\sum_{n=1}^N p_n - 1\right).
\end{equation}
Maximizing $\mathscr{L}$ with respect to $p_n$:
\begin{equation}
-\ln p_n - 1 - \lambda = 0 \quad \therefore \quad p_n = e^{-\lambda - 1}.
\end{equation}
Using the normalization condition, it is found that $p_n =1/N$. This is the microcanonical {\it ensemble}, in which, for a closed statistical system with fixed energy, all accessible microstates are equally probable.

For the canonical ensemble, the system is considered in thermal contact with a reservoir, so its total energy is not fixed, but the average energy $ \langle E \rangle $ must remain constant. Thus, in addition to the normalization of the probability distribution, the following constraint also applies:

\begin{equation}
\sum_{n=1}^N p_n E_n = \langle E \rangle,
\label{canonical constraint}
\end{equation}
where $ E_n $ represents the energy of the system's state $ n $.

The functional to be maximized, with the two constraints, is given by:
\begin{equation}
\mathscr{L} = -\sum_{n=1}^N p_n \ln p_n - \lambda \left(\sum_{n=1}^N p_n - 1\right) - \beta \left(\sum_{n=1}^N p_n E_n - \langle E \rangle\right),
\end{equation}
where $\lambda$ and $\beta$ are Lagrange multipliers associated with the normalization and average energy constraints, respectively.

To maximize $\mathscr{L}$ with respect to $p_n$, the partial derivative of $\mathscr{L}$ with respect to $p_n$ is taken, yielding:
\begin{equation}
\ln p_n = \lambda - 1 - \beta E_n \quad \therefore \quad p_n = e^{-\lambda - 1} e^{-\beta E_n}.
\end{equation}
Using the normalization condition, the constant $Z=e^{-\lambda - 1}$ is determined, known as the canonical partition function:
\begin{equation}
Z = \sum_{n=1}^N e^{-\beta E_n},
\end{equation}
where $\beta=1/k_BT$ in this equation, so the probability distribution can be expressed as:
\begin{equation}
p_n = \frac{e^{-\beta E_n}}{Z}.
\end{equation}
This is the well-known {\it Boltzmann distribution}, which describes the canonical ensemble, where the system's average energy is maintained constant in contact with a thermal reservoir.

This procedure of entropy maximization to determine the probability distributions of statistical ensembles was introduced by Edwin T. Jaynes \cite{jaynes1957information1}, and following this methodology, it is also possible to show the relation of Shannon entropy to the {\it grand canonical ensemble}. Additionally, Shannon's formula in the continuous limit becomes the expression for Gibbs entropy\cite{gibbs1902elementary}. In the continuous limit, where the system's variables form a continuous state space, replacing the sum with an integral and the probability $ p_i $ with the probability density $ \rho(x) $, where $ x $ represents the system's continuous state variables, the entropy becomes:
\begin{equation}
S = - \int \rho(x) \ln \rho(x) \, dx.
\label{gibs formula}
\end{equation}
This expression is known in Statistical Mechanics as the \textit{Boltzmann-Gibbs entropy}. It measures the uncertainty or dispersion of the probability density distribution $ \rho(x) $ in a continuous state space. Therefore, through the application of the maximum entropy principle in Shannon's expression, it is possible to establish a connection between Information Theory and Statistical Mechanics.

\section{The von Neumann Entropy}
\label{chapter 2: Section 5 - The von Neumann Entropy}

In the framework of Quantum Mechanics, the appropriate notion of entropy is attributed to John von Neumann. Here, the averages refer to the concept of ensemble averages, meaning systems {\it a priori} identically prepared. After a measurement is performed, a statistical characterization of the constituents of the total final state is obtained, composed of all the subsystems where the measurement was conducted. For example, after performing a {\it Stern-Gerlach} experiment\cite{gerlach1989experimentelle}, it is known that the physical state of the silver atom beam, after interaction with the external magnetic field, has 50\% of its atoms collapsed into an up-spin state, and the remaining portion, also composed of 50\%, has a down-spin state. However, upon exiting the furnace, or, in other words, before the measurement, it is not possible to characterize the physical states of the atoms in the beam: the individual spin of each atom could be pointing in any direction; in general terms, the physical state is {\it random}.

In the case of physical systems where no measurement has occurred, it is known that they are composed of a finite number of constituents, so it is possible to assign a weight to their relative population in a given particular state, $p_m$, where $1\leq m\leq N$, associated with the $m$-th state $\ket{m}$, and $N$ is the number of individuals in the ensemble, or the number of identically prepared systems. In this case, care must be taken not to confuse the number of individuals composing the system with the dimension of the space generated by the eigenvectors of a given observable: the parameter $N$ generally exceeds the dimension of the eigenspace of a given operator. Since it is a fractional population, the sum of the weights must equal unity, similar to Eq.~[\ref{microcaninical bound}].

Furthermore, no geometric information about the kets is available before the measurement: they may very well be orthogonal to each other or not; they may be eigenvectors of a common operator, or they may not be, and it is not determined whether the operators representing them are compatible or not. Thus, it is possible to infer the statistical nature of this set: before a measurement is performed in a system composed of a population of physical states, and considering that more than one $p_m$ is different from zero, the system is said to constitute a {\it mixed ensemble}. Now, after a measurement is performed, it is possible to fully analyze the fraction of the population characterized by a specific physical state, i.e., the collection of physical systems represented by a single ket. This latter case is referred to as a {\it pure ensemble}. In other words, a mixed {\it ensemble} consists of a collection of pure {\it ensembles}.

When aiming to obtain the measure of an observable, this is only possible through an ensemble average. Considering, for instance, the observable $\hat G$, which in the formal construction of quantum mechanics is an operator, its average $\expval{G}$ is obtained as:
\begin{equation}
\begin{split}
   \expval{ G}&=\sum_{m=1}^{N}p_m\bra{m}\hat G \ket{m} =\sum_{m=1}^{N}p_m\bra{m}\hat G \mathbbm{1}\ket{m} \\
    &=\sum_{m=1}^{N}\sum_g p_m \bra{m} \hat G \ket{g} \bra{g}\ket{m}.   
\end{split}
\end{equation}
Using the eigenvalue equation $\hat G\ket{g}=g\ket{g}$, $\expval{G}$ becomes:
\begin{equation}
   \expval{ G} = \sum_{m=1}^{N}\sum_g p_m |\bra{g}\ket{m}|^2 g.
\end{equation}
From this result, it should be noted that two independent statistical constructions are involved in obtaining a single measurement: the population weights of each physical state form a statistical approach mediating the ensemble average of the quantum predictions, which themselves constitute a statistical framework.

The quantum formalism allows for as many basis changes as needed, utilizing the completeness relation given by:
\begin{equation}
    \sum_i\ket{i}\bra{i}=\mathbbm{1},
\end{equation}
where $\mathbbm{1}$ is the identity operator, enabling basis changes to be expressed in a very compact form. Thus, the expected value can be evaluated as follows:
\begin{equation}
\begin{split}
   \expval{ G}&=\sum_{m=1}^{N}p_m\bra{m}\mathbbm{1}\hat G \mathbbm{1}\ket{m} =\sum_{m=1}^{N}p_m \sum_i\sum_j \bra{m} \ket{i}\bra{i}\hat G \ket{j}\bra{j}\ket{m} \\
    &=\sum_i\sum_j \left ( \sum_{m=1}^N p_m \bra{j}\ket{m}\bra{m}\ket{i}\right )\bra{i}\hat G \ket{j}.   
\end{split}
\end{equation}
The term highlighted in parentheses is defined as the matrix element of a Hermitian operator, called the {\it density operator} $\hat \rho$,
\begin{equation}
   \rho_{ij}\equiv \bra{i}\hat \rho \ket{j} = \sum_{m=0}^N p_m\bra{i}\ket{m}\bra{m}\ket{j}.
\end{equation}
Reconciling the matrix representation of quantum mechanics with this operator, the general expression of the density operator is defined as:
\begin{equation}
\hat \rho \equiv \sum_{m=0}^N p_m\ket{m}\bra{m}.
\end{equation}

Considering this construction, the expression for $\expval{ G}$ takes a more compact form:
\begin{equation}
\begin{split}
   \expval{ G}&=\sum_i\sum_j\bra{j}\hat \rho \ket{i}\bra{i} \hat G \ket{j} = \sum_j \bra{j} \hat \rho \underbrace{\sum_i \ket{i}\bra{i}}_{\mathbbm{1}}\hat G\ket{j} \\
    &=\sum_j\bra{j}\hat \rho \hat G \ket{j}=\Tr[\hat \rho \hat G].   
\end{split}
\end{equation}
where the operation $\Tr[\hat \rho \hat G]$ corresponds to the trace of the operator resulting from the calculation of $\hat \rho \hat G$, making explicit the generalized power of this construction: {\it the trace is representation-independent}.

Summarizing, the ensemble average of an observable $\hat G$ is given by:
\begin{equation}
    \expval{G} = \Tr[\hat \rho \hat G].
\end{equation}
Now, analyzing the trace of the identity operator separately:
\begin{equation}
\begin{split}
   \Tr[\hat \rho]&=\sum_j\sum_{m=0}^N p_m\bra{j}\ket{m}\bra{m}\ket{j}=\sum_{m=0}^Np_m\bra{m}\underbrace{\left(\sum_j \ket{j}\bra{j}\right)}_{\mathbbm{1}}\ket{m} \\
    &=\sum_{m=0}^Np_m\underbrace{\bra{m}\ket{m}}_{1}=1,   
\end{split}
\end{equation}
for a pure {\it ensemble}, where the relative population becomes total, with $p_1=1$, the density matrix $\hat \rho_P$ is:
\begin{equation}
    \hat \rho_P = \sum_{m=1}^N p_m \ket{m} \bra{m}= \ket{m}\bra{m}.
\end{equation}
Hence,
\begin{equation}
    \hat \rho_P \hat \rho_P = \hat \rho_P^2 = \ket{m}\underbrace{\bra{m}\ket{m}}_1\bra{m}=\ket{m}\bra{m}=\hat \rho_P,
\end{equation}
meaning $\hat \rho_P$ is a projector:
\begin{equation}
    \hat \rho_P^2=\hat \rho_P.
\end{equation}
Only for a pure state:
\begin{equation}
   \Tr[\hat \rho_P^2]=1.
\end{equation}
Thus, the eigenvalues associated with the density operator of pure {\it ensembles} must always be zero or one. When diagonalizing the density matrix $\hat \rho_P$, one expects to find a mathematical object in the form:
$$
    \hat \rho_P\doteq
    \begin{pmatrix}
    0       &  ...   & 0& 0& 0   & ...  & 0        \\ 
    \vdots  &  ...   & 0& 0& 0   & ...  & 0        \\
    0       &  ...   & 0& 1& 0   & ...  & 0        \\
    \vdots  &  ...   & 0& 0& 0   & ...  & 0        \\ 
    0       &  ...   & 0& 0& 0   & ...  & 0
    \end{pmatrix}.
$$
In contrast, a completely mixed {\it ensemble} has a density matrix $\hat \rho_M$ structured as:
$$
    \hat\rho_M\doteq\frac{1}{N}
    \begin{pmatrix}
    1       &  ...   & 0& 0& 0   & ...  & 0        \\ 
    \vdots  &  ...   & 1& 0& 0   & ...  & 0        \\
    0       &  ...   & 0& 1& 0   & ...  & 0        \\
    \vdots  &  ...   & 0& 0& 1   & ...  & 0        \\ 
    0       &  ...   & 0& 0& 0   & ...  & 1
    \end{pmatrix}=\frac{1}{N}\mathbbm{1}_N,
$$
where $\mathbbm{1}_N$ is the $N$-dimensional identity matrix. Thus, two diagonal matrices, subject to the same normalization condition, represent diametrically opposed physical objects. It is convenient, therefore, to define a quantity that distinguishes the intrinsic physical properties of each object. In this spirit, the {\it von Neumann entropy} is defined\cite{von2018mathematical}:
\begin{equation}
    S_{vN}\equiv-k_B\Tr[\hat \rho \ln{\hat \rho}].
    \label{svn}
\end{equation}
This entropy measures the system's deviation from a pure state, the amount of degraded information, and, once again, chaos. 

To deepen the analysis of von Neumann entropy, there are two fundamental characteristics: (1) the von Neumann entropy of a pure quantum state is zero, and (2) for mixed states described by classical probabilities, von Neumann entropy recovers the form of Shannon entropy.

For a pure state, when diagonalizing $\hat \rho_P$, the associated eigenvalues will be $1$ for the occupied state $\ket{\psi}$ and $0$ for all other states. Thus:
\begin{equation}
    \Tr(\hat \rho_P \ln \hat \rho_P) = 1 \cdot \ln(1) + 0 \cdot \ln(0) = 0.
\end{equation}
Therefore, the von Neumann entropy of a pure quantum state is zero. This characteristic reflects the fact that a pure state contains maximum information about the system, with no associated uncertainty. For a mixed state, the density operator is given by a combination of pure states with associated probabilities $p_i$, which are the probabilities of finding the system in the state $\ket{\psi_i}$, with $\sum_i p_i = 1$. By diagonalizing $\hat \rho$, a diagonal matrix with eigenvalues $p_i$ is obtained. Thus, the expression for $S_{vN}$ becomes:
\begin{equation}
    S_{vN} = -k_B \sum_i p_i \ln p_i=k_BH.
\end{equation}
In other words, it is exactly the form of Shannon entropy in the natural base multiplied by $k_B$. This equivalence demonstrates that von Neumann entropy generalizes Shannon entropy to the quantum context, recovering it in situations where the system is described by classical probabilities.

Moreover, in Quantum Mechanics, the Rényi entropy is given by\cite{calabrese2009entanglement}:
\begin{equation}
S_\alpha = \frac{1}{1-\alpha}\ln \Tr[\hat \rho^\alpha]
\label{quantum renyi}
\end{equation}
In this case, for $\alpha\rightarrow1$, it recovers von Neumann entropy:
\begin{equation}
\lim_{\alpha\to1}\frac{1}{1-\alpha}\ln\Tr[\hat \rho^\alpha]=-\Tr[\hat \rho \ln \hat \rho]=\frac{1}{k_B}S_{vN}
\label{von neumann and renyi}
\end{equation}

Illustratively, consider the case of the one-dimensional harmonic oscillator. In this case, the Boltzmann distribution is given by:
\begin{equation}
p_n = \frac{e^{-\beta E_n}}{\sum_{n=0}^\infty e^{-\beta E_n}} = \frac{e^{-n \beta \omega \hbar}}{\sum_{n=0}^\infty e^{-n \beta \omega \hbar}} = (1 - e^{-\beta \omega \hbar}) e^{-n \beta \omega \hbar},
\label{harmonic_oscilator}
\end{equation}
where the expression for the eigenvalues of the one-dimensional harmonic oscillator (Eq. [\ref{OHE}]) and properties of the hypergeometric series were used. Substituting this result into the von Neumann entropy:
\begin{equation}
S = -k_B (1 - e^{-\beta \omega \hbar}) \left[\ln (1 - e^{-\beta \omega \hbar}) \sum_{n=0}^\infty e^{-\beta \omega \hbar} - \beta \omega \hbar \sum_{n=0}^\infty n e^{-n \beta \omega \hbar}\right].
\end{equation}
Using some manipulations, substituting $\beta = 1 / k_B T$, and multiplying by \( N \) oscillators, the expression for the entropy of the Einstein solid developed in the first section is recovered:
\begin{equation}
S(N, T) = N k_B \left[\frac{\omega \hbar}{k_B T} \frac{e^{-\frac{\omega \hbar}{k_B T}}}{1 - e^{-\frac{\omega \hbar}{k_B T}}} - \ln \left(1 - e^{-\frac{\omega \hbar}{k_B T}}\right)\right].
\end{equation}
It is interesting to note that, in the first case, which used more rudimentary counting methods, the fundamental principle of statistical mechanics and Boltzmann’s expression led to exactly the same result for the von Neumann entropy—a concept typical of quantum theory. This example was included to demonstrate how the different notions of entropy are related. Throughout this work, in the models evaluated, it is common to derive one form of entropy from manipulations of another.

\begin{figure}[ht]
\centering
\includegraphics[width=0.7\linewidth]{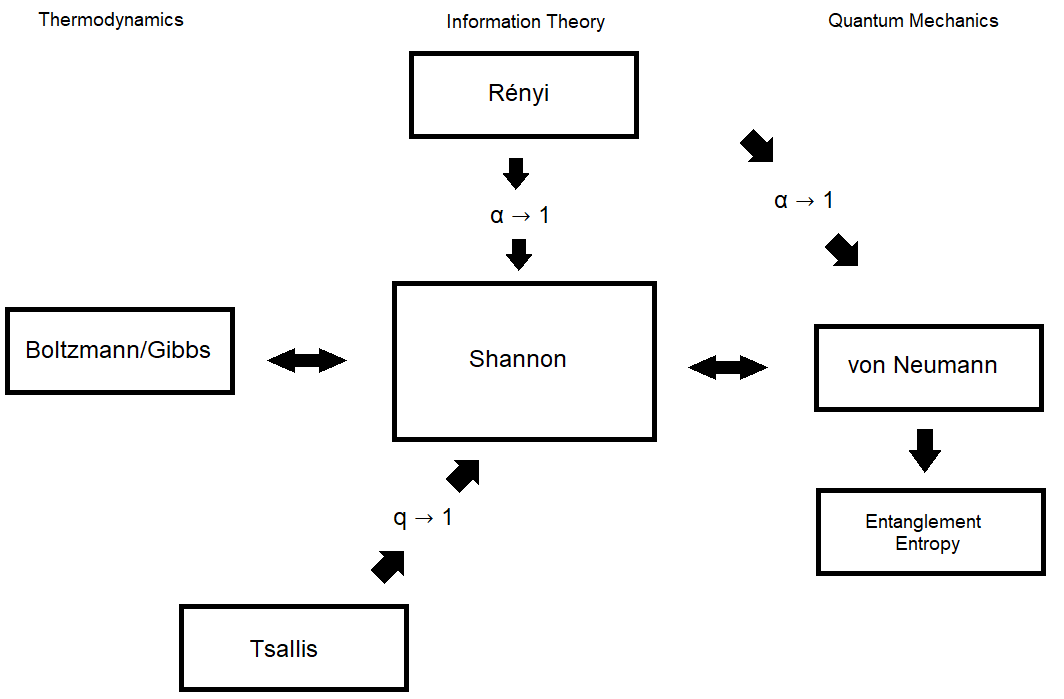}
\caption{Relationship between the notions of entropy studied in this chapter. Entanglement entropy will be studied in Chapter [\ref{chapter 4: Entanglement Entropy}].}
\label{entropyrelation}
\end{figure}

For instance, in the calculation of entanglement entropy in elastic collisions \cite{peschanski2016,peschanski2019}, it is necessary to first obtain Rényi entropy, calculate the limit as $\alpha \to 1$, and then determine an entanglement entropy written in the Shannon form, making it essential to understand the relationship between the different notions of entropy. In any case, Fig.~[\ref{entropyrelation}] illustrates the relationship between the notions of entropy evaluated in this work.

\section{Mutual Information}
\label{chapter 2: Section 6 - Mutual Information}

Mutual information, also called the Kullback–Leibler Relative Entropy, is a fundamental measure to evaluate the divergence between two probability distributions, $p$ and $p^{(0)}$, where $p^{(0)}$ represents a reference distribution\cite{tsallis2009introduction}. To quantify this discrepancy, various metrics can be adopted, with the Kullback–Leibler divergence being one of the most relevant. Formally, the mutual information between $p$ and $p^{(0)}$ is defined as:
\begin{equation}
I(p, p^{(0)}) = \int dx \, p(x) \ln \left[\frac{p(x)}{p^{(0)}(x)}\right].
\label{mutual information}
\end{equation}
Mutual information is non-negative and equals zero if and only if $p(x) = p^{(0)}(x)$. The non-negativity property makes the Kullback–Leibler divergence useful for comparing distributions by indicating the proximity of $p$ to the reference $p^{(0)}$.

Now, consider a change of variables in a continuous probability distribution $p(x)$, such that $y = f(x)$. In this case, $dx=|f'(x)|dy$. Applying this transformation to the probability distribution in the Boltzmann-Gibbs entropy (Eq.~[\ref{gibs formula}]):
\begin{equation}
S= - \int dy \, p(y) \ln p(y) = - \int dx \, p(x) \left( \ln p(x) + \ln |f'(x)| \right).
\label{variance bg}
\end{equation}
On the other hand, after the same variable transformation in mutual information, one obtains:
\begin{equation}
\begin{split}
I(p, p^{(0)}) &= \int dy \, p(y) \ln \left[ \frac{p(y)}{p^{(0)}(y)} \right] \\
&= \int dy \frac{dx}{dy} \, p(x) \ln \left[ \frac{p(x) |f'(x)|}{p^{(0)}(x) |f'(x)|} \right]\\
&= \int dx \, p(x) \ln \left[ \frac{p(x)}{p^{(0)}(x)} \right].
\end{split}
\end{equation}
This is an important characteristic of $I(p, p^{(0)})$, its invariance under variable transformations that preserve the measure. This means that even when performing a change of variables, the divergence between $p$ and $p^{(0)}$ remains constant. In many calculations, this property makes mutual information preferable for manipulation compared to Boltzmann-Gibbs entropy (Eq.~[\ref{gibs formula}]), as in the latter case, invariance is lost under certain transformations, as shown in Eq.~[\ref{variance bg}].

\section{Non-Equilibrium Statistical Mechanics}
\label{chapter 2: Section 7 - Non Equilibrium Mechanical Statistics}

In general, the study of Thermodynamics and Statistical Mechanics considers processes and transformations between equilibrium states, described by state variables such as entropy $S$, internal energy $U$, the number of particles $N$, or any other extensive variable. This equilibrium is necessary for measuring intensive variables, such as pressure $P$ and temperature $T$. Broadly speaking, temperature can be interpreted as the average kinetic velocity of the particles that make up the gas, while pressure is given by the sum of forces applied over the area of a certain piston. Without equilibrium, there is no satisfactory average of forces or velocities for intensive variables to be properly defined, as in the previous cases. 

The second law of thermodynamics describes the fundamental limitation of possible transitions between equilibrium states; however, the understanding of non-equilibrium systems remains primitive. It is precisely in this context that the so-called {\it Non-Equilibrium Statistical Mechanics} comes into play. This is a broad and developing field, but in this work, only two topics will be briefly addressed: the Jarzynski Identity and the Hatano-Sasa Relation, which will be discussed next.

\subsection{Jarzynski Identity}
\label{chapter 2: Section 7 - Subsection 1: Jarzynski Identity}

The Jarzynski identity\cite{jarzynski1997nonequilibrium} is a fluctuation relation that connects the work performed during a thermodynamic process between equilibrium states with the respective change in free energy. This identity establishes an expression involving the stochastic distribution of thermodynamic work, $\rho(W)$, allowing the study of processes far from equilibrium.

In isolated systems, it is well known in statistical physics that the average work $\expval{W}$ performed in a thermodynamic system satisfies inequalities related to the second law of thermodynamics. Thus, for a transition between two equilibrium states, the following relation holds:
\begin{equation}
\Delta S = \frac{\expval{W}-\Delta F}{k_BT} \geq 0 \equiv \frac{\expval{W_{\text{Diss}}}}{k_BT} \geq 0,
\end{equation}
Here, $\Delta E \equiv \expval{W}$ represents the total energy variation, and $\Delta F$ is the change in the system's free energy. The term $W - \Delta F = W_{\text{Diss}}$ corresponds to the dissipated work, which is zero only in reversible processes.

The Jarzynski identity generalizes this principle by directly connecting the average work to fluctuations outside equilibrium, especially in cases of fluctuations where $W_{\text{Diss}} < 0$. It is given by the expression:
\begin{equation}
\expval{e^{-W/k_BT}} = e^{-\Delta F/k_BT},
\label{jarzinski identity}
\end{equation}
As demonstrated in the earlier sections of this chapter, the second law of thermodynamics has a statistical nature, and therefore, very rarely, fluctuations occur where $-W < \Delta F$\footnote{Using the convention where $-W$ represents the work done by an external force on the system, and $W$ is the work done by the system on the environment.}. These fluctuations may be rare, but with large $W$ (i.e., strongly negative $-W$), their contribution to the average of $e^{-W/k_BT}$ can be significant. Thus, the Jarzynski formula includes all fluctuations, including those that violate $-W \geq \Delta F$.

This identity can be written as:
\begin{equation}
\Delta F = - k_B T \ln \expval{ e^{ - W / k_B T}} = - k_B T \ln \int dW \rho ( W ) e^{ - W / k_B T},
\label{jarzinsky 2}
\end{equation}
where $\rho ( W )$ is called the stochastic distribution of thermodynamic work, describing the probability that the work performed on a system during a thermodynamic process takes a specific value. In a non-equilibrium thermodynamic process, the system is driven from an initial equilibrium state to a final state by a perturbation, such as the application of a mechanical force altering the volume or an electric field changing the internal energy. The work performed, in this case, depends on the system's microscopic trajectories, i.e., the specific paths that particles follow due to thermal fluctuations during the process. Hence, problems involving this identity aim to determine this distribution, with examples including the one-dimensional piston \cite{lua2005practical} and the adiabatic compression of a dilute gas \cite{crooks2006work}.

\subsection{Hatano-Sasa Identity}
\label{chapter 2: Section 7 - Subsection 2: Hatano-Sasa Identity}

Compared to the Jarzynski expression, the Hatano-Sasa identity\cite{hatano2001steady} does not have a variable equivalent to temperature. This identity applies to non-equilibrium steady-state systems and is defined based on a dynamic parameter $\lambda$. In this case, for each value of $\lambda$, there is a steady-state spectrum in phase space with a probability distribution $P_{\text{Stat}}(z; \lambda)dz$, where the variable $z$ describes the system's phase space. Under this configuration, the following equality holds:
\begin{equation}
\left\langle \exp \left(-\int_{\tau_1}^{\tau_2} d\tau \, \frac{d\lambda}{d\tau} \, \frac{\partial}{\partial \lambda} \ln P(z; \lambda, \tau) \right) \right\rangle_{\tau_2} = 
\int dz \, \exp \left(-\int_{\tau_1}^{\tau_2} d\tau \, \frac{d\lambda}{d\tau} \, \frac{\partial }{\partial \lambda} \ln P(z; \lambda, \tau)\right) \, P(z; \lambda_2, \tau_2) = 1,
\label{hatano-sasa}
\end{equation}
In this equation, the transition between a non-equilibrium steady state occurs with the change in the dynamic variable $\lambda(\tau)$, such that $\lambda(\tau_1)=\lambda_1$ evolves to $\lambda(\tau_2)=\lambda_2$, and $\tau$ is the transition time between two distinct non-equilibrium steady states. Thus, $P(z; \lambda, \tau) \equiv P_{\text{Stat}}(z(\tau); \lambda(\tau))$ represents the steady-state solution for the value $\lambda(\tau)$ in phase space variables “frozen” at time $\tau$. Note that this identity holds independently of the arbitrary “history” of $d\lambda/d\tau$ in the non-equilibrium mechanism.

\chapter{Entropy of QCD Dense States}
\label{chapter 3: QCD dense states}

The main objective of this chapter is to investigate the production of entropy in high-energy physics related to the so-called QCD dense states. As previously discussed, the creation of entropy due to quantum entanglement and decoherence will also be addressed. However, since the application of these quantum phenomena will be associated with fundamental characteristics of high-energy physics, this chapter also serves as an introduction to the respective chapters dedicated to each of the aforementioned phenomena.

Furthermore, the rapid thermalization of the QGP suggests significant entropy creation in the early stages of its formation; hence, some basic characteristics of this plasma will also be presented. Ultimately, the central goal of the chapter will be established with the presentation of the so-called QCD dynamic entropy, a proposal to evaluate the entropy density per unit of rapidity defined from observables of saturation physics.

Given the mathematical complexity inherent to Quantum Field Theory (QFT), it is essential to establish the notation conventions used in this work: bi- and three-dimensional vectors will be indicated with an overhead arrow: for vector $A$, for example, $\vec{A}$ will be used; four-vectors will have an index or sub-index with a Greek character in the form $A^{\mu}$ and may also be represented simply as $A$, where disambiguation will be determined by the operation in which the object is involved. From now on, natural units will be used, where
\begin{equation}
    \hbar = c = k_B = 1.
\end{equation}
Thus, the relationship between units of measurement is given as
\begin{equation}
    [\text{mass}] = [\text{energy}] = [\text{time}]^{-1} = [\text{length}]^{-1} = \text{GeV},
\end{equation}
with these quantities subject to the conversion factor,
\begin{equation}
    1 \, \text{GeV} = 5.0677 \, \text{fm}^{-1}.
\end{equation}

As this work involves quantum field theories, with special relativity as a prerequisite for the mathematical manipulations established here, it is ideal to treat quantities that are Lorentz invariant. The reference frame where the high-energy regime is most relevant is called the {\it Breit frame}, which will be further discussed in the next chapter, focusing on some fundamental characteristics of high-energy physics.

\section{The Parton Model}
\label{chapter 3: QCD dense states - Section 1; The Parton Model}

In 1911, {\it Ernest Rutherford} initiated a scientific revolution with the discovery of the atomic nucleus\cite{ruther}. This breakthrough was crucial for advancing atomic models and theories, rendering the conceptions of {\it Thomson} and {\it Dalton} obsolete and providing a new explanation for the structure of matter. In 1913, {\it Bohr} refined the theory by introducing quantized orbits, addressing the problem of matter collapse predicted by {\it Maxwell}'s equations in {\it Rutherford}'s model. Shortly thereafter, the existence of an exact analytical solution to {\it Schrödinger}'s equation for the hydrogen atom paved the way for the birth of quantum mechanics. By the late 1950s, the groundwork was laid for the formulation of QFT, enabling satisfactory derivations for three of the fundamental forces: electromagnetic, weak, and strong; while the gravitational force still lacks a quantum theory.

By properly applying QFTs, it is possible to derive expressions for cross-sections with good results. In addition to the development of these theories, experimental setups also evolved drastically: larger accelerators were designed and subsequently built. Current examples include the LHC and RHIC, among others, which rely on the collaboration of researchers and engineers from around the world to further explore not only the nuclear realm but also the behavior of matter at even more fundamental levels — specifically, the structure of nucleons (hadrons that make up atomic nuclei, {\it protons}, and {\it neutrons}), including quarks and gluons. 

\begin{wrapfigure}{r}{0.5\textwidth}
\begin{center}
\includegraphics[width=0.9\linewidth]{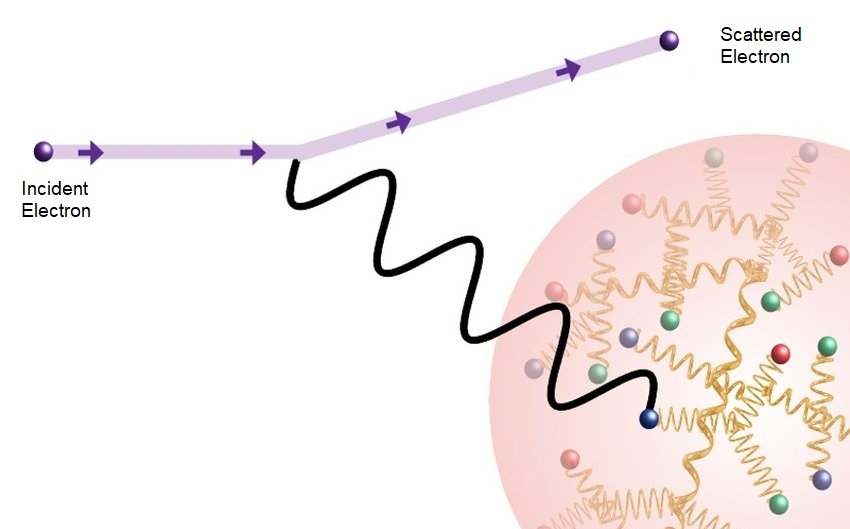}
\end{center}
\caption{Representation of DIS. Adapted from \cite{abt2021azimuthal}.}
\label{DISfigure}
\end{wrapfigure}

A formalism of paramount importance in this research is the {\it Parton Model}\cite{feynmanpartons}. In 1968, a series of experiments was conducted at the {\it Stanford Linear Accelerator Center} (SLAC), involving Deep Inelastic Scattering (DIS) of leptons off nucleons\cite{invencaodos}. These experiments consist of the scattering of a lepton, with the measurement of the deflection angle revealing aspects of the process's nature. Specifically, in {\it inelastic} processes, the target absorbs part of the collision's kinetic energy from a high-energy projectile, which can be associated with a small wavelength, allowing the probing of so-called {\it deep} regions (Fig.~[\ref{DISfigure}]). During these experiments, theoretical physicist {\it Richard Feynman} met with members of the responsible experimental group and gained access to the first obtained results, where the cross-section was parametrized following suggestions by {\it James D. Bjorken}\cite{bjorken}.

To understand {\it Bjorken}'s contribution, it is necessary first to establish the following set of definitions:

\begin{enumerate}
  \item[$\blacksquare$] $q$ is used to label both the quarks and the transferred 4-{\it momenta}, $q^{\mu}=k^{\mu}-k'^{\mu}$, with disambiguation applied in context.
    
  \item[$\blacksquare$] $k^{\mu}$ is used for the 4-{\it momentum} of the incident electron.
    
  \item[$\blacksquare$] $k'^{\mu}$ is used for the 4-{\it momentum} of the scattered electron.
    
  \item[$\blacksquare$] $P^{\mu}$ is used for the 4-{\it momentum} of the proton.
    
  \item[$\blacksquare$] $p^{\mu}$ is used for the 4-{\it momentum} of the parton.
    
  \item[$\blacksquare$] $Q^2$ quantifies the deviation of the virtual photon from the mass shell, referred to as the {\it photon virtuality}.
    
  \item[$\blacksquare$] $x$ is the {\it Bjorken} kinematic variable, defined as:
    \begin{equation}
        x\equiv \frac{Q^2}{2P\cdot q};
    \end{equation}
    
  \item[$\blacksquare$] $\xi$ represents the fraction of the $i$-th parton's {\it momentum} relative to the proton's {\it momentum}:
    \begin{equation}
    p_i^{\mu}=\xi P^{\mu};
    \end{equation}
        
  \item[$\blacksquare$] $\nu$ is a kinematic variable given by:
    \begin{equation}
    \nu=\frac{P\cdot q}{M},
    \end{equation}
where $M$ is the nucleon mass, although the mass of the proton, $m_p$, is frequently used in this chapter.
\end{enumerate}

\begin{wrapfigure}{r}{0.3\textwidth}
\begin{center}
\includegraphics[width=0.8\linewidth]{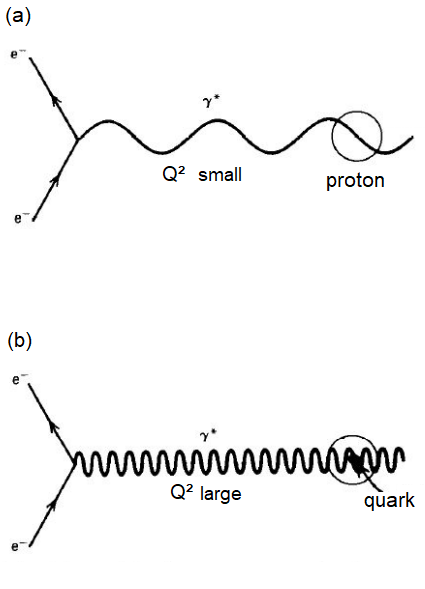}
\end{center}
\caption{Two experiments where an electron collides with a proton. In (a), little energy is transferred in the collision, so the virtual photon $\gamma^*$ has a wavelength on the order of the proton's size. In (b), sufficient energy is transferred in the collision, so the wavelength of the virtual photon can probe the proton's internal structure.}
\label{measurefigure}
\end{wrapfigure}

Using current algebra, {\it Bjorken} noted that in the limit where the momentum and energy transferred in the impact tend to infinity, $q^2 \to \infty$ and $\nu = (E - E')_{\text{lab}} \to \infty$, in the reference frame called the {\it Breit Frame}, a fixed ratio is established, known as {\it Bjorken's} $x$:
\begin{equation}
    x = \frac{Q^2}{2 P \cdot q} = -\frac{q^2}{2M\nu}.
    \label{bjorken x}
\end{equation}
Moreover, {\it Bjorken} observed that, in this limit, the structure functions, analogous to the form factors present in elastic collisions, depend solely on $x$, exhibiting scaling behavior. {\it Feynman} interpreted this scaling behavior in collisions as an indication that the nucleon possesses charged constituents, allowing the short wavelength of the projectile lepton to probe these constituents (Fig.~[\ref{measurefigure}]), known as {\it Partons} (from “part of hadrons”). When publishing his investigations on DIS, {\it Feynman} introduced for the first time the distinction between {\it exclusive} and {\it inclusive} scattering. In the exclusive case, the processes involve knowledge of which particles will be produced. In contrast, the inclusive case studies the behavior of a specific particle in the final state across various kinematic ranges of longitudinal and transverse momentum, without specifying the other particles involved.

Practically, an example of an exclusive process is given by:
$$
p^+ + p^+ \rightarrow p^+ + n^0 + \pi^+.
$$
The particles involved in this example are the proton $p^+$, the neutron $n^0$, and the positively charged pion $\pi^+$. An inclusive process can be described as:
$$
p^+ + p^+ \rightarrow \pi^+ + X,
$$
where $X$ represents an unspecified final state, and in this example, the analyzed particle would be the {\it pion}. {\it Feynman} argued that, in a high-energy regime, inclusive cross-sections should exhibit scaling behavior when the total energy is very large and the transverse momentum of the final state particle remains limited.

Following the methodology proposed by Feynman and Bjorken, the cross-section in a DIS is parametrized in terms of $q^{\mu}$ and $P^{\mu}$:
\begin{equation}
\left(\frac{d\sigma}{d\Omega dE'}\right) = \frac{\alpha^2}{4\pi m_p q^4} \frac{E'}{E} L^{\mu\nu} W_{\mu\nu}.
\end{equation}
In this equation, $\alpha \approx 1/137$ is the electromagnetic coupling constant, $L_{\mu\nu}$ is the {\it leptonic tensor}, and $W_{\mu\nu}$ is the hadronic tensor. For an unpolarized scattering, $L_{\mu\nu}$ is given by:
\begin{equation}
L_{\mu\nu} = 2(k'_{\mu} k_{\nu} + k_{\mu} k'_{\nu} - k \cdot k' g_{\mu\nu}),
\end{equation}
and the most general form of the hadronic tensor is given by:
\begin{equation}
W^{\mu\nu} = W_1 \left(-g^{\mu\nu} + \frac{q^{\mu} q^{\nu}}{q^2}\right) + W_2 \left(P^{\mu} - \frac{P \cdot q}{q^2} q^{\mu}\right)\left(P^{\nu} - \frac{P \cdot q}{q^2} q^{\nu}\right).
\end{equation}

The scalar quantities on which $W_1$ and $W_2$ may depend must be Lorentz invariants, namely, $P^2 = M^2$, $q^2$, and $P \cdot q$. We use $Q \equiv \sqrt{-q^2} > 0$, which is the energy scale in the collision, and, in the LAB frame, $P \cdot q / M = (E - E')$. Thus, contracting the leptonic tensor with the hadronic tensor:
\begin{equation}
\left(\frac{d\sigma}{d\Omega dE'}\right) = \frac{\alpha^2}{8\pi E^2 \sin^4 \frac{\theta}{2}} \left[\frac{M}{2} W_2(x, Q) \cos^2 \frac{\theta}{2} + \frac{1}{M} W_1 \sin^2 \frac{\theta}{2}\right].
\label{eita}
\end{equation}
In the inelastic case, this cross-section reveals all the necessary characteristics.

To test the Parton Model, it is necessary to verify the form factors by considering that the electron elastically scatters off the proton's constituents with mass $m_q$. To perform this evaluation, the initial and final 4-{\it momentum} of the parton, $p_i^{\mu}$ and $p_f^{\mu}$, respectively, are considered. Using {\it momentum} conservation, $p_i^{\mu} + q^{\mu} = p_f^{\mu}$, it follows that:
\begin{equation}
\frac{Q^2}{2 p_i \cdot q} = 1.
\end{equation}
However, the parton's 4-{\it momentum} is not directly measurable. Thus, it is assumed that the parton carries a fraction $\xi$ of the proton's {\it momentum}:
\begin{equation}
p_i^{\mu} = \xi P^{\mu},
\end{equation}
hence,
\begin{equation}
x = \frac{\xi Q^2}{2 p_i \cdot q} = \xi.
\end{equation}
Therefore, the measurement of $x$ reveals the fraction of {\it momentum} that the parton carries from the parent proton.

To calculate the elastic scattering $e^-q$, it is assumed that the partons do not interact significantly with each other. Thus, the form factors exhibit only a weak logarithmic dependence on $Q^2$\footnote{Similar to the $e^-\mu^-$ scattering, well established in Quantum Electrodynamics.}, with the initial partonic {\it momentum} fixed, meaning $x$ remains constant. The cross-section approximately independent of $Q^2$ for fixed $x$ is known as {\it Bjorken scaling}.

The parton model also uses the probabilities $f_i(\xi)d\xi$ of the photon interacting with the $i$-th parton carrying a fraction $\xi$ of the proton's {\it momentum}. These $f_i(\xi)$'s are known as {\it parton distribution functions} (PDFs). The model predicts that the cross-section for the scattering $e^-P^+ \rightarrow e^-X$, $\sigma_T$, is given by $e^-p_i \rightarrow e^-X$, $\sigma_n$, where $p_i$ is the parton with {\it momentum} $p_i^{\mu} = \xi P^{\mu}$, integrated over all $\xi$:
\begin{equation}
\sigma_T = \sum_i \int_0^1 d\xi f_i(\xi) \sigma_n.
\end{equation}
Assuming the partons are free except for electromagnetic interactions, the electron scatters only off charged particles, which, in the proton, are quarks. For a given quark with {\it momentum} $p_i$, the partonic cross-section $e^-q \rightarrow e^-q$, $\sigma_p$, is approximated by a point-like scattering in Quantum Electrodynamics, given by the {\it Rosenbluth} formula with $F_1 = 1$ and $F_2 = 0$, such that:
\begin{equation}
\left(\frac{d\sigma_p}{d\Omega}\right)_{LAB} = \frac{\alpha_e^2 Q_i^2}{4E^2 \sin^4(\theta/2)}\left[\cos^2 \frac{\theta}{2} + \frac{Q^2}{2m_q^2} \sin^2 \frac{\theta}{2}\right] \delta\left(E - E' - \frac{Q^2}{2m_q}\right),
\label{Rosenquark}
\end{equation}
where $Q_i$ is the quark charge.

To obtain the DIS cross-section, it is necessary to integrate this expression over the momentum of the incident parton. Thus, the result is:
\begin{equation}
\left(\frac{d\sigma_p}{d\Omega}\right)_{LAB}=\sum_i f_i(x)\frac{\alpha_e^2Q_i^2}{4E^2\sin ^4(\theta/2)} \left(\frac{2m_p}{Q^2}x^2\cos ^2 \frac{\theta}{2}+\frac{1}{m_p}\sin ^2\frac{\theta}{2}\right).
\label{secao2}
\end{equation}
Comparing [\ref{eita}] with [\ref{secao2}], the following relations are obtained:

\begin{equation}
\begin{cases}
W_1(x,Q)=2\pi\sum_iQ_i^2f_i(x);\\
W_2(x,Q)=8\pi \frac{x^2}{Q^2}\sum_i Q_i^2f_i(x).
\end{cases}
\label{relation}
\end{equation}
These relations provide a solid foundation for the experimental prediction of {\it Bjorken} scaling, as the quantities $W_1(x,Q^2)$ and $Q^2W_2(x,Q)$ exhibit weak dependence on $Q$ at fixed $x$, as confirmed experimentally.

From this perspective, partons are interpreted as bare particles of a new fundamental interaction—the strong force—being investigated through DIS in the Breit Frame. In this reference frame, the hadron undergoes longitudinal spatial contraction, making its interaction with the lepton (electromagnetic or weak, depending on the lepton involved) nearly instantaneous. Thus, the internal state of hadrons is "frozen", and the interactions among partons can be neglected, allowing them to be treated as quasi-free particles. Later, partons were identified with the quarks introduced by {\it Gell-Mann}\cite{quark}, establishing the {\it Standard Model}, and the methodology of {\it Bjorken} and {\it Feynman} became fundamental for the analysis of {\it Perturbative Quantum Chromodynamics}.

To better understand the use of the Parton Model in investigating hadronic structures, it is useful to briefly review the theory that describes elementary particle physics: the {\it Standard Model}. This model organizes fundamental particles into 12 spin-$1/2$ fermions, divided into two main groups. The first group includes quarks, which constitute hadronic matter: {\it up}, {\it down}, {\it charm}, {\it strange}, {\it top}, and {\it bottom}. The second group consists of leptons: the {\it electron}, {\it muon}, {\it tau}, and their respective neutrinos. Together, these fermions form all ordinary matter\footnote{There is still much to discover about other forms of matter, such as {\it dark matter} and {\it dark energy}, which are currently the focus of several research programs. Information about these forms of matter comes from gravitational theories, and they are estimated to account for about 95\% of all the matter in the universe.}.

\begin{figure}[ht]
\centering
\includegraphics[width=0.55\linewidth]{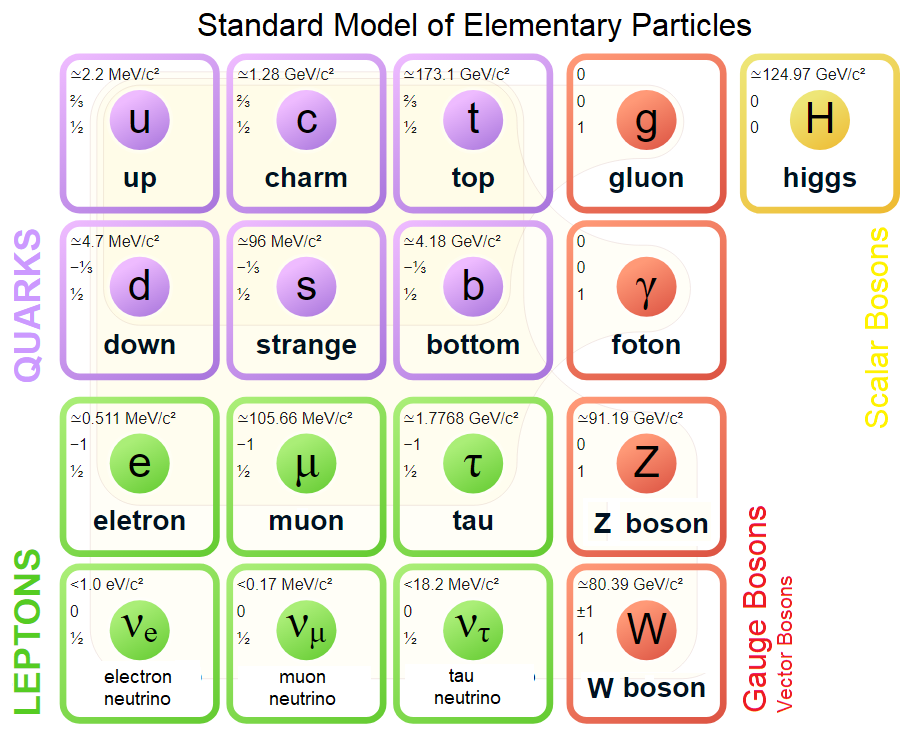}
\caption{Organization of the experimentally confirmed particles in the Standard Model. For each particle, its mass, electric charge, and spin values are also shown. Reproduced from \cite{modelo}.}
\label{standart model}
\end{figure}

The interactions between these particles are mediated by the {\it gauge bosons}, spin-$1$ particles associated with the fundamental forces: the photon (electromagnetic force), the gluon (strong nuclear force), and the $W^{\pm}$ and $Z^0$ bosons (weak force). Additionally, the Standard Model includes a scalar boson, the {\it Higgs}, whose existence was experimentally confirmed and is fundamental to the mechanism that gives mass to other particles. Fig.~[\ref{standart model}] illustrates some details about the constituents of the Standard Model.

Particles interact with different forces depending on the charges they possess. For instance, quarks carry a color charge, which can be red, blue or green. This charge allows them to form "colorless" hadrons, such as {\it mesons} and {\it baryons}. Mesons are bound states of a quark and an antiquark, while baryons consist of three quarks. In addition to the color charge, quarks have fractional electric charges, enabling them to interact electromagnetically and be detected in processes such as DIS.

Thus, the Parton Model provides a clearer understanding of quarks and gluons by describing the composition of nucleons, which, in DIS, possess a fraction $\xi$ of the parent hadron's {\it momentum}, maintaining collinearity with it. At low energies, the so-called {\it valence quarks} constitute the nucleons in triads. However, with increasing energy in DIS, the need to consider PDFs arises. These distributions obey certain relations, known as {\it sum rules}\cite{schaw}. For instance, to ensure the conservation of the proton's quantum numbers, its PDFs must satisfy the following relations:

\begin{equation}
\int_0^1 d\xi \left[ f_d(\xi) - f_{\bar{d}}(\xi) \right] = 1, \quad \int_0^1 d\xi \left[ f_u(\xi) - f_{\bar{u}}(\xi) \right] = 2, \quad \int_0^1 d\xi \left[ f_s(\xi) - f_{\bar{s}}(\xi) \right] = 0,
\label{sum rule}
\end{equation}
with the latter relation also applicable to bottom and charm quarks. Although there is no specific conservation rule for the number of gluons, the sum of their PDFs with those of other quarks must satisfy the condition:

\begin{equation}
\sum_j \int \xi f_j(\xi) d\xi = 1.
\label{general sum rule}
\end{equation}

\begin{figure}[ht]
\centering
\includegraphics[width=0.65\linewidth]{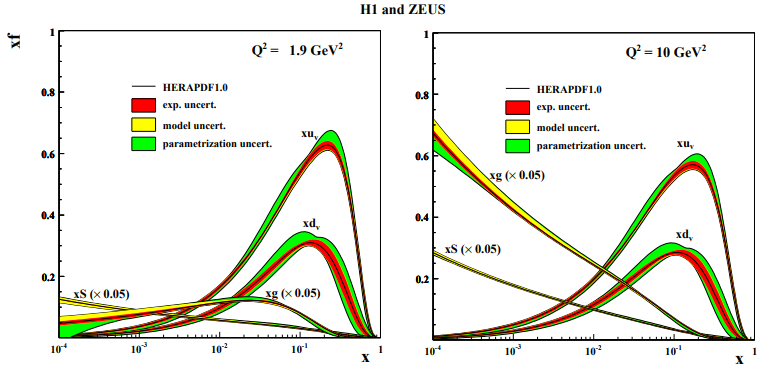}
\caption{Behavior of PDFs as a function of the {\it Bjorken} variable for virtualities $Q^2=1.9$ GeV$^2$ (left) and $Q^2=10$ GeV$^2$ (right). The functions shown here include those for up quarks, $xu_v$, down quarks, $xd_v$, gluons, $xg$, and sea quarks, $xS=2x(\bar{U}+\bar{D})$. The gluon and sea distributions have been reduced by a factor of 20 to allow evaluation on the same scale. Reproduced from \cite{pdfss}.}
\label{pdfs}
\end{figure}

In the case of the proton, the presented sum rules indicate that only $38\%$ of the proton's {\it momentum} is carried by the valence quarks (up and down). On the other hand, gluons carry a fraction ranging from $35\%$ to $50\%$, depending on the scale. Quarks not accounted for in these distributions are referred to as {\it sea quarks}. Fig.~[\ref{pdfs}] illustrates the behavior of these distribution functions in a DIS involving an electron and a proton for two values of virtuality ($Q^2 = -q^2$): for high values of $x$, corresponding to low-energy processes, the distributions describe the valence quarks; whereas in the small-$x$ regime, in high-energy processes, the proton's composition is dominated by gluons, a region known as {\it saturation physics}, which will be addressed in more detail throughout this work.

\subsection{Saturation Physics}
\label{chapter 3: QCD dense states - Section 1; High Energy Physics / subsection 1. Saturation Physics}

\begin{wrapfigure}{r}{0.4\textwidth}
\begin{center}
\includegraphics[width=0.9\linewidth]{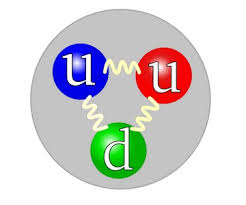}
\end{center}
\caption{Schematic representation of the proton at low energies with its three valence quarks. Image from reference \cite{quarks}.}
\label{proton}
\end{wrapfigure}

Atomic nuclei are composed of two types of particles called nucleons: protons and neutrons. At low energies, these particles are essentially a combination of three quarks with distinct colors, arranged in such a way that the nucleon's quantum numbers are balanced. For instance, the proton (Fig.~\ref{proton}) consists of two up quarks with a positive electric charge of $+2/3 \, e$ and one down quark with an electric charge of $-1/3 \, e$, resulting in a net charge of $+e$. Quarks are spin-$1/2$ particles, meaning the proton's spin wave function is a superposition of its constituents' various spin configurations, arranged so that the total spin equals $1/2$. This combination does not violate \textit{Pauli's exclusion principle} due to the presence of an additional quantum number, the color charge of the strong interaction, which can be red, blue, or green, along with their respective anti-color charges.

This type of description works well to understand the physical composition of nucleons. However, it is not possible to make robust theoretical predictions or predict collision behaviors with satisfactory results because, in perturbation theory, the strong coupling constant $\alpha_s$ is of the order of unity, making calculations infeasible. Fortunately, as the energy increases, the coupling constant decreases, and perturbative calculations become effective. Physically, it becomes possible to probe smaller regions to obtain the cross-section, and new phenomena begin to emerge.

Before providing further details, it is worth clarifying some differences between electrodynamics and quantum chromodynamics: although both are quantum field theories and can have their properties derived from abstract groups — U(1) for electrodynamics and SU(3) for chromodynamics — a key difference lies in the fact that the generators\footnote{In abstract algebra, a generating set of a group is a subset that is not contained in any proper subgroup of the group. Equivalently, a generating set of a group is a subset such that every element of the group can be expressed as a combination (under the group operation) of finite elements of the subset and their inverses.} of the latter group do not commute as those of the former do. Hence, quantum chromodynamics is described as a {\it non-Abelian} theory\footnote{In abstract algebra, an abelian group, also known as a commutative group, is a group $(G,*)$ where $a * b = b * a$ for any $a$ and $b$ $\in G$.}. This characteristic of the symmetry group leads to the fact that the mediators of the strong force, the gluons, carry color charge and therefore interact with each other, a striking contrast to the photons in Quantum Electrodynamics.

Generally, the resolution at which the hadronic structure is evaluated is described in terms of two variables: the {\it virtuality} $Q^2$, representing the {\it momentum} transferred in the interaction, and Bjorken's $x$, which in the high-energy regime can be approximated as $x \approx Q^2 / s$, where $s$ is the squared energy of the system in the center-of-mass frame. Before the collision, it is assumed that the nucleon carries a certain 4-{\it momentum} $P^{\mu}$. Consequently, by conservation, the 4-{\it momenta} of the quarks must be fractions of the total, $p_i^{\mu} = xP^{\mu}$. The Bjorken variable represents the fraction of the parent nucleon's {\it momentum}. The virtuality of the photon, $Q^2$, is related to the square of the transferred 4-{\it momentum}, $Q \equiv \sqrt{-q^2}$.

\begin{wrapfigure}{r}{0.5\textwidth}
\begin{center}
\includegraphics[width=0.8\linewidth]{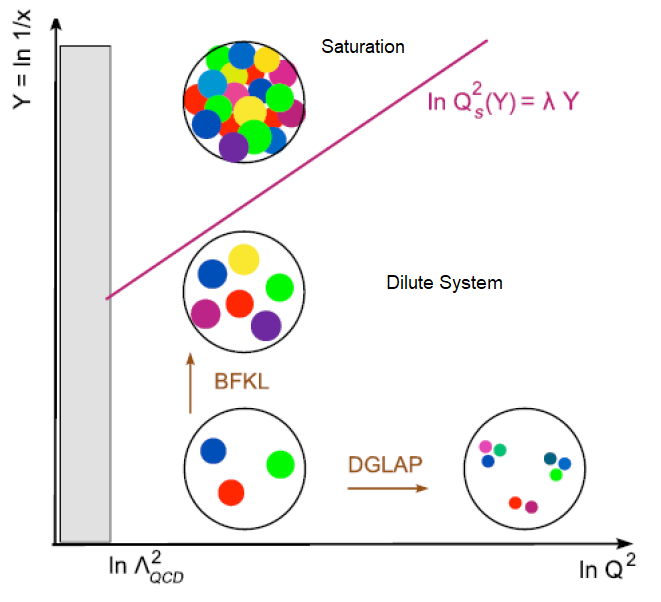}
\end{center}
\caption{The characterization of the nucleon's state changes depending on the variable being analyzed. Considering the evolution with respect to $x$, the system becomes saturated, with its evolution described by the BFKL equation. On the other hand, analyzing only the increase in virtuality $Q^2$, the system becomes more dilute, described by the DGLAP equations. Figure adapted from \cite{iancu_qcd}.}
\label{nucleon}
\end{wrapfigure}

Based on these considerations, the analysis of the partonic structure can be elaborated by considering the phenomenon of {\it bremsstrahlung}, where partons can emit gluons, which can fluctuate into quark-antiquark pairs or emit additional gluons. In the high-energy regime, $x \ll 1$ (and within the first-order approximation in $\alpha_s$), the differential probability $P_g$ for gluon emission is given by\cite{iancu_qcd}:
\begin{equation}
P_{\text{bremm}} \propto \alpha_s(k^2) \frac{d^2p}{k^2} \frac{dx}{x},
\end{equation}
where $k$ is the 4-{\it momentum} of the emitted parton, with $p = (\sqrt{k^2 + k_z^2}, \vec{k}, k_z = xP_z)$. Thus, in the limit where the transverse {\it momentum} approaches a small value, the likelihood of generating a gluon collinear with a fraction of {\it momentum} $x$ increases. Physically, the size of the hadron remains nearly constant as the energy increases, and the gluon population within this state of matter has a transverse {\it momentum} that can be associated with an area (using the uncertainty principle), also transverse, which is on the order of $\sim 1 / Q^2$.

In the high-energy regime, it becomes possible to characterize the nucleon's constituents by analyzing their dependence on Bjorken's $x$ variable (or equivalently the rapidity $Y$, since $Y = \ln 1/x$) and the virtuality $Q^2$. In Fig.~[\ref{nucleon}], it is possible to observe how the parton distribution within the nucleon changes with variations in $x$ and $Q$. As virtuality increases, the system begins to dilute, and this description is given by the set of DGLAP equations (named after the works of {\it Dokshitzer, Gribov, Lipatov, Altarelli, and Parisi})\cite{altarelli,dokshitzer,gribov}:
\begin{equation}
Q^2\frac{\partial}{\partial Q^2}\begin{pmatrix}
f_q(x,Q^2) \\ 
f_g(x,Q^2) \\
\end{pmatrix}=\frac{\alpha_s}{2\pi}\begin{pmatrix}
P_{qq} & P_{qg} \\ 
P_{gq} & P_{gg}   \\
\end{pmatrix}
\otimes
\begin{pmatrix}
f_q(Q^2) \\ 
f_g(Q^2) \\
\end{pmatrix} (x).
\label{DGLAP}    
\end{equation}
In this equation, $P_{qq}$, $P_{qg}$, $P_{gq}$, and $P_{gg}$ are the {\it splitting functions}, and the $f_i(x)$'s are the PDFs. At leading order, the splitting functions that can be obtained are listed below:
\begin{equation}
P_{qq}(z)=C_F\left[(1+z^2)\left[\frac{1}{1-z}\right]_++\frac{3}{2}\delta(1-z)\right], 
\label{pqq}
\end{equation}
\begin{equation}
P_{qg}(z)=N_c[z^2+(1-z)^2],
\label{pqg}
\end{equation}
\begin{equation}
P_{gq}(z)=C_F\frac{1+(1-z)^2}{z},
\label{pgq}
\end{equation}
and
\begin{equation}
P_{gg}(z)=2N_c\left[\frac{z}{[1-z]_+}+\frac{1-z}{z}+z(1-z)\right]+\frac{11N_C-2N_f}{6}\delta(1-z),
\label{pgg}
\end{equation}
where in these expressions, $N_c$ is the number of quark flavors, and $z=Q^2/(2p_i\cdot q)$. Qualitatively, as virtuality increases, partons occupy a smaller transverse area due to the proportionality relation given by $1/Q^2$ (Fig.~[\ref{nucleon}]), resulting in a more diluted system.

Now, the evolution with respect to the variable $x$ presents a different scenario. The emission of gluons with small {\it momentum} becomes increasingly favored as $x$ decreases, essentially occupying the same transverse area. In this case, the leading-order evolution equation is given by:
\begin{equation}
    \frac{\partial}{\partial (\ln1/x)}f_g(x,k^2)=\frac{N_c\alpha_s}{\pi}k^2\int_0^{\infty}\frac{dk'^2}{k'^2}\left[\frac{f_g(x,k'^2)-f_g(x, k^2)}{|k'^2-k^2|}+\frac{f_g(x,k^2)}{\sqrt{4k'^2+k^2}}\right],
    \label{BFKL}
\end{equation}
and it is called the BFKL equation, named after the works of {\it Balitsky, Fadin, Kuraev}, and {\it Lipatov} \cite{bfkl1,bfkl2}. In the BFKL equation, $f_g(x,k)$ is the gluon density and can be written in terms of the unintegrated gluon distribution function,
\begin{equation}
xf_g(x,Q^2)=\int^{Q^2}d{k^2}\frac{f_g(x,k^2)}{k^2}.
\label{non integraded gluon pdf}
\end{equation}
In this regime, beyond a certain value of $x$, the growth of the gluon population increases the likelihood of their mutual interaction in the form of recombination, initiating processes where $gg \rightarrow g$, which compensate for {\it bremsstrahlung} emissions and saturate the gluon density in the hadronic system, i.e., the {\it saturation regime}.

A way to determine whether the physical system is in a dense or dilute state is provided by the phenomenological physical model of {\it Golec-Biernat-Wüsthoff} (GBW) \cite{golec}, which effectively delineates the saturation regime through the {\it saturation scale} $Q_s^2$,

\begin{equation}
    Q_s^2(x) = (x_0/x)^{\lambda},
    \label{saturation scale}
\end{equation}
where, in this work, the values $x_0=4.2\times10^{-5}$ and $\lambda=0.248$ were generally adopted, based on experimental DIS data from $ep$ collisions \cite{valores}. Some key properties of the GBW model include the {\it geometric scaling} (gs), in which the DIS cross sections can be expressed as a function of a single variable, $\tau=Q^2/Q^2_s$.

The geometric scaling allows extending the application of the GBW model to proton-nucleus collisions. Specifically, the relationship between the cross-section of a virtual photon interacting with a nucleus and the target's transverse area can be rewritten as a function dependent on nuclear saturation:
\begin{equation}
    \frac{\sigma^{\gamma^*A}(\tau_A)}{\pi R_A^2} = \frac{\sigma^{\gamma^*p}(\tau)}{\pi R_p^2},
\end{equation}
where $R_A = (1.12A^{1/3} - 0.86A^{-1/3}) \, \text{fm}$ represents the nuclear radius, and $A$ is the nucleus mass number. To achieve this equivalence, the transverse cross-section is adapted as $\sigma_0 \rightarrow \sigma_A$, and the saturation scale as $Q_s^2(Y) \rightarrow Q_{s,A}^2(Y)$. Thus, the nuclear saturation scale $Q_{s,A}(Y)$ can be expressed as:
\begin{equation}
    Q_{s,A}^2(Y) = \left(\frac{R_p^2A}{R_A^2}\right)^\Delta Q_s^2(Y),
    \label{gsprocedure01}
\end{equation}
where $\Delta \approx 1.27$ and $R_p \approx 3.56 \, \text{GeV}$ \cite{Armesto:2004}. This formulation enables adjusting the model to account for the increased particle density in interactions with nuclei, resulting in a saturation scale tailored to the nuclear context. Hence, the property of geometric scaling provides a means to extrapolate predictions made for isolated protons to interactions involving nuclei, with the nuclear saturation scale $Q_{s,A}^2$ reflecting the increased gluon density in nuclear systems.

\subsection{The Colour Glass Condensate}
\label{chapter 2 - section 3: the saturation physics and the cgc; subsection 1 - The Colour Glass Condensate}

The model that satisfactorily describes the behavior of the QGP during its initial stages is the {\it Colour Glass Condensate} (CGC) \cite{gelis}, which translates literally to the {\it color glass condensate} model. It describes the matter associated with a high density of gluons through the wave function of a hadron at high energies.

\begin{wrapfigure}{r}{0.4\textwidth}
\begin{center}
\includegraphics[width=0.65\linewidth]{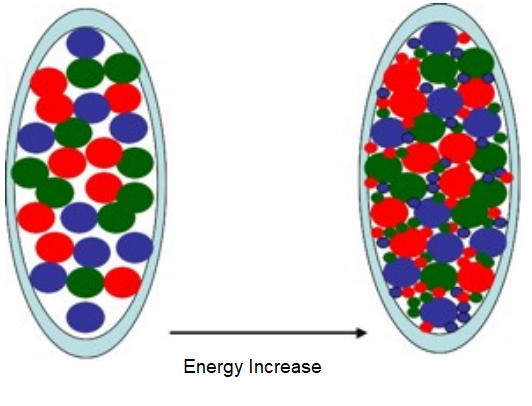}
\end{center}
\caption{For a given energy value $E_1$, gluons fill the proton. For $E_2>E_1$, even smaller gluons are generated, organizing themselves in previously unoccupied spaces. Adapted from \cite{mclerran2008brief}.}
\label{brief2008}
\end{wrapfigure}

The Fig.~[\ref{pdfs}] illustrates how, in the regime of high-energy physics or, equivalently, at small-$x$, the gluon population grows indefinitely. However, as energy increases, the size of the proton remains nearly constant. Thus, if more and more gluons are added, these particles will fill all the \textquoteleft empty' regions of the proton. If gluons had a fixed size, there would be a determined value to fully occupy the proton; this is not observed. On the other hand, if newly generated gluons are smaller, they can be arranged in previously unoccupied regions, which also become progressively smaller, in a process that can continue indefinitely (Fig.~[\ref{brief2008}]), akin to the paradox of Achilles never overtaking the tortoise in a world where convergent series do not exist.

The saturation scale $Q_s$ is inversely proportional to the space occupied by a gluon, $R_s=1/Q_s$. Thus, if the scale $Q_s$ is fixed, only a finite number of gluons can be arranged, as the entire proton region is occupied in a {\it condensed} system.
    Upon their generation, each gluon is incorporated into the wave function describing the hadron, such that:
\begin{equation}
    \ket{h}=\ket{qqq}+\ket{qqqg}+...+\ket{qqqg...q\bar{q}ggg},
\end{equation}

The terminology {\it glass} is related to the time scale and the order of the fields associated with this state of matter: they evolve much more slowly compared to natural time scales. In this model, gluons at high energies are described by classical fields produced by lower-energy gluons or, initially, by the valence quarks, understood here as the set of quarks that composed the hadron at low energies.

This mechanism causes the \textquoteleft daughter' gluons, also called {\it wee partons}, to have their time evolution scale dilated relative to the \textquoteleft parent' gluon. As a result, the older a gluon is within this hierarchy, the more it is perceived by others as a static classical field. Thus, the different configurations of gluons contributing to the hadronic wave function can be treated as an {\it ensemble} of non-interacting fields. Finally, the term {\it color} refers to the particles forming the CGC, primarily gluons, which carry color in QCD.

There exists a coordinate system better suited to handle the mathematical objects in CGC, known as the {\it Light Cone Coordinate System} (Appendix \ref{apendice A: light cone variables}). To work within this framework, it is necessary to quantize the theory in this set of variables, where the initial value is taken on the temporal surface of the light cone $x^+ = (t+z)/\sqrt{2} = 0$. This approach offers the significant advantage of simplifying the vacuum state, which coincides for both the interacting theory and the free theory\footnote{The boost operation in light cone quantization commutes with the light cone Hamiltonian in QCD. This property is not satisfied in the usual boost operation, which leads to particle creation.}, while also enabling the calculation of hadronic wave functions as an expansion of {\it Fock} states. Thus, the QCD light cone Hamiltonian, $P^-_{QCD}$, can be expressed as:
\begin{equation}
P^-_{QCD} = P^{-0}_{QCD} + V_{QCD}.
\end{equation}
Each {\it wee parton} in a densely populated configuration carries a small fraction $x = k^+/P^+$ of the total hadron momentum $P^+$. From this, it is possible to observe the exotic temporal characteristics of this model, as the typical timescale for reactions involving valence quarks, $t_{q}$, is much larger than that for reactions involving {\it wee partons}:
\begin{equation}
    t_{wee} = \frac{1}{k^-} = \frac{2xP^+}{k^2},
\end{equation}
\begin{equation}
    t_q \approx \frac{2P^+}{k^2},
\end{equation}
and since $x \ll 1$, it follows that $t_{wee} \ll t_q$.

Thus, the CGC is an effective field theory based on the separation of degrees of freedom into two categories: \textit{frozen color sources} and \textit{dynamic color fields}. These two types of degrees of freedom are separated by a renormalization group equation, known as the JIMWLK Equation, which ensures the independence of physical quantities from the cutoff, defined by a scale $\Lambda^+$ that demarcates the division between the two categories.

Fast gluons, which act as color sources, have a longitudinal momentum $k^+ > \Lambda^+$ and remain "frozen" due to Lorentz time dilation, forming a color current described by the color charge density $\rho_a(x^-, x_\perp)$. On the other hand, {\it wee partons} with $k^+ < \Lambda^+$ are described by the usual QCD gauge fields, $A^\mu$. The interaction between these two types of gluons is coupled in an eikonal manner\footnote{The term "eikonal" originates from wave theory and optics, specifically from the concept of the eikonal approximation, which describes the behavior of waves in media where phase variations are rapid compared to amplitude variations. The word "eikonal" derives from the Greek \textit{eikōn}, meaning "image" or "appearance." In high-energy physics and the CGC framework, this idea was adapted to describe particle interactions in scenarios where one component moves at extremely high speeds. In this context, "eikonal" refers to a situation where the "fast" particles act as a fixed field source, creating a sort of "frozen image" for the "slow" particles.}, meaning that the fast gluons act as sources for the slow gluons. While the color density $\rho_a$ remains constant during a collision, it varies from event to event, generating a probabilistic distribution $W_{\Lambda^+}[\rho]$, known as the \textit{CGC weight function}, which encodes all correlations of the color charge density at the cutoff scale $\Lambda^+$.

The network of charges with which the \textit{wee partons} couple is represented by a classical color charge density per unit of transverse area, $\rho$, in a random distribution, such that:
\begin{equation}
    \expval{\rho^a(x_T)} = 0; \quad \expval{\rho^a(x_T)\rho^b(y_T)} = \mu_A^2\delta^{ab}\delta^{(2)}(x_T-y_T),
\end{equation}
where
\begin{equation}
    \mu_A^2 = \frac{g_s^2 A}{2\pi R_A^2},
\end{equation}
is the squared color charge per unit of transverse area.

The CGC weight function describes the statistical distribution of $\rho$ and characterizes the correlations of the color charge density in the system. Observables of interest for a \textit{wee parton} are constructed from a classical field $A_a^{\mu}$ and are denoted by $O[A]$. Thus, the expected value of such an observable is obtained by averaging over all possible configurations of $\rho$:
\begin{equation}
    \expval{O[A]}_Y = \int W_Y[\rho]O[A[\rho]]d\rho.
    \label{observables on CGC}
\end{equation}
The evolution of $W_{\Lambda^+}[\rho]$ with $\Lambda^+$ is governed by the JIMWLK functional equation:
\begin{equation}
    \frac{\partial W_{\Lambda^+}[\rho]}{\partial \ln(\Lambda^+)} = -H_{JIMWLK}\left[\rho, \frac{\delta}{\delta \rho}\right] W_{\Lambda^+}[\rho],
\label{JIMWLK equation}
\end{equation}
where $H$ is the JIMWLK Hamiltonian, named after the work of \textit{Jalilian-Marian, Iancu, McLerran, Weigert, Leonidov, and Kovner}. The solution to this equation is numerical and can be expressed as a path integral or through the Balitsky hierarchy. The Balitsky-Kovchegov equation \cite{balitsky1996,kovchegov1999} (BK, described in more detail in Appendix \ref{apendice C: bk equation}) represents a mean-field approximation of this evolution, valid in the limit of a large number of colors, $N_c \rightarrow \infty$.

The \textit{McLerran-Venugopalan} model \cite{mclerran} (MV) provides a physical initial condition for the JIMWLK evolution, especially useful for studying gluon distributions in nuclei. In this model, the color charge distribution $W_{\Lambda_0^+}[\rho]$ is Gaussian in $\rho$:
\begin{equation}
    W_{MV}[\rho] = \mathscr{N} e^{-\int_k \frac{1}{2\mu^2(k)}\rho_a(k)\rho_a(-k)},
    \label{MV weight function}
\end{equation}
where $\mathscr{N}$ is a normalization factor. However, this hypothesis is inadequate for describing the evolution of observables at small-$x$, as it does not depend on rapidity. Nevertheless, this approach allows for both the theoretical motivation of the effective theory and direct phenomenological studies in collisions. Regarding the application of CGC, this model will be used in this thesis.
\newpage

\section{Generation and Phases of QGP}
\label{chapter 2: Section 8 - QGP Generation and Phases}

This section focuses on describing the processes that generate QGP from the collision of two heavy ions and summarizes its different phases up to the final stage, called freeze-out, where the plasma cools and the resulting hadrons are detected. Each of these phases is described by a different number of degrees of freedom, and an illustration of the stages can be seen in Fig.~[\ref{phases on a qgp}]:

\begin{wrapfigure}{r}{0.3\textwidth}
\begin{center}
\includegraphics[width=0.85\linewidth]{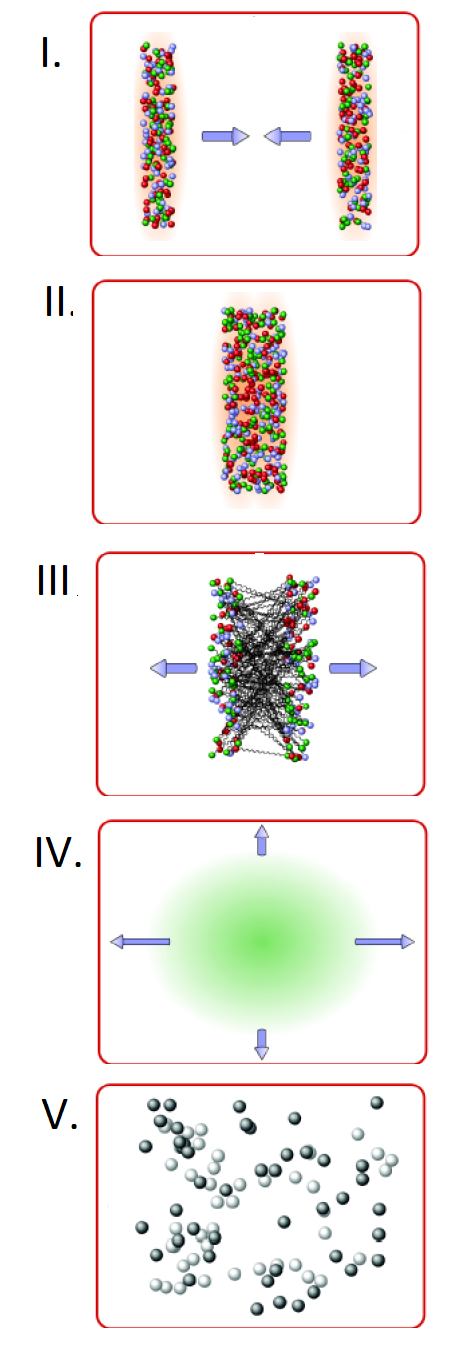}
\end{center}
\caption{Representation of the phases of QGP. Image adapted from \cite{epelbaum}.}
\label{phases on a qgp}
\end{wrapfigure}

I. Here, the center-of-mass frame is considered. The ions are accelerated until they reach an ultra-relativistic velocity (with a {\it Lorentz} factor of the order of 100), and it is reasonable to approximate that the ions are distributed in a spherical shape. The effect of spatial contraction, which manifests in the direction of propagation, causes the matter to be distributed almost entirely in a longitudinal disk relative to the direction of motion. Saturation physics is evident, and the ions are composed mainly of gluons, with a small fraction of {\it momenta} ($x\ll1$). This system is well described by the {\it Colour Glass Condensate} model.

II. The collision occurs, marking the initial time $\tau_0=0$ fm/c. The nuclei pass through each other almost instantaneously but leave residues whose size varies depending on the centrality of the collision. The first processes that occur are the so-called {\it hard} processes, which involve a large transfer of {\it momenta}. These processes are responsible for creating the necessary conditions for the subsequent generation of heavy quarks, vector mesons, hadronic jets, and direct photons, particles that play an important role in the final state of the system.

III. The third stage occurs at approximately $\tau=0.2$ fm/c. Most of the partons are released by the collision, creating an extremely dense and out-of-equilibrium medium, about ten times denser than the atomic nucleus. In this phase, gluon fields dominate the system, and the interactions between partons are highly non-linear. This state is called {\it glasma}, an intermediate phase between the initial state of the collided ions and the formation of the QGP itself.

IV. This phase is marked by collective effects that indicate strong interactions among partons, such as {\it elliptic flow}\cite{jia}. In this phase, the system thermalizes in an impressively short time: $\tau=1$ fm/c. Quantum field theories are unable to explain such a short time, which is obtained from transport theory, more precisely, from relativistic hydrodynamics.

V. The plasma matter continues to expand and cool, hadronizing when it reaches temperatures below the critical value, at approximately $\tau=10$ fm/c. The resulting hadronic medium is still relatively dense, and energy exchange among its constituents maintains thermal equilibrium. At this stage, at about $\tau=20$ fm/c, the system consists of a hot and dense gas of hadrons. When inelastic processes, which convert hadrons of one species into others, cease, the hadronic abundances stabilize, and the system undergoes {\it chemical freeze-out}. Then, {\it thermal freeze-out} occurs, the stage where the momentum of the particles in the medium no longer changes, that is, when all elastic and inelastic collisions cease. From these points onward, the particles in the medium become free and move smoothly to the final detectors.

\section{The Effective Theories for Each Phase}
\label{chapter 2 - section 2: the phase effective theories}

\begin{figure}[ht]
\centering
\includegraphics[width=0.7\linewidth]{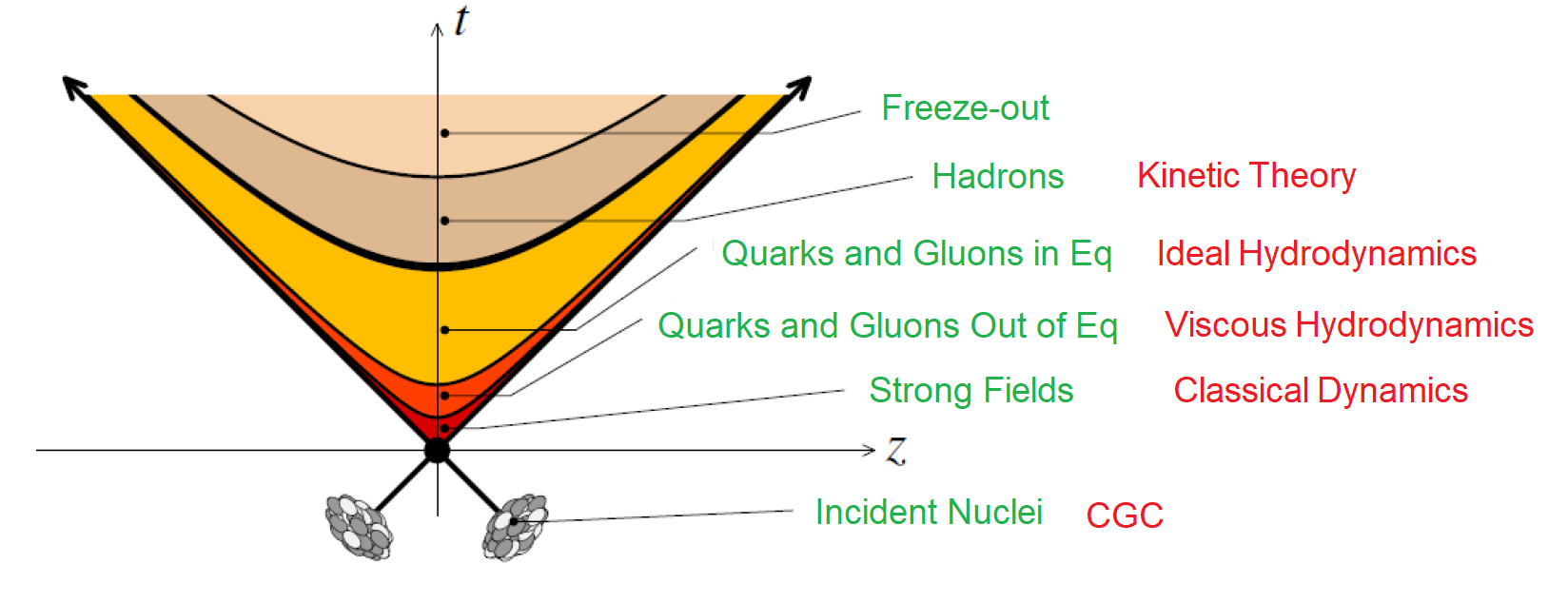}
\caption{Representation of the QGP phases. As time evolves, the system's degrees of freedom change, making different physical theories convenient for describing each phase. Image adapted from \cite{iancu_qcd}.}
\label{lc}
\end{figure}

As discussed earlier, the QGP undergoes various phases throughout its evolution. Some of the properties that arise in these different stages include non-equilibrium situations, short-range effects, non-homogeneity, $N$-body phase space, resonance or particle production, collective dynamics, and freeze-out. This range of characteristics can be effectively studied using transport phenomena. 

Thus, it is possible to employ microscopic, macroscopic (hydrodynamic), or hybrid transport theories to provide a comprehensive temporal description of the QGP. Accordingly, the following subsections are devoted to a brief summary of the two main theories used to construct and characterize each stage of the system: {\it Relativistic Hydrodynamics} and {\it Kinetic Theory}, excluding the CGC, which has already been discussed and accounts for the early stages of the QGP.

\subsection{Relativistic Hydrodynamics}
\label{chapter 2 - section 2 - subsection 2: relativistic hydrodynamics}

The modern conception describes hydrodynamics as an effective low-energy theory of quantum field theory. This theory effectively describes the intermediate phases of the QGP. Its fundamental equations describe the conservation of current, energy, and {\it momentum},

\begin{equation}
    \partial_{\mu}T^{\mu\nu}=0, \quad \partial_{\mu}j_i^{\mu}=0,
    \label{hidro}
\end{equation}
where $j_i^{\mu}$, with $i=B,S,Q$, represents the conserved current, and $T^{\mu \nu}$ is the energy-{\it momentum} tensor. The local flow velocity $u^{\mu}$ and the rank-2 tensor perpendicular to the flow, $\Delta^{\mu\nu}=g^{\mu\nu}-u^{\mu}u^{\nu}$, can be used to decompose the tensor $T^{\mu\nu}$ and the conserved currents into {\it space-like} and {\it time-like} components, such that:
\begin{equation}
\begin{split}
T^{\mu\nu}&=\epsilon u^{\mu}u^{\nu}-p\Delta^{\mu\nu}+W^{\mu}u^{\nu}+W^{\nu}u^{\mu} +\pi^{\mu\nu}, \\   
j_i^{\mu}&=n_iu^{\mu}+V_i^{\mu}. 
\end{split}
\end{equation}
In these equations, $\epsilon=u_{\mu}T^{\mu\nu}u_{\nu}$ is the energy density, $p=p_s+\Pi=-\frac{1}{3}\Delta_{\mu\nu}T^{\mu\nu}$ is the sum of the hydrostatic and volumetric pressures, $W^{\mu}=\Delta^{\mu}_{\alpha}T^{\alpha\beta}u_{\beta}$ is the energy current, $n_i=u_{\mu}j^{\mu}_i$ is the charge density, $V_i^{\mu}=\Delta^{\mu}_{\nu}j_i^{\nu}$ is the charge current, and $\pi^{\mu\nu}\equiv\expval{T^{\mu\nu}}$ is the shear stress tensor. The {\it brackets} in the definition of the shear stress tensor indicate the following mathematical operation,

\begin{equation}
    \expval{A^{\mu\nu}}=\left[\frac{1}{2}(\Delta^{\mu}_{\alpha}\Delta^{\nu}_{\beta}+\Delta^{\mu}_{\beta}\Delta^{\nu}_{\alpha})-\frac{1}{3}\Delta^{\mu\nu}\Delta_{\alpha\beta}\right]A^{\alpha\beta}.
\end{equation}

To simplify the discussion of interest, we now consider the case of a conserved charge and denote the baryonic current by $j_{\mu}=j_B^{\mu}$. Then, each term of the current and the energy-momentum tensor can be explicitly separated into \textit{ideal} and \textit{dissipative} components:
\begin{equation}
\begin{split}
T^{\mu\nu}&=T^{\mu\nu}_{\text{id}}+T^{\mu\nu}_{\text{dis}}=[\epsilon u^{\mu}u^{\nu}-p_s\Delta^{\mu\nu}]_{\text{id}}+[-\Pi\Delta^{\mu\nu}+W^{\mu}u^{\nu}+W^{\nu}u^{\mu}+\pi^{\mu\nu}]_{\text{dis}},\\
j_{\mu}&=j_{\text{id}}^{\mu}+N_{\text{dis}}^{\mu}=[nu^{\mu}]_{\text{id}}+[V^{\mu}]_{\text{dis}}.
\end{split}
\end{equation}
Neglecting the dissipative terms constitutes \textit{ideal hydrodynamics}. In this case, the solutions to the hydrodynamic equation [\ref{hidro}] with a given set of initial conditions describe the spatio-temporal evolution of six variables, three of which are state variables, $\epsilon(x)$, $p(x)$, and $n(x)$. The remaining three are the spatial components of the flow velocity $u^{\mu}$. However, the conservation equations [\ref{hidro}] consist of only 5 independent equations. A sixth equation relating $p$ and $\epsilon$ must be added to solve the problem. A good description is provided by the equation of state derived from the thermodynamic calculation of QCD at high temperatures with low chemical potentials\cite{bazavov}.
\begin{equation}
    \epsilon-3p=-\frac{T}{V}\frac{d\ln \Xi}{d\ln a},
\end{equation}
where $T$ is the temperature, $V$ is the pressure, $\Xi$ is the grand canonical partition function, and $a$ is the lattice spacing\footnote{Using calculations in the lattice-QCD model, a non-perturbative approach to chromodynamics. It consists of a gauge theory in discretized form, where the spacetime points correspond to the points of a finite 4-dimensional lattice.}. From this equation of state, it is still possible to describe the strongly interacting matter below the deconfinement temperature $T_c$, where all thermodynamic quantities are well described by a hadronic resonance gas initially proposed by {\it Hagedorn}\cite{hagedorn},

\begin{equation}
    \epsilon-3p=\sum_{m_i\leq m_{max}}T^4\frac{d_i}{2\pi^2}\sum_{k=1}^{\infty}\frac{(-\eta_i)^{k+1}}{k}\left(\frac{m_i}{T}\right)^3K_1\left(\frac{km_i}{T}\right),
\end{equation}
where $K_1(km_i/T)$ is the modified Bessel function, having species of different particles with mass $m_i$, degeneracy factor $d_i$, and $\eta_i=\pm 1$, with $+1$ for fermions and $-1$ for bosons. The sum runs over all particles up to the resonance mass $m_{max}=2.5 \, \text{GeV}$.
 
In the literature, two definitions of flow can be found; one related to the energy flow due to {\it Landau}\cite{landau}, and the other proposed by {\it Eckart}\cite{eckart}, which refers to the conserved charge flow. Respectively,

\begin{equation}
    \begin{split}
    u_L^{\mu}&=\frac{T^{\mu}_{\nu}u_L^{\nu}}{\sqrt{u_L^{\alpha}T_{\alpha}^{\beta}T_{\beta\gamma}u_L^{\gamma}}}=\frac{1}{e}T^{\mu}_{\nu}u_L^{\nu}, \\
    u_E^{\nu}&=\frac{j^{\nu}}{\sqrt{j_{\nu}j^{\nu}}}.
    \end{split}
\end{equation}

In the {\it Landau} definition, $W^{\nu}=0$, while in {\it Eckart}: $V^{\nu}=0$. When ideal hydrodynamics is considered, the two definitions become equivalent. Particularly, in heavy-ion collisions, the evolution of matter is described in a region with zero baryon decomposition number, $j=0$, as in the mid-rapidity regions at the LHC and the top energy of RHIC, making the {\it Landau} definition more appropriate for the case of interest.

To solve the hydrodynamic equations without discounting the dissipative terms, it is common to introduce two phenomenological definitions, also known as {\it constitutive equations}, for the stress tensor and the bulk pressure\cite{hirano}:
\begin{equation}
    \begin{split}
    \pi^{\mu\nu}&=2\eta\expval{\nabla^{\mu}u^{\nu}}, \\
    \Pi&=-\zeta\partial_{\mu}u^{\mu}=-\zeta\nabla_{\mu}u^{\mu}.
    \end{split}
    \label{NS}
\end{equation}
The new coefficients $\eta$ and $\zeta$ are called shear viscosity and bulk viscosity, respectively.

For the {\it Bjorken} flow invariant under {\it boosts}\cite{bjorken1983}, with a velocity in the $z$ direction, $v_z$, we have:

\begin{equation}
    u_{BJ}^{\mu}=\frac{x^{\mu}}{\tau}=\frac{t}{\tau}\left(1,0,0,\frac{z}{t}\right),
\end{equation}

where $\tau$ is the proper time. With this definition, it is possible to find the equation of motion\cite{hirano2009}:

\begin{equation}
    \frac{d\epsilon}{d\tau}=-\frac{\epsilon+p_s}{\tau}\left(1-\frac{4}{3\tau T}\frac{\eta}{s}-\frac{1}{\tau T}\frac{\zeta}{s}\right),
\label{hidroe}
\end{equation}

where $s$ is the entropy density. If the two terms on the left are disregarded, the {\it Bjorken} solution for ideal hydrodynamics\cite{bjorken1983} is obtained. These last two terms describe the compression of the energy density due to viscous corrections. The first term is related to the shear viscosity, $\eta/s$, while $\zeta/s$ reflects an intrinsic property of fluids.

Hydrodynamics provides an effective description of a system in local thermal equilibrium and can be derived from the kinetic description through expansions in series of the entropy 4-current $S^{\mu}=su^{\mu}$ in gradients of the local thermodynamic variables. Zero-order gradients reflect ideal hydrodynamics, making higher-order terms correspond to dissipative quantities arising due to irreversible thermodynamic processes in the fluid, such as the frictional energy dissipated between two fluid elements in relative motion. 

Thus, the {\it Navier-Stokes} equation [\ref{NS}], which includes only linear dependencies on the velocity gradient, results in some issues: the thermodynamic flux in $\pi^{\mu\nu}$ or $\Pi$, which is purely a local function of the velocity gradient, either disappears or manifests instantaneously, leading to non-causal influences and causing numerical instabilities.  

To address this problem, the inclusion of second-order terms in the gradients must be considered in the implementation of a relativistic dissipative fluid, resulting in relaxation-type equations for $\pi^{\mu\nu}$ and $\Pi$ with macroscopic relaxation times $\tau_{\pi}\equiv 2\eta \alpha$ and $\tau_{\Pi}=\zeta\beta$. Qualitatively, these times reflect the difference in duration between the manifestation of thermodynamic gradients driving the system out of local equilibrium and the establishment of dissipative fluxes in response to these gradients, restoring causality.

\subsection{Kinetic Theory}
\label{chapter 2 - section 2 - subsection 3 - kinetic theory}

Kinetic theory fits well in the final stages of the plasma, when the system becomes more dilute and begins to hadronize towards freeze-out and, finally, the detection of the multiplicity of final particles. This model accounts for a wide range of interactions between particles, including both elastic and inelastic collisions. 

The relevant quantities in kinetic theory are described through phase-space densities, which must be calculated for each particle species embedded in the QGP. This quantity is dimensionless and measures the number of particles of a given type per phase-space unit divided by the number of choices for each possible discrete degree of freedom. Taking the case of gluons, which dominate the early stages of the QGP, the density $f$ is defined as:

\begin{equation}
f\equiv \frac{1}{2(N_c^2-1)}\frac{dN_g}{d^3xd^3p}.
\end{equation}
Here, $2(N_c^2-1)$ is the degeneracy factor for gluons, such that $f(\vec{p}, \vec{x}, t)d^3xd^3p$ represents the average number of gluons within the volume $d^3x$ around the point $\vec{x}$ with a momentum between $\vec{p}$ and $\vec{p}+d^3p$ at time $t$. The time evolution of the distribution of a given set of particles follows the Boltzmann equation \cite{mermin2006}:

\begin{equation}
    \left ( \frac{\partial}{\partial t}+\vec{v_c}\cdot \nabla+\vec{F}_{ext}\cdot \nabla_{p} \right)f(\vec{x},\vec{p},t)=C[f].
\end{equation}
In this equation, $f$ is the phase-space density of a given type of particle in the QGP, $\vec{v}_p=\vec{p}/E_p$ is the velocity of gluons, and $\vec{F}_{ext}$ is a generic external force.
 
\subsection{Some Results from Effective Theories}
\label{chapter 2 - section 3: some results}

The particle multiplicity in the final state of the collision is a key point in understanding the phases where processes occurred, generating a myriad of complexities ranging from hadronic jets to heavy quarks. This brief subsection focuses mainly on the successes achieved during the intermediate phases described by hydrodynamics, which inspire greater confidence due to their predictive power from the earliest RHIC experiments to the LHC.

\begin{wrapfigure}{r}{0.45\textwidth}
\begin{center}
\includegraphics[width=0.75\linewidth]{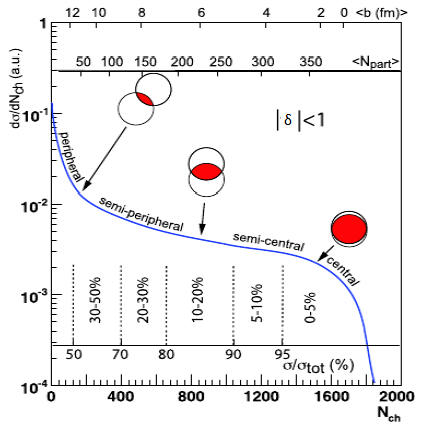}
\end{center}
\caption{The definition of the centrality of charged particle multiplicity in the final state $N_{ch}$ and its correlation with the average impact parameter $\expval{b}$ and the average number of participating nuclei $\expval{N_p}$. Image adapted from reference \cite{sarkar2009}.}
\label{glau}
\end{wrapfigure}

In the initial description of the hadronic or nuclear interaction state of the type $(A+B)$, the {\it de Broglie} wavelength of the incident nucleus is much smaller than the intra-nuclear distances of the target nucleus. For each incident nucleon, the positions of the nucleons in the target appear frozen in time. After each nucleon-nucleon (NN) collision, whether elastic or inelastic, both participating nucleons acquire a transverse momentum relative to the incidence direction, which in most cases is much smaller than the longitudinal component of the same quantity, making the momenta before and after the collision approximately the same $p_z \approx p_z'$. Now, high incidence energies coupled with small scattering angles indicate interactions dominated by large orbital momentum $\ell$. In this case, it is convenient to modify the partial wave expansion of the scattering amplitude by introducing an impact parameter $b=(1+\ell)/p$. Under these circumstances, the semi-classical approximation model proposed by {\it Glauber}\cite{glauber2006} comes into play, treating nuclear collisions as multiple NN interactions\cite{joachain1974}. 

The nucleons that experienced at least one NN collision are called {\it participants}, and those that felt none are the {\it spectators}. The total number of spectators and participants follows the rule $N_{es}+N_p=A+B$. Additionally, there is a restriction on the number of collisions $N_{col}$, which must satisfy the inequality $N_{col} \leq N_p/2$.

Experimentally, the number of charged particles $N_{ch}$ is measured, and, broadly speaking, the ingredients of the model can be arranged in the form:
\begin{equation}
    \frac{dN_{ch}}{d\delta} \propto (N_p \leftrightarrow N_{col}) \propto b \rightarrow \delta. 
\end{equation}
In this equation, the derivative on the left-hand side represents the experimental observable, known as the {\it multiplicity}\footnote{This should not be confused with the concept of multiplicity $\Omega$ developed in the study of entropy, although this observable is also used to estimate entropy in collisions.}, which is proportional to the number of participants or collisions according to the {\it Glauber} model. The impact parameter $b$ is determined through simulations aimed at establishing the centrality $\delta$ of the collision (Fig.~[\ref{glau}]).

\begin{figure}[ht]
\centering
\includegraphics[width=0.75\linewidth]{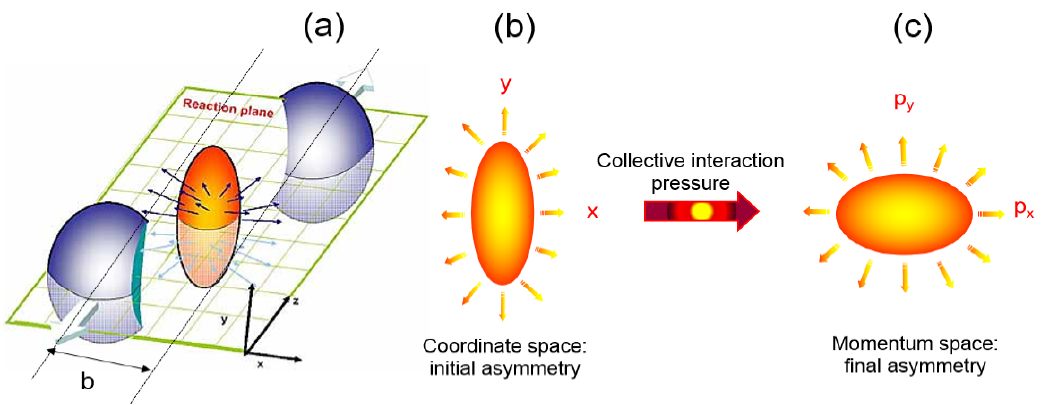}
\caption{A {\it non-central} collision of two nucleons generates an interaction region {\bf (a)}, where the spatial anisotropy of the generated region induces pressure gradients in the transverse plane {\bf (b)}. This process ultimately results in a momentum anisotropy in the generated particles. Figure reproduced from reference \cite{nouicer2016}.}
\label{v2}
\end{figure}

One of the most well-established predictions regarding the collective behavior of matter created in ultra-relativistic heavy-ion collisions occurs in {\it non-central} collisions, due to transverse evolutions caused by pressure gradients stemming from the spatial anisotropy of the initial density (Fig.~[\ref{v2}])\cite{heinz2010}. This anisotropy is quantified using {\it Fourier} coefficients\cite{poskanzer1998}:

\begin{equation}
    v_{n}=\expval{\cos[n(\phi-\Psi_n)]},
\end{equation}

where $\phi$ represents the particle's azimuthal angle, $\Psi_n$ the angle of symmetry of the initial state plane, and $n$ the harmonic order. For the case of interest, a {\it non-central} collision of heavy ions, the beam axis and the impact parameter define the azimuthal reaction plane $\Psi_{RP}$. For a nucleus with a smooth matter distribution, the symmetry plane corresponds to the reaction plane, $\Psi_n=\Psi_{RP}$, which results in odd Fourier coefficients being null due to symmetry.

\begin{figure}[ht]
\centering
\includegraphics[width=0.85\linewidth]{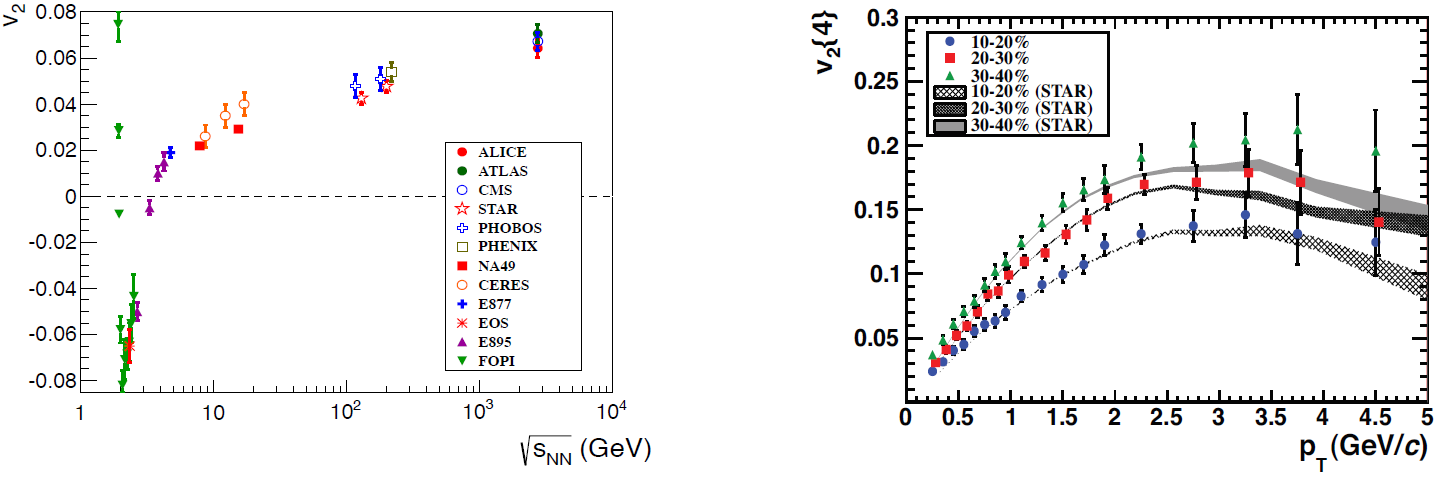}
\caption{{\bf (Left)} Experimental results for the $v_2$ coefficient as a function of beam energy, reproduced from reference \cite{heinz2013}. {\bf (Right)} Elliptic flow as a function of transverse {\it momenta} for charged particles. Reproduced from reference \cite{aysto2010}.}
\label{v2exp}
\end{figure}

However, fluctuations in the matter distribution cause the symmetry plane to vary event by event around the reaction plane. This plane is determined by the participating nucleons and is referred to as the participant plane, $\Psi_{PP}$\cite{manly2006}. Since the symmetry planes $\Psi_n$ are not experimentally measured, the anisotropic flow coefficients are estimated based on correlations observed among detected particles\cite{aysto2010}, as shown in Fig.~[\ref{v2exp}]. In this figure, the dominant term $v_2$ is expressed as a function of various beam center-of-mass energies (left) and for different values of transverse {\it momenta} of charged particles.

\begin{figure}[ht]
\centering
\includegraphics[width=0.5\linewidth]{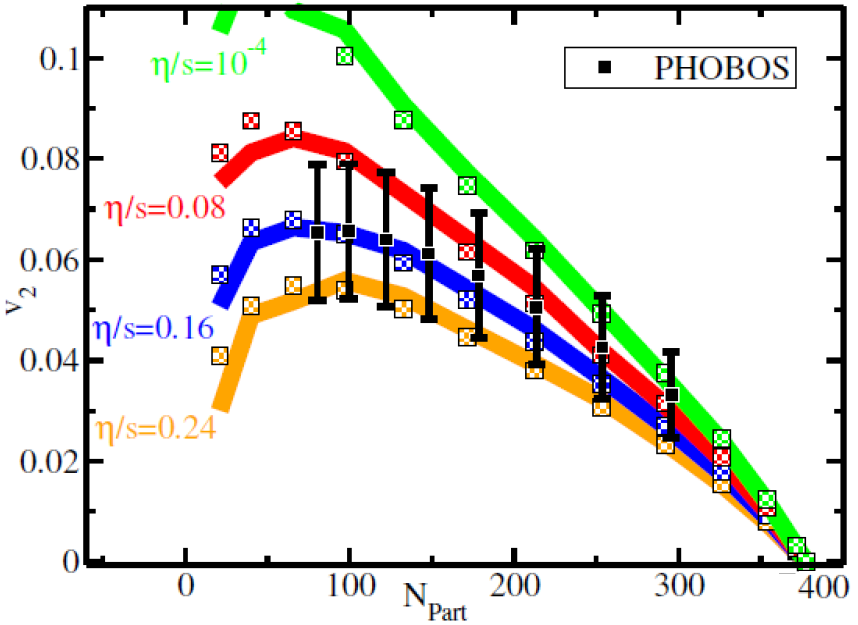}
\caption{Data modeled by hydrodynamic simulations with $\eta/s$ values in the range [0.08-0.2]. The black points represent experimental results. Results from \cite{policastro2002}.}
\label{v2exp2}
\end{figure}

From this perspective, the results involving hydrodynamic calculations gain prominence due to their ability to predict a relatively low viscosity per unit entropy $\eta/s$ (Fig.~[\ref{v2exp2}]). These results can also be compared with studies involving conformal field theories\footnote{Quantum field theory that is invariant under conformal transformations, which may be classical or quantum. Such transformations are performed on an arbitrary metric covariant with a Weyl transformation ($g_{ab}\rightarrow e^{-2\omega(x_{\mu})}g_{ab}$). In the quantum case, this invariance leaves the partition function $\Xi$ of the system unchanged.} (which is not the case for QCD). In these theories, the variation of the system's action with respect to the adjacent metric is proportional to the stress tensor, and a variation in the metric will be proportional to the trace of the tensor. Consequently, the trace of the stress tensor must vanish, which does not always happen; thus, the phenomenon is called conformal or trace anomaly. As a result, in these theories, $\eta/s$ would a priori be 0, but considering the trace anomaly in gluonic matter for terms of order $\eta/s=\mathcal{O}(0.1-1)$ in lattice $SU(3)$ gauge theory yields a value of $\eta/s=1/4\pi$\cite{kovtun2005}, aligning with the simulation results in Fig.~[\ref{v2exp2}].

By examining expression [\ref{hidroe}], it is observed that the theory can also predict the proper time, which, during the flow generation phases, corresponds to the system's thermalization time $\tau_T$. The obtained value is $\tau_T=1 \hbox{fm/c}$\cite{adcox2005}, meaning that during plasma and flow generation, the system organizes and thermalizes within a short period. However, when employing theories that consider the system's microscopic effects, such as QCD, it is not possible to predict such a short time, remaining an open research area regarding the QGP. Physically, this short thermalization time is associated with abrupt entropy creation during the initial stages of collisions. Determining how this entropy is generated is the primary objective of this work. Henceforth, models aiming to address this objective will be presented, starting with the dynamic entropy of QCD dense states.

\section{Dynamic Entropy in QCD}
\label{chapter 5 - section 2: QCD Dynamic Entropy}

The first notion of entropy studied in this work is now presented, related to the dense states of QCD. These states can occur in different physical contexts but generally imply a high concentration of gluons and other partons in the early stages of a collision. For an appropriate description, the hadronic medium is well-explained by the CGC, characterized by saturation physics in a dense configuration. This description can also be applied to $pA$ collisions in the high-energy regime. Thus, these states can be theoretically described in QCD in a weak-coupling regime through nonlinear energy evolution, where the initial conditions are described by a dense state.

QCD evolution occurs with respect to rapidity, $Y$, which, as it grows, increases the density of partons until the saturation regime, delimited by $Q_s$, is reached, resulting in a CGC. Since $Y=-\ln x$, the saturation scale described in Eq.~[\ref{saturation scale}] can be rewritten as:
\begin{equation}
    Q_s^2(Y) = (x_0/x)^{\lambda}=x_0^\lambda e^{\lambda Y},
    \label{Y saturation scale}
\end{equation}
Thus, the CGC state is characterized, among other properties, by a limiting transverse size $R_s \sim 1/Q_s$. Therefore, the parton size $R_s$ decreases as rapidity increases. With increasing energy, the parton density becomes high enough for recombination reactions involving gluons to occur, establishing the saturation regime.

In this situation, properties such as geometric scaling are crucial, making the Unintegrated Gluon Distributions (UGD's), $\phi(k,Y)$, functions of a single variable $\tau=k^2/Q_s^2=k^2R_s$, where $k^2$ is the squared transverse momentum carried by the gluons. Thus:
\begin{equation}
    \phi(k,Y)d^2k \sim \phi(\tau)R_s^2d^2k.
\end{equation}
Initially, a state can be associated with a rapidity value $Y_1$, which, due to rapidity evolution in the dense regime, will lead to a new value $Y_2$. This evolution describes how the system increases its parton density, forming a dense medium of gluons in the CGC. Thus, the initial value, $Y_1$, must correspond to a state that can already be described by saturation physics. For this purpose, in this work, we use $Y\approx4.6$ ($x=10^{-2}$). This transition to a higher density regime implies a behavior in which gluons branch and recombine at balanced rates, characterizing the saturation state. The variation in rapidity represents a dynamic parameter that alters the system's energy. Although it is an out-of-equilibrium process, it generates a stable saturation state, allowing the system to evolve into a denser CGC medium as the energy increases with rapidity.

In Non-Equilibrium Statistical Mechanics, using the Hatano-Sasa identity (Sub-section [\ref{chapter 2: Section 7 - Subsection 2: Hatano-Sasa Identity}]), a distribution of steady states, $P(z;\lambda)$, can describe the probability of finding the system in a specific configuration in phase space $z$ for a given value of the dynamic parameter $\lambda$. The transition from $\lambda_1$ to $\lambda_2$ represents a change in the dynamic parameter over time, resulting in a new equilibrium configuration, $P_2(z;\lambda_2)$. In high-energy physics, considering the following comparisons:
\begin{itemize}
    \item[$\blacksquare$] The phase-space variable $z$ corresponds to the transverse momentum of the partons $k$;
    \item[$\blacksquare$] The dynamic variable $\lambda$ corresponds to the rapidity $Y$;
    \item[$\blacksquare$] The steady-state distribution $P(z;\lambda)$ corresponds to $P(k, Y)$.
\end{itemize}
Thus, it is possible to define a probability distribution for the transverse momentum of gluons $P(k,Y)$ between $k^2$ and $k^2 + d^2k$, written in terms of the UGD's and given by:
\begin{equation}
P(k,Y)d^2k=\frac{\phi(k,Y)}{\int \phi(k,Y)d^2k}d^2k,
\label{kdist}
\end{equation}
subject to the normalization condition:
\begin{equation}
 \int P(k,Y)d^2 k=\int P(\tau)d\tau=1.
\end{equation}
Using the Hatano-Sasa identity (Eq.~[\ref{hatano-sasa}]), we have:
\begin{equation}
\begin{split}
\expval{\exp \left[ - \int_{Y_1}^{Y_2} \frac{d}{dY}\ln P(k, Y) dY \right]}_{Y_2}&=\expval{\exp \left[ - \int_{Y_1}^{Y_2}\frac{dP}{P(k, Y)}\right]}_{Y_2}= \int P(k, Y_2) \, e^{- \ln \frac{P(k, Y_2)}{P(k, Y_1)}} d^2k \\
&\equiv \int P(k, Y_1) \, d^2k \equiv 1.
\end{split}
\label{dynamical hatano-sasa}
\end{equation}
Here, $\expval{...}_{Y_2}$ represents an average calculated over the probability distribution at state $Y_2$, and the term $\int_{Y_1}^{Y_2} \frac{d}{dY}\ln P(k, Y) dY$ refers to the variation in the logarithm of the probability distribution over the rapidity interval from $Y_1$ to $Y_2$. Now, using arguments from statistical mechanics, the increase in rapidity $Y$, which causes the size $R_s(Y)$ to decrease, i.e., $R_s(Y_1)>R_s(Y_2)$ for $Y_2>Y_1$, can be interpreted as a compression. This compression alters the parton probability distribution, $P(k, Y)$, in a manner that aligns this distribution with the Hatano-Sasa identity.

Using these concepts, it is possible to define the {\it QCD dynamical entropy}\cite{pescha}, $\Sigma^{Y_1\rightarrow Y_2}$, in a medium described by the CGC with rapidity $Y_2$, arising from the QCD evolution from $Y_1\rightarrow Y_2$. The dynamical entropy is defined as:
\begin{equation}
\Sigma^{Y_1\rightarrow Y_2}=\expval{\ln\frac{P(k,Y_2)}{P(k,Y_1)}}_{Y_2}\equiv \int d^2k P(k,Y_2)\ln\left[\frac{P(k,Y_2)}{P(k,Y_1)}\right].
\label{qcd dynamical entropy}
\end{equation}
This quantity measures the amount of disorder created in the CGC medium due to the evolution of rapidity. It can be noted that it is mathematically equivalent to mutual information (Eq.~[\ref{mutual information}]), such that variable transformations involving the {\it geometric scaling} $\tau=k^2/Q_s^2$ will not alter the value of the dynamical entropy. Hence, the following sections will present some characteristics of QCD dynamical entropy, particularly its positivity, the relation of geometric scaling in terms of Non-Equilibrium Statistical Mechanics, and the dynamical entropy of a CGC state.

\subsection{Positivity of QCD Dynamical Entropy}
\label{chapter 5 - section 2: QCD Dynamic Entropy; subsecion 1: QCD Dynamical Entropy Positivity}

Initially, consider {\it Jensen's Inequality}, a fundamental result applicable to convex functions. In simple terms, it states that for a convex function and a random variable $X$, the average of the function is always greater than or equal to the function of the average, i.e.:
\begin{equation}
e^{\expval{X}} \leq \expval{e^X}.
\label{jansen desigualdade}
\end{equation}
From this inequality, we have:
\begin{equation}
\ln e^{\expval{X}} \leq \ln \expval{e^X} \quad \therefore \quad  \expval{X} \leq \ln \expval{e^X}.
\end{equation}
Using $X=-\ln P(k,Y_2)/P(k, Y_1)$, this result can be related to the dynamical entropy, as:
\begin{equation}
\begin{split}
\Sigma^{Y_1\rightarrow Y_2} &= \expval{\ln \frac{P(k,Y_2)}{P(k, Y_1)}}_{Y_2}\geq -\ln \expval{e^{-\ln \left[\frac{P(k,Y_2)}{P(k, Y_1)}\right]}}_{Y_2}\\
&= - \ln \int d^2k P(k,Y_2)\frac{P(k,Y_1)}{P(k, Y_2)}=- \ln \underbrace{\int d^2k P(k,Y_1)}_1 = 0.
\end{split}
\end{equation}
This holds for all $Y_2 \leq Y_1$, ensuring the positivity condition for any increase in rapidity.

\subsection{Geometric Scaling in Terms of Non-Equilibrium Thermodynamics}
\label{chapter 5 - section 2: QCD Dynamic Entropy; subsection 2: Geometric scaling in Terms of Non-Equilibrium Mechanical Statistics}

Considering $\tau_n = k^2/Q_s^2(Y_n)$ and $R_n = R_s^2(Y_n)$, Eq.~[\ref{dynamical hatano-sasa}] can be rewritten in terms of the scaling variable $\tau$. Using the normalization condition from Eq.~[\ref{kdist}]:
\begin{equation}
P(k,Y) = \frac{\phi(k,Y)}{\int \phi(k,Y) d^2k} = \frac{\phi(\tau) R_s^2}{\int \phi(\tau) R_s^2 d^2k} = \frac{R_s^2}{\pi}P(\tau),
\label{normalization 2}
\end{equation}
where the variable substitution $d\tau = dk 2kR_s^2$ was applied. Now, evaluating the Hatano-Sasa identity in QCD (Eq.~[\ref{dynamical hatano-sasa}]) in terms of the scaling variable $\tau$:
\begin{equation}
\expval{\exp\left[-\ln\frac{P(k, Y_2)}{P(k, Y_1)}\right]}_{Y_2} = \expval{\exp\left[-\ln\frac{P(\tau_2) R_2^2}{P(\tau_1) R_1^2}\right]}_{Y_2} = \underbrace{\int d^2 k P(k,Y_2)}_1 e^{-\ln \frac{R_2^2}{R_1^2}} = e^{-\ln \frac{R_2^2}{R_1^2}}.
\label{dynamical hatano-sasa 2}
\end{equation}
Thus:
\begin{equation}
\expval{\exp\left[-\ln\frac{P(\tau_2) R_2^2}{P(\tau_1) R_1^2}\right]}_{Y_2} = e^{-\ln \frac{R_2^2}{R_1^2}}.
\label{dynamical hatano-sasa 3}
\end{equation}
This equation can be related to the Jarzynski identity (Eq.~[\ref{jarzinski identity}]). The Jarzynski identity connects the stochastic distribution of thermodynamic work during the process $A \rightarrow B$ to the free energy difference $\Delta F$ between two equilibrium states $A \rightarrow C$. An interesting feature is that the amount of dissipative work $W_{\text{Dif}} \equiv W - \Delta F$ performed during the process $A \rightarrow B$ is related to the entropy production $\Delta S = \frac{\expval{W} - \Delta F}{T} \geq 0$, provided the state $B$ can relax to temperature $T$, keeping the driving parameter constant.

Comparing the Jarzynski identity with Eq.~[\ref{dynamical hatano-sasa 3}], it becomes clear that the term $\ln \frac{R_2^2}{R_1^2}$ corresponds to the logarithm of the ratio of the available phase space dimensions $R_2^2$ relative to $R_1^2$. This term represents the change in the free energy of a particle in an ideal gas confined within a two-dimensional "box" when its size decreases from $R_1$ to $R_2 < R_1$. Hence, a thermodynamic interpretation of the QCD relation derived in Eq.~[\ref{dynamical hatano-sasa 3}] is that the rapidity shift $Y_1 \rightarrow Y_2$ induces a change in the set of CGC states with a reduced saturation size $R_2$, generating entropy upon subsequent relaxation.

This analysis leads to the following heuristic comparisons between the variables of dynamical entropy and thermodynamic variables:
\begin{equation}
\begin{split}
\ln\frac{P(\tau_2)R_2^2}{P(\tau_1)R_1^2} &\rightarrow \frac{W}{T},\\
\ln \frac{R_2^2}{R_1^2} &\rightarrow \frac{\Delta F}{T},\\
\ln \frac{P(k, Y_2)}{P(k, Y_1)} &\rightarrow \frac{W - \Delta F}{T} \equiv \frac{W_{\text{Dif}}}{T},\\
\Sigma_{Y_1 \rightarrow Y_2} = \expval{\ln \frac{P(k, Y_1)}{P(k, Y_2)}}_{Y_2} &\rightarrow \Delta S.
\end{split}
\label{heuristic}
\end{equation}
where $\Delta S$, in the thermodynamic context, represents the entropy production (due to the gluon degrees of freedom) caused by the compression $R_1 \rightarrow R_2$ when the system relaxes to a state at the same initial temperature $T$ but confined within the reduced size $R_2$.

\subsection{Dynamical Entropy of a CGC State}
\label{chapter 5 - section 2: QCD Dynamic Entropy; Dynamical Entropy of a CGC State}

Considering the evolution $Y_1 \rightarrow Y_2$ and $R_1 \rightarrow R_2$, the addition of individual contributions in the calculation of the dynamical entropy density $dS/dy$ for a final CGC state involves summing over all degrees of freedom, referred to as the color multiplicity $N_c^2 - 1$ and the gluon occupancy number in the longitudinal coordinate space $\sim 1/4\pi N_c \alpha_s$. For the transverse degrees of freedom, it is necessary to account for the average number of "transverse cells" $R_T^2/R_2^2$ with an initial rapidity $Y_1$, where $R_T$ represents the size of the hadronic target, and the average number of gluon degrees of freedom within a cell is $\mu$. Thus, the following expression is obtained:
\begin{equation}
    \frac{dS}{dy}=\frac{C_F}{2\pi\alpha_s}\frac{R_T^2}{R_1^2}\mu \Sigma^{Y_1 \rightarrow Y_2}.
    \label{dynamical entropy density}
\end{equation}
In this study, correction factors due to gluon correlation effects are neglected under the ideal gas approximation.

Regarding the average number of gluon degrees of freedom, $\mu$, Gaussian models for the UGD in the CGC, such as the one presented in \cite{golec}, can be used to obtain $dS/dy$ and compare the result with equation (25) in \cite{kutak2011gluon}, where the saturation related to this UGD introduces thermodynamic entropy with $Q_s^2(Y) = 2\pi T$. Based on this relation, the average number of gluon degrees of freedom is identified as $\mu = \frac{3 \pi}{2}$, which is the value adopted in this thesis.

\chapter{Entanglement Entropy}
\label{chapter 4: Entanglement Entropy}

This chapter addresses entropy production due to the phenomenon of quantum entanglement. To understand how this phenomenon is related to entropy generation in high-energy physics, a brief introduction will be provided, emphasizing its main characteristics.

Subsequently, the concept of entanglement entropy will be defined, along with a set of strategies that can be employed to calculate it. Finally, calculations of entanglement entropy will be presented in three scenarios involving high-energy physics: (1) entanglement between measured and unmeasured spatial regions in a DIS; (2) entanglement between incident and scattered particles in elastic collisions; and (3) entanglement between valence quarks and {\it wee} partons in the CGC.

\section{Quantum Entanglement}
\label{chapter 4 - section 1: quantum entanglement}

In 1935, A. Einstein, B. Podolsky, and N. Rosen published a paper challenging the completeness of Quantum Mechanics\cite{einstein1935}. In this publication, the physicists proposed a thought experiment called the {\it EPR paradox}, which aimed to prove that the only sustainable interpretation in the quantum universe is the {\it realist} one\footnote{In this interpretation, the measurement in a quantum system reflects the physical state of the system just before its realization. If this interpretation is correct, then quantum mechanics is an incomplete theory, as it \textquoteleft fails' to determine the measurement, providing only probabilities.} in complete contrast to the {\it Copenhagen} school\footnote{Also called the orthodox interpretation. It asserts that, before measurement, the physical system was not in any defined state but rather in a {\it superposition} of states. The act of measurement forces the system to \textquoteleft choose' one of these superposed states, i.e., \textquoteleft compels' nature to decide the state.}.

David Bohm proposed a simplification of the EPR paradox, focusing on the spin measurement of a particle. In this formulation, the decay of the pion meson into an electron and a positron is considered,
$$
\pi^0 \rightarrow e^- + e^+
$$
If the pion is at rest, the electron and positron move in opposite directions due to the conservation of linear {\it momentum}. Additionally, the pion has spin 0, so the conservation of angular {\it momentum} requires that the positron and electron are in a singlet configuration.

Thus, if the electron has an upward spin, the positron has a downward spin, and vice versa. The theoretical framework of quantum mechanics is unable to predict which combination will be obtained in a measurement; it only predicts that, on average, there will be half of each case. In this experiment, the distance the particles travel is arbitrary, meaning that upon measuring the spin of the electron, for instance, as upward, the positron's spin is determined as downward through angular {\it momentum} conservation, without any measurement process being inferred on the second particle, regardless of whether it is meters or light-years away.

Now, the realist school's argument with this experiment is that the electron indeed had an upward spin (and the positron a downward spin) at the moment they were created. The orthodox interpretation of this problem comes at a cost: for angular {\it momentum} conservation to hold, the wave function collapse caused by the experimenter measuring the electron's spin would have to travel faster than the speed of light, which {\it Bohm} later called {\it non-locality}. Since quantum theory does not predict the result of a measurement with certainty and locality was not verified, the authors of the EPR experiment claimed that quantum mechanics, in its current form, is incomplete.

Classical mechanics also assumes certain well-behaved statistical aspects. For instance, when flipping a fair coin, there is an equal probability of heads or tails; however, knowing the force applied to the coin, the local gravitational acceleration, air viscosity, the height at which the coin is flipped, and the air temperature, among other variables, it is possible to determine with precision which side the coin will show upon hitting the ground. The quantum analogue of these additional variables is referred to as {\it hidden variables}.

The decay analyzed in this experiment is the most traditional example of a purely quantum phenomenon, {\it entanglement}\cite{schrodinger1935}, which occurs between the electron and the positron. It consists of the description of a quantum system composed of two or more particles, where the defined characterization of one of its entities reveals that of the others in an inseparable manner, regardless of the distance separating them, precisely as in the EPR paradox. Today, this phenomenon is central to research in fields such as cryptography based on {\it Bell}'s inequalities\cite{bell1964,schrodinger1935,ekert1992}, teleportation\cite{bennett1993}, and other applications.

\subsection{Characterizing Entanglement from the State Function}
\label{chapter 4 - section 1 - subsection 1: Characterizing Entanglement from the State Function}

Quantum entanglement is an intrinsic characteristic of composite systems. Therefore, for entanglement to exist, the total system must be divisible into at least two subsystems. This means that the total Hilbert space $\mathscr{H}$ can also be subdivided. In the simplest case, where the division occurs into two subspaces, the total composite state $\ket{\psi} \in \mathscr{H}$ can be represented in terms of the eigenvectors of the subspaces $A$ and $B$, with $\ket{a_i} \in \mathscr{H}_A$ and $\ket{b_i} \in \mathscr{H}_B$, such that the total space is $\mathscr{H} = \mathscr{H}_A \otimes \mathscr{H}_B$. Thus, the physical state $\ket{\psi}$ of a composite system can be written as:
\begin{equation}
    \ket{\psi} = \sum_{i,k} c_{ik} \ket{a_i} \otimes \ket{b_k},
    \label{emaranhado}
\end{equation}

A physical state $\ket{\psi}$ in $\mathscr{H} = \mathscr{H}_A \otimes \mathscr{H}_B$ is called \textit{separable} when it can be written as the direct product of the states of the subsystems \(A\) and \(B\). In other words, $\ket{\psi}$ is separable if there exist states $\ket{\phi_A} \in \mathscr{H}_A$ and $\ket{\phi_B} \in \mathscr{H}_B$ such that:
\begin{equation}
    \ket{\psi} = \ket{\phi_A} \otimes \ket{\phi_B} = \sum_i \alpha_i \ket{a_i} \otimes \sum_k \beta_k \ket{b_k}.
    \label{estado_separavel}
\end{equation}
Here, the coefficients $c_{ik}$ can be factored as $c_{ik} = \alpha_i \beta_k$, where $\alpha_i$ and $\beta_k$ are the amplitudes associated with the states $\ket{a_i}$ and $\ket{b_k}$, respectively. For example, consider the following physical state:
\begin{equation}
    \ket{\psi} = \left(\frac{1}{\sqrt{2}} \ket{a_1} + \frac{1}{\sqrt{2}} \ket{a_2}\right) \otimes \ket{b_1}.
\end{equation}
In this case, the state of $A$ is a superposition of $\ket{a_1}$ and $\ket{a_2}$, while the state of $B$ is fixed at $\ket{b_1}$. The coefficients $c_{ik}$ are:
\begin{equation}
c_{ik} = 
\begin{cases}
    \frac{1}{\sqrt{2}}, & \text{if } i = 1 \text{ or } i = 2 \text{ and } k = 1, \\
    0, & \text{otherwise}.
\end{cases}
\end{equation}
Although the state of \(A\) is a superposition, the total state is still separable, as there is no dependence between subsystems \(A\) and \(B\). Thus, composition is a necessary condition for entanglement, but it is not sufficient.

For entanglement to exist, a necessary and sufficient condition is that the state be \textit{non-separable}, also called \textit{entangled}. In this case, it is any state for which the representation of Eq.~[\ref{estado_separavel}] does not hold, i.e.:
\begin{equation}
    \ket{\psi} \neq \sum_i \alpha_i \ket{a_i} \otimes \sum_k \beta_k \ket{b_k}.
    \label{estado_emaranhado}
\end{equation}
Considering the pion decay discussed at the beginning of this section, the only possible configuration of the system's physical state, $\ket{\psi}$, before any measurement is the singlet. Labeling with the subscript \(A\) the quantities inherent to the electron and \(B\) to the positron, Eq.~[\ref{emaranhado}] takes the form:
\begin{equation}
    \ket{\psi} = c_{12} \ket{a_1} \otimes \ket{b_2} + c_{21} \ket{a_2} \otimes \ket{b_1}.
    \label{intermediario_singleto}
\end{equation}
The wave function must be normalized, $|c_{12}|^2 + |c_{21}|^2 = 1$. Additionally, by imposing $|c_{12}|^2 = |c_{21}|^2$, we have $c_{12} = c_{21} = \pm 1/\sqrt{2}$, yielding:
\begin{equation}
    \ket{\psi} = \frac{1}{\sqrt{2}} \left( \ket{a_1} \otimes \ket{b_2} - \ket{a_2} \otimes \ket{b_1} \right).
    \label{intermediario_singleto_2}
\end{equation}
This equation states that after the pion decay, upon measuring the electron's spin direction, the positron's spin direction is instantly determined, i.e., measuring the state $\ket{a_1} \in \mathscr{H}_A$ reveals the configuration of the state $\ket{b_2} \in \mathscr{H}_B$, without it undergoing the measurement process.

Finally, it is important to highlight that quantum entanglement is inherently dependent on the choice of partition of the physical system, i.e., on the representation of the subspaces. For instance, consider the pure state of the hydrogen atom, composed of an electron and a proton. When the wave function of this state is described in terms of the electron's and proton's coordinates, it cannot be separated as the product of two wave functions, each depending solely on the coordinates of one of the particles. Thus, in this partition, the electron and proton are entangled. 

\subsection{The Reduced Density Matrix}
\label{chapter 4 - section 1 - subsection 2: The Reduced Density Matrix}

The density matrix of an entangled state $\hat{\rho}$ is given by:
\begin{equation}
    \hat{\rho} = \sum_{i,j,k,l} c_{ik} c_{jl}^* \ket{a_i}\bra{a_j} \otimes \ket{b_k}\bra{b_l},
\end{equation}
The method used to extract the system's information contained in the subspace $\mathscr{H}_A$ from the density matrix $\hat{\rho}$ is the \textit{partial trace}, performed over the basis of $\mathscr{H}_B$, resulting in the \textit{reduced density matrix} $\hat{\rho}_A$:
\begin{equation}
\hat{\rho}_A = \Tr_B [\hat{\rho}] = \sum_r \bra{b_r} \hat{\rho} \ket{b_r}.
\end{equation}
Roughly speaking, this operation is analogous to integrating out a specific variable, such as $x$, in a two-variable function $\rho = \rho(x,y)$:
\begin{equation}
\rho'(y) = \int dx \, \rho(x,y).
\end{equation}
Thus, the reduced density matrix, $\hat{\rho}_A$ or $\hat{\rho}_B$, can be determined from $\hat{\rho}_{A,B} = \Tr_{B,A}[\hat{\rho}]$.

In general, the density matrix of an entangled state is given by the expression:
\begin{equation}
\begin{split}
\hat{\rho}_A = \tr_B \hat{\rho} &= \sum_m \bra{b_m} \sum_{i,j,k,l} c_{ij} c_{kl}^* \ket{a_i} \bra{a_k} \otimes \ket{b_j} \underbrace{\bra{b_l} \ket{b_m}}_{\delta_{lm}}, \\
&= \sum_{i,j,k,l} c_{ij} c_{kl}^* \ket{a_i} \bra{a_k} \otimes \sum_m \bra{b_m} \ket{b_j} \delta_{lm}, \\
&= \sum_{i,j,k,l} c_{ij} c_{kl}^* \ket{a_i} \bra{a_k} \delta_{lj}.
\end{split}
\label{delta_aplication}
\end{equation}
Therefore:
\begin{equation}
\hat{\rho}_A = \sum_{i,j,k} c_{ij} c_{kj}^* \ket{a_i} \bra{a_k}.
\label{2.1.7}
\end{equation}
From this expression, it is evident that $\ket{a_i} \bra{a_k} \in \mathscr{H}_A$; however, if the state is entangled, the constants $c_{ij} c_{kj}^*$ carry information about the $\mathscr{H}_B$ state contained within the subspace $A$.

For example, the density matrix of the singlet state is given by:
\begin{equation}
\begin{aligned}
\hat{\rho} = \frac{1}{2} \Big(
    & \ket{a_1} \otimes \ket{b_2} \bra{a_1} \otimes \bra{b_2} 
    - \ket{a_1} \otimes \ket{b_2} \bra{a_2} \otimes \bra{b_1} \\
    & - \ket{a_2} \otimes \ket{b_1} \bra{a_1} \otimes \bra{b_2} 
    + \ket{a_2} \otimes \ket{b_1} \bra{a_2} \otimes \bra{b_1}
\Big).
\end{aligned}
\end{equation}
Considering the following representation:
\begin{equation}
\ket{a_1} \equiv \begin{bmatrix} 1 \\ 0 \end{bmatrix}, \quad \ket{a_2} \equiv \begin{bmatrix} 0 \\ 1 \end{bmatrix},\\
\ket{b_1} \equiv \begin{bmatrix} 1 \\ 0 \end{bmatrix}, \quad \ket{b_2} \equiv \begin{bmatrix} 0 \\ 1 \end{bmatrix}.\\
\label{representation}
\end{equation}
The density matrix of the singlet state is given by:
\begin{equation}
\hat \rho = \frac{1}{2}
\begin{bmatrix}
0 & 0 & 0 & 0 \\
0 & 1 & -1 & 0 \\
0 & -1 & 1 & 0 \\
0 & 0 & 0 & 0
\end{bmatrix}.
\end{equation}
Thus, the reduced density matrix for subspace \( A \) is:
\begin{equation}
\hat{\rho}_A=\Tr_B[\hat \rho]= \frac{1}{2}\mathbbm{1}_2.
\label{singlete reduced density matrix}
\end{equation}
where \( \mathbbm{1}_2 \) is the two-dimensional identity matrix.

Lastly, it is possible to apply the {\it Gram-Schmidt} procedure (Appendix \ref{Apendice B: Gram-Schmidt Procedure}) to the entangled state given by Eq.~[\ref{emaranhado}]. This technique allows for a simpler mathematical form at the cost of performing a basis transformation. With this, the entangled state is given by:
\begin{equation}
\ket{\psi} = \sum_i \alpha_i \ket{\phi_i^A} \otimes \ket{\phi_i^B},
\label{gram_schmidt_entangled}
\end{equation}
where $\ket{\phi_i^A} \in \mathscr{H}_A$ and $\ket{\phi_i^B} \in \mathscr{H}_B$. Thus, the total density matrix can be rewritten as:
\begin{equation}
\hat \rho = \sum_{n,m} \alpha_n \alpha_m^* \ket{\phi_n^A} \bra{\phi_m^A} \otimes \ket{\phi_n^B} \bra{\phi_m^B}.
\end{equation}
The reduced density matrix $\hat \rho_A$ is given by:
\begin{equation}
\hat \rho_A = \Tr_B \hat \rho = \sum_s \bra{\phi_s^B} \sum_{n,m} \alpha_n \alpha_m^* \ket{\phi_n^A} \bra{\phi_m^A} \otimes \ket{\phi_n^B} \underbrace{\bra{\phi_m^B} \ket{\phi_s^B}}_{\delta_{ms}}.
\end{equation}
Similarly to the operations performed in Eq.~[\ref{delta_aplication}], the reduced density matrix can be expressed as:
\begin{equation}
\hat \rho_A = \sum_n \alpha_n^2 \ket{\phi_n^A} \bra{\phi_n^A}.
\label{optimal_entanglement_entropy}
\end{equation}
Furthermore, from Eq.~[\ref{gram_schmidt_entangled}], the reduced density matrix $\hat \rho_B$ is given by:
\begin{equation}
\hat \rho_B = \sum_n \alpha_n^2 \ket{\phi_n^B} \bra{\phi_n^B}.
\label{B_reduced_density_matrix}
\end{equation}

\section{Entanglement Entropy}
\label{chapter 4 - section 1: the entanglement entropy}

The Entanglement Entropy is given by:
\begin{equation}
S(\hat \rho_A) = -\Tr\left[\hat \rho_A \ln \hat \rho_A\right].
\label{entanglement entropy}
\end{equation}
In other words, it is the mathematical expression of von Neumann entropy (Eq.~[\ref{svn}]), where the reduced density matrix $\hat \rho_A$ is used instead of the total density matrix $\hat \rho$. Thus, the entanglement entropy can also be calculated using Shannon entropy:
\begin{equation}
S(p_n) = -\sum _n p_n \ln p_n.
\label{shannon entanglment entropy}
\end{equation}
In this equation, $p_n = \alpha_n^2$, obtained from Eq.~[\ref{optimal_entanglement_entropy}].

Entanglement entropy is a measure of the degrees of entanglement in the system. Therefore, the greater the entropy, the more entangled the system is. Thus, the entanglement entropy with respect to substate $A$, $S(\hat \rho_A)$, is identical to the entanglement entropy with respect to substate $B$, $S(\hat \rho_B)$:
\begin{equation}
S(\hat \rho_A) = -\Tr[\hat \rho_A \ln \hat \rho_A] = -\sum_n |\alpha_n|^2 \ln |\alpha_n|^2 = -\Tr[\hat \rho_B \ln \hat \rho_B] = S(\hat \rho_B).
\end{equation}
This occurs because, as this entropy measures the degree of entanglement between the two subsystems, if evaluated in $A$, it must yield the same result as in $B$, since it does not measure internal characteristics of the subsets, but rather how they are related.

\begin{figure}[ht]
\centering
\includegraphics[width=0.5\linewidth]{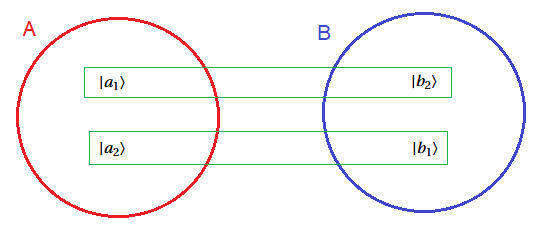}
\caption{Maximum entanglement in the singlet state. By making a measurement in $A$, the observer knows all the information about $B$.}
\label{singlete figure}
\end{figure}

For example, the singlet state, with a reduced density matrix described in expression [\ref{singlete reduced density matrix}], has an entanglement entropy such that:
\begin{equation}
S(\mathbbm{1}_2/2) = -\frac{1}{2} \Tr\left[\mathbbm{1}_2 \ln\frac{\mathbbm{1}_2}{2}\right].
\label{singlete entanglement entropy}
\end{equation}
To calculate the entanglement entropy, problems generally boil down to calculating the logarithm of the operator. In this case, it has a simple resolution:
\begin{equation}
\ln \hat \rho_A = -\mathbbm{1}_2 \sum_{n=1}^\infty \frac{1}{n 2^n} = -\mathbbm{1}_2 \ln 2 .
\end{equation}
That is, the entanglement entropy of the singlet state is given by:
\begin{equation}
S(\mathbbm{1}_2/2) = \frac{\ln 2}{2} \Tr[\mathbbm{1}_2] = 1 \, \text{bit}.
\end{equation}
In this case, the entropy is maximal, and the system is said to be maximally entangled. This happens because, upon determining the state of the system in $A$, all information about $B$ is automatically obtained (Fig.~[\ref{singlete figure}]).

This relationship is not always maximal. To verify this, let us now consider the entangled state $\ket{W}$:
\begin{equation}
\ket{W} = \frac{1}{\sqrt{3}} \left( 
    \ket{1}_A \ket{0}_B \ket{0}_C + \ket{0}_A \ket{1}_B \ket{0}_C + \ket{0}_A \ket{0}_B \ket{1}_C.
\right)
\label{W state}
\end{equation}
This state is divided into three subsets, such that ${\ket{0}_A, \ket{1}_A} \in \mathscr{H}_A$, ${\ket{0}_B, \ket{1}_B} \in \mathscr{H}_B$, and ${\ket{0}_C, \ket{1}_C} \in \mathscr{H}_C$. Physically, it can represent the case where a quantum is distributed among a set of three identical and entangled particles, where $\ket{1}$ represents the particle in the first excited level and $\ket{0}$ represents the particle in the ground state.

\begin{figure}[ht]
\centering
\includegraphics[width=0.6\linewidth]{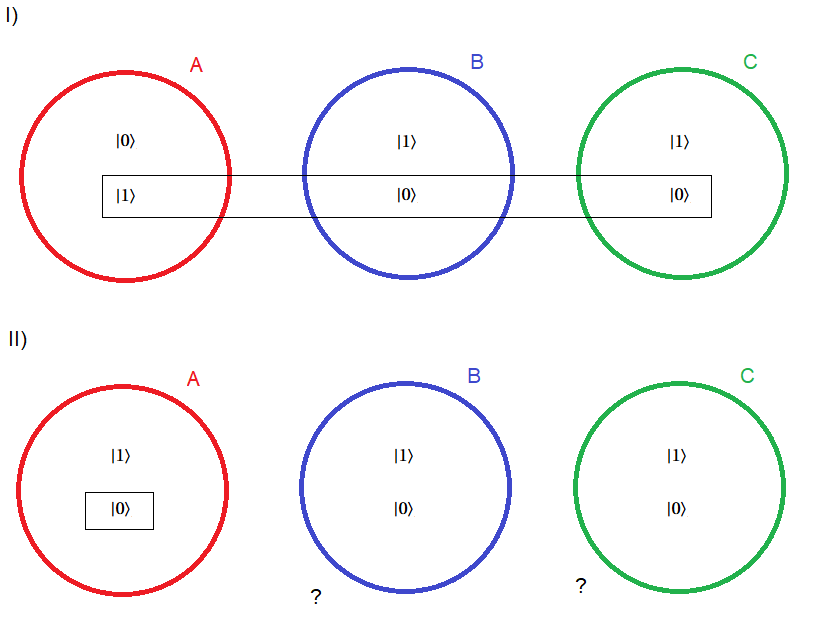}
\caption{Entanglement between the states of subsets $A$, $B$, and $C$: {\bf (I.)} If the state in $A$ is $\ket{1}$, it is automatically known that the state in $B$ and $C$ is $\ket{0}$. {\bf (II.)} Upon measuring $\ket{0}$ in $A$, it is not possible to determine the configuration of $B$ and $C$.}
\label{w state figure}
\end{figure}

The reduced density matrix is given by:
\begin{equation}
\hat \rho_A = \Tr_{BC}\hat \rho = \frac{1}{3} \ket{1}_A \bra{1}_A + \frac{2}{3} \ket{0}_A \bra{0}_A.
\label{W reduced density matrix}
\end{equation}
Calculating the entanglement entropy, we see that it is not maximal:
\begin{equation}
S(\rho_A) = -\left( \frac{1}{3} \ln \frac{1}{3} + \frac{2}{3} \ln \frac{2}{3} \right) \approx 0.92 \, \text{bits}.
\end{equation}
Fig.~[\ref{w state figure}] aids in interpreting this result. For this, two hypotheses are considered: (I.) The observer, having access only to subspace $A$, measures the energy and finds the eigenvalue corresponding to state $\ket{1}$. In this case, upon measuring in $A$, they know that subspaces $B$ and $C$ must be in state $\ket{0}$; in hypothesis (II.), the observer measures state $\ket{0}$ for particle $A$. In this case, they cannot determine the states of particles in subspaces $B$ and $C$, such that they remain in a superposition state. Note that in the case of the singlet, this ambiguity does not occur.

There are some useful strategies for calculating entanglement entropy. The first is to obtain the probability distribution $p_n$ of the entangled set and then use Shannon entropy, given by expression [\ref{shannon entanglment entropy}], linked to entanglement via $p_n = |\alpha_n|^2$. The second strategy is to evaluate the definition of the logarithmic function, such that:
\begin{equation}
\operatorname{ln} \hat \rho_A = \lim_{\epsilon \rightarrow 0} \frac{1}{\epsilon} \left(\hat \rho_A^{\epsilon} - \mathbbm{1}_d \right).
\label{logaritm identity}
\end{equation}
In the equation above, $\mathbbm{1}_d$ is the identity operator in a space of dimension $d$, equivalent to the dimension of subspace $A$. Substituting this expression into the definition of entanglement entropy:
\begin{equation}
S(\hat \rho_A) = \lim_{\epsilon \rightarrow 0} \frac{1}{\epsilon} \left( 1 - \Tr[\hat \rho_A^{1+\epsilon}] \right).
\label{logaritm property entanglement entropy}
\end{equation}

The Rényi entanglement entropy is given by the expression:
\begin{equation}
S(\hat \rho_A) = \frac{1}{1-\alpha} \ln \Tr[\hat \rho_A^\alpha].
\label{renyi entanglement entropy}
\end{equation}
In some cases, it is easier to obtain an expression for the Rényi entropy. Thus, it is possible to use the fact that this entropy reduces to the von Neumann expression when $\alpha \to 1$. Therefore:
\begin{equation}
\lim_{\alpha \to 1} \frac{1}{1-\alpha} \ln \Tr[\hat \rho_A^\alpha] = -\Tr[\hat \rho_A \ln \hat \rho_A] = S(\hat \rho_A).
\label{renyi entanglement entropy limit}
\end{equation}

In summary, three strategies for calculating entanglement entropy will be presented in this thesis:
\begin{itemize}
    \item[$\blacksquare$] {\bf Strategy I}: Obtain an expression for $p_n$ and then calculate the Shannon entanglement entropy given by Eq.~[\ref{shannon entanglment entropy}], subject to the condition $p_n = |\alpha_n|^2$, obtained in Eq.~[\ref{optimal_entanglement_entropy}];
    \item[$\blacksquare$] {\bf Strategy II}: Obtain an expression for $\Tr[\hat \rho_A^{1+\epsilon}]$, substitute it into Eq.~[\ref{logaritm property entanglement entropy}], and then calculate the limit as $\epsilon \rightarrow 0$;
    \item[$\blacksquare$] {\bf Strategy III}: Find the Rényi entanglement entropy given by expression [\ref{renyi entanglement entropy}] and then calculate the limit as $\alpha \to 1$ to obtain the entanglement entropy.
\end{itemize}
Subsequently, these strategies will be applied in three distinct models of entanglement entropy in high-energy physics. Strategy I will be used to calculate entanglement entropy in a DIS using the {\it Kharzeev-Levin Model} (KL)\cite{karkar}. Then, starting from an expression for the Rényi entropy, the entanglement entropy in elastic collisions will be obtained, confirming the program of Strategy II. Finally, Strategy III will be used to calculate the entanglement entropy in the CGC.

\section{Entanglement Entropy in the Kharzeev-Levin Model}

The KL Model\cite{karkar} considers quantum entanglement in a DIS. To this end, two spatial regions are considered: $A$, the region probed by the DIS with a Hilbert space $\mathscr{H}_A$, and $B$, the unprobed region. In this model, the physical states in $A$ are entangled with those in $B$. From this, it is possible to characterize the entangled wave function using Eq.~[\ref{gram_schmidt_entangled}].

\begin{figure}[ht]
\centering
\includegraphics[width=0.45\linewidth]{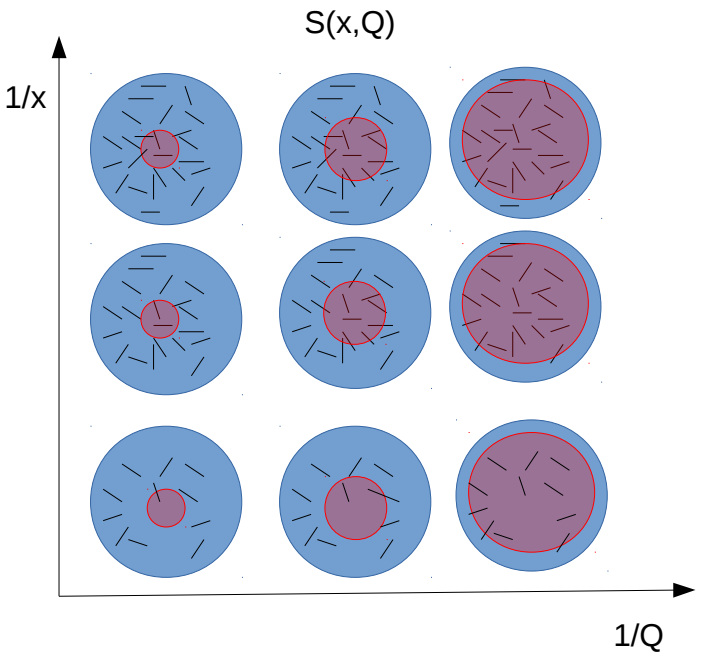}
\caption{Regions probed in a DIS and their dependence on the $1/x$ and $1/Q$ scales. Adapted from \cite{hentschinski2022maximally}.}
\label{henta}
\end{figure}

In the dipole model, the appropriate system evolution equation is the BK equation (Appendix \ref{apendice C: bk equation}), so that, applying the Strategy I program, it is possible to reconcile an expression for $p_n$ with the BK equation. Furthermore, in a DIS, the probed region $A$ has its size determined by a transverse area of the order of $\sim 1/Q^2$. According to reference \cite{hentschinski2022maximally}, the entanglement between regions $A$ and $B$ will be established by the presence of color dipoles at the boundary between the two regions (Fig.~[\ref{henta}]).

Thus, in the following subsections, the methods for obtaining $p_n$ to calculate the entanglement entropy in the KL Model will be presented for two cases, (1+1)-dimensional and (3+1)-dimensional, which describe the dipole problem initially with only one spatial dimension and with three dimensions, respectively.

\subsection{Entanglement Entropy: (1+1)-dimensional Model}
\label{chapter 4 - section 1 - subsection 1: entanglement entropy: (1+1)-dimensional model}

The evolution of partonic distributions can be modeled in the dipole representation\cite{karkar}, where partons are represented by a set of color dipoles. In the $(1+1)$-dimensional model, it depends only on one spatial direction and rapidity $Y$. Thus, the information that different dipoles have distinct sizes is neglected.

From this, the following definitions are considered:
\begin{itemize}
    \item[$\blacksquare$] $p_n(Y)$: Probability of finding $n$ dipoles with rapidity $Y$;
    \item[$\blacksquare$] $\omega_0$: Constant probability of a dipole decaying into two.
\end{itemize}

The equation relating the variation of probability with rapidity\footnote{It also relates to energy, as the expression $e^{\omega_0Y} = x^{-\omega_0}$ holds.} is given by:
\begin{equation}
\frac{dp_n}{dY} = -\omega_0 n p_n + (n-1)\omega_0 p_{n-1}.
\label{2.2.1}
\end{equation}
The first term on the right-hand side of Eq.~[\ref{2.2.1}] refers to the decrease in the probability of finding $n$ dipoles due to their division into $n + 1$ dipoles, while the second term describes the increase due to the division of $n - 1$ dipoles into $n$.

In addition to the color dipole model, the generating function $G(Y,u)$ is introduced to solve the problem. This resolution technique was first used by A. Mueller in\cite{mumumu}. Thus, we have:
\begin{equation}
G(Y,u) = \sum_{n=1}^{\infty} p_n u^n.
\label{2.2.2}
\end{equation}
Unlike infinite series frequently used, generating functions can diverge, meaning that $G(Y,u)$ is not always the true function, and the variable being sought may, in reality, be indeterminate.

One of the boundary conditions for the problem is the case where $u=1$:
\begin{equation}
G(Y,1) = \sum_{n=1}^{\infty} p_n 1^n = \sum_{n=1}^{\infty} p_n = 1.
\end{equation}
Additionally, for $Y=0$, we have $P_1(0)=1$ (a single dipole) and $P_{n>1}=0$, indicating a pure state. From this, the boundary conditions are completely defined:
\begin{equation}
\begin{cases}
G(0,u) = u; \\
G(Y,1) = 1.
\label{2.2.4}
\end{cases}             
\end{equation}

It is necessary to connect the generating function formalism with the evolution of $p_n(Y)$ [\ref{2.2.1}]. For this, derivatives are considered:
\begin{equation}
\frac{\partial}{\partial Y} G(Y,u) = \sum_{n=1}^{\infty} \left[\left(\frac{dp_n}{dY}\right) u^n\right],
\label{2.2.5}
\end{equation}
and
\begin{equation}
\frac{\partial}{\partial u} G(Y,u) = \sum_{n=1}^{\infty} p_n n u^{n-1}.
\label{2.2.6}
\end{equation}
Substituting [\ref{2.2.1}] into [\ref{2.2.5}]:
\begin{equation}
\begin{split}
\frac{\partial}{\partial Y} G(Y,u) &= \omega_0 \sum_{n=1}^{\infty} \left[-n p_n(Y) + (n-1) p_{n-1}\right] u^n, \\
&= -\omega_0 \underbrace{\sum_{n=1}^{\infty} n p_n u^n}_{u \frac{\partial}{\partial u} G(Y,u)} + \omega_0 \sum_{n=1}^{\infty} (n-1) p_{n-1} u^n,
\end{split}
\label{2.2.7}
\end{equation}
For the second term of Eq.~[\ref{2.2.7}]:
\begin{equation}
\sum_{n=1}^{\infty} (n-1) p_{n-1} u^n \underbrace{=}_{n-1=m} \sum_{m=0}^{\infty} m p_m u^{m+1} = 0 + \sum_{m=1}^{\infty} m p_m u^{m+1} = u^2 \frac{\partial}{\partial u} G(Y,u).
\label{2.2.8}
\end{equation}

Observing [\ref{2.2.7}] and [\ref{2.2.8}], it can be observed that the partonic evolution in the color dipole model with the generating function is modeled by the partial differential equation:
\begin{equation}
\frac{\partial}{\partial Y} G(Y,u) = \omega_0 u (u-1) \frac{\partial}{\partial u} G(Y,u).
\label{2.2.9}
\end{equation}
The general solution to this equation, under the conditions [\ref{2.2.4}], is given by:
\begin{equation}
G(Y,u) = \frac{u e^{-\omega_0 Y}}{1+u(e^{-\omega_0 Y}-1)} = u e^{-\omega_0 Y} \underbrace{\frac{1}{1+u(e^{-\omega_0 Y}-1)}}.
\label{2.2.10}
\end{equation}
Now, the term highlighted in [\ref{2.2.10}], considering $|u(1-e^{-\omega_0 Y})| < 1$, yields:
\begin{equation}
\frac{1}{1+u(e^{-\omega_0 Y}-1)} = \sum_{j=0}^{\infty} u^j [1-e^{-\omega_0 Y}]^j \underbrace{=}_{j+1=n} \sum_{n=1}^{\infty} u^{n-1} (1-e^{-\omega_0 Y})^{n-1}.
\end{equation}
Thus, the general solution can be expressed as:
\begin{equation}
G(Y,u) = e^{-\omega_0 Y} \sum_{n=1}^{\infty} (1-e^{-\omega_0 Y})^{n-1} u^n.
\label{2.2.12}
\end{equation}

Comparing equation [\ref{2.2.12}] with the definition of the generating function in [\ref{2.2.2}], the probability $p_n(Y)$ is determined:
\begin{equation}
p_n(Y) = e^{-\omega_0 Y} (1-e^{-\omega_0 Y})^{n-1}.
\label{2.2.13}
\end{equation}

Substituting $p_n(Y)$ obtained in [\ref{2.2.13}] into the entanglement entropy expression [\ref{entanglement entropy}]:
\begin{equation}
\begin{split}
S(Y) &= -\sum_n \left\{ e^{-\omega_0 Y} (1-e^{-\omega_0 Y})^{n-1} \ln\left[e^{-\omega_0 Y} (1-e^{-\omega_0 Y})^{n-1}\right] \right\} \\
&= -e^{-\omega_0 Y} \sum_n \left\{ (1-e^{-\omega_0 Y})^{n-1} \ln\left[-\omega_0 Y + (n-1)\ln(1-e^{-\omega_0 Y})\right] \right\} \\
&= \omega_0 Y \sum_n \underbrace{e^{-\omega_0 Y} (1-e^{-\omega_0 Y})^{n-1}}_{p_n} - \sum_n e^{-\omega_0 Y} (1-e^{-\omega_0 Y})^{n-1} (n-1)\ln(1-e^{-\omega_0 Y}) \\
&= \omega_0 Y \underbrace{\sum_n p_n}_1 + \ln(1-e^{-\omega_0 Y})^{n-1} \left[\sum_n \underbrace{e^{-\omega_0 Y} (1-e^{-\omega_0 Y})^{n-1}}_{p_n} - \sum_n n \underbrace{e^{-\omega_0 Y} (1-e^{-\omega_0 Y})^{n-1}}_{p_n}\right].
\label{2.2.14}
\end{split}
\end{equation}
Thus:
\begin{equation}
S(Y) = \omega_0 Y + \ln(1-e^{-\omega_0 Y})(1-\sum_n n p_n).
\label{formageral}
\end{equation}
In the limit of interest, which is high energies, $Y \rightarrow \infty$, we have:
\begin{equation}
\lim_{Y \rightarrow \infty} \ln(1-e^{-\omega_0 Y}) = 0.
\end{equation}
Therefore, the entropy takes the simple form:
\begin{equation}
S(Y) \approx \omega_0 Y.
\label{uni}
\end{equation}

It is possible to relate the obtained entropy with the gluon density. To do this, the average number of partons is defined as\footnote{Currently, the state-of-the-art proposes that the number of particles $\expval{x}=xf_g(x)+xf_{\text{sea}}(x)$, where $xf_{\text{sea}}(x)$ is the sea quark PDF. However, this contribution will not be considered in this work. For a discussion, consider references \cite{hentschinski2022maximally} and \cite{kharzeev2021deep}.}:
\begin{equation}
\expval{n}\equiv xf_g(x),
\label{1+1 ocupation}
\end{equation}
In this expression, $xf_g(x)$ is the gluon distribution for a given value of $x$, enabling the evaluation of entropy per unit rapidity. Thus,
\begin{equation}
xf_g(x)=\expval{n}=\sum_{n=0}^{\infty}np_n.
\label{2.2.17}
\end{equation}
The distribution appears in the far-right term of [\ref{formageral}]. For it,
\begin{equation}
\begin{split}
\sum_{n=0}^{\infty}np_n &= e^{-\omega_0Y}\sum_{n=1}^{\infty}n(1-e^{-\omega_0Y})^{n-1}, \\
&= \frac{1}{\omega_0}\frac{d}{dY}\sum_{n=1}^{\infty}(1-e^{-\omega_0Y})^n.
\end{split}
\end{equation}
Then, for $|1-e^{-\omega_0 Y}|<1$, we have:
\begin{equation}
\sum_{n=1}^{\infty}np_n =\frac{1}{\omega_0}\frac{d}{dY}\frac{1}{1-(1+e^{-\omega_0 Y})}=e^{\omega_0 Y}.
\label{2.2.19}
\end{equation}

Now, for large values of $x$, more precisely, in the regime where $x\gg 1$, the rapidity $Y$ is related to the Bjorken scale in the form $Y=-\ln x$. Adding this information to the results of Equations [\ref{2.2.19}] and [\ref{2.2.17}], we obtain:
\begin{equation}
xf_g(x)=e^{\omega_0Y}.
\end{equation}
Thus, the von Neumann entropy in the asymptotic limit is given by:
\begin{equation}
S(x)=\ln [xf_g(x)].
\label{2.2.21}
\end{equation}
It is noted that, for the small $x$ regime, the entropy [\ref{2.2.21}] emerges in the limit where all probabilities $p_n$ become equiprobable. Observing Eq.~[\ref{2.2.13}] under these circumstances, we have the probabilistic equipartition:
\begin{equation}
p_n=e^{-\omega_0Y}=\frac{1}{\expval{n}}.
\label{arrumando}
\end{equation}

The postulates of equilibrium statistical mechanics lead to entropy maximization. Thus, Equation [\ref{arrumando}] describes a maximally entangled physical state, meaning that in such a hadronic state, it is impossible to predict how many partons will be detected, as all microstates are equally probable.

\begin{figure}[ht]
\centering
\includegraphics[width=0.65\linewidth]{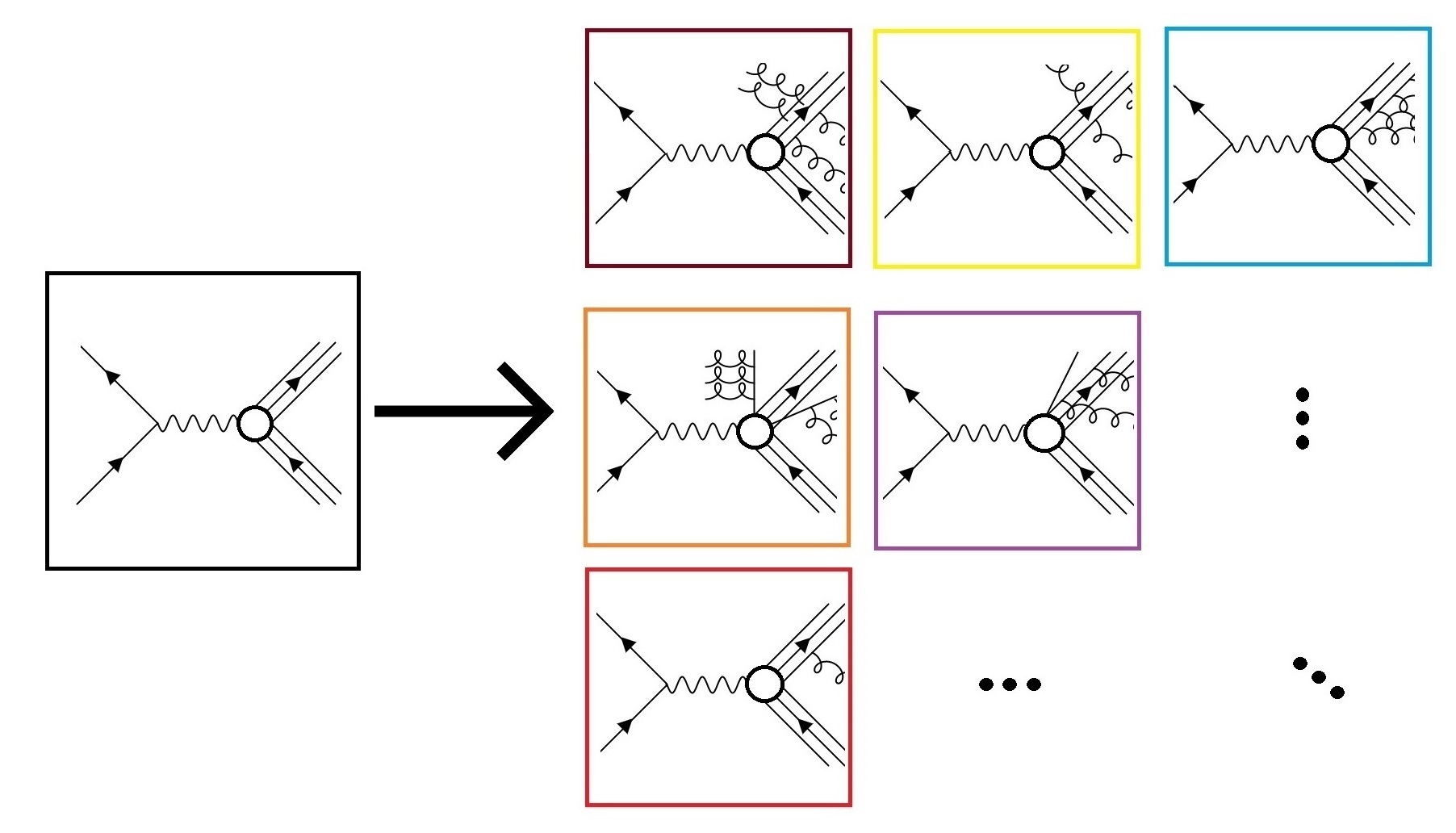}
\caption{Representation of equally probable quantum states in the von Neumann entanglement entropy for partons in the (1+1) model. Initially, at low energies, the parent hadron does not exhibit internal structure, and the scattering can be analyzed through QED. As the energy increases, reducing the wavelength of the virtual photon, the internal structure is revealed. The case becomes maximally entropic in the high-energy limit, where it is no longer possible to establish the system's final state, maximizing chaos.}
\label{Entropia de Emaranhamento equiprovavel}
\end{figure}

\subsection{Entanglement Entropy: (3+1)-dimensional Model}
\label{chapter 4 - section 1 - subsection 2: entanglement entropy: (3+1)-dimensional model}

The (3+1)-dimensional case introduces significant complications: now the transverse degrees of freedom, the size of the dipoles, and the impact parameter vector $\vec{b}$ are considered. Initially, the variables are defined as:
\begin{equation}
\begin{split}
\{\vec{a_i}\} &= \vec{a}_1, \vec{a}_2, ..., \vec{a}_i; \\
\vec{b}_{\text{in}} &= \vec{b}_i + \frac{1}{2}\vec{r}_i = \vec{b}_n - \frac{1}{2}\vec{r}_i.
\end{split}
\label{abreviations}
\end{equation}

The partonic cascade equation for the (3+1) case is given by:
\begin{equation}
\begin{split}
\frac{\partial}{\partial Y}P_n(Y, \{\vec{b}_n, \vec{r}_n\}) = &-\sum_{i=1}^n \omega_G(r_i) P_n(Y, \{\vec{b}_n, \vec{r}_n\}) \\
&+ \frac{\bar{\alpha}_s}{2\pi} \sum_{i=1}^{n-1} \frac{(\vec{r}_i+\vec{r}_n)^2}{r_i^2 r_n^2} P_{n-1}(Y, \vec{r}, \vec{b}; \vec{r}_1, \vec{b}_1, ..., (\vec{r}_i+\vec{r}_n), \vec{b}_{\text{in}}, ..., \vec{r}_{n-1}, \vec{b}_n).
\end{split}
\label{31cascate}
\end{equation}
In this equation, $\bar{\alpha}_s = N_c \alpha_s / \pi$. In the (3+1) case, $P_n(Y, \{\vec{b}_n, \vec{r}_n\})$ is the probability of having $n$ dipoles of size $r_i$, with impact parameter $b_i$, at rapidity $Y$.

The initial conditions for this problem in a DIS scattering are given by:
\begin{equation}
\begin{cases}
P_1(Y=0, \vec{r}, \vec{b}; \vec{r}_1, \vec{b}_1) = \delta^{(2)}(\vec{r}-\vec{r}_1)\delta^{(2)}(\vec{b}-\vec{b}_1); \\
P_{n>1}(Y=0; \{r_i\}) = 0,
\end{cases}
\label{initialconditions31}
\end{equation}
that is, at the initial moments when $Y=0$, there is only one dipole. The probabilities follow the usual normalization rule:
\begin{equation}
\sum_{n=1}^{\infty} \int \prod_{i=1}^n d^2r_i d^2b_i P_n(Y; \{\vec{r}_i, \vec{b}_i\}) = 1.
\label{normalization31}
\end{equation}

The dipole generating function in the (3+1)-dimensional model is given by
\begin{equation}
    G(Y, \vec{r}, \vec{b};[u_i])=\sum_{n=1}^\infty\int P_n(Y, \{\vec{b}_n, \vec{r}_n\})\prod_{i=1}^nu(\vec{r}_i,\vec{b}_i)d^2r_id^2b,
    \label{31geratriz}
\end{equation}
where $u(\vec{r}_i,\vec{b}_i)\equiv u_i$ is an arbitrary function. The boundary conditions [\ref{initialconditions31}] and the normalization [\ref{normalization31}] restrict the generating function as follows:
\begin{equation}
\begin{cases}
G(Y=0, \vec{r}, \vec{b};[u_i])=u(\vec{r}, \vec{b});\\
G(Y, r, [u_i=1])=1.
\label{initialconditions31Z}
\end{cases}             
\end{equation}

Multiplying both sides of Eq.~[\ref{31cascate}] by $\prod_{i=1}^n u_i$ and integrating over the variables $r_i$ and $b_i$, we obtain the expression:
\begin{equation}
\frac{\partial}{\partial Y}G(Y, \vec{r}, \vec{b};[u_i])=\int d^2r'K(\vec{r}', \vec{r}-\vec{r}'|\vec{r})\left[-u(r,b)+u\left(\vec{r}',\vec{b}+\frac{1}{2}(\vec{r}-\vec{r}')\right)\right]u\left(\vec{r}-\vec{r}', \vec{b}-\frac{1}{2}\vec{r}'\right)\frac{\delta G}{\delta u(r,b)}.
\label{31quasiBK}
\end{equation}
In this equation:
\begin{equation}
\begin{cases}
K(\vec{r}', \vec{r}-\vec{r}'|\vec{r})=\frac{\bar{\alpha}_s}{2\pi}\frac{r^2}{r'^2(\vec{r}-\vec{r}')^2}\equiv K, \\
\omega_G(\vec{r})=\int d^2r'K.
\label{definitionsquasiBK}
\end{cases}             
\end{equation}

A fim de encontrar soluções na forma $G([u(r_i, b_i,Y)])$ com as condições [\ref{initialconditions31Z}], é possível reescrever [\ref{31quasiBK}] na forma:
\begin{equation}
\frac{\partial}{\partial Y}G(Y, \vec{r}, \vec{b};[u_i])=\int d^2r'K\left[G(r',\vec{b}+\frac{1}{2}(\vec{r}-\vec{r}');[u_i])G(\vec{r}-\vec{r'},\vec{b}-\frac{1}{2}\vec{r'};[u_i])-G(Y, \vec{r}, \vec{b};[u_i])\right].
\label{quasiquasi31bk}
\end{equation}
Agora, definindo a amplitude de espalhamento:
\begin{equation}
    N(Y, r,b)=\sum_{n=1}^\infty \frac{(-1)^{n-1}}{n!}\int \prod_{i=1}^v\left[d^2 r_i\gamma(r_i,b)\frac{\delta}{\delta u_i}\right]G(Y,r,b,[u_i])|_{u_i=1}.
    \label{Namplitude31}
\end{equation}
Here, $\gamma(r_i,b)$ is the scattering amplitude for the dipole interactions at low energies. From these definitions, the {\it Balitsky-Kovchegov} equation is recovered (Appendix \ref{apendice C: bk equation}). The central objective of this development is to find the solution to [\ref{31cascate}]. For that, we define: 
\begin{equation}
    \tilde{P}_n(Y,r)\equiv \int P_n(Y, \{\vec{b}_n, \vec{r}_n\}) \prod_{i=1}^n d^2 r_i d^2b'.
    \label{alternativeP}
\end{equation}
In this definition, $\tilde{P}_n$ is the probability of finding $n$ dipoles with all possible sizes for the same values of the impact parameter. The conditions [\ref{initialconditions31}] and [\ref{normalization31}] constrain $\tilde{P}_n$ in the form of the following expressions:
\begin{equation}
\begin{cases}
\tilde{P}_1(Y=0,r,b)=1;\\
\tilde{P}_{n>1}(Y=0,r,b)=0;\\
\sum_{n=1}^\infty \tilde{P}_n(Y,r,b)=1.
\label{alternativePconditions}
\end{cases}   
\end{equation}

With these conditions, it is possible to solve Eq.~[\ref{31cascate}] recursively: for $n=1$, we have:
\begin{equation}
    \frac{\partial }{\partial Y}P_1(Y, r,b, r_1, b_1)=-\omega_G(r_1)P_1(Y, r,b, r_1, b_1).
    \label{31cascaten=1}
\end{equation}

For $\tilde{P}_1$, the equation takes the form:
\begin{equation}
    \frac{\partial}{ \partial Y}\tilde{P}_1(Y,r,b)=-\omega_G(r)\tilde{P}_1(Y,r,b),
    \label{31cascatetilden=1}
\end{equation}
with the solution:
\begin{equation}
    \tilde{P}_1(Y,r,b)=e^{-\omega_G(r)Y}.
    \label{tidelP1}
\end{equation}

For $P_2$, Eq.~[\ref{31cascate}] is given by:
\begin{equation}
    \frac{\partial}{\partial Y} P_2(Y,r,b;\{r_2, b_2\})=-[\omega_G(r_1)+\omega_G(r_2)]P_2(Y,r,b;\{r_2, b_2\})+\frac{\bar{\alpha}_s}{2\pi}\frac{(\vec{r}_1+\vec{r}_2)^2}{r_1^2r_2^2}P_1(Y,r,b;\vec{r}_1+\vec{r}_2,b').
    \label{31cascatetilden=2} 
\end{equation}

Initially, the value of $\omega_G(r)$ is estimated from Eq.~[\ref{definitionsquasiBK}] as:
\begin{equation}
    \begin{split}
        \omega_G(r)&=\frac{\bar{\alpha}_s}{2\pi}\int d^2r'\frac{r^2}{r'^2(\vec{r}-\vec{r}')^2}=\frac{\bar{\alpha}_s}{\pi}\int d^2r'\frac{r^2}{r'^2[r'^2+(\vec{r}-\vec{r}')^2]}\\
        &=\frac{\bar{\alpha}_s}{\pi}\int_{r_0}^{r} d^2r'\frac{r^2}{r'^2[r'^2+(\vec{r}-\vec{r}')^2]}+\frac{\bar{\alpha}_s}{\pi}\int_r^{\infty} d^2r'\frac{r^2}{r'^2[r'^2+(\vec{r}-\vec{r}')^2]}\\
        &=\underbrace{\bar{\alpha}_s\ln(r^2/r_0^2)}_{r'\leq r}+\underbrace{0}_{r'\geq r}=\int_{r_0^2}^{r^2}\frac{dr'^2}{r'^2},
    \end{split}
\label{omegaG}
\end{equation}
so only dipoles smaller than $r$ contribute to the value of $\omega_G(r)$.

The authors of reference \cite{gotsman2020} suggest that Eq.~[\ref{31cascatetilden=2}] has a solution in the form of:
\begin{equation}
    \int d^2bP_2(Y,r,b;r_1, b', r_2,b')=\frac{1}{r_1^2r_2^2}\Theta(r-r_1)\Theta(r-r_2)p_2(r,b),
    \label{sugestion}
\end{equation}
where $\Theta(z)$ is the Heaviside function:
\begin{equation}
    \begin{cases}
    \Theta(z)=1, \quad \hbox{for} \quad z>0, \\
    \Theta(z)=0, \quad \hbox{for} \quad z<0.
    \end{cases}
    \label{stepfunction}
\end{equation}

To obtain the solution of Eq.~[\ref{sugestion}], it is possible to derive the equation for $\tilde{P}_2$ by integrating both sides of [\ref{31cascatetilden=2}] over $b'$, $r_1$, and $r_2$:
\begin{equation}
    \frac{\partial}{\partial Y}\tilde{P}_2(Y,r)=2\omega_G(r)\tilde{P}_2(Y,r)+\omega_G(r)\tilde{P}_1(Y,r).
    \label{31cascatetilden=2 II}
\end{equation}
Using the solution [\ref{tidelP1}]:
\begin{equation}
    \tilde{P}_2(Y,r)=e^{-\omega_G(r)Y}(1-e^{-\omega_G(r)Y}).
\label{tildeP2}
\end{equation}
The condition $\tilde{P}_2(Y=0,r)=0$ is satisfied, corresponding to Eq.~[\ref{alternativePconditions}]. For $\omega_G(r)Y\ll 1$, there should be only two terms in the partonic cascade: $\tilde{P}_1$ and $\tilde{P}_2$, so that Eq.~[\ref{normalization31}] reduces to:
\begin{equation}
\tilde{P}_1+\tilde{P}_2\rightarrow 1 - \omega_G(r)Y+\omega_G(r)Y=1.
\label{redution}
\end{equation}
This equation suggests that the $P_n$ are negligible for large sizes, i.e., $r_i>r$, because they have a low probability of relevance.

Thus, the general solution of Eq.~[\ref{31cascate}] has the form:
\begin{equation}
P_n(Y, \{\vec{b}_n, \vec{r}_n\})=\prod_{i=1}^n \Theta(r-r_i)\frac{1}{r_i^2}p_n(Y,r).
\label{sugestion2}
\end{equation}
For this solution, using Eq.~[\ref{31cascate}], $\tilde{P}_n$ is given by the following differential equation:
\begin{equation}
    \frac{\partial}{\partial Y}\tilde{P}_n(Y,r)=-n\omega_G(r)\tilde{P}_n(Y,r)+(n-1)\omega_G(r)\tilde{P}_{n-1}(Y,r),
    \label{31tildePn I}
\end{equation}
with the solution:
\begin{equation}
 \tilde{P}_n(Y,r)=e^{-\omega_G(r)Y}(1-e^{-\omega_G(r)Y})^{n-1}.
 \label{31tildePn II}
\end{equation}
Analogous to Eq.~[\ref{2.2.19}]:
\begin{equation}
    \expval{n} = \sum_{n=1}^\infty n \tilde{P}_n(Y,r)=e^{\omega_G(r)Y}.
    \label{ocupation31}
\end{equation}
Thus, the entanglement entropy for high energies in this case will reproduce the results obtained in Eq.~[\ref{2.2.14}], with the change that the probability of a dipole decaying into two is no longer constant, but depends on its size $r$:
\begin{equation}
    S\approx \omega_G(r)Y = \ln \expval{n}.
\label{(3+1)-dimensional entanglemente entropy I}
\end{equation}
This result is similar to the one obtained by the authors in the previous work for the $(3+1)$-dimensional case in \cite{karkar}, leading to the expression:
\begin{equation}
    S\approx \bar{\alpha}_s\ln(r^2Q_s^2)Y.
\label{(3+1)-dimensional entanglemente entropy II}
\end{equation}

Alternatively, it is possible to extend the studies of the $(3+1)$-dimensional case with other equations compatible with the BK evolution and the generating function formalism. In reference \cite{liu2022}, for example, the authors investigate the behavior of the entanglement entropy for the 4-dimensional formulation of QCD with the aim of studying its divergences. Thus, it is stated that the evolution of the generating function, $Z(Y,u,b)$, defined from Eq.~[\ref{2.2.2}], is given by:
\begin{equation}
Z(Y,u,b)=e^{-mbY}+um\int_0^YdY_1e^{-mb(Y-Y_1)}\int_0^bdb'Z(b-b',Y_1,u)Z(b',Y_1,u),
\label{zizi31cascate}
\end{equation}
where $m$ and $b$ are parameters of the integral equation. The solution to this equation, considering the probability conservation $Z(b,Y,1)=1$, has the form of the Poisson distribution:
\begin{equation}
    Z(Y,u,b)= e^{-mbY}\sum_{n=0}^{\infty}u^n\frac{(mbY)^n}{n!},
\end{equation}
that is:
\begin{equation}
    p_n=e^{-mbY}\frac{(mbY)^n}{n!}.
    \label{zizipnsolution}
\end{equation}
Thus:
\begin{equation}
    \begin{split}
        \expval{n}&=\sum_{n=1}^\infty np_n =e^{-mbY}\sum_{n=1}^\infty n \frac{(mbY)^n}{n!},\\
        &= mbYe^{-mbY}\sum_{n=1}^\infty \frac{(mbY)^{n-1}}{(n-1)!}\underbrace{=}_{i=n+1}e^{-mbY}mbY\underbrace{\sum_{i=0}^\infty \frac{(mbY)^i}{i!}}_{e^{mbY}}=mbY\equiv N.
    \end{split}
    \label{expvaln31zizi}
\end{equation}

Inserting [\ref{zizipnsolution}] into the entanglement entropy (Eq.~[\ref{entanglement entropy}]),
\begin{equation}
S(Y)=-\sum_{n=1}^{\infty}p_n\ln\left[e^{-mbY}\frac{(mbY)^n}{n!}\right]=-\sum_n \frac{N^n}{n!}e^{-N} \ln\left(\frac{N^n}{n!}e^{-N}\right).
\label{entropyzizi31}
\end{equation}
With the help of the Stirling approximation and the integral representation of the logarithmic function,
\begin{equation}
    \ln n = \int_{0}^{\infty}\frac{ds}{s}(e^{-s}-e^{-ns}),
\end{equation}
the sum in Eq.~[\ref{entropyzizi31}] can be done analytically, with an asymptotic validity for $N\gg 1$:
\begin{equation}
    S=\frac{1}{2}\left[\ln (2\pi eN)-\frac{1}{6N}+\mathscr{O}(1/N^2)\right].
    \label{realentropy1}
\end{equation}

\section{Entanglement Entropy of an Elastic Scattering}
\label{chapter 4 - section 1 - subsection 3: entanglement entropy of elastic scattering}

In this subsection, the focus is on the entanglement entropy generated by the interaction of particles in a collision. Following this line of investigation, the underlying dynamics is governed by non-perturbative QCD or, in {\it Regge} phenomenology, by non-perturbative Pomeron physics (soft Pomeron).

In particular, hadron-hadron interactions in scatterings with strong interactions are described for both the elastic channel ($A+B\rightarrow A+B$) and the inelastic channel ($A+B\rightarrow X$), using the scattering matrix formalism $S$. To this end, we strictly follow the steps developed in Refs.~\cite{peschanski2016,peschanski2019}, where the reduced density matrix, $\hat{\rho}_A$, of the final state of two particles that underwent elastic scattering is written in terms of the partial wave expansion of two-body states.

It is used the partial wave expansion of physical observables, such as the total, elastic, and inelastic cross sections of the collision ($\sigma_T$, $\sigma_{\text{el}}$, and $\sigma_{\text{inel}}$), as well as the differential elastic cross section, $d\sigma_{\text{el}}/dt$, where $t$ is the Mandelstam variable associated with the transferred momentum. Using Strategy 2, proposed earlier, the entanglement entropy will be obtained from the Rényi entropy:
\begin{equation}
S_E = -\lim_{\alpha\rightarrow 1} \frac{\partial}{\partial \alpha} \Tr[\hat{\rho}_A^\alpha] = -\ln \Omega,
\label{elastic entanglement entropy I}
\end{equation}
\begin{equation}
\Omega = 1 - \left(\frac{\sigma_{\text{el}} - \frac{4}{f_V} \frac{d \sigma_{\text{el}}}{dt}\big|_{t=0}}{\pi f_V - \sigma_{\text{inel}}}\right).
\label{capital omega function}
\end{equation}

In the equations above, $f_V = V/k^2$ with $V = \sum_{\ell} (\ell + 1)$ being the total volume of the phase space. Such a volume is formally divergent because the complete {\it Hilbert} space encompasses all partial waves up to $\ell \rightarrow \infty$. In reference \cite{peschanski2019}, the physical origin of this divergence is identified, and its regularization is treated correctly.

Initially, the elastic scattering of two non-interacting particles, $A$ and $B$, will be considered. Before the scattering, particle $A$ has a 3-{\it momentum} $\vec{k}$, while particle $B$ has $\vec{l}$. After the interaction, $A$ and $B$ have 3-{\it momenta} $\vec{p}$ and $\vec{q}$, respectively.

In elastic collisions, the density matrix of the system is given by:
\begin{equation}
\hat \rho = \frac{1}{N} \int \frac{d^3p}{2E_A(\vec{p})} \frac{d^3q}{2E_B(\vec{q})} \frac{d^3p'}{2E_A(\vec{p}')} \frac{d^3q'}{2E_B(\vec{q}')} 
\ket{\vec{p}, \vec{q}} \bra{\vec{p}, \vec{q}} S \ket{\vec{k}, \vec{l}} \bra{\vec{k}, \vec{l}} S^\dagger \ket{\vec{p}', \vec{q}'} \bra{\vec{p}', \vec{q}'}.
\end{equation}
The reduced density matrix is constructed in terms of the $S$ matrix, projecting the initial state of the two bodies onto the final state: with $\hat Q$ being the projection operator,
\begin{equation}
    \ket{\psi_f}=\hat Q \hat S \ket{\psi_i},
    \label{scattering projection}
\end{equation}
such that $\ket{\psi_{f,i}}$ are the final and initial states, respectively. By calculating the partial trace of the total density matrix $\hat \rho$ with respect to the vectors of the Hilbert space B, $\mathscr{H}_B$\footnote{With $\mathscr{H}_{B}\otimes\mathscr{H}_{B}=\mathscr{H}$, where $\mathscr{H}$ is the Hilbert space that contains the entire system.},
\begin{equation}
    \hat \rho_A=\rho_0\int \frac{d^3p}{2E_{A,p}}\delta(p-k)\frac{|\bra{\vec{p},-\vec{p}}\hat S\ket{\vec{k},-\vec{k}}|^2}{4k(E_{A,k}+E_{B,k})}\ket{\vec{p}}\bra{\vec{p}},
    \label{elastic reduced density matrix}
\end{equation}
where,
\begin{equation}
    \rho_0^{-1}=\delta^3(0)\int d^3p\delta(p-k)\frac{|\bra{\vec{p},-\vec{p}}\hat S\ket{\vec{k},-\vec{k}}|^2}{4k(E_{A,k}+E_{B,k})},
    \label{inverse of rho 0}
\end{equation}
for which the normalization condition is satisfied,
\begin{equation}
    \Tr_A[\hat \rho_A]=\Tr_B[\hat \rho_B]=1.
    \label{elastic scattering normalization}
\end{equation}
This condition is responsible for the delta function in Eq. [\ref{inverse of rho 0}] and is a possible source of divergence in the entropy. In the above equations, $p=|\vec{p}|$ and $k=|\vec{k}|$ with, $\cos\theta=\vec{p}\cdot\vec{k}/(pk)$.

To calculate the trace, following the form of Eq. [\ref{elastic entanglement entropy I}],
\begin{equation}
    \Tr_A[\hat \rho_A]^n=\int d^3p\delta^{(3)}(0)\left(\rho_0\delta(p-k)\frac{|\bra{\vec{p},-\vec{p}}\hat S\ket{\vec{k},-\vec{k}}|^2}{4k(E_{A,k}+E_{B,k})}\ket{\vec{p}}\bra{\vec{p}}\right)^n,
    \label{trace over rho of A in n power}
\end{equation}
such that the extra delta function appears due to the trace calculation over the 3-momentum of particle $A$. Consider the definition,
\begin{equation}
    \bra{\vec{p},\vec{q}}\hat S\ket{\vec{k},\vec{l}}\equiv \delta^{(4)}(P_{p+q}-P_{k+l})\bra{\vec{p},\vec{q}}s\ket{\vec{k},\vec{l}},
    \label{expected value identidy}
\end{equation}
with the notation $P$ for the 4-vector of the center of mass, and $s=1+2\mathbbm{i}t$ for the reduced $\hat S$ matrix, and $t$ for the reduced transfer matrix.

By expanding the elements of the matrix $s$ and the scattering amplitude in partial waves,
\begin{equation}
\bra{\vec{p},-\vec{p}}s\ket{\vec{k},-\vec{k}}=\frac{E_{A,k}+E_{B,k}}{(\pi k/2)}\left[\delta(1-\cos \theta)+\frac{\mathbbm{i}\mathcal{A}}{16\pi}\right],
\label{expected value reduced s}
\end{equation}
where the scattering amplitude is given by,
\begin{equation}
    \mathcal{A}(s,t)=16\pi\sum_{\ell=0}^{\infty}(2\ell+1)\tau_{\ell}P_{\ell}(\cos\theta),
    \label{elastic scattering amplitude}
\end{equation}
the trace of Eq. [\ref{trace over rho of A in n power}] can be calculated. In Eq. [\ref{elastic scattering amplitude}], the variable $s_{\ell}=1+2\mathbbm{i}\tau_{\ell}$ refers to the $\ell$-th partial wave of the two-body $\hat S$ matrix. A total phase space volume can be defined as,
\begin{equation}
    V\equiv 2\delta(0)=\sum_{\ell=0}^{\infty}(2\ell+1),
    \label{phase-space volume}
\end{equation}
which is related to the three-dimensional delta functions in the form $V=4\pi k^2\sigma^{(3)}(0)/\sigma(0)$.

After integrating over the 3-momentum, writing Eq. [\ref{expected value reduced s}] in terms of the scattering angle $\theta$ and factoring out the constant factors, it is possible to write,
\begin{equation}
 \Tr_A[\hat \rho_A]^n=\left(\frac{V}{2}\right)^{1-n}\int_{-1}^1d\cos\theta[\mathcal{P}(\theta)]^n,
 \label{trace over rho of A in n power II}
\end{equation}
with,
\begin{equation}
\mathcal{P}(\theta)=\delta(1-\cos\theta)\left(1-\frac{2\sum_{\ell}(2\ell+1)|\tau_{\ell}|^2}{V/2-\sum_{\ell}(2\ell+1)f_{\ell}}\right)+\frac{|\sum_{\ell}(2\ell+1)\tau_{\ell}P_{\ell}(\cos\theta)|^2}{V/2-\sum_{\ell}(2\ell+1)f_{\ell}},
\label{function P}
\end{equation}
where, in this equation, $f_{\ell}$ are the partial wave components of the inelastic cross-section related to the elastic components $\tau_{\ell}$ through the unitarity relation,
\begin{equation}
    f_{\ell}=2(\Im \tau_{\ell}-|\tau_{\ell}|^2).
    \label{unitary relation II}
\end{equation}

The next step is to express the $\mathcal{P}(\theta)$ function in terms of physical observables, $\sigma_{\text{el}}$, $\sigma_{\text{inel}}$, $\sigma_{T}$, and $d\sigma_{\text{el}}/dt=|\mathcal{A}|^2/(256\pi k^4)$, which are commonly described in terms of the partial wave components $\tau_{\ell}$ and $f_{\ell}$. We obtain,
\begin{equation}
    \mathcal{P}(\theta)=\delta(1-\cos\theta)\left(1-\frac{\sigma_{\text{el}}}{\pi V/k^2-\sigma_{\text{inel}}}\right)+\frac{2k^2}{\sigma_{\text{el}}\frac{d\sigma_{\text{el}}}{dt}}\frac{\sigma_{\text{el}}}{\pi V/k^2-\sigma_{\text{inel}}},
    \label{function P for physical observables}
\end{equation}
with the Mandelstam variable $t=2k^2(\cos\theta-1)$ being the squared momentum transfer. Finally, the entanglement entropy $S$ [\ref{elastic entanglement entropy I}] is,
\begin{equation}
    S=\ln \frac{V}{2}-\int_{-1}^1d\cos\theta\mathcal{P}(\theta)\ln \mathcal{P}(\theta).
   \label{elastic entanglement entropy II} 
\end{equation}

As discussed earlier, divergences are identified in the expression for entanglement entropy due to the presence of the total volume in phase space $V$. The authors of reference \cite{peschanski2019} point out three regularization options: (i) volume regularization, (ii) cutoff regularization with a Heaviside function [\ref{stepfunction}], and (iii) cutoff regularization with a Gaussian function. This is possible because, for a given energy, the first term of Eq. [\ref{function P for physical observables}] comes from the part of one of the two bodies in the Hilbert space of the final states that do not correspond to the interacting states. The natural way to remove these non-interacting states is by regularizing the phase space volume so that the first term of Eq. [\ref{function P for physical observables}] disappears.

To carry out these procedures, the operations are defined such that $\sigma_{\text{el}}/[(\pi\tilde{V}/k^2)-\sigma_{\text{inel}}]=1$. Using the fact that $\sigma_T=\sigma_{\text{el}}+\sigma_{\text{inel}}$, we obtain $\tilde{V}=k^2\sigma_{T}/\pi$, and,
\begin{equation}
    \tilde{\mathcal{P}}(\theta)=\frac{2k^2}{\sigma_{\text{el}}}\frac{d\sigma_{\text{el}}}{dt}.
    \label{function P regularized}
\end{equation}
This is considered the volume regularization hypothesis. With this procedure, the entanglement entropy will be,
\begin{equation}
    S=-\int_{0}^{\infty}d|t|\frac{1}{\sigma_{\text{el}}}\frac{d\sigma_{\text{el}}}{dt}\ln\left(\frac{4\pi}{\sigma_T \sigma_{\text{el}}}\frac{d\sigma_{\text{el}}}{dt}\right),
    \label{volume regularization entanglement entropy}
\end{equation}
which depends only on measurable observables.

The authors of \cite{peschanski2019} made estimates for Eq.~[\ref{volume regularization entanglement entropy}] in such a way that, in their derivation, they assumed the diffractive peak approximation in high-energy hadron-hadron scattering. In this case, the differential elastic cross-section is given by,
\begin{equation}
    \frac{d\sigma_{\text{el}}}{dt}=\frac{\sigma_T^2}{16\pi}e^{-B_{\text{el}}|t|},
    \label{diferential elastic cross section in volume regularization}
\end{equation}
with,
\begin{equation}
    \sigma_{\text{el}}=\int_0^{\infty}d|t|\frac{d\sigma_{\text{el}}}{dt}=\frac{\sigma_T^2}{16\pi B_{\text{el}}},
    \label{total elastic cross section in volume regularization}
\end{equation}
where $B_{\text{el}}(\sqrt{s})$ is the elastic slope parameter.

Now, the procedures for the cutoff with step and Gaussian functions are followed. With these regularizers, the authors of \cite{peschanski2016} rewrote the scattering amplitude $\mathcal{A}$ in the impact parameter representation as,
\begin{equation}
    a(s,b)=\frac{1}{2\pi}\int d^2 qe^{-\mathbbm{i}\vec{q}\cdot \vec{b}}f(s,t),
    \label{scattering amplitude in the impact-parameter representation}
\end{equation}

\begin{equation}
    f(s,t)=\frac{1}{2\pi}\int d^2be^{\mathbbm{i}\vec{q}\cdot \vec{b}} a(s,b),
    \label{reverse scattering amplitude in the impact-parameter representation}
\end{equation}
such that $f(s,t)=\mathscr{A}/(16\pi k^2)$ and then, $\sigma_T=2\int d^2b\Im a(s,b)$ and $\sigma_{\text{el}}=\int d^2b|a(s,b)|^2$, with $t=-\vec{q}^2$.

The following prescription is used to approximately obtain the Hilbert space. Identifying that $bk\sim \ell$, the region for a large impact parameter does not contribute to the scattering amplitude $a(s,b)$. The regularization procedure is performed by truncating the modes with a large impact parameter by introducing a cutoff function $C(b)$ that vanishes as $b\rightarrow\infty$. Thus, the regulated quantities are given by,
\begin{equation}
\hat \sigma_T=2\int_0^{\infty}d^2bC^2(b)\Im a(s,b),
\label{total cross section with cutoff step regularization}
\end{equation}

\begin{equation}
\hat \sigma_{\text{el}}=\int_{0}^{\infty}d^2bC^2(b)|a(s,b)|^2,
\label{elastic cross section with cutoff step regularization}    
\end{equation}

\begin{equation}
 \frac{d \hat \sigma_{\text{el}}}{dt}=\frac{1}{4\pi}\Bigm\lvert \int_0^{\infty}d^2be^{\mathbbm{i}\vec{q}\cdot\vec{b}}C(b)a(s,b) \Bigm\lvert ^2.
\label{diferential elastic cross section with cutoff step regularization}   
\end{equation}
Thus, the volume of the regularized Hilbert space is given by $\tilde{V}\approx\hat{V}=k^2\hat \sigma_{T}/\pi$ and as a consequence,

\begin{equation}
\hat{\mathcal{P}}(\theta)=\frac{2k^2}{\hat \sigma_{\text{el}}}\frac{d \hat \sigma_{\text{el}}}{dt}.
\label{function P regularized II}
\end{equation}
The simplest functions for applying this method are precisely the step function and the Gaussian,
\begin{equation}
 C(b)=\begin{cases}
 1, \quad \text{for} \quad b\leq 2\Lambda, \\
 0, \quad \text{for} \quad b> 2\Lambda,
 \end{cases}
 \label{step function regulator}
\end{equation}

\begin{equation}
C(b)=e^{-\frac{1}{2}\frac{b^2}{4\Lambda^2}}.
\label{gaussian regulator}
\end{equation}
With the cutoff applied, the entanglement entropy is given by,
\begin{equation}
    \hat{S}=-\int_0^{\infty}d|t|\frac{1}{\hat{\sigma}_{el}}\frac{d\hat{\sigma}_{el}}{dt}\ln \left(\frac{4\pi}{\hat{\sigma}_{T}\hat{\sigma}_{el}}\frac{d\hat{\sigma}_{el}}{dt}\right).
    \label{regularized entanglement entropy}
\end{equation}
Both cutoffs perform the regularization above in the infinite Hilbert space volume because now $\ell$ has a defined upper limit given by $\ell_{max}\equiv 2\Lambda k$ and now $\hat V=2k^2\int_0^{\infty}C^2(b)=4k^2...
\Lambda$. De qualquer forma, a condição que determina o cutoff é $\Lambda^2=\hat{\sigma}_T/4\pi$. 

\section{Entanglement Entropy in the CGC}
\label{chapter 4 - section 2: CGC entanglement entropy}

In this model, the characterization of wave functions in the CGC formalism is used to investigate entanglement entropy. To do this, we first address the structure of the hadronic wave function in a DIS at high energies in order to obtain the reduced density matrix of the system, and then calculate the entanglement entropy.

At high energies, the hadronic wave function has a significant contribution from the soft gluons (which carry small momentum), which have enough energy to scatter a hadronic target, within a rapidity interval,
\begin{equation}
    0<T<\Delta Y,
    \label{rapidity interval}
\end{equation}
with $Y\sim 1/\alpha_s$.

In the CGC model, the hadronic wave function takes the form,
\begin{equation}
    \Psi[a,A]=\psi[A]\chi [a,\rho],
    \label{hadronic wave function CGC}
\end{equation}
where $a$ are the modes of the soft gluons, $A$ are the modes of the valence quarks (with rapidity $Y>\Delta Y$), and $\rho^a(x)$ is the color charge density as a function of the coordinate $x$. For $\rho \ll 1/\alpha_s$, the wave function of the gluons is given by a coherent state,
\begin{equation}
    \chi[a,\rho]=e^{\mathbbm{i}\int_kb_a^i(k)[a^{\dagger i}_a(k)+a_a^i(-k)]}\ket{0},
    \label{soft gluon wave function}
\end{equation}
with the Weizsacker-Williams field,
\begin{equation}
    b_a^i(k)=g\rho_a(k)\frac{\mathbbm{i}k^i}{k^2},
    \label{WW gloun field}
\end{equation}
where $g$ is the coupling of the strong interaction. The creation and annihilation operators in Eq.~[\ref{soft gluon wave function}] are integrated over rapidity,
\begin{equation}
    a_i^a(k)\equiv \frac{1}{\sqrt{\Delta Y}}\int _{Y<\Delta Y}\frac{d Y}{2\pi}a_i^a(Y,k).
    \label{creation and aniquilation gluon operators}
\end{equation}

For a fixed energy, observables that depend only on the color charge density $O[A]$ are calculated from Eq.~[\ref{observables on CGC}]. In this model, the McLerran-Venugopalan model \cite{mclerran} will be used, with the weight functional given by Eq.~[\ref{MV weight function}].

The reduced density matrix $\hat \rho_r$ for the soft gluons in the MV model is,
\begin{equation}
    \hat \rho_r=\mathscr{N}\int D[\rho]e^{-\int_k\frac{1}{2\mu^2(k)\rho_a(k)\rho_a(-k)}}e^{\mathbbm{i}\int_q b_b^i(q)\phi_b^i(-q)}\ket{0}\bra{0}e^{-\mathbbm{i}\int_pb_c^j(p)\phi_c^j(-p)},
    \label{reduced gluon density matrix MV 1}
\end{equation}
where,
\begin{equation}
    \phi_a^i(k)=a_a^i(k)+a_a^{\dagger i}(-k).
    \label{phi field abreviation}
\end{equation}
The integral over the charge density in [\ref{reduced gluon density matrix MV 1}] results in,
\begin{equation}
    \hat \rho_r=\sum_n \frac{1}{n!}e^{-\frac{1}{2}\phi_iM_{ij}\phi_j}\left[\prod_{m=1}^n M_{i_mj_m}\phi_{i_m}\ket{0}\bra{0}\phi_{j_m}\right]e^{-\frac{1}{2}\phi_iM_{ij}\phi_j},
    \label{reduced gluon density matrix MV 2}
\end{equation}
where the compact notations are considered,
\begin{equation}
    \begin{cases}
    \phi_i\equiv [a_i^{\dagger a}(x)+a_i^a(x)]; \\
    M_{ij}\equiv \frac{g^2}{4\pi^2}\int_{u,v}\mu^2(u,v)\frac{(x-u)_i(y-v)_j}{(x-u)^2(y-v)^2}\delta^{ab},
    \end{cases}
    \label{CGC abreviations}
\end{equation}
and the matrix $\hat M$ has two polarizations, two colors, and two coordinate indices, collectively denoted by the pair $\{i,j\}$.

Thus, in this section, the goal is to find an expression for $\Tr[\hat \rho^{1+\epsilon}]$ in order to establish the program proposed in Strategy III. To do so, we first calculate $\Tr[\hat \rho_r^N]$ for an arbitrary $N$, and then take the limit $N\rightarrow 1+\epsilon$. For $\Tr[\hat \rho_r^2]$,
\begin{equation}
    \Tr[\hat \rho_r^2]=\sum_{n,n'}\frac{1}{n!n'!}\bra{0}e^{-\phi_iM_{ij}\phi_j}
     \left(\prod_{m=1}^n\prod_{m'=1}^{n'}M_{i_mj_m}M_{i_{m'}j_{m'}}\phi_{j_{m'}}\phi_{i_m}\ket{0}\bra{0}\phi_{j_m}\phi_{i_{m'}}\right)e^{-\phi_iM_{jm}\phi_j}\ket{0}.
    \label{trace density matrix squared}
\end{equation}
This expression is used to compute the Rényi entropy in the reference \cite{cgcvenu}. The generalization of Eq.~[\ref{trace density matrix squared}] for $\hat \rho_r^N$ involves the product of $N$ elements of the vacuum matrix operators that depend on the field $\phi$. Each matrix element is computed separately, so the fields entering the expression can be considered independent. Thus, we define the replicated field multiplet $\phi_i^{\alpha}$, with $\alpha=1,2,...,N$.
\begin{equation}
    \Tr[\hat \rho_r^N]=\bra{0}e^{-\sum_{\alpha=1}^N\phi_i^{\alpha}M_{ij}\phi_j^{\alpha}+\sum_{\alpha=1}^N\phi_i^{\alpha}M_{ij}\phi_j^{\alpha+1}}\ket{0},
    \label{trace density matrix N squared 1}
\end{equation}
where now $\ket{0}$ is the vacuum in the light-cone of all the replicated fields $\phi^{\alpha}$. Note that the nearest neighbor of the \textquoteleft interaction' in the replicated fields follows the periodic boundary condition,
\begin{equation}
    \phi^{N+1}=\phi^1.
    \label{neighborhood rule}
\end{equation}

Thus, it is possible to rewrite Eq.~[\ref{trace density matrix N squared 1}] as,
\begin{equation}
    \Tr[\hat \rho_r^N]=\left(\frac{\det[\hat \pi]}{2\pi}\right)^{N/2}\int \prod_{\alpha=1}^N [D\phi^{\alpha}]e^{-\frac{\pi}{2}\sum_{\alpha=1}^N\phi_i^{\alpha}\phi_i^{\alpha}-\frac{1}{2}\sum_{\alpha=1}^N(\phi_i^{\alpha}-\phi_{i}^{\alpha+1})M_{ij}(\phi_j^{\alpha}-\alpha_j^{\alpha+1})},
    \label{trace density matrix N squared 2}
\end{equation}
where, in this equation,
\begin{equation}
    \hat \pi=\pi \delta_{ij}\delta^{ab}\delta^2(x-y),
    \label{pi matrix}
\end{equation}
The \textquoteleft action' is diagonalized by a Fourier transform in $\alpha$:
\begin{equation}
    \begin{cases}
    \tilde{\phi}^n=\frac{1}{N}\sum_{\alpha=1}^Ne^{\mathbbm{i}\frac{2\pi}{N}\alpha n}\phi^{\alpha}; \\
    \phi^{\alpha}=\sum_{n=0}^{N-1}e^{-\mathbbm{i}\frac{2\pi}{N}\alpha n}\tilde{\phi}^n,
    \end{cases}
    \label{discrete fourier transform}
\end{equation}
with the periodicity relation,
\begin{equation}
    \tilde{\phi}^{N-n}=\tilde{\phi}^{*n}.
    \label{periodicity}
\end{equation}

The relation between neighbors in Fourier space results in,
\begin{equation}
    (\phi_i^{\alpha}-\phi_i^{\alpha+1})(\phi_j^{\alpha}-\phi_j^{\alpha+1})=\sum_{n,m}(e^{-\mathbbm{i}\frac{2\pi}{N}n}-1)(e^{-\mathbbm{i}\frac{2\pi}{N}m}-1)e^{-\mathbbm{i}\frac{2\pi}{N}\alpha(n+m)}\tilde{\phi}_i^n\tilde{\phi}_j^m
    \label{neighborhood fouries space rule}
\end{equation}
Using,
\begin{equation}
    \sum_{\alpha}e^{-\mathbbm{i}\frac{2\pi}{N}\alpha(n+m)}=N\delta_{(n+m),N},
    \label{fourier identidy}
\end{equation}
we obtain,
\begin{equation}
    \begin{split}
        \sum_{\alpha}(\phi_i^{\alpha}-\phi_i^{\alpha+1})(\phi_j^{\alpha}-\phi_j^{\alpha+1})&=N\sum_n (e^{-\mathbbm{i}\frac{2\pi}{N}n}-1)(e^{-\mathbbm{i}\frac{2\pi}{N}n}-1)\tilde{\phi}_i^n\tilde{\phi}_j^{*n} \\
        &=4N\sum_n\sin^2\left(\frac{\pi n}{N}\right)\tilde{\phi}_i^n\tilde{\phi}_j^{*n}.
    \end{split}
    \label{fourier identidy use}
\end{equation}
Then,
\begin{equation}
    \Tr[\hat \rho_r^N]=N^{N/2}\left(\frac{\det[\hat \pi]}{2\pi}\right)^{N/2}\int \prod_n [D\tilde{\phi}^n]e^{-\frac{N}{2}\sum_{n=0}^{N-1}\tilde{\phi}_i^n[\hat \pi +4\hat M\sin^2(\pi n/N)]_{ij}\tilde{\phi}_j^{*n}}.
    \label{trace density matrix N squared 2: I}
\end{equation}
where $N^{N/2}$ is the Jacobian of the transformation [\ref{discrete fourier transform}]. Thus, we have,
\begin{equation}
    \hat \pi +4\hat M \sin^2\frac{\pi n}{N}=\hat \pi + 2\hat M \left(1-\cos \frac{2\pi n}{N}\right),
    \label{pi and M matrix manipulation}
\end{equation}
The Gaussian integral in Eq.~[\ref{trace density matrix N squared 2: I}] is,
\begin{equation}
        \Tr[\hat \rho_r^N]=\det[\hat \pi]^{N/2}\det\left\{\prod_{n=0}^{N-1}\left[\hat \pi +2\hat M \left(1-\cos\frac{2\pi n}{N}\right)\right]^{-1/2}\right\}.
        \label{trace density matrix N squared 2: II}
\end{equation}
Using the tabulated result (1.396) from reference \cite{zwillinger2007},
\begin{equation}
    \begin{split}
        \prod_{n=0}^{N-1}\left [\hat \pi +2\hat M\left(1-\cos\frac{2\pi n}{N}\right)\right ] &= (2\hat M)^N\prod_{n=0}^{N-1}\left(1-\cos\frac{2\pi n}{N}+\frac{\hat M^{-1}\hat \pi}{2}\right)\\
        &=2\hat M^N\left\{\cosh\left[N\cosh^{-1}\left(1+\frac{\hat M^{-1}\hat \pi}{2}\right)\right]\right\}.
    \end{split}
    \label{table idendity}
\end{equation}
Thus,
\begin{equation}
    \Tr[\hat \rho_r^N]=\exp\left\{-\frac{1}{2}\ln 2-\frac{N}{2}\Tr[\ln\hat \pi^{-1}\hat M]-\frac{1}{2}\Tr[\ln (\cosh(N\cosh^{-1}\left(1+\frac{\hat M^{-1}\hat \pi}{2}\right))-1)]\right\}.
    \label{trace density matrix N squared 2: III}
\end{equation}

The entropy is obtained by taking $N=1+\epsilon$ and keeping the linear terms in $\epsilon$, so that,
\begin{equation}
    \Tr[\hat \rho_r^{1+\epsilon}]\approx 1+\frac{\epsilon}{2}\Tr\left[\ln(\hat \pi^{-1}\hat M)-\sqrt{1+4\hat \pi^{-1}\hat M}\ln\left[1+\frac{\hat{M}^{-1}\hat \pi}{2}\left(1+\sqrt{1+4\hat \pi^{-1}\hat M}\right)\right]\right].
    \label{trace density matrix 1 + epsilon}
\end{equation}
Thus, the entanglement entropy is,
\begin{equation}
    S=\frac{1}{2}\Tr\left\{\ln(\hat \pi^{-1}\hat M)+\sqrt{1+4\hat \pi^{-1}\hat M}\ln\left[1+\frac{\hat M^{-1}\hat \pi}{2}\left(1+\sqrt{1+4\hat \pi^{-1}\hat M}\right)\right]\right\}.
    \label{CGC entanglement entropy I}
\end{equation}
To understand this equation, consider the case invariant under translation, where the matrix $\hat M$ is diagonal in momentum space, 
\begin{equation}
    M_{ij}^{ab}(p)=g^2\mu^2(p^2)\frac{p_ip_j}{p^4}\delta^{ab}.
    \label{M matrix weak coumpling}
\end{equation}

In the original MV model, $\mu^2$ is a constant and does not depend on the \textit{momentum}. The contribution to the entropy with high transverse \textit{momentum} modes can be obtained by expanding Eq.~[\ref{CGC entanglement entropy I}] to leading order in $\hat M$, since for large \textit{momenta} ($g^2\mu^2<p^2$), the eigenvalues of $\hat M$ are small. The expression for entropy in the weak coupling limit is,
\begin{equation}
    S_{E}^{M\ll 1}=\Tr \left[\hat \pi^{-1}\hat M\ln \hat M^{-1}\hat \pi e\right].
    \label{CGC entanglement entropy weak coupling}
\end{equation}
Thus, the dominant ultraviolet contribution is,
\begin{equation}
    S_{E}^{UV}\approx-\frac{g^2}{\pi}(N_c^2-1)S\int \frac{d^2p}{(2\pi)^2}\frac{\mu^2(p^2)}{p^2}\ln\left[\frac{g^2\mu^2(p^2)}{e\pi p^2}\theta\left(p^2-\frac{g^2}{\pi}\mu^2(p^2)\right)\right],
    \label{CGC entanglemente entropy UV I}
\end{equation}
where, in this equation, $S$ is the total area of the projectile. In the original MV model with $\mu$ independent of the \textit{momentum}, the expression [\ref{CGC entanglemente entropy UV I}] is logarithmically divergent. Introducing the ultraviolet cutoff $\Lambda$,
\begin{equation}
    S_{E}^{UV}\approx\frac{Q_s^2}{4\pi g}(N_c^2-1)S\left[\ln^2\frac{g^2\Lambda^2}{Q_s^2}+\ln\frac{g^2\Lambda}{Q_s^2}\right],
      \label{CGC entanglemente entropy UV II}
\end{equation}
so that the saturation scale was identified in the usual way in the model,
\begin{equation}
    Q_s^2=\frac{g^4\mu^2}{\pi}.
        \label{MV saturation scale}
\end{equation}

The contribution from infrared modes can also be calculated. For momentum ranges $p^2<Q_s^2/g^2$, it is possible to expand $\hat M^{-1}$, so that,
\begin{equation}
    S_{E}^{M\rightarrow \infty}\approx\frac{1}{2}\Tr [\ln e^2\hat \pi^{-1}\hat M].
    \label{CGC entanglemente entropy IR I}
\end{equation}
Thus,
\begin{equation}
    S_{E}^{IR}\approx\frac{1}{2}(N_c^2-1)S\int\frac{d^2p}{(2\pi)^2}\ln \frac{e^2g^2\mu^2(p^2)}{\pi p^2}\theta(Q_s^2-g^2p^2)=\frac{3}{8\pi g^2}(N_c^2-1)SQ_s^2.
    \label{CGC entanglemente entropy IR II}
\end{equation}

Combining the two expressions, the CGC entanglement entropy in the MV model is given by,
\begin{equation}
    S_{E}\approx S_{E}^{IR}+S_{E}^{UV}=\frac{SQ_s^2}{4\pi g^2}(N_c^2-1)\left(\ln^2\frac{g^2\Lambda^2}{Q_s^2}+\ln\frac{g^2\Lambda^2}{Q_s^2}+\frac{3}{2}\right).
    \label{CGC entanglement entropy II}
\end{equation}

In contrast to the field representation, in the reference \cite{duan2020}, the same computation of the entropy is performed in the number operator representation,
\begin{equation}
    \expval{N}=\Tr\left[\int \frac{d^2k}{(2\pi)^2a^{\dagger}_ka_k\hat \rho_r}\right].
    \label{number operator}
\end{equation}
yielding a coincident expression,
\begin{equation}
    S_{E}^{CGC}\approx\frac{1}{2}SC_F\int_0^{\infty}\frac{d^2k}{(2\pi)^2}\left[\ln\left(\frac{g^2\mu^2}{k^2}\right)+\sqrt{1+4\frac{g^2\mu^2}{k^2}}\ln\left(1+\frac{k^2}{2g^2\mu^2}+\frac{k^2}{2g^2\mu^2}\sqrt{1+4\frac{g^2\mu^2}{k^2}}\right)\right].
    \label{CGC entanglemente entropy in number representation}
\end{equation}

For phenomenological purposes, saturation can evolve with rapidity following the GBW model. Moreover, the ultraviolet regulator can be identified as the virtuality of the photon in DIS $Q^2$, with the arbitrary choice $Q^2=g^2\Lambda^2$. Thus, analytically,
\begin{equation}
    S_{E}^{CGC}=\frac{1}{2}S\frac{C_F}{4\pi}\tilde{Q}_s^2\left[\tau \ln(\tau^{-1})+\tau\sqrt{1+4\tau^{-1}}\ln\left(\frac{\sqrt{1+4\tau^{-1}}+1}{\sqrt{1+4\tau^{-1}}-1}\right)+\ln^2\left(\frac{\sqrt{1+4\tau^{-1}}+1}{\sqrt{1+4\tau^{-1}}-1}\right)\right],
    \label{CGC entanglement entropy III}
\end{equation}
where $\tau=Q^2/\tilde{Q}_s^2$ and $\tilde{Q}_s^2=(9/4)Q_s^2$.

The parametric behaviors of equations [\ref{CGC entanglemente entropy UV II}] and [\ref{CGC entanglemente entropy IR II}] are obtained for $\tau=1$ ($Q^2=\tilde{Q}_s^2$), so,
\begin{equation}
    S_{E}^{CGC}\sim S\tilde{Q}_s^2,
    \label{CGC entanglement entropy IV}
\end{equation}
On the other hand, for large values of $\tau$, $\sqrt{1+4\tau^{-1}}\approx1+2\tau^{-1}$, so,
\begin{equation}
    S_{E}^{CGC}\sim S\tilde{Q}_s^2(2\ln \tau +\ln^2\tau).
    \label{CGC entanglement entropy V}
\end{equation}
For numerical calculations, it is possible to use,
\begin{equation}
    S=\pi R_p^2 =\sigma_0/2,
    \label{hadron area}
\end{equation}
where $R_p$ is the radius of the proton.

\chapter{Quantum Decoherence Entropy}
\label{chapter 5: Decoherence and Entropy}

The last form of entropy generation studied in this thesis will be due to quantum decoherence. Thus, this brief chapter will address the characterization of the phenomenon, followed by the calculation of decoherence entropy in CGC states.

\section{Quantum Decoherence}

The beginning of the 20th century brought with it physical theories that challenged the well-established classical conception, especially the theories of Relativity and Quantum Mechanics, which not only explain results where the old conceptions fail, but also predict new phenomena. However, it would be naive to abandon all the predictions and results obtained by classical theories, so today we know that each of these theories has its own domain of dominance. For example, the Lorentz transformations of Special Relativity for low speeds reduce to the classical Galilean transformation, so it is safe to state that relativity is a theory that deals with objects at high speeds.

On the other hand, the boundaries between the Classical and Quantum regimes of dynamics do not have such a simple correspondence. Some equations may indeed recover, in their own way, those established by Classical Mechanics in the mean values of observables through Ehrenfest's theorem. However, this does not explain the disappearance of the commutation algebra between observables, and in some cases, it comes with several restrictions. In the search for delimiting the theories, Bohr introduced the correspondence principle, where the idea was to study the limit in which Planck's constant tends to zero ($\hbar\rightarrow 0$). Thus, energy would have a continuous spectrum, equivalent to the classical case. However, explanations were still lacking for the aforementioned problem of the discrepancies between algebras proposed by quantum and classical mechanics.

Another unprecedented quantum phenomenon with no classical counterpart is the {\it Wave Function Collapse}, where, to understand it, it is necessary to elucidate the basic modes of operation of quantum dynamics: given a system composed of a particle of mass $m$ (it could be a set of particles, but this would bring some difficulties not relevant to the discussion) immersed in the influence of a potential field $V(\vec{r},t)$, the evolution of the system is given by the {\it Schrödinger equation},
\begin{equation}
\mathbbm{i}\hbar\frac{\partial}{\partial t}\ket{\psi(\vec{r},t)}=\left[-\frac{\hbar^2}{2m} \nabla^2+V(\vec{r},t) \right]  \ket{\psi(\vec{r},t)}, 
\end{equation}
$\ket{\psi}$ is the solution to this equation: the {\it quantum state}. However, it does not inherently carry a physical meaning, unlike, for example, the equation of a wave on a stretched string, which reveals its vertical displacement, or the pressure difference in air molecules during the propagation of a sound wave, or even the fields in electromagnetic waves. The true physical significance of the wave function was established by the German physicist {\it Max Born}: {\it the squared modulus of the projection of the quantum state in the position representation, {\it i.e.}, the wave function $\psi(\vec{r},t)=\bra{\vec{r}}\ket{\psi}$, represents the probability density of finding the particle at position $\vec{r}$ within $\vec{r}+d\vec{r}$ at time $t$}, that is,
\begin{equation}
 P(\vec r,t)d^3 r=|\psi(\vec r,t)|^2d^3 r   .  
\end{equation}
Therefore, if one is interested in analyzing the probability of finding the particle between points $a$ and $b$, denoted by $P_{ab}$, the probability, in the one-dimensional case, is given by
\begin{equation}
P_{ab}=\int_a^b|\psi( x,t)|^2dx.
\end{equation}

Once the {\it Schrödinger} equation is solved, i.e., the wave function is obtained, it can be expressed as a linear combination of the eigenstates that form the basis of the system in question,
\begin{equation}
\ket{\psi}=\sum_{i=1}^{N}c_i\ket{i},
\label{sobre}
\end{equation}
with the set $\{\ket{i}\}$ forming an $N$-dimensional vector space called the {\it Hilbert space} $\mathcal{H}$, the vector space of square-integrable finite functions. Here, the set of scalars $\{c_i\}$ consists of complex numbers. In this way, it is said that the physical state (wave function) is a linear combination of independent states, i.e., it is a system in {\it superposition}. Therefore, if a measurement is made on the system, it will instantly collapse to the measured state, and all the other states that constituted the superposition will disappear, so that the superposition is {\it never} measured. This is the wave function collapse.

It is possible to treat an $N$-dimensional vector space. However, for simplicity, we consider here the simplest case, where $N=2$, so that Eq.~[\ref{sobre}] takes the form,
\begin{equation}
    \ket{\psi}=c_1\ket{1}+c_2\ket{2}.
    \label{sobre2}
\end{equation}
Born's interpretation reveals that the probability of measuring the state $\ket{1}$ or $\ket{2}$ follows the rule,
\begin{equation}
\begin{cases}
\text{Probability of measuring} \quad \ket{1}:\quad  c_1c_1^*; \\ 
\text{Probability of measuring} \quad \ket{2}:\quad  c_2c_2^*;
\end{cases}
\end{equation}
Thus, for a two-state system (which is a general result that can be adapted for $N$ states), we have,
\begin{equation}
    \sum_{i=1}^Nc_ic_i^*=1 \quad \therefore \quad c_1c_1^*+ c_2c_2^*=1,
\label{impo}
\end{equation}
i.e., for a $50\%$ probability of measuring state $\ket{1}$, it is possible to have the coefficient $c_1=1/\sqrt{2}$ and, according to the constraints in [\ref{impo}], $c_2=1/\sqrt{2}$. However, it has been said that the coefficients of the linear combination [\ref{sobre2}] can be complex numbers, and indeed, it is possible to assign a phase $\theta$ to the coefficients without loss of physical meaning, for example, $c_2=e^{\mathbbm{i}\theta}/\sqrt{2}$. Since, 
\begin{equation}
    c_2c_2^*=\left[\frac{1}{\sqrt{2}}e^{\mathbbm{i}\theta}\right]\left[\frac{1}{\sqrt{2}}e^{-\mathbbm{i}\theta}\right]=\underbrace{e^{\mathbbm{i}\theta}e^{-\mathbbm{i}\theta}}_1\frac{1}{2}, =\frac{1}{2},
\end{equation}
i.e., the addition of a real phase $\theta$ does not alter the physics of the problem. That said, consider now a material object, which can have either microscopic dimensions (an electron, for example) or macroscopic dimensions (a grain of sand), which can be in two states, $\ket{1}$ and $\ket{2}$, with energies $E_1$ and $E_2$. The time evolution of each state is,
\begin{equation}
    \ket{n(t)}\rightarrow e^{-\mathbbm{i}E_nt/\hbar}\ket{n},
\end{equation}
therefore, the time evolution of the representation [\ref{sobre2}] is such that,
\begin{equation}
    \ket{\psi(t)}\rightarrow e^{-\mathbbm{i}E_1t/\hbar}c_1\ket{1}+e^{-\mathbbm{i}E_2t/\hbar}c_2\ket{2}.
\end{equation}
Defining $\omega\equiv(E_2-E_1)/\hbar$, the time evolution of the system (which can also be done relative to $E_2$ if preferred) can be represented as,
\begin{equation}
    \ket{\psi(t)}\rightarrow e^{-\mathbbm{i}E_1 t/\hbar}\left[c_1\ket{1}+e^{-\mathbbm{i}\omega t}c_2\ket{2}\right].
\end{equation}

Now, considering that the object in question is subjected to the Earth's gravitational field, with $\omega=mg\Delta z/\hbar$, we have, for an electron with wave packet separations of $\Delta z= 1$ nm or $\Delta z= 1$ m, $\omega=10^{-4}$ Hz or $\omega=10^{5}$ Hz; for a mass of $1$ g, $\omega=10^{23}$ Hz or $\omega=10^{32}$ Hz. For comparison, $10^{-22}$ s is the time that light takes to cross an atomic nucleus. Thus, when the phases are changing so rapidly, only one of the states can be measured.

It is possible to calculate the density matrix of the state being analyzed, with $\rho_{mn}=c_mc_n^*$, we have,
\begin{equation}
    \begin{cases}
    \rho_{11}=c_1e^{-\mathbbm{i}E_1t/\hbar}c_1^*e^{\mathbbm{i}E_1t/\hbar}=c_1c_1^*=|c_1|^2, \\
    \rho_{12}=c_1e^{-\mathbbm{i}E_1t/\hbar}c_2^*e^{\mathbbm{i}E_1t/\hbar}e^{\mathbbm{i}\omega t}=c_1c_2^*e^{\mathbbm{i}\omega t}, \\
    \rho_{21}=c_2e^{-\mathbbm{i}E_1t/\hbar}c_1^*e^{\mathbbm{i}E_1t/\hbar}e^{-\mathbbm{i}\omega t}=c_2c_1^*e^{-\mathbbm{i}\omega t}, \\
    \rho_{22}=c_2e^{-\mathbbm{i}E_1t/\hbar}c_2^*e^{\mathbbm{i}E_1t/\hbar}=|c_2|^2, \\
    \end{cases}
\end{equation}
so that,
\begin{equation}
\hat \rho=
    \begin{pmatrix}
    |c_1|^2    &  c_1c_2^*e^{\mathbbm{i}\omega t} \\ 
    c_2c_1^*e^{-\mathbbm{i}\omega t}    &  |c_2|^2
    \end{pmatrix}.
    \label{forma1}
\end{equation}

\begin{figure}[ht]
\centering 
\includegraphics[width=1\linewidth]{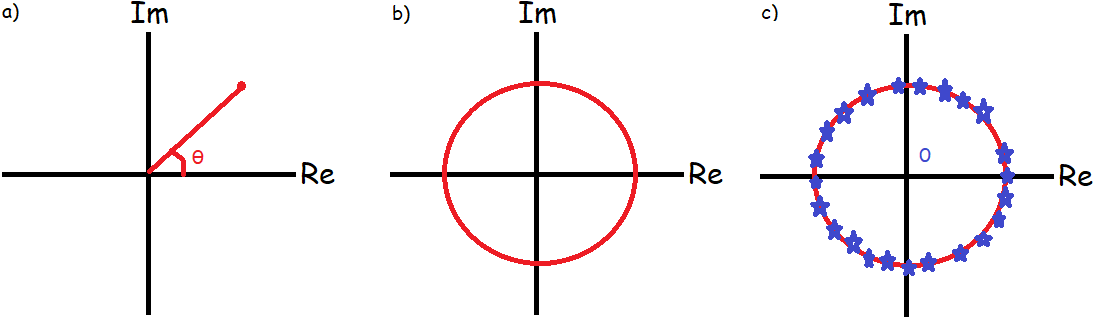} 
\caption{{\bf(a)} Polar representation of a complex number in the real-imaginary plane; {\bf(b)} for values of $\theta$ between $0$ and $2\pi$, a unit circle centered at the origin is formed; {\bf(c)} if a large number of interactions of the system with its environment or detector are considered, where its manifestation is due to a random change in phase, the average value of the complex number is zero.} 
\label{fases} 
\end{figure}

Every measurement requires a finite time $T$, although in everyday values it can be considered instantaneous ($10^{-12}$s). Thus, it is possible to analyze what happens to the density matrix of a system subjected to a measurement by averaging it over $T$,
\begin{equation}
\frac{1}{T}\int_{0}^T\hat \rho dt=
    \begin{pmatrix}
    c_1c_1^*    &  c_1c_2^*s(T) \\ 
    c_2c_1^*s(T)    &  c_2c_2^*
    \end{pmatrix}.
    \label{forma2}
\end{equation}
with,
\begin{equation}
    \lim_{\omega T\rightarrow \infty}s(T)=\lim_{\omega T\rightarrow \infty} e^{\mathbbm{i}\omega T/2}\frac{\sin (\omega T/2)}{\omega T/2}=0,
\end{equation}
Therefore,
\begin{equation}
\hat \rho=
    \begin{pmatrix}
    |c_1|^2     &  0 \\ 
    0               &  |c_2|^2
    \end{pmatrix}.
\label{decoro2}
\end{equation}
For example, for $\omega=10^{32}$Hz and $T=10^{-16}$s, $|s(T)|\sim 10^{-16}$. These cases would require great experimental precision for the superposition to be detected. After performing the measurement, it is elucidated that the system must interact with the environment and with the detector, which will amplify the signal, i.e., the particle that composed the system will interact several times with other particles before being detected. This is the description of {\it quantum decoherence}, the constant interactions that the system in superposition must endure will result in a phase shift of the off-diagonal elements. Each time the system interacts with any object, whether from the environment or the detector, irreversible changes will occur due to a random phase shift $\theta$, so the key to understanding decoherence lies in calculating the average of the complex number that accompanies the phase (Fig.~[\ref{fases}]), that is, {\it zero}: It is possible to represent the complex number $e^{\pm\mathbbm{i}\theta}$ in the so-called {\it polar form} where the coordinate axes are composed of the set of real and imaginary numbers. A change in the phase $\theta$ of the exponential, depending on the direction, will cause the slope of the line connecting the origin and the point on the unit circle to change, and for a large number of perturbations, the average value is zero.

The matrix [\ref{decoro2}] no longer corresponds to any physical system, meaning that it is not possible to find the wave function $\ket{\psi}$ that generated it, because by analyzing the form [\ref{forma1}], if the off-diagonal elements are null, the diagonal elements will also be null. However, the disappearance of the non-diagonal terms in the system's density matrix is the effect of quantum decoherence, and without these terms, the system loses its ability to interfere. Finally, it is also worth highlighting that decoherence preserves the unitarity of quantum theory, unlike wave function collapse, which does not maintain this property. While decoherence describes the loss of coherence due to interaction with the environment, collapse implies a non-unitary and irreversible change in the quantum state.

\section{Theoretical Formalism of Decoherence}
\label{chapter 5 - section 2; the decoherence theoretical formalism}

\begin{figure}[ht]
\centering 
\includegraphics[width=0.45\linewidth]{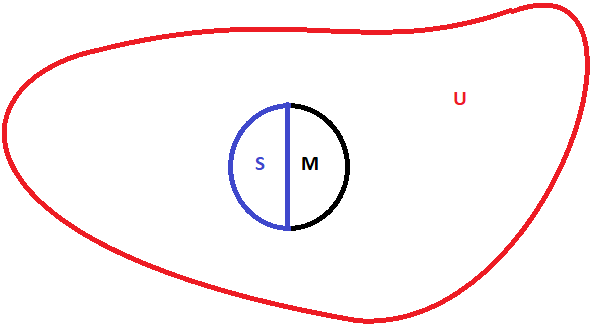} 
\caption{Division of the regions analyzed in the study of decoherence: $S$ is the studied system, which could be a hydrogen atom, a harmonic oscillator, etc.; a physical system of interest; $M$ is the measurement apparatus that will collapse the system's wave function. In order to perform a measurement, it must necessarily interact with $S$; $U$ is the environment, which can never be isolated from the system, and could be photons illuminating the sample, other atoms surrounding it, or the cosmic microwave background radiation.} 
\label{separa} 
\end{figure}

Physically, the basic idea in the computation of decoherence effects is to consider the effect of the environment and the measurement on the quantum system, leading to a separation between the system of interest $S$, the measurement apparatus $M$, and an open relation for interactions with the environment $U$ (Fig. \ref{separa}). Some of the effects of measurement were discussed earlier, and in this section, only the effects of the environment on the system of interest are considered. Thus, the system $S$ has its eigenvector basis $\ket{i}$ which generates the Hilbert space $\mathcal{H}_S$. The environment, in turn, is described by the basis $\ket{n}$ of the space $\mathcal{H}_U$. In the interaction between these two systems, the general state will be written in a basis $\ket{i}\otimes\ket{n}\doteq \ket{i,n}$ that generates the composite space $\mathcal{H}=\mathcal{H}_S\otimes \mathcal{H}_U$. It is now assumed that the density matrices of the composite space are not independent, that is, the matrices in the space $\mathscr{H}=\mathscr{H}_S\otimes\mathscr{H}_U$ are more general than $\hat \rho_S\otimes\hat \rho_U$.

In the Schrödinger picture ($H=H_U+H_S$), it is known that the temporal evolution of the system's density matrix follows the von Neumann equation \cite{breuer2002}
\begin{equation}
    \mathbbm{i}\hbar\frac{\partial}{\partial t}\hat \rho_s = [H_S,\hat \rho_s].
\end{equation}
Considering that it is possible to separate the internal dynamics of the system from the interaction with the environment, where the duration of a scattering process with the medium is short compared to the internal time scale, the von Neumann equation will have a small modification introduced by the scattering interaction with the environment \cite{joos2013}, considering an appropriate scattering matrix $\hat S$:
\begin{equation}
    \mathbbm{i}\hbar\frac{\partial}{\partial t}\hat \rho_s = [H,\hat \rho_s] + \mathbbm{i}\hbar\frac{\partial}{\partial t}\hat \rho_s\bigg|_{\text{scattering}}.
\end{equation}

In most cases, a broad sequence of scatterings can be treated as damping of the off-diagonal elements, so that,
\begin{equation}
\frac{\partial}{\partial t}\rho_{nm}^s\bigg|_{\text{scattering}}=-\lambda\rho_{nm}^s(t),
\end{equation}
where the constant $\lambda$ is given by,
\begin{equation}
    \lambda=\Gamma(1-\bra{n_0}\hat S_m^{\dagger}\hat S_n\ket{n_0}),
\end{equation}
where $\Gamma$ is the collision rate and $\ket{n_0}$ is the environment state at a given initial condition.

Next, two examples related to the phenomenon of quantum decoherence will be discussed: the spatial localization of material objects and the Quantum Zeno effect.

\subsection{Spatial Localization}
\label{chapter 3 - section 2 - subsection 1: Spatial Localization}

The first example to be explored is the localization of macroscopic objects. It may not seem so at first glance, but the classical perception we have of the localization of macroscopic objects, formed by microscopic entities governed by quantum mechanics, where the physical state does not have a well-defined position and momentum, is not a trivial problem. Not only the perception but also the use of classical theories that assertively describe, in certain limits, the physical localization of these objects. The hypothesis is considered that different spatial configurations of a system must undergo a decoherence process very quickly due to a strong influence by scattering processes.

A formal treatment of the effect can be given. For this, $\ket{x}$ is the position eigenstate of a macroscopic object, and $\ket{a}$ is the state of the incident particle. Thus, the effect of the interactions with the system in the temporal evolution can be expressed as:
\begin{equation}
    \ket{x}\ket{a}\overbrace{\rightarrow}^t\ket{x}\ket{a_x}=\ket{x}\hat S_x\ket{a},
\end{equation}
so that the scattering is calculated from the appropriate scattering matrix $\hat S_x$. The same representation for an initial wave packet state is given by:
\begin{equation}
    \int d^3x\phi(x)\ket{x}\ket{a}\overbrace{\rightarrow}^t\int d^3x\phi(x)\ket{x}\hat S_x\ket{a}.
\end{equation}
The reduced density matrix that describes the changes in the object is given by:
\begin{equation}
    \hat \rho(x,x')=\phi(x)\phi^*(x')\bra{a}\hat S_x^{\dagger}\hat S_x\ket{a}.
\end{equation}

Obviously, a single scattering process will not localize the object, so the elements of the matrix above are close to unity. However, the wide occurrence of scatterings causes an exponential damping of spatial coherence:
\begin{equation}
    \hat \rho (x,x',t)=\hat \rho (x,x',0)e^{-\Lambda(x-x')^2}.
\end{equation}
Thus, the contribution of this effect is described by a single parameter $\Lambda$, called the {\it localization rate}, given by:
\begin{equation}
    \Lambda=\frac{k^2Nv\sigma_{\text{ef}}}{V},
\end{equation}
where $k$ is the wave number of the incident particles, $Nv/V$ is the flux, and $\sigma_{\text{ef}}$ is of the order of the total cross section. Some values of $\Lambda$ are given in Table [\ref{tab1}]. Most of the numbers in the table are quite large, revealing how strong the connection is between macroscopic objects, the size of dust particles, and the environment. Even the intergalactic space would not be excluded from this effect due to the cosmic background radiation.

\begin{table}[ht]
\centering
\caption{Localization rate $\Lambda$ in $cm^{-2}s^{-1}$ for three particle sizes in various types of scatterings. This quantity measures how quickly the interference between different positions disappears as a function of distance over time. Data provided by \cite{joos2013}.}
\label{tab1}
\begin{tabular}{|c|c|c|c|c|}
\hline

Size                                         & $10^{-3}$cm & $10^{-5}$cm & $10^{-6}$cm    \\ \hline
Cosmic background radiation                  & $10^{6}$    & $10^{-6}$   & $10^{-12}$     \\
Photons at 300K                               & $10^{19}$   & $10^{12}$   & $10^{6}$       \\
Solar light (on Earth)                        & $10^{21}$   & $10^{17}$   & $10^{13}$      \\
Molecules in the air                         & $10^{36}$   & $10^{32}$   & $10^{30}$      \\
Laboratory vacuum ($10^3$ particles/cm$^3$)  & $10^{23}$   & $10^{19}$   & $10^{17}$      \\ \hline
\end{tabular}
\end{table}

\begin{figure}[ht]
\centering 
\includegraphics[width=0.8\linewidth]{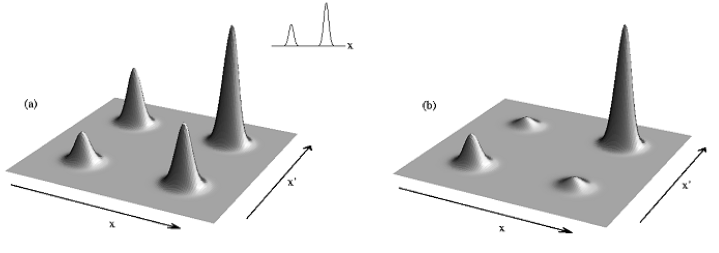} 
\caption{{\bf (a)} Density matrix of two Gaussian wave packets. The coherence between the two packets is represented by the off-diagonal elements. {\bf (b)} The density matrix after losing its coherence. Figure adapted from \cite{joos2013}}
\label{pacotes} 
\end{figure}

In the case of decoherence in the superposition of two wave packets, we analyze the distance between the packets. Fig.~[\ref{pacotes}] {\bf a)} represents the density matrix well, illustrating four peaks: two around the main diagonal and two off-diagonal. The latter contributions represent the coherence between the two parts. However, if the off-diagonal terms are damped, decoherence occurs and the peaks disappear, as shown in Fig.~[\ref{pacotes}] {\bf b)}.

\subsection{Quantum Zeno Effect}

Macroscopically, it is not desirable that the act of measurement interferes with the object of interest; or rather, in most cases, it is possible to perform a measurement without disturbing the system being measured, obtaining information about it. However, it is known that this verification is not analogous to the quantum case, where the wave function collapses. Furthermore, it is not possible to immediately obtain all the measurement values of the observables of a system, so we always work with the idea of ensemble averages.

One phenomenon that addresses the peculiarities of measurement in quantum systems is the {\it Zeno} effect\cite{angioni2009}, and now its mathematical derivation is presented: let $\hat H$ be the Hamiltonian of a quantum system, where $\ket{\Psi_0}$ is its state at the initial time, and its time evolution is given by
\begin{equation}
    \ket{\Psi(t)}=e^{-\mathbbm{i}\hat H t}\ket{\Psi_0}.
\end{equation}
Thus, it is possible to calculate the transition rate $P(t)$, from a physical state evolving over a short time interval $\delta t$ to the initial state:
\begin{equation}
P(t)=|\bra{\Psi_0}e^{-\mathbbm{i}\hat H t}\ket{\Psi_0}|^2,
\end{equation}
Now, expanding this probability in a Taylor series:
\begin{equation}
P(t)=\left|\sum_{n=1}^{\infty}\frac{(\mathbbm{i}\delta t)^n}{n!}\expval{\hat H^n}\right|^2=1-\delta t^2(\expval{\hat H}^2-\expval{\hat H^2})+\mathcal{O}(\delta t^3)\approx 1-\delta t^2\sigma_H^2,
\end{equation}
here, $\sigma_H=\sqrt{\expval{\hat H}^2-\expval{\hat H^2}}$ is the standard deviation of the expected value associated with the Hamiltonian. Thus, the transition probability depends on $t^2$, and if measurements are performed at regular time intervals, say $\tau=t/N$, where $N$ is the number of measurements performed, we have:
\begin{equation}
P(t)=(1-\tau^2\sigma_H^2)^N=\left(1-\frac{\frac{t^2}{N}\sigma_H^2}{N}\right)^N\approx e^{-\frac{t^2}{N}\sigma_H^2},
\end{equation}
making it clear that for a large number of measurements:
\begin{equation}
    \lim_{N\rightarrow\infty}P(t)=1,
\end{equation}
a dramatic result, where if a large number of measurements and short time intervals are made, the probability of the system remaining in the initial state is $100\%$, meaning that the continuous measurement process forces the system to remain in the initial state.

{\it Aristotle} wrote in his {\it Physics}\cite{angioni2009} a series of paradoxical arguments, attributed to the pre-Socratic philosopher {\it Zeno of Elea}, where the proposal was to dialectically prove inconsistencies in the philosophical concepts of the time, such as {\it multiplicity}, {\it divisibility}, and {\it motion}. Among these arguments, the famous story of the race between {\it Achilles} and the tortoise appears: since the animal would have a total disadvantage due to the difference in speeds, it is allowed to start from a more advanced position ahead of the hero of the {\it Iliad}. The paradox proposed by {\it Zeno} is the demonstration that {\it Achilles} would never surpass the tortoise, for when he reaches the position where it started, say A, the tortoise will be at B, and when he reaches B, the animal will have moved to C, and this process would continue indefinitely, yet the hero would never win.

\begin{wrapfigure}{r}{0.4\textwidth}
\begin{center} 
\includegraphics[width=0.8\linewidth]{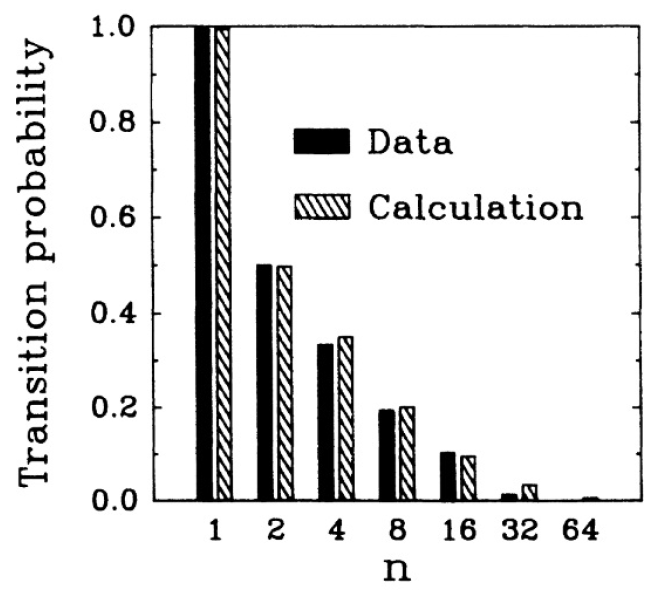} 
\end{center} 
\caption{Calculations and experimental data of the transition rate of a three-level quantum system as a function of the number of measurements $n$ performed during the transition processes. For large values of $n$, the transition rate drops to 0. Reproduced from \cite{itano1990}.} 
\label{zeno} 
\end{wrapfigure}

At that time, nothing was known about infinitesimal calculus or Newtonian reference frames; however, by introducing the two competitors, the use of reference frames is intuitively suggested, where it could be the velocity of {\it Achilles} relative to the tortoise and vice versa, where both objects have independent velocities. Now, if there is a relationship between the runners, say, {\it Achilles} restricts his movement to a constant observation of where the tortoise is, so that whenever he reaches point C, he checks the tortoise at D, continuing this process indefinitely, like a pattern to determine his motion. An artificial situation is created in which {\it Achilles} is governed by the tortoise's space, exactly what was proposed in the calculation of the transition rate probability, and hence the name of the phenomenon is the quantum {\it Zeno} effect. One way to explain the classical solution to the {\it Zeno} paradox involves introducing concepts such as {\it limit}, {\it convergence}, and {\it infinitesimal}, where the philosopher's proposal fails in classical dynamics by assuming that the sum of infinite time intervals is always infinite; however, it is known that there is a possibility for an infinite sum of terms to result in finite values. But what about the quantum case? In 1900, {\it Itano, Bollinger}, and {\it Wineland}\cite{itano1990} experimentally observed the effect in a three-level quantum system (Fig.~[\ref{zeno}]).

\begin{figure}[ht]
\centering 
\includegraphics[width=0.7\linewidth]{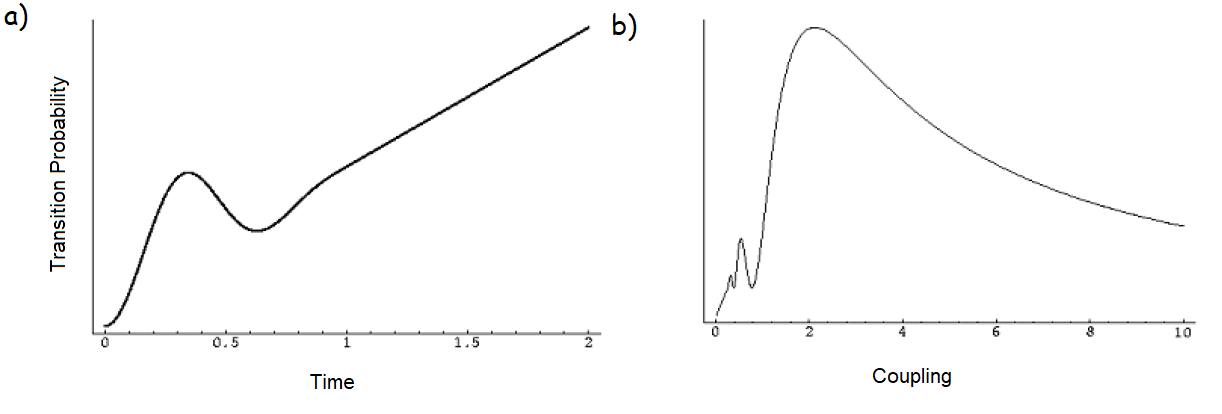} 
\caption{{\bf (a)} Temporal dependence of the probability of finding the system in state $\ket{2}$, where at $t=0$ it was in state $\ket{1}$, for a fixed coupling constant. {\bf (b)} Dependence on the coupling constant of the probability of finding the system in state $\ket{2}$, where at $t=0$ it was in state $\ket{1}$, calculated at fixed time. Image adapted from \cite{itano1990}.} 
\label{final} 
\end{figure}

Decoherence enters the effect precisely to elucidate the difference between the quantum and classical cases, with a simple model being used here, justifying its means in the problem. For this, a simplified model of {\it von Neumann's measurement theory} is considered, where a {\it pointer} (measurement apparatus) is coupled to a two-level system, $\ket{1}$ and $\ket{2}$, described by the Hamiltonian:

\begin{equation}
\hat H = \hat H_0 + \hat H_{int} =V(\ket{1}\bra{2}+\ket{2}\bra{1})+E\ket{2}\bra{2}+\gamma\hat p (\ket{1}\bra{1}-\ket{2}\bra{2}),
\end{equation}

where the transitions between the two levels are induced by the potential $V$ and monitored by the {\it pointer}, with its intensity measured by the constant $\gamma$. Thus, it is possible to calculate the transition rate as a function of time (Fig.~[\ref{final}(a)]) and the coupling constant (Fig.~[\ref{final}(b)]). As previously predicted, for a small time interval, i.e., the measurement is made in a very short time, a quadratic dependence on time is obtained. However, for longer times, this evolution becomes linear. Furthermore, if the system and the measurement apparatus are strongly coupled, the transition rate starts to smooth out.

In this chapter, basic concepts of decoherence and quantum entanglement were discussed, with examples, without exhausting the subject, only elucidating their basic characteristics and fundamental elements used in the research in question. In the next section, these concepts will be used in the formulation of much of the entropic notions studied in the analysis of entropy creation in high-energy particle collisions.

\section{Decoherence Entropy in Heavy Ion Collisions}
\label{chapter 5 - section 3: the decoherence entropy on the heavy ions scattering}

In the initial stages of ultra-relativistic heavy ion collisions, the available states are characterized by a coherent configuration of gluon fields. These fields are generated by the quasi-static color charges of the valence quarks of the nuclei and can be approximated with semi-classical color fields randomly oriented in a CGC medium.

The phenomenon of decoherence can play a significant role in the entropy production in these reactions. Initially, the color fields in the reactions of ultra-relativistic heavy ion collisions can be described by classical coherent fields,
\begin{equation}
    \ket{\Psi[J]}=\prod_{\vec{k}}\exp(\mathbbm{i}\alpha_{\vec{k},\lambda}a_{\vec{k},\lambda}^{\dagger}-\mathbbm{i}\alpha_{\vec{k},\lambda}^*a_{\vec{k},\lambda})\ket{0},
\end{equation}
where $\vec{k}$ is the {\it momentum}, $\lambda$ is the polarization, and the amplitude $\alpha_{\vec{k},\lambda}$ is determined by the classical current creation field $\vec{J}$,
\begin{equation}
    \alpha_{\vec{k},\lambda}=\frac{\vec{\epsilon}_{\vec{k},\alpha}\cdot \vec{J}(\vec{k},\omega_{\vec{k}})}{\sqrt{\hbar \omega_{\vec{k}}V}}.
\end{equation}
For simplicity, consider here only a mode with $\vec{k}$ and $\lambda$. The coherent state can be written as a superposition of the number states $\ket{n}$ with eigenvalue $\alpha$,
\begin{equation}
    \ket{\alpha}=e^{-|\alpha|^2/2}\sum_{n=0}^{\infty}\frac{\alpha^n}{\sqrt{n!}}\ket{n}.
\end{equation}
Since this is a pure state, it is associated with a density matrix whose elements are given by,
\begin{equation}
    \rho_{mn}=\bra{m}\ket{\alpha}\bra{\alpha}\ket{n},
\end{equation}
which satisfies the projection relation $\rho^2=\rho$ and has a zero {\it von Neumann} entropy, $S=-\Tr\{\hat\rho\ln\hat\rho\}=0$.

The complete decoherence of this state corresponds to an amortization of the off-diagonal elements of the associated density matrix, so that,
\begin{equation}
    \rho_{mn}^{dec}=|\bra{n}\ket{\alpha}|^2\delta_{mn}=e^{|\alpha|^2}\frac{|\alpha|^2}{(n-1)!}\delta_{mn}.
\end{equation}
The number of particles in the mixed state can be characterized by a Poisson distribution with an average number of particles $\bar{n}=|\alpha|^2$. The entropy that contains the mixed states is given by,
\begin{equation}
S_{dec}^{(cs)}= e^{-\bar{n}}\sum_{n=0}^{\infty}\frac{\bar{n}^n}{n!}(n\ln \bar{n}-\bar{n}-\ln n!),
\label{cs}
\end{equation}
where the subscript $(cs)$ indicates that the result is the same for a coherent state. This equation is identical to Eq.~[\ref{realentropy1}], meaning that the entropy will be given by the same expression as Eq.~[\ref{realentropy1}], and for $\bar n\gg1$,
\begin{equation}
  S_{dec}^{(cs)}=\left[\ln (2\pi e\bar n)-\frac{1}{6\bar n}+\mathscr{O}(1/\bar n^2)\right].
    \label{decoherence entropy}
\end{equation}
Thus, the number of density matrix elements that contribute to the entropy calculation is given by $\Delta n=\sqrt{\bar{n}}$ due to the Poisson distribution.

\newpage

The energy of a single quantum harmonic oscillator at equilibrium temperature $T$ is given by,
\begin{equation}
    S_{eq}=\ln (\bar{n}+1)+\bar{n}\ln \left(1+\frac{1}{\bar{n}}\right),
    \label{equilibrium entropy}
\end{equation}
\begin{wrapfigure}{r}{0.5\textwidth}
\centering
\includegraphics[width=0.8\linewidth]{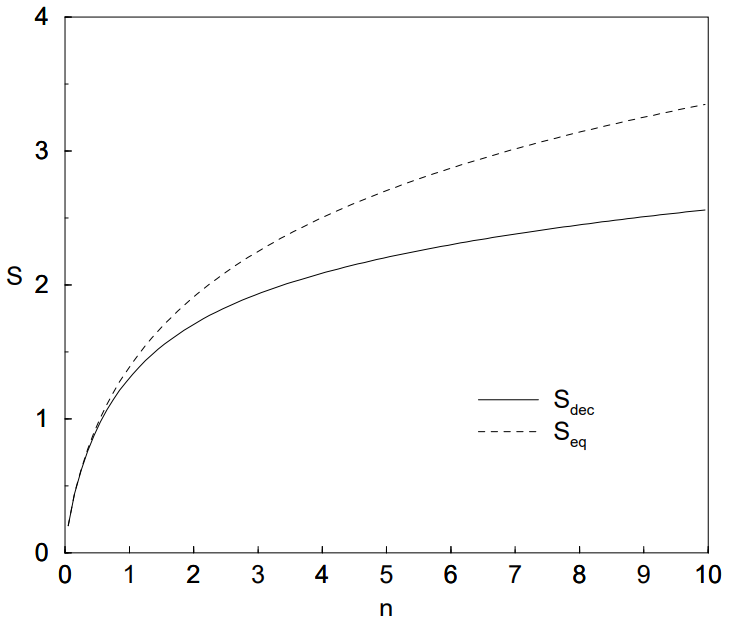}
\caption{Decoherence entropy $S_{dec}$ for a single mode $\vec{k},\lambda$ and equilibrium entropy $S_{eq}$ for the same average energy and occupation number $\bar{n}$. Reproduced from \cite{sdeco}.}
\label{deco}
\end{wrapfigure}
so that the average particle occupation number is given by $\bar{n}=(e^{\omega/T}-1)^{-1}$. Considering a large $\bar{n}$, asymptotically, one obtains $S_{eq}\approx 2S_{dec}^{(cs)}$ (Fig.~[\ref{deco}]), and the thermal entropy becomes twice as large as the decoherence entropy. However, for intermediate or average values of the occupation number, $S_{dec}^{(cs)}/S_{eq}$ is close to unity ($\sim 0.75$ for $\bar{n}=10$). In this theoretical model, it is stated that, for not so large values of the average occupation, the decoherence process is considerably responsible for a large fraction of the created entropy. Thus, the difference in entropy created in the reaction is due to processes that occur during the temporal evolution of the system generated in the initial heavy ion scattering. Clearly, decoherence is a phenomenon that occurs rapidly compared to thermal equilibrium processes; thus, these results imply that the high entropy generation rate observed in heavy ion collisions is primarily due to the effects of decoherence of the initial color fields.

\chapter{Results and Conclusion}
\label{chapter 6: Results}

In this chapter, the results developed in this research are presented, involving phenomenological investigations into the properties of entanglement entropy, its version in the CGC model, dynamic entropy in QCD, and decoherence entropy.

The first results concern the comparison between the entanglement entropy models in the KL and CGC models in the investigation of collisions where nucleons are the targets\cite{ramos2020a}. The entanglement entropy in the KL model is compared with data from hadronic entropy in $pp$ collisions at the LHC and $ep$ from the H1 Collaboration\cite{ramos2024investigating}. This work is extended to nuclear targets. Continuing the research on entanglement entropy, the case of elastic collisions is investigated using the model-independent Lévy femtoscopy method, obtaining results for typical RHIC, Tevatron, and LHC energies\cite{ramos2020b}. Finally, using different UGD models, it is possible to compute the dynamic entropy of QCD for nucleons\cite{ramos2022} and for $eA$ collisions. These results are compared with decoherence entropy. At the end of the chapter, the conclusions of this doctoral thesis will be presented.

\section{Entanglement Entropy at High Energies in DIS for pp and ep Collisions}
\label{chapter 6 - section 1: entanglement entropy of protons and nuclei}

The expressions for entanglement entropy given by [\ref{2.2.21}] and [\ref{(3+1)-dimensional entanglemente entropy I}] depend on the gluon PDFs. The PDF model from reference \cite{golec1999} contains a phenomenological analytical expression for this distribution, valid both for large values of photon virtuality ($Q^2 \leq 50$ GeV$^2$) and small values ($Q^2 \ll 1$ GeV$^2$). This model is advantageous compared to the usual PDFs extracted from fitting with initial conditions at approximately $Q^2 = Q_0^2 \approx 2$ GeV$^2$. Another convenience of this expression is that it is an explicit function of the saturation scale $Q_s$. Starting from the GBW saturation model, it is possible to obtain a non-integrated gluon distribution:
\begin{equation}
    \alpha_s\mathcal{F}(x,k) = \frac{N_0 k^2}{Q_s^2} e^{-k^2/Q_s^2},
    \label{gbw ugd}
\end{equation}
with $N_0 = 3\sigma_0/4\pi^2$. The PDF can be obtained from the integral:
\begin{equation}
\begin{split}
    \expval{n} \equiv x f_g(x,Q^2) &= \int_0^{Q^2} dk^2 \mathcal{F}(x,k) \\
    &= \frac{3\sigma_0}{4\pi^2 \alpha_s} Q_s^2 \left[1 - \left(1 + \frac{Q^2}{Q_s^2}\right) e^{-\frac{Q^2}{Q_s^2}}\right].
\end{split}
\label{analytical gluon distribution by gbw}
\end{equation}
In this expression, the saturation scale $Q_s$ is given by Eq.~[\ref{saturation scale}], and the values used for the parameters are $\sigma_0 = 27.32\ \text{mb}$, $\lambda = 0.248$, and $x_0 = 4.2 \times 10^{-5}$, as adjusted in reference \cite{valores}.
\begin{wrapfigure}{r}{0.57\textwidth}
\begin{center} 
\includegraphics[width=0.85\linewidth]{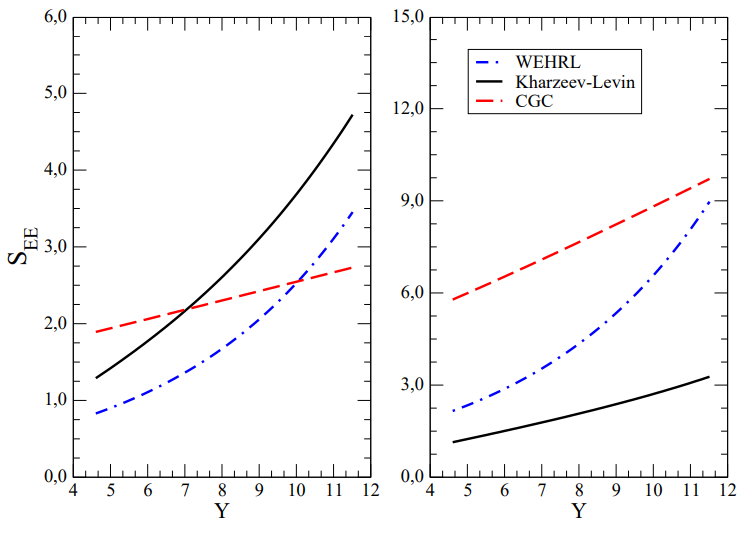} 
\end{center} 
\caption{Comparison of different models for partonic entropy at high energies. The entropy is expressed as a function of Bjorken $x$ for virtualities of $Q^2 = 2$ GeV$^2$ (left) and $Q^2 = 10$ GeV$^2$ (right) in a proton DIS. The results are obtained for entanglement entropy in the KL Model, CGC model, and Wehrl model. Reproduced from \cite{ramos2020a}.} 
\label{result1figure1}
\end{wrapfigure}

Initially, analyzing the dependence of the entropic concepts studied with rapidity $Y$, the entropy in the KL model, specifically Eq.~[\ref{(3+1)-dimensional entanglemente entropy II}], is evaluated and compared with the expressions for Wehrl entropy [\ref{final wehrl entropy}] (studied in the dissertation \cite{dissertacaogabriel} and briefly exposed in Appendix \ref{apendice E: qcd wehrl entropy}) and the CGC model [\ref{CGC entanglement entropy III}]. The results are shown in Fig.~[\ref{result1figure1}]. The value for the strong coupling constant is $\alpha_s=0.25$. Analyzing the different entropy models, it is observed that both for the CGC entanglement model and the partonic version of Wehrl entropy, they are proportional to the transverse area of the target. This is an intrinsic property of an extensive observable like entropy. This property is not accounted for in the entanglement entropy expression in the KL model [\ref{2.2.21}]. This expression behaves as $S\sim Y^2$ with a logarithmic suppression in $1/Q^2$. The choice $r^2=4/Q^2$ for the dipole's mean size is used for the product inside the logarithm, such that $Q_s^2r=4Q_s^2/Q^2+e$ (the second term is used to avoid negative values when $Q_s^2\ll Q^2$). On the other hand, the Wehrl entropy behaves as $S_W\sim e^{\lambda Y}$ and grows with $Q^2$ with the simplification performed in the $k$ integration, which is sufficient for this phenomenological analysis. Finally, the entanglement entropy in the CGC model behaves as $S_{CGC}\sim e^Y(\ln^2Q^2-2\lambda Y)$.

\begin{figure}[ht]
\centering
\includegraphics[width=\linewidth]{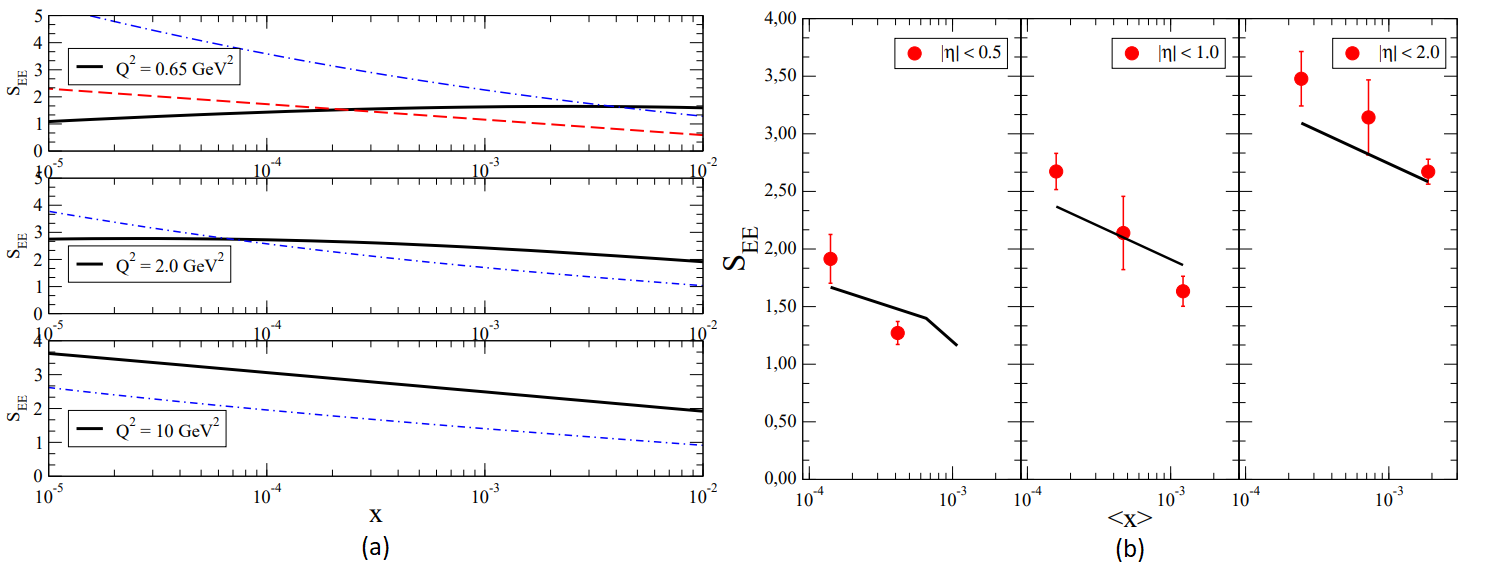}
\caption{{\bf (a)}: Entanglement entropy as a function of the Bjorken $x$ for the virtualities $Q^2=0.63$, $2$, $10$ GeV$^2$ in a proton DIS. The maximum entropy for the $Q^2=0.63$ scale is represented by the red line (dashed line). The parametric expression [\ref{(3+1)-dimensional entanglemente entropy II}] is also presented (dotted lines).
{\bf (b)}: Entanglement entropy in $pp$ collisions compared with the final hadronic entropy determined for different pseudo-rapidity domains at the LHC (the points with $|\eta|<0.5,1.0,2.0$) obtained from \cite{tu2020}. The numerical result is represented by solid lines. Adapted from \cite{ramos2020a}.}
\label{result1figure2}
\end{figure}

In Fig.~[\ref{result1figure2}], the results for entanglement entropy are presented using the analytical gluon PDF in high energies [\ref{gbw ugd}] as a function of $x$ ($10^{-5}\leq x \leq 10^{-2}$) for some specific values of photon virtuality. The analytical model allows an analysis at the soft scale, $Q^2=0.65$ GeV$^{-2}$, a regime that cannot be evaluated by DGLAP evolution [\ref{DGLAP}], since this equation generally starts to be evaluated at $Q_0^2\sim 2$ GeV$^2$. Results for the virtualities $Q^2=2$ GeV$^2$ and $Q^2=10$ GeV$^2$ are also presented. The transition between soft and hard scales is quite clear. The advantage of using the analytical model for $xf_g$ is the ability to check previous behaviors in terms of the scale variable $\tau=Q^2/Q_s^2$. For $\tau \ll 1$, a series expansion results in $xf_g\propto Q^4/Q_s^2$, so $S\propto-\ln Q_s^2$. For small values of $x$, one obtains $S\sim \lambda \ln x$. When $\tau=1$, $xf_g\propto [1-2/e]Q_s^2$, with the entropy $S\sim -\ln(x)-1$, causing the curve to change its inflection point in the transition region $Q^2\approx Q_s^2$. In the hard regime, when $Q^2\ll Q_s^2$, the asymptotic behavior is given by $xf_g \propto Q_s^2$ and $S\sim-\lambda \ln(x)$. This is seen for large values of $x$ in the $Q^2=2$ GeV$^2$ graph and for all $x$ in the $Q^2=10$ GeV$^2$ case.

The determination of entanglement entropy from experimental data is carried out in \cite{tu2020}. For a small-$x$ proton DIS in the energy range $\sqrt{s_{ep}}\approx 225$ GeV for DESY-HERA. The authors used the Monte Carlo method for the multiplicity distribution in order to obtain the entropy of the hadrons in the final stages with the hadronic entropy $S_{h}$, comparing it with the entanglement entropy, showing that both are uncorrelated at $Q^2=2$ and $Q^2=10$ GeV$^2$. In both virtualities, the result is independent of $\expval{x}$ with $S_h \approx 1.5$, in contrast to the power-law behavior of the entanglement entropy. The result of the model proposed in this thesis, observed in {\bf (b)} of Fig.~[\ref{result1figure2}] for $Q^2=2$ GeV$^2$, is similar to the one obtained by the authors of \cite{tu2020}. Table [\ref{tabela 1 artigo 1}] presents the entanglement entropy given by Eq.~[\ref{2.2.21}] using the scale $Q^2=Q_s^2(x)$ and employing the same procedures as in \cite{tu2020} to compare with the entropy $S_h$. The choice for the hadron rapidity $y$ is made based on the different experimental cuts of the hadron multiplicity distribution with pseudo-rapidity $\eta$. Thus, $S_h$ is obtained from experimental data from the CMS collaboration \cite{khachatryan2010}, which is consistent with ATLAS and ALICE.

\begin{table}[ht]
\centering
\caption{Entanglement entropy in $pp$ collisions at the LHC predicted by the gluon saturation PDF using the procedures from \cite{tu2020}. Some values extracted from CMS data are also presented (in parentheses) \cite{tu_private}. Adapted from \cite{ramos2020a}.}
\label{tabela 1 artigo 1}
\begin{tabular}{|c|c|c|c|c|c|}
\hline

$\sqrt{s_{pp}}$(TeV)  & $|y|<0.5$                  & $|y|<1.0$                  & $|y|<1.5$ & $|y|<2$                   & $|y|<2.4$  \\ \hline
$7.00$                & $1.668 (1.914 \pm 0.212)$  & $2.368 (2.673 \pm 0.157)$  & $2.787$   & $3.093 (3.478 \pm 0.236)$ & $3.291$ \\
$2.36$                & $1.398 (1.271 \pm 0.099)$  & $2.100 (2.139 \pm 0.318)$  & $2.517$   & $2.823 (3.142 \pm 0.326)$ & $3.022$ \\
$0.90$                & $1.160$                    & $1.860 (1.633 \pm 0.130)$  & $2.277$   & $2.584 (2.671 \pm 0.108)$ & $2.784$ \\ \hline

\end{tabular}
\end{table}

Thus, the analytical expression for the entanglement entropy is given by,
\begin{equation}
    S(Q^2=Q_s^2)=\ln[Q_s^2(x)]+S_0,
    \label{final entanglement entropy}
\end{equation}
with $S_0=\ln[3(e-2)R_p^2/4e\pi\alpha_s]\approx 2$ for $\alpha_s=0.2$ and $S=S_0$ when $Q_s^2=1$ GeV$^2$. In Fig.~[\ref{result1figure2}] {\bf (b)}, the results show good agreement between the entanglement entropy obtained and the entropy reconstructed from the hadronic multiplicity for small-$x$ values.

Finally, the KL entanglement entropy was calculated using the GBW gluon PDF [\ref{analytical gluon distribution by gbw}], comparing it with the H1 collaboration data \cite{H1} for $ep$ collisions. The data provide the hadronic entropy in the final state, derived from the charged multiplicity distributions for pseudo-rapidity $\eta$ in the hadronic center-of-mass frame, restricted to the range $0<\eta<4$. Furthermore, the H1 collaboration measured the hadronic entropy in four photon virtuality ranges: $5<Q^2<10$, $10<Q^2<20$, $20<Q^2<40$, and $40<Q^2<100 \operatorname{GeV^2}$.

\begin{figure}[ht]
\centering
\includegraphics[width=0.8\linewidth]{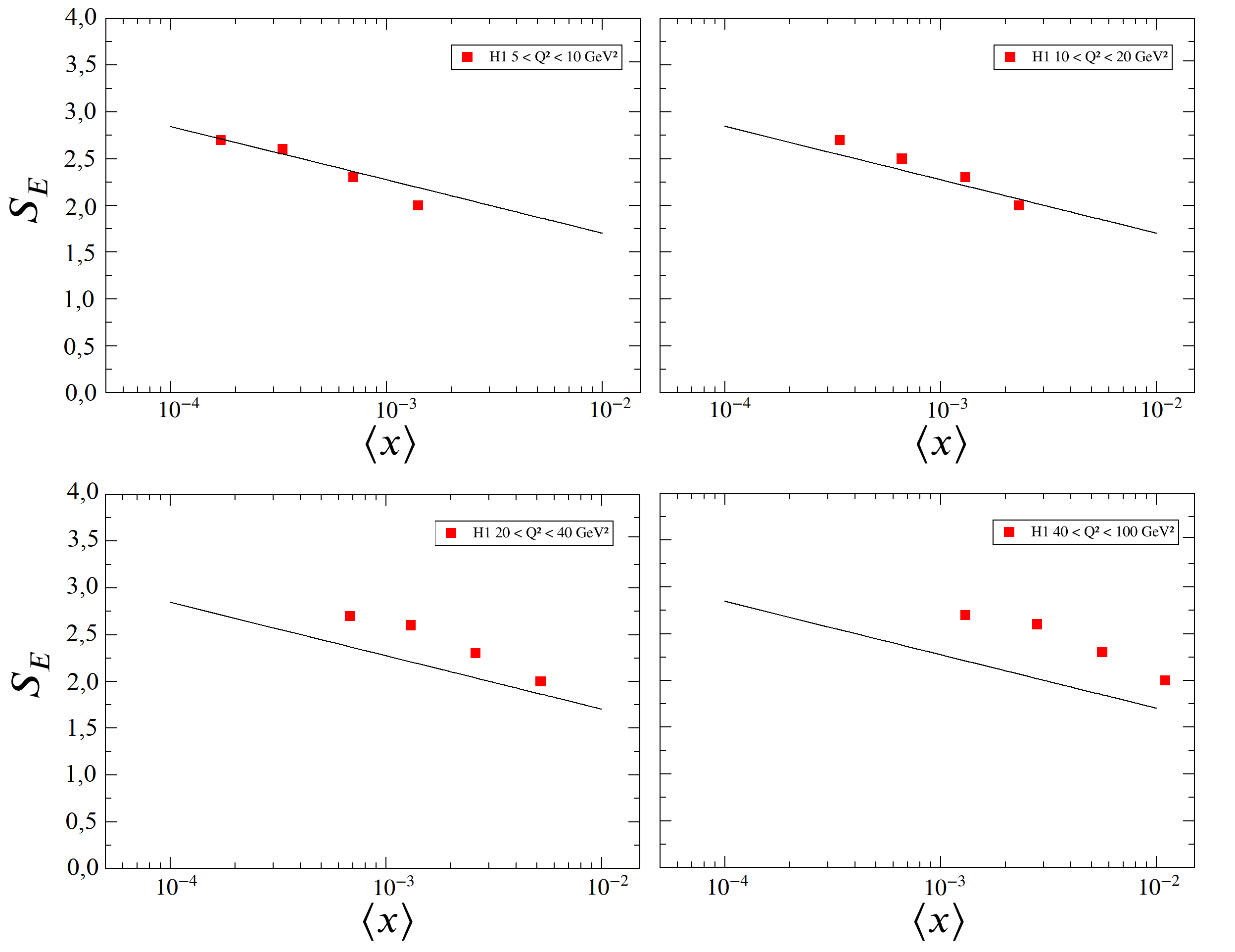}
\caption{Partonic entanglement entropy versus Bjorken-$x$. The results are contrasted with the hadronic entropy derived from the charged multiplicity distributions measured by the H1 collaboration \cite{H1}. The numerical results from this work are represented by solid lines. Reproduced from \cite{ramos2024investigating}.}
\label{figureh1}
\end{figure}

To compare the KL expression with the data, it is necessary to adapt the entanglement entropy formula to include the contribution from $Q^2$. One way to perform this procedure is provided in \cite{hentschinski2022maximally}, where the authors relate the KL entanglement entropy to the entropy of final states, pointing out several uncertainties in the current comparison with the data. In particular, they highlight the global normalization, the relation between charged hadron multiplicity versus total multiplicity in comparison with experimental results, as well as different methods to determine the number of partons in a DIS. They also included the sea quark PDF, $xf_{sea}(x,Q^2)$, modifying Eq.~(\ref{1+1 ocupation}) to $\expval{n}=xf_g(x,Q^2)+xf_{sea}(x,Q^2)$. Here, we will only keep the gluon contribution to make use of the analytical GBW expression for the gluon PDF. Following \cite{hentschinski2022maximally}, the treatment for the measurements in $Q^2$ ranges will be given by:
\begin{equation}
\expval{n(x,Q^2)}_{Q^2}=\frac{1}{Q^2_{\text{max}}-Q^2_{\text{min}}}\int_{Q^2_{\text{min}}}^{Q^2_{\text{max}}} dQ^2 xf_g(x,Q^2)
\label{kutak PDF}
\end{equation}
After the modifications in Eq.~(\ref{2.2.21}), the final expression to evaluate the KL entanglement entropy, to be compared with the H1 data analysis, is:
\begin{equation}
\expval{S(x,Q^2)}_{Q^2}=\ln\expval{n(x,Q^2)}_{Q^2}
\label{efective entanglement entropy}
\end{equation}
The results are presented in Fig.~[\ref{figureh1}]. In general, the results fit well, except for the data in the range $40<Q^2<100 \operatorname{GeV^2}$. The reason for this is that the analytical GBW formula is valid up to $50 \operatorname{GeV^2}$. A DGLAP evolution is needed in this kinematic region.

\section{Entanglement Entropy at High Energies in DIS for Nuclei}
\label{chapter 6 - section 2: entanglement entropy of nuclei}

In this section, the partonic entanglement entropy for nuclear targets is discussed. To simplify the analysis of nuclear DIS, the geometric scaling property is considered in the partonic saturation approximation. Thus, the cross-section of an $eA$ DIS at small-$x$ is directly related to the cross-section for a proton target. Nuclear effects are absorbed by the nuclear saturation scale $Q_{s,A}^2$,
\begin{equation}
Q_{s,A}^2(x,A)=\left(\frac{A\pi R_p^2}{\pi R_A^2}\right)^{\Delta}Q_s^2(x) \sim A^{4/9}Q_s^2(x),
\label{nuclear saturation scale}
\end{equation}
with $\Delta\approx 1.27$\cite{armesto2005}, and the rescaled cross-section in relation to the $ep$ case is given by the substitution,
\begin{equation}
    \sigma_A \rightarrow \frac{\pi R_A^2}{\pi R_p^2}\sigma_0 \sim A^{2/3}\sigma_0,
    \label{nuclear normalization}
\end{equation}
where the nuclear radius is given by $R_A\approx 1.12 A^{1/3}$ fm. Thus, the simplest extension of the nuclear gluon distribution is,
\begin{equation}
    xf_{g,A}(x,Q^2)=\frac{3R_A^2}{4\pi \alpha_s}Q_{s,A}^2\left[1-\left(1+\frac{Q^2}{Q_{s,A}^2}\right)e^{-\frac{Q^2}{Q_{s,A}^2}}\right]
    \label{analytical gluon distribution by gbw for nuclei}
\end{equation}

\begin{figure}[ht]
\centering
\includegraphics[width=0.8\linewidth]{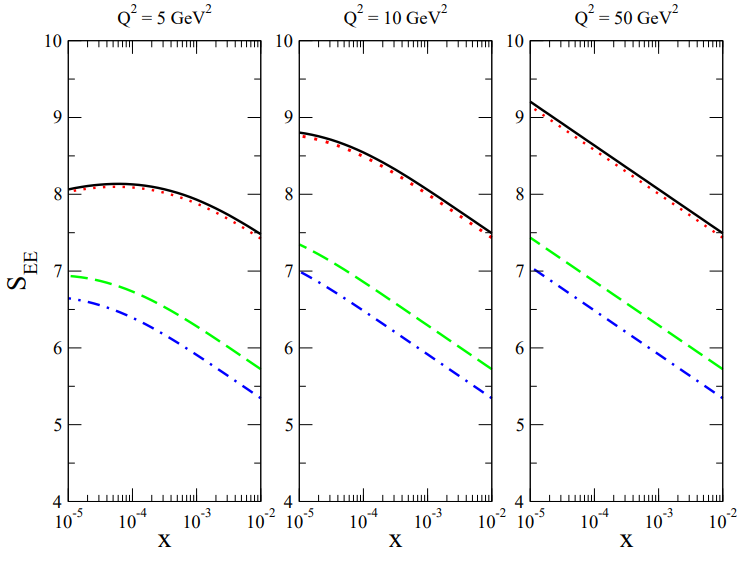}
\caption{Nuclear entanglement entropy as a function of $x$ for virtualities $Q^2=5$, $10$, $50$ GeV$^2$ in a nuclear DIS. For each virtuality, the following nuclei were examined: $Pb$ (solid lines), $Au$ (dashed lines), $Ca$ (long dashed lines), and $Si$ (dotted lines). Reproduced from \cite{ramos2020a}.}
\label{result1figure3}
\end{figure}

In Fig.~[\ref{result1figure3}], the results for the nuclear entanglement entropy are presented based on the parametrization for the nuclear PDF [\ref{analytical gluon distribution by gbw for nuclei}]. The virtualities $Q^2=5$, $10$, $50$ GeV$^2$ were considered, and the following nuclei: lead ($Pb$), gold ($Au$), calcium ($Ca$), and silicon ($Si$). The $Pb$ and $Au$ nuclei will be investigated in future electron-ion colliders such as LHeC and eRHIC. The case $Q^2=2$ is interesting because the saturation scale is enhanced by a factor of $A^{4/9}$ compared to the proton case. This factor is $10$ for lead ($A=208$) and $5$ for calcium ($A=40$). Therefore, in the model used, the scale $Q_{s,A}^2$ is on the order of $2$ GeV$^2$ for $x\approx 10^{-2}$ for $Pb$ and $x\approx 10^{-3}$ for $Ca$, while for the proton case this happens at $x\sim 10^{-5}$ (Fig.~[\ref{result1figure2}] {\bf (a)}). This means that the entanglement entropy reaches its maximum value at a higher $x$ value compared to nuclear DIS due to the rapid saturation in the nuclear case.

\section{Entanglement entropy in elastic scattering using hadronic femtoscopy}
\label{chapter 6 - section 3: entanglement entropy in elastic scattering}

In references \cite{peschanski2016,peschanski2019}, the calculation of entanglement entropy in elastic collisions was performed as developed in Section [\ref{chapter 4 - section 1 - subsection 3: entanglement entropy of elastic scattering}]. However, in order to obtain the expression for entanglement entropy [\ref{volume regularization entanglement entropy}], the authors, in addition to performing a volume regularization, had to make some assumptions regarding the dependence of physical observables on the Mandelstam variable $t$, in this case, the diffraction peak approximation in hadron-hadron scattering.

It is possible to obtain an expression for entanglement entropy \cite{peschanski2016,peschanski2019} without considering any assumptions for the dependence on $t$, such as the diffraction peak. However, for this, model-independent femtoscopy models are used, more specifically, the model-independent {\it Lévy} femtoscopic method for elastic collisions (Appendix \ref{apendice D: model-independent femtoscopic lévy imaging for elastic scattering}).

The objective of this section is to present the results obtained for the entanglement entropy using the {\it Lévy} femtoscopy model \cite{ramos2020b}. Not only high-energy data were considered, but also lower-energy regimes. {\it Lévy} expansions up to the fourth order were used for the $pp$ scattering data measured in the energy domain of the ISR ($\sqrt{s}=23.5$, $30.7$, $44.7$, $52.8$, and $62.5$ GeV). Furthermore, for $p\bar{p}$ collisions, the {\it Lévy} expansions used were up to the second degree for energies $\sqrt{s}=53$ GeV (ISR) and $\sqrt{s}=1960$ GeV (D0, Tevatron), and expansions up to the third order for $\sqrt{s}=546$ GeV and $\sqrt{s}=630$ GeV (UA4). For the LHC energies, the {\it Lévy} expansions were performed up to the fourth order for all differential cross-section measurements in elastic $pp$ collisions for $7$ and $13$ TeV. The parameters for the expansion, $R$, $\alpha$, and the complex coefficients $c_i$ are available in Appendices A and B of the reference \cite{apendix}. In any case, typically $\alpha\approx0.9$ and $R\approx 0.6$-$0.7$ fm are used.

A part of the results is presented in Table [\ref{tabela 1 artigo 2}], using the three regularization methods proposed in reference \cite{peschanski2016}, originally with $\sqrt{s}=1.8$, $7$, $8$, and $13$ TeV. The measured values for the total and elastic cross-sections are also presented. Predictions were added for RHIC energies, $0.2$ TeV, and LHC with $2.76$ TeV, as well as recent results for $\sigma_T$ and $\sigma_{\text{el}}$ in $pp$ collisions at RHIC with $\sqrt{s}=200$ GeV.

\begin{figure}[ht]
\centering
\includegraphics[width=0.7\linewidth]{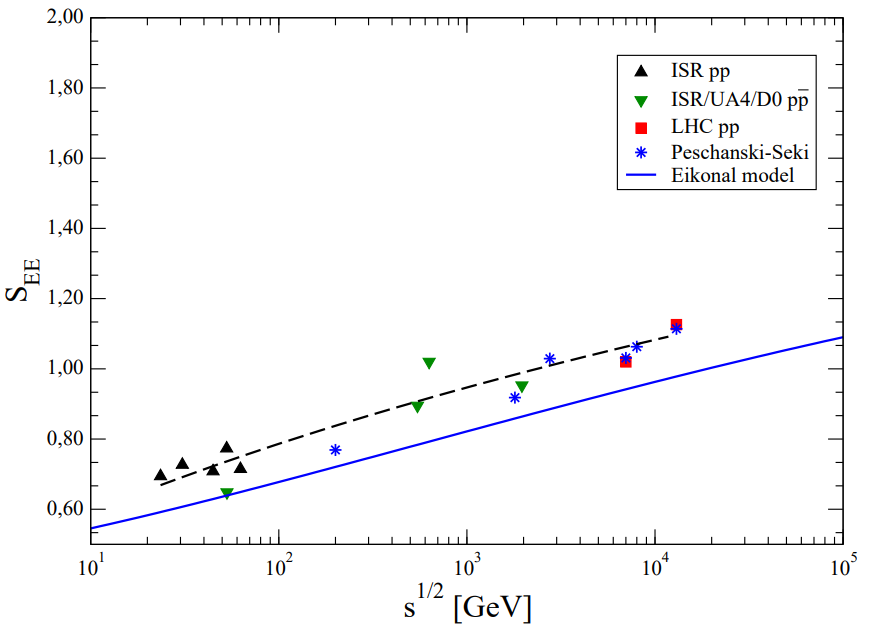}
\caption{Entanglement entropy for elastic collisions as a function of the center-of-mass collision energy $\sqrt{s}$. The extraction obtained using the {\it Lévy} method is presented at both low and high energies and compared with the results from \cite{peschanski2019}. The values for LHC, Tevatron, and RHIC are given in Table [\ref{tabela 1 artigo 2}]. The prediction for the diffractive peak approximation using a channel in the eikonal model is presented (solid line). A fit based on the single pole contribution of {\it Regge} for the light Pomeron is also presented (dashed line). Reproduced from \cite{ramos2020b}.}
\label{result2figure1}
\end{figure}

In Fig.~[\ref{result2figure1}], the results for the entanglement entropy extracted as a function of the center-of-mass energy of the collision are presented using volume regularization [\ref{volume regularization entanglement entropy}], with the {\it Lévy} femtoscopy methodology. The $pp$ collision data for low energies in the ISR are labeled with upward triangles, while the $p\bar{p}$ collision data from ISR, UA4, and D0 are represented by inverted triangles. The TOTEM-LHC data with energies of $7$ and $13$ TeV, represented by squares, are displayed along with the entanglement entropy from reference \cite{peschanski2019}, marked as stars in the graph.

The calculation of entanglement entropy using the {\it Lévy} methodology resulted in high values using volume regularization due to the additional contribution at large $t$, which is suppressed in the diffractive peak approximation. However, the deviation is not as large for the small $t$ approximation, and it can be considered a compatible extraction of entanglement entropy. It is reiterated that the regularization using the step function consumed more machine time due to the oscillating integrand in Eq.~[\ref{diferential elastic cross section with cutoff step regularization}].

\begin{table}[ht]
\centering
\caption{The entanglement entropy determined using the model-independent {\it Lévy} method compared with the diffractive peak approximation presented in \cite{peschanski2019}. Results are also shown for three types of regularization (volume regularization, and step function/gaussian cutoffs). Predictions for $0.2$ TeV (RHIC) and $2.76$ TeV (LHC), which do not originally appear in \cite{peschanski2019}, are calculated. Adapted from \cite{ramos2020b}.}
\label{tabela 1 artigo 2}
\begin{tabular}{|c|c|c|c|c|c|}
\hline

$\sqrt{s_{pp}}$ (TeV) & {\it Lévy} & Volume Reg. & Experimental Data [$\sigma_T$, $\sigma_{\text{el}}$] (mb) & Heaviside & Gaussian   \\ \hline
$13.00$               & $1.126$    & $1.114$        & [$110.6 \quad 3.4, \quad 31.0 \quad 1.7$  ]          & $1.212$            & $0.8621$       \\
$8.00$                & -          & $1.063$        & [$101.7 \quad 2.9, \quad 27.1 \quad 1.4$  ]          & $1.197$            & $0.7965$       \\
$7.00$                & $1.020$    & $1.031$        & [$98.0  \quad 2.5, \quad 25.1 \quad 1.1$  ]          & $1.192$            & $0.7539$       \\
$2.76$                & -          & $1.029$        & [$84.7  \quad 3.3, \quad 21.8 \quad 1.4$  ]          & $1.144$            & $0.7509$       \\ 
$1.80$                & $0.953$    & $0.918$        & [$72.10 \quad 3.3, \quad 16.6 \quad 1.6$  ]          & $1.193$            & $0.6009$       \\ 
$0.20$                & -          & $0.769$        & [$54.67 \quad 1.89,\quad 10.85\quad 1.103$]          & $1.103$            & $0.3909$       \\ \hline

\end{tabular}
\end{table}

\newpage

\section{The Dynamical Entropy of QCD}
\label{chapter 6 - section 4: QCD dynamical entropy in high energy collisions}

The dynamical entropy of dense states in high energy $pp$ collisions can be studied using phenomenological models for the UGDs. In this way, it is possible to obtain the transverse momentum probability distributions evaluated in terms of rapidity. The dynamical entropy is evaluated as a function of $\Delta Y=Y_0-Y$, where $Y_0$ is the initial rapidity. 

Four UGDs were analyzed in the investigation of dynamical entropy: the Gaussian CGC distribution\cite{kutak2011} $\phi_{\text{Gaus}}(Y, k)$, the phenomenological proposal considering the production of charged hadrons in $pp$ collisions combined with Tsallis-type distributions\cite{moriggi2020} $\phi_{\text{MPM}}(Y, k)$, the model that considers both the initial CGC conditions by McLerran-Venugopalan and the Levin-Tuchin solutions in a broad transverse momentum distribution domain\cite{siddiqah2017} $\phi_{\text{LT}}(T,k)$, and the numerical proposal from reference \cite{sapeta2018} $\phi_{\text{KS}}(Y,k)$. However, initially, it is necessary to obtain the probability distributions for each of the aforementioned UGDs.

\subsection{The Probability Distributions of the QCD Dynamical Entropy}
\label{chapter 6 - section 4 - subsection 1: the probability distribution of qcd dynamical entropy}

Now, we demonstrate in detail the procedures used for calculating the dynamical entropy in the case of the Gaussian UGD, which can be solved analytically. Except for the $\phi_{\text{KS}}(Y,k)$ proposal, the other distributions were obtained using a similar methodology, so this demonstration follows a more pedagogical approach.

Initially, to obtain the probability distributions of the {\it momentum}, it is necessary to normalize the UGD, because, from Eq.~[\ref{kdist}],
\begin{equation}
    P(\tau)=\frac{\phi(\tau)}{\int d\tau \phi(\tau)}=\frac{\phi(\tau)}{N} \quad \therefore \quad N=\int d^2k \phi (\tau).
    \label{ugd probalistic distribution}
\end{equation}
Remembering that $\tau=k^2/Q_s^2(x)$.

The Gaussian QCD UGD from reference \cite{golec} is given by,
\begin{equation}
    \phi(\tau)=\frac{C_F A_{T}}{4\pi^2\alpha_s}\tau e^{-\tau/2}\equiv C\tau e^{-\tau/2} \quad \therefore \quad C\equiv \frac{C_F A_T}{4\pi^2\alpha_s}.
    \label{gaussian ugd}
\end{equation}
In this equation, $A_T$ is the transverse area of the proton. Thus, the normalization factor can be obtained through a trivial variable substitution,
\begin{equation}
\begin{split}
N &= \int_{-\infty}^{+\infty}\int_{-\infty}^{+\infty}dk_xdk_y\phi(k^2,x)=2\pi\int_{0}^{\infty}dkk\phi(\tau)  \\
&=\pi Q_s^2(x)\int_{0}^{\infty}\phi(\tau)d\tau=\pi Q_s^2(x)C\underbrace{\int_{0}^{\infty}\tau e^{-\tau/2}d\tau}_4 \\
\end{split}
\label{gaussian normalization}
\end{equation}
Thus, $N=4\pi C Q_s^2(x)$. Therefore, the distribution for the Gaussian case $P_{\text{gaus}}(\tau)$ is given by,
\begin{equation}
    P_{\text{gaus}}(\tau)=\frac{\tau e^{-\tau/2}}{4\pi Q_s^2}.
    \label{gaussian ugd probalistic distribution}
\end{equation}

For the case of the UGD $\phi_{\text{MPM}}$, given by,
\begin{equation}
\phi_{\text{MPM}}=\frac{3\sigma_0}{4\pi^2\alpha_s}\frac{\tau \beta(\tau)}{(1+\tau)^{1+\beta(\tau)}},
\label{mpm ugd}
\end{equation}
in this equation, $\alpha_s=0.2$, with $Q_s^2(Y)=k_0^2e^{0.33Y}$ and $k_0^2=\bar{x}_0^{0.33}$ GeV$^2$. The power-law behavior of the gluons produced in the high {\it momentum} spectrum is determined by the function $\beta(\tau)=a\tau^b$. The set of parameters $\sigma_0$, $\bar{x}_0$, $a$, and $b$ is fitted from data obtained from a small-$x$ DIS \cite{moriggi2020},
\begin{equation}
    \begin{cases}
    \sigma_0=20.47 \, \text{mb};\\
    \bar{x}_0=3.52\times10^{-5};\\
    a=0.055;\\
    b=0.204.
    \end{cases}
    \label{mpm fitting}
\end{equation}

Analogously to the procedure followed for obtaining the Gaussian distribution, for the case of $\phi_{\text{MPM}}$, the distribution is obtained as,
\begin{equation}
    P_{\text{MPM}}(\tau)=\frac{1}{\pi Q_s^2\xi}\frac{\tau \beta(\tau)}{(1+\tau)^{1+\beta(\tau)}},
    \label{mpm ugd probabilistic distribution}
\end{equation}
where $\xi=4.34618$ is a constant derived from numerical integration [\ref{ugd probalistic distribution}].

In order to analyze information from the transverse momentum distributions (TMD) of the gluon, which carry more information about the correct theoretical behavior for both large and small momenta, the distribution $\phi_{\text{LT}}$ is considered. It is derived from a general solution form $\phi(Y,k)$ that reproduces both the initial conditions of the MV model and the Levin-Tuchin (LT) solutions in their appropriate limits. This distribution smoothly connects both limits and more closely approximates the first-order numerical solution of the BK equation, especially in the saturation region. In this limit, the TMD for the gluon tends to $0$. Initially, with the TMD of gluons for a small transverse momentum from the LT matrix $S$ solution, the UGD takes the following form in the region $Q_s > k > \Lambda_{QCD}$,
\begin{equation}
    \phi_{\text{LT}}^{\text{sat}}(Y,k)=-\frac{N_c A_T \epsilon}{\pi^3 \alpha_s}\ln \left(\frac{\tau}{4}\right)e^{-\epsilon \ln^2\left(\frac{\tau}{4}\right)},
    \label{lt satured ugd}
\end{equation}
so that $\phi_{\text{LT}}^{\text{sat}}(Y,k)$ is obtained for a small transverse momentum in terms of a series of Bell polynomials. This expression corresponds to the dominant logarithmic approximation for the resummed series with a constant $\epsilon \approx 0.2$, which arises due to the saddle point condition around the saturation edge. Outside the saturation boundaries ($k > Q_s$), but close to the saturation line, the QCD dipole amplitude in transverse space takes the form $N(r,Y)\approx (r^2Q_s^2)^{\gamma_s}$. In this limit, the TMD can be written as,
\begin{equation}
    \phi_{\text{LT}}^{\text{dil}}(Y,k) \propto \frac{N_c A_T \epsilon}{\pi^3 \alpha_s}\tau^{-\gamma_s}.
    \label{lt diluted ugd}
\end{equation}

Using the normalization procedures [\ref{ugd probalistic distribution}] and considering the details of the physical region of interest, the following distribution is proposed:
\begin{equation}
    P_{\text{LT}}(Y,k)=\begin{cases}
    -B\ln \left(\frac{\tau}{4}\right)e^{-\epsilon\ln^2(\tau/4)}, \quad \text{for} \quad \tau<1,\\
    B(d\tau)^{-\gamma_s}e^{-\epsilon^2(\tau/4)}, \quad \text{for} \quad \tau \leq 1,
    \end{cases}
    \label{lt ugd probabilistic distribution}
\end{equation}
where $d=(\ln 4)^{-1/\gamma_s}$ and $B\approx0.1/\pi Q_s^2$ are the normalization parameters.

Finally, the non-linear UGD model was used, based on the algorithm provided by the author in the reference \cite{sapeta2018}, applied in this work to verify the theoretical uncertainties associated with the other models used. This UGD is treated numerically with the mathematical formalism of dynamic entropy using algorithms made available by one of the authors.

\begin{figure}[ht]
\centering
\includegraphics[width=1\linewidth]{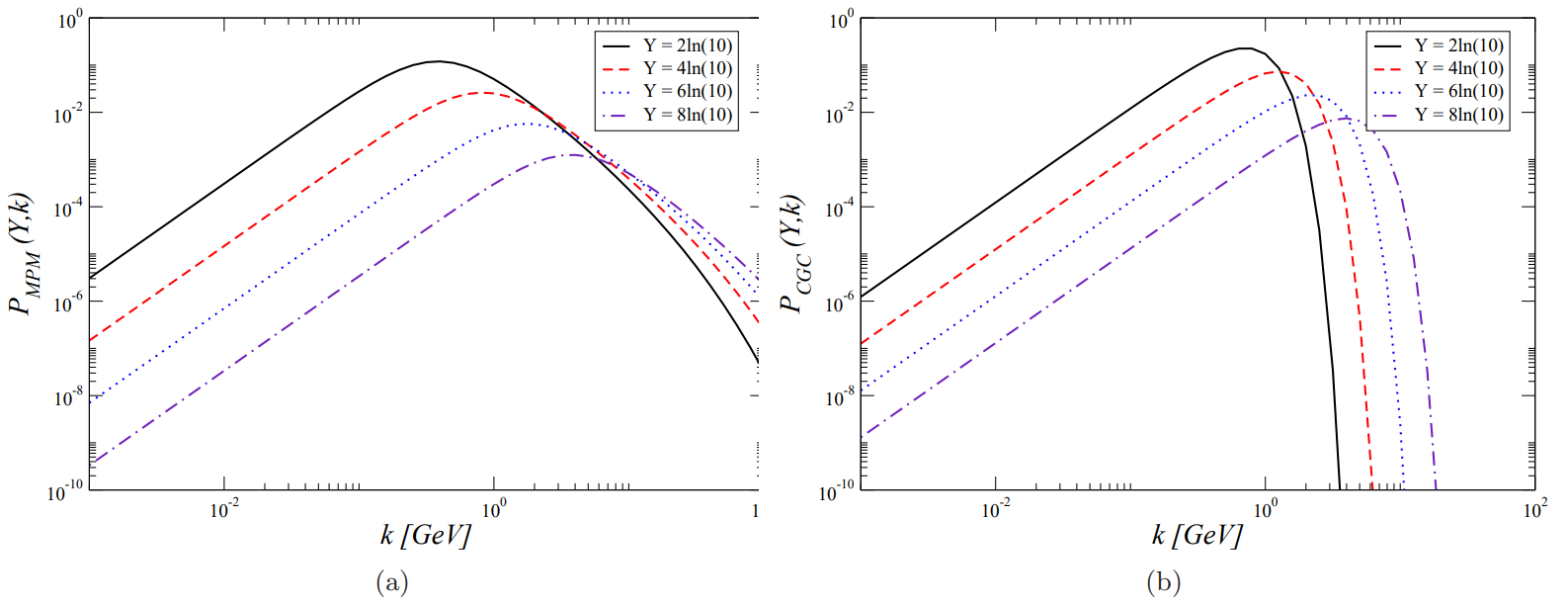}
\caption{The probability distribution, $P(Y,k)$, as a function of $k$ for fixed values of $Y=-\ln x$ ($x=10^{-8}$-$10^{-2}$). Results for the MPM model (left) and Gaussian CGC model (right). Reproduced from \cite{ramos2022}.}
\label{result3figure1}
\end{figure}

In Fig.~[\ref{result3figure1}], the transverse momentum distributions of gluons are shown for the MPM model (a) and the Gaussian CGC model (b). The dependence on $k$ is presented for various values of $Y$ ($2$, $4$, $6$, $8$, $\ln 10$), corresponding to the longitudinal momentum fractions of the gluons, $x=10^{-8}$-$10^{-2}$. Both models exhibit the geometric scaling property, $\phi(Y,k)\sim \phi(\tau=k^2/Q_s^2)R_s$, and their peak occurs at a transverse momentum proportional to the saturation scale. However, in each model, this peak is reached for different values: for the Gaussian CGC, it occurs at $k^{max}=\sqrt{2}Q_s(Y)$, while for the MPM model, $k^{max}\approx \sqrt{0.954}Q_s(Y)$.

\subsection{Results for QCD Dynamical Entropy and Decoherence Entropy}
\label{chapter 6 - section 4 - subsection 2: results for qcd dynamical entropy and the decoherence entropy}

\begin{figure}[ht]
\centering
\includegraphics[width=0.99\linewidth]{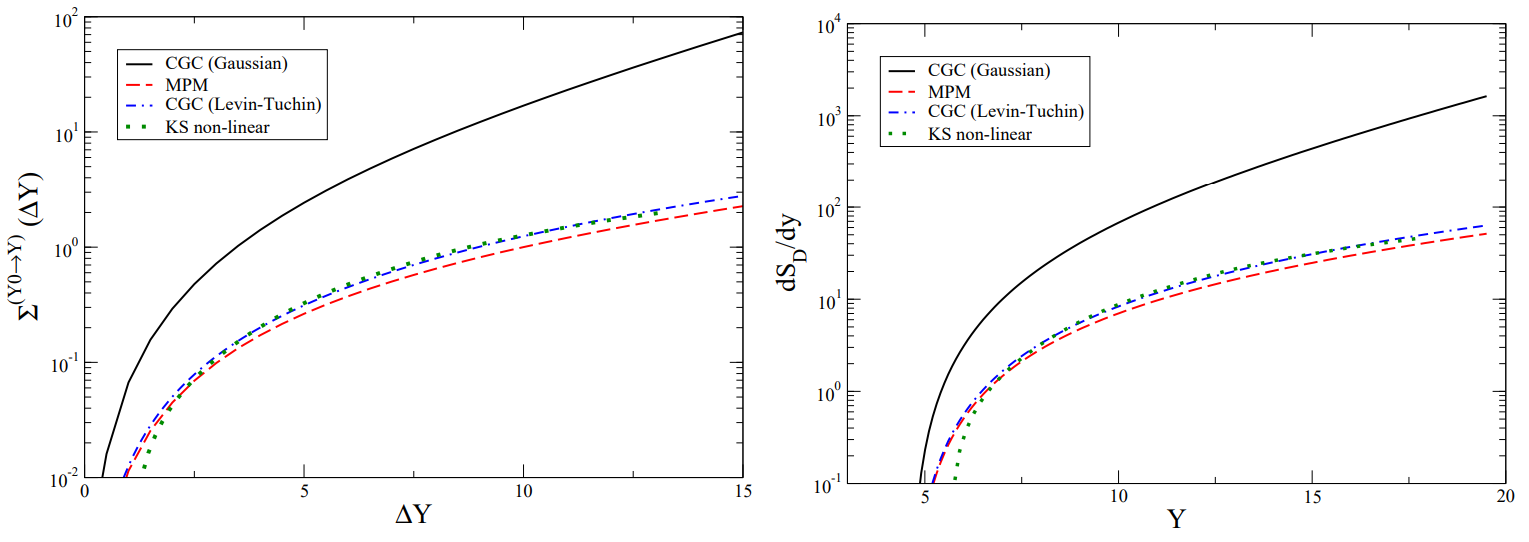}
\caption{({\bf left}): Dynamical entropy corresponding to a QCD evolution from rapidity $Y_0\rightarrow Y$, with $\Delta Y=Y-Y_0$. The initial rapidity is $Y_0=-\ln x_0$, with $x_0=10^{-2}$. Numerical results for the MPM model (dashed line), Gaussian CGC model (solid line), CGC LV (dashed line), and KS model (dots). ({\bf right}): Total dynamical entropy in a $pp$ collision for a QCD evolution from rapidity $Y_0\rightarrow Y$, in the domain $\Delta Y=[0,15]$. Reproduced from \cite{ramos2022}.}
\label{result3figure2}
\end{figure}

In this section, the results obtained for the QCD dynamical entropy with a rapidity $Y$ subject to the evolution $Y_0\rightarrow Y$ are presented. An initial rapidity of $Y_0\approx 4.6$ was considered. The values of $x\leq x_0$ correspond to the validity limit for the phenomenological application of the UGDs considered in this work. For this initial rapidity, the partons populate a transverse area proportional to the size $R_0(Y_0)=1/Q_s(Y_0)$.

The case of the Gaussian UGD model of the CGC can be treated analytically. Using the expression [\ref{qcd dynamical entropy}] with the Gaussian UGD given by [\ref{gaussian ugd probalistic distribution}],
\begin{equation}
    \Sigma_{gaus}^{Y_0\rightarrow Y}=\int 2\left[\left(\frac{Q_s^2(Y)}{Q_s^2(Y_0)}-1\right)-\ln \left[\frac{Q_s^2(Y)}{Q_s^2(Y_0)}\right]\right].
    \label{gaussian qcd dynamical entropy I}
\end{equation}
Using the expression for the saturation scale as a function of rapidity, $Q_s^2(Y)=Q_s^2(Y_0)e^{\lambda \Delta Y}$, we obtain:
\begin{equation}
\Sigma_{\text{Gauss}}^{Y_0\rightarrow Y}=2(e^{\lambda \Delta Y}-1-\lambda \Delta Y),
\label{gaussian qcd dynamical entropy II}
\end{equation}

For the MPM model case, the entropy is parametrized as:
\begin{equation}
\Sigma_{\text{MPM}}^{Y_0\rightarrow Y}=(1+\gamma_s)(e^{\sigma \Delta Y}-g-\sigma\Delta Y),
\label{mpm qcd dynamical entropy I}
\end{equation}
where $\sigma\approx 0.088$ and $g\approx 0.95$ in the region $\Delta Y \ll 5$.

In Fig.~[\ref{result3figure2}], the dynamical entropy is shown for all the UGD models studied: MPM models (dashed line), Gaussian CGC (solid line), LT (dashed line), and nonlinear KS (dots). The MPM and LV models are practically coincident, meaning that the phenomenology performed in the MPM model correctly mimics the theoretical behavior of the LT UGD in the saturation region. The Gaussian model presents an entropy of higher magnitude compared to the others.

The dynamical entropy density [\ref{dynamical entropy density}] is also calculated and presented in Fig.~[\ref{result3figure2}], with the same notation as the lines on the left side of the same figure. In Eq.~[\ref{dynamical entropy density}], partonic correlations are not included \cite{pescha}. The results were obtained in the domain $\Delta Y=[0,15]$, which corresponds to a QCD evolution from $x=10^{-2}$ to $x=10^{-8}$. The numerical computation used $\alpha_s=0.2$ and $R_p=0.8414$ fm. For the MPM and LT models, the magnitude is very close. The KS UGD model does not exhibit geometric scaling, especially for large $k$. This model mimics the LV and MPM UGDs very well, especially for large $\Delta Y$.

\begin{figure}[ht]
\centering
\includegraphics[width=0.75\linewidth]{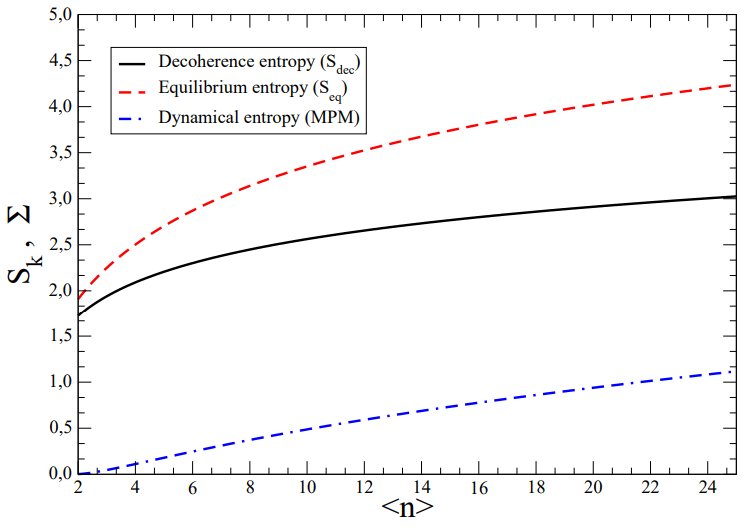}
\caption{The dynamical entropy $\Sigma$ as a function of the average occupancy number (dashed line). Comparison with the decoherence entropy (solid line) and equilibrium entropy (dotted line) for coherent states of a single mode. Reproduced from \cite{ramos2022}.}
\label{result3figure3}
\end{figure}

Finally, in Fig.~[\ref{result3figure3}], the dynamical entropy by the average gluon occupancy number $\expval{n}$ is presented in comparison with the decoherence entropy of a single mode [\ref{decoherence entropy}] and the equilibrium entropy [\ref{equilibrium entropy}]. To obtain the results for dynamical entropy as a function of the average occupancy, the instructions from reference \cite{karkar} were used, where $\expval{n}=xf_g(x)$. The resolution scale is $Q^2=Q_s^2(Y)$, so the gluon occupancy density is given at this scale by:
\begin{equation}
    xf_g(x=e^{Y},Q_s^2)=CQ_s^2(Y),
    \label{gluon population em dynamical entropy}
\end{equation}
with $C=3\pi R_p^2(1-2/e)/4\pi^2\alpha_s$. The result is an expression for the MPM model. Comparing the results from Fig.~[\ref{result3figure3}], it can be observed that, for a large number of occupancies, the equilibrium entropy behavior is very similar to that of dynamical entropy. However, it is noted that different definitions for $\expval{n}$ were used in each case.

\subsection{Dynamical Entropy in pA Collisions}

To adapt the calculation of Dynamical Entropy, it is necessary to find nuclear UGDs. This type of mathematical object is scarce in the literature, so two strategies were established to adapt the UGDs already studied in the case of protons to the case of {\it nuclei}. The first one uses the geometric scaling property proposed in reference \cite{Armesto:2004}, as shown in Eq.~[\ref{gsprocedure01}], where the transverse area of the target can be absorbed by the saturation scale depending on the atomic mass \( A \), i.e., \( \sigma^{\gamma^*A}(\tau_A)/\pi R_A^2=\sigma^{\gamma^*p}(\tau)/\pi R_p^2 \), with \( R_A=(1.12A^{1/3}-0.86A^{-1/3}) \) fm being the nuclear radius. Thus, it is necessary to adapt the transverse cross section \( \sigma_0 \rightarrow \sigma_A \) and the saturation scale \( Q_s^2(Y) \rightarrow Q_{s,A}^2(Y) \). Specifically, for the nuclear saturation scale \( Q_{s,A}(Y) \):
\begin{equation}
Q_{s,A}^2(Y)=\left(\frac{R_p^2A}{R_A^2}\right)^\Delta Q_s^2(Y),
\end{equation}
where \( \Delta \approx 1.27 \) and \( R_p \approx 3.56 \) GeV.

Now, in the calculation of dynamical entropy, the normalization procedure is identical to the case of protons, as shown in previous results, where all dependencies on the cross section \( \sigma_0 \) are negligible due to the normalization process. Therefore, performing the operation \( Q_s^2(Y) \rightarrow Q_{s,A}^2(Y) \), the transverse momentum probability distributions for the GBW, MPM, and LV models in the proton-nucleus case are:
\begin{equation}
P_{\mathrm{GBW}}^A (\tau_A)=\frac{\tau_A e^{-\tau_A/2}}{4\pi Q_{s,A}^2},
\label{Pgbw}
\end{equation}
\begin{equation}
P_{\mathrm{MPM}}^A(\tau_A)=\frac{1}{\pi \xi Q_{s,A}^2}\frac{\tau_A(1+a\tau_A^b)}{(1+\tau_A)^{2+a\tau_A^b}},
\label{Pmpm}
\end{equation}
\begin{equation}
P_{\mathrm{LV}}^A(\tau_A)=
\begin{cases}
-\frac{\ln\left(\frac{\tau_A}{4}\right)}{8\pi Q_{s,A}^2} e^{-\epsilon \ln^2\left(\frac{\tau_A}{4}\right)}, & \text{for } \tau_A < 1; \\
\frac{(d\tau_A)^{-\gamma_s}}{8\pi Q_{s,A}^2} e^{-\epsilon \ln^2\left(\frac{\tau_A}{4}\right)}, & \text{for } \tau_A \geq 1.
\end{cases}
\label{Plv}
\end{equation}
In these equations, the scaling variable is now \( \tau_A = k^2/Q_{s,A}^2 \), and \( \xi = 4.346 \) is the normalization factor for the MPM model.

The second method to obtain a nuclear UGD uses the Glauber-Gribov formalism, where the total dipole cross section of the proton, $\sigma_{\mathrm{dip}}(r,Y)$, is replaced by the nuclear cross section $\sigma_{dA}(x,r)=\int d^2b \, \sigma_{dA}(x,r,b)$, with:
\begin{equation}
\sigma_{\mathrm{dA}}(Y,r,b)=2\left[1-\exp\left(-\frac{1}{2}T_A\sigma_{\mathrm{dip}}(Y,r)\right)\right].
\end{equation}
In this equation, $T_A(b)$ is the nuclear thickness function, defined as the nuclear profile function $T_A=\int_{-\infty}^{+\infty}\rho_A(z,\vec{b})$, normalized to the atomic mass, $\int d^2b T_A(b)=A$. In this work, the Woods-Saxon parametrization was used for the nuclear density $\rho_A$.

The nuclear UGD is given by the expression:
\begin{equation}
\varphi_A(Y,k)=-\frac{N_c k^2}{4\pi^2\alpha_s}\int \frac{d^2bd^2r}{2\pi}e^{i\vec{k}\cdot \vec{r}}\sigma_{\mathrm{dA}}(Y,r,b).
\end{equation}

In particular, for the GBW model in the small-$x$ regime, the proton dipole cross section can be used, and the nuclear UGD is \cite{armesto2002simple}:
\begin{equation}
\varphi_A^{\text{GBW}}(x,k)=\frac{N_c}{\pi^2 \alpha_s}\frac{k^2}{Q_s^2}\int d^2b\sum_{n=1}^{\infty}\frac{(-B)^n}{n!}
\sum_{\ell=1}^n C_{\ell}^n\frac{(-1)^{\ell}}{\ell}e^{-k^2/\ell Q_s^2},
\label{glaubergribovpratical}
\end{equation}
where, in this equation, $C_{\ell}^n$ is the binomial coefficient and $B=\frac{1}{2}T_A(b)\sigma_0$.

\begin{figure}[ht]
\centering
\includegraphics[width=1\linewidth]{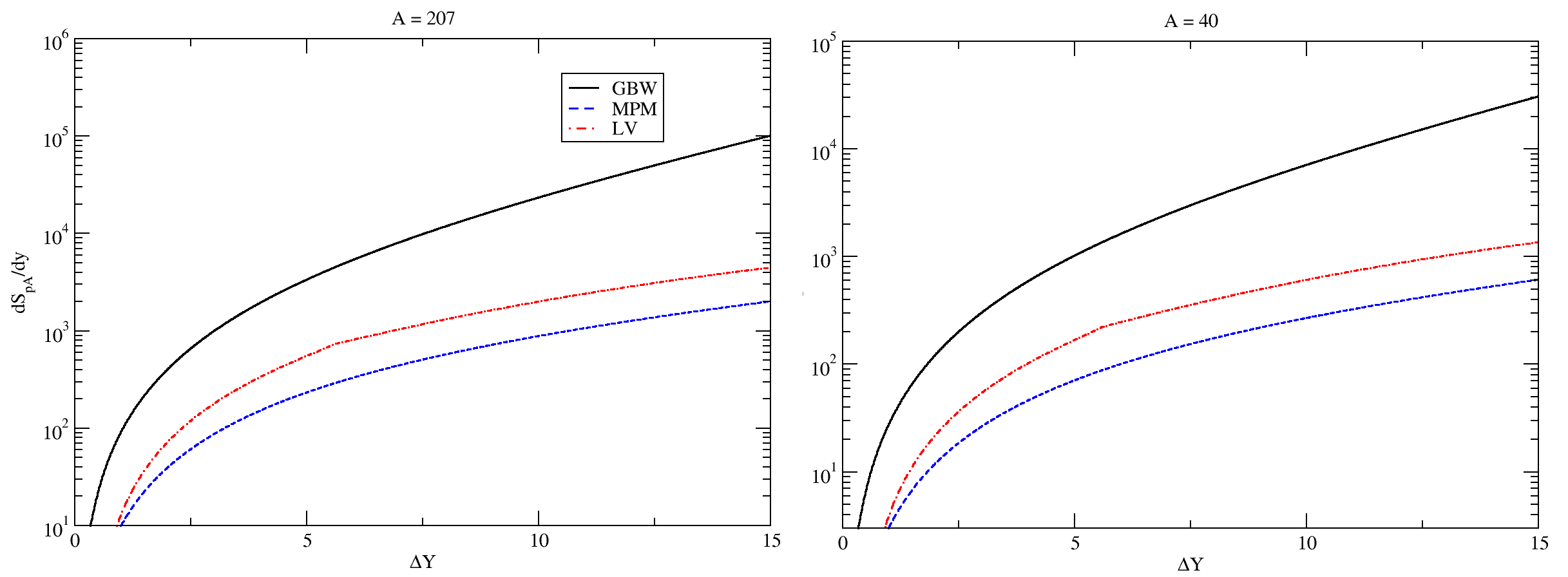}
\caption{Total QCD dynamical entropy, $d\Sigma_d/dy$, produced in a $pA$ collision as a function of $\Delta Y = Y - Y_0$, with $Y_0 \approx 4.6$, for the UGD models GBW (solid lines), MPM (dashed lines), and LV (dashed-dotted lines), for lead (left) and calcium (right).}
\label{newarticle}
\end{figure}

By calculating the nuclear dynamical entropy using the geometric scaling strategy, and using the nuclear transverse momentum probability distributions [\ref{Pgbw}]-[\ref{Plv}], the same result for the proton shown in Fig.~[\ref{result3figure2}] was obtained. To understand this, the dynamical entropy can be evaluated as follows:
\begin{equation}
\Sigma^{Y_0\rightarrow Y}=\pi Q_{s,A}^2(Y)\int_0^{\infty}d\tau_A P(\tau_A)\ln\left[\frac{P(\tau_A)}{P(\tau_{A}^0)}\right],
\label{dynamicalentropy02}
\end{equation}
with $\tau_{A}^0=k^2/Q_{s,A}^2(Y_0)$. From this, $k^2=\tau_AQ_{s,A}^2(Y)=\tau_A^0Q_{s,A}^2(Y_0)$, and it is useful to define the ratio,
\begin{equation}
\frac{Q_{s,A}^2(Y)}{Q_{s,A}^2(Y_0)}=\frac{\left(\frac{R_p^2A}{R_A^2}\right)^{\Delta}Q_s^2(Y)}{\left(\frac{R_p^2A}{R_A^2}\right)^{\Delta}Q_s^2(Y_0)}=e^{\lambda\Delta Y}\equiv s,
\label{sdefinition}
\end{equation}
where $\Delta Y=Y-Y_0$ with $Y_0\approx 4.6$ ($x_0=0.01$). This initial rapidity is considered because values of $x \leq x_0$ correspond to the validity limit for the application of the phenomenological UGD models considered here. Thus, initially, the partons occupy a transverse area proportional to the initial color correlation size $R_0(Y_0) = 1/Q_s(Y_0)$.

From the expression [\ref{dynamicalentropy02}], the ratio between \( P(\tau_A) \) and \( P(s\tau_A) \) can be analyzed for the different expressions of the nuclear transverse momentum probability distributions in the geometric scaling adaptation strategy [\ref{Pgbw}]-[\ref{Plv}]. For the GBW model, it can be observed:
\begin{equation}
\ln\left[\frac{P_{\mathrm{GBW}}(\tau_A)}{P_{\mathrm{GBW}}(s\tau_A)}\right]=2\lambda\Delta Y+k^2[R_s^2(Y_0)-R_s^2(Y)].
\label{ratioGBW}
\end{equation}
Substituting this result into the expression [\ref{qcd dynamical entropy}], we obtain an expression equivalent to that obtained in Eq.~[14] of Ref. \cite{pescha}, recovering the dynamical entropy of the proton:
\begin{equation}
\Sigma_{\mathrm{GBW}}^{Y_0\rightarrow Y}(\Delta Y)=2\left(e^{\lambda \Delta Y}-1-\lambda\Delta Y\right),
\label{protongbwdynamicalentropy}
\end{equation}

A similar procedure can be carried out for the dynamical entropy of the MPM model, \( \Sigma_{\mathrm{MPM}}^{Y_0\rightarrow Y} \), since it exhibits the geometric scaling property:
\begin{equation}
\Sigma^{Y_0\rightarrow Y}_{\mathrm{MPM}}=\frac{1}{\xi }\int_0^{\infty}d\tau_A\frac{\tau_A(1+a\tau_A^b)}{(1+\tau_A)^{2+a\tau_A^b}}\ln\left[\frac{(1+a\tau_A^b)(1+s\tau_A)^{2+as^b\tau_A^b}}{s^2(1+as^b\tau_A^b)(1+\tau_A)^{2+a\tau_A^b}}\right].
\label{protonMPMdynamicalentropy}
\end{equation}
This expression also recovers the dynamical entropy of the proton shown in Fig.~[\ref{result3figure2}]. The normalization procedure [\ref{ugd probalistic distribution}] eliminates all nuclear dependence on the transverse size of the target, $S_{\perp}^{A} = \pi R_A^{2}$. The same effect can be demonstrated for the case of the LV UGD, both for the diluted and saturated contributions.

In Fig.~[\ref{newarticle}], the entropy density [\ref{dynamical entropy density}] is calculated for lead (left panel) and calcium (right panel) for all models based on the phenomenon of geometric scaling: GBW (solid line), MPM (dashed line), and LV (dashed-dotted line) in the range $\Delta Y=[0,15]$. Although the nuclear dynamical entropy is independent of $A$, its density is related to the nuclear radius size as $\frac{d S_D}{dy} \sim R_A^2$. In the definition proposed in reference \cite{pescha}, the ratio between all available unit cells in the CGC medium, $\sim \frac{\pi R_A^2}{\pi R_0^2}$, should be taken into account.

\begin{figure}[ht]
\centering
\includegraphics[width=0.7\linewidth]{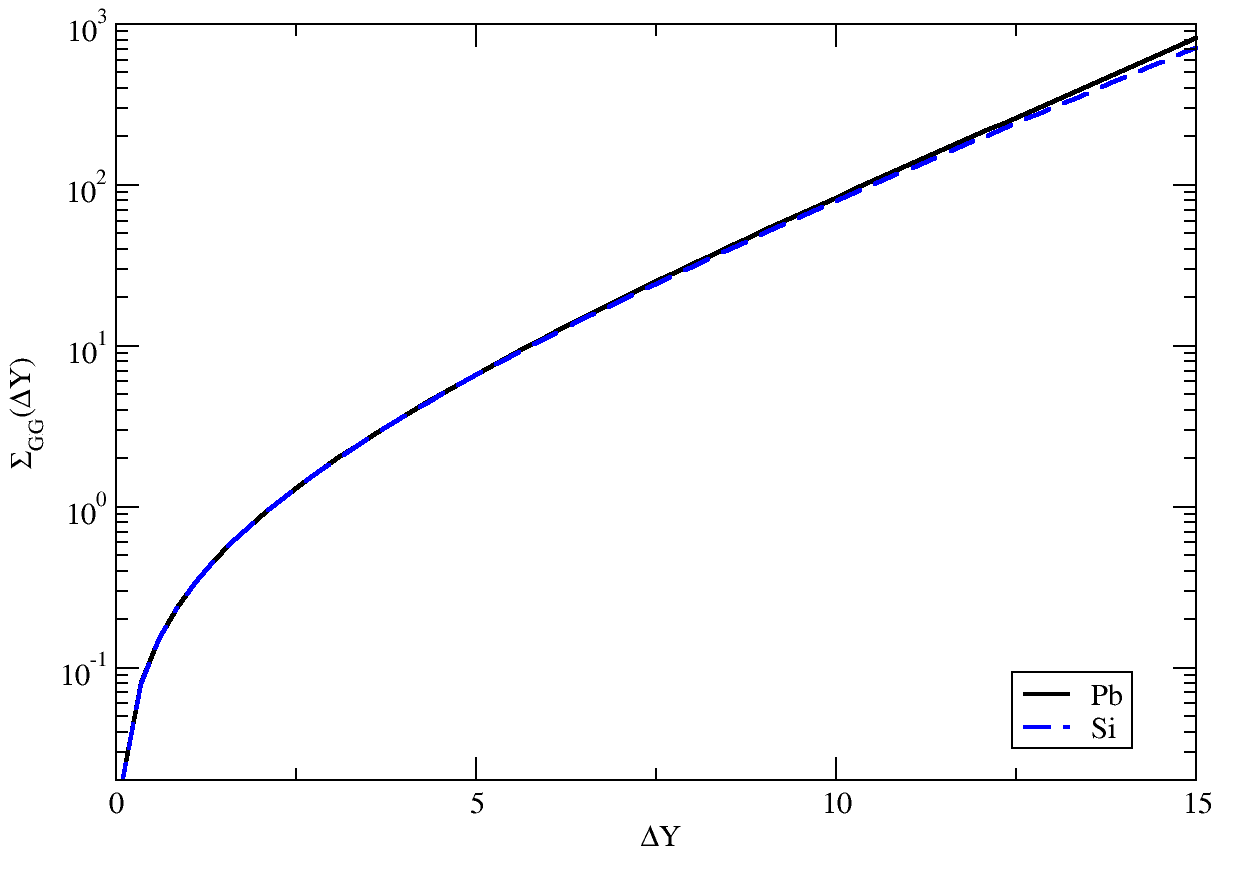}
\caption{Nuclear QCD dynamical entropy in proton-nucleus collisions corresponding to the QCD evolution in rapidity, $Y_0\rightarrow Y$, in the range $\Delta Y=[0,15]$ in the Glauber-Gribov approach. The entropy is calculated for lead (solid line) and silicon (dashed lines).}
\label{newarticle2}
\end{figure}

Finally, the results obtained for the nuclear dynamical entropy via the Glauber-Gribov formalism are given in Fig.~[\ref{newarticle}2], where the dynamical entropy again shows independence from $A$, also plotted as a function of $\Delta Y$ in the range $[0,15]$. Although obtaining the UGD involves a more complex process via Eq.~[\ref{glaubergribovpratical}], it seems that the geometric scaling and the normalization procedure also eliminate the $A$ dependence in the dynamical entropy. A notable difference in relation to the geometric scaling strategy is that this result does not reduce to the proton case and is approximately ten times larger.

\section{Conclusion}

In this thesis, the entanglement entropy in DIS processes for $pp$, $ep$, and $pA$ was investigated, as well as in elastic scatterings. The theoretical framework was based on the entanglement entropy using von Neumann’s expression, written in terms of the number of gluons as a function of Bjorken $x$ and the photon virtualities, $Q^2$. Analytical expressions for the gluon density related to parton saturation physics were used, with models based on the color dipole representation. The analysis included both integrated and unintegrated densities, allowing the description of fundamental observables in DIS for small $x$ and intermediate values of $Q^2 \sim 50 \operatorname{GeV}^2$. Additionally, an extrapolation was performed using geometric scaling properties to obtain the nuclear gluon density, with results consistent with experimental data.

The investigation of entanglement entropy in elastic scattering processes in $pp$ and $\bar{p}p$ collisions was conducted using the $S$-matrix formalism and partial wave expansion. The model-independent extraction, based on the Lévy image method, allowed the systematic analysis of entanglement in the hadronic final states. It was found that, at high energies, the entropy for elastic scattering saturates at asymptotic energies, with behavior parametrized as $S \sim 1 + \ln(2) - \ln(\ln(s))$.

The QCD dynamical entropy was studied in $pp$ and $pA$ collisions, based on various UGD models. For $pp$ collisions, analytical models such as MPM were used, which accurately describe the charged particle spectra at the LHC, as well as CGC models based on Gaussian distributions and the Levin-Tuchin law. The results showed that, in all cases, the peak of the distribution occurs around $k \sim Q_s$, highlighting the geometric scaling property. The total dynamical entropy and its density were calculated, showing a strong dependence on $\Delta Y$, especially in the case of the Gaussian CGC model. For $pA$ collisions, strategies based on geometric scaling and the Glauber-Gribov formalism were employed, confirming that the dynamical entropy is independent of the atomic number $A$, due to the normalization procedure.

The results compared the dynamical entropy with the decoherence entropy and the equilibrium entropy of a single mode, highlighting similarities for large average gluon occupancy numbers. The detailed analysis of these entropies, using analytical and phenomenological tools, offers a new perspective on understanding the dynamics of the initial states in heavy ion collisions and the production of multiple particles at high energies.

Finally, the robust analysis of various entropy notions confronts the good results obtained by the entanglement entropy in the LV Model. In this case, since the reduced density matrix can be written as a function of the virtuality, which, in turn, is related to the spatial portion of the hadron investigated in a DIS, and this is entirely $A_h$, the idea is proposed that for low values of the virtuality, in the order of $A_h \sim 1/Q^2$, we will have $\hat{\rho}_A \sim \hat{\rho}$, meaning the density matrix is similar to the total system’s density matrix, so the entanglement entropy is equivalent to the hadronic entropy.

\appendix

\chapter{Light Cone Variables}
\label{apendice A: light cone variables}

The light cone coordinates are the usual system in high-energy particle physics. Traditionally, there are two ways of treating this system, which may change depending on the reference studied. The first is called the Lepage-Brodsky (LB) convention, and the second, the Kogut-Soper (KS) convention. For example, when discussing this system of variables, in the section where the CGC formalism is presented [\ref{chapter 3: QCD dense states - Section 1; High Energy Physics / subsection 1. Saturation Physics}], the reproduced results used the KS convention; in contrast, when performing manipulations to obtain the Wehrl entropy in this appendix [\ref{apendice E: qcd wehrl entropy}], the LB convention was used.

Thus, in this appendix, the foundations of both conventions are briefly presented, since for each treatment of the theories and formalisms addressed in this work, the choices and conventions of each author were maintained. For a comprehensive discussion of the light cone variables, reference \cite{lightfront} is recommended.

\subsection*{Lepage-Brodsky Convention}

The contravariant 4-vectors of position $x^{\mu}$ are written as:
\begin{equation}
    x^{\mu}=(x^+,x^-,x^1,x^2)=(x^+,x^-,\vec{x}_T).
    \label{4 position in LB}
\end{equation}
The time-like and space-like components are given by:
\begin{equation}
    x^+=(x^0+x^3) \quad \text{and} \quad x^-=(x^0-x^3),
    \label{time and space light vectors in LB}
\end{equation}
respectively, and are called the {\it light cone time} and {\it light cone position}. The covariant vectors are obtained using $x_{\mu}=g_{\mu\nu}x^{\nu}$, with the metric tensors:

\begin{equation}
    g^{\mu\nu}= \begin{pmatrix}
    0 & 2 & 0  & 0 \\ 
    2 & 0 & 0  & 0\\
    0 & 0 & -1 & 0\\
    0 & 0 & 0  & -1
    \end{pmatrix}
\label{covariant metric in LB}
\end{equation}

\begin{equation}
    g_{\mu\nu}= \begin{pmatrix}
    0   & 1/2 & 0  & 0 \\ 
    1/2 & 0   & 0  & 0\\
    0   & 0   & -1 & 0\\
    0   & 0   & 0  & -1
    \end{pmatrix}
\label{contravariant metric in LB}
\end{equation}

The scalar product is given by:
\begin{equation}
    x \cdot p=x^{\mu}p_{\mu}=x^+p_++x^-p_-+x^1p_1+x^2p_2=\frac{1}{2}(x^+p^-+x^-p^+)-\vec{x}_T\cdot\vec{p}_T.
    \label{scalar product in LB}
\end{equation}

\subsection*{\bf Kogut-Soper Convention}

{\it Kogut} and {\it Soper} used the following expressions for the time-like and space-like components:
\begin{equation}
    x^+=\frac{1}{\sqrt{2}}(x^0+x^3) \quad \text{and} \quad x^-=\frac{1}{\sqrt{2}}(x^0-x^3),
    \label{time and space light vectors in KS}
\end{equation}
The metric tensors are:

\begin{equation}
    g^{\mu\nu}= g_{\mu\nu}= \begin{pmatrix}
    0 & 1 & 0  & 0 \\ 
    1 & 0 & 0  & 0\\
    0 & 0 & -1 & 0\\
    0 & 0 & 0  & -1
    \end{pmatrix}
\label{metric in KS}
\end{equation}
The scalar product is given by:
\begin{equation}
    x \cdot p=x^{\mu}p_{\mu}=x^+p_++x^-p_-+x^1p_1+x^2p_2=x^+p^-+x^-p^+-\vec{x}_T\cdot\vec{p}_T.
    \label{scalar product in KS}
\end{equation}

Finally, it is demonstrated that the ratio $p^+/p^-$ provides a measure of the Lorentz {\it boost} a particle undergoes relative to its rest frame. The {\it rapidity} $Y$ is a quantity defined in relation to this ratio, given, both in KS and in LB, by:
\begin{equation}
Y=\frac{1}{2}\ln \left(\frac{p^+}{p^-}\right)= \frac{1}{2}\ln \left(\frac{x_0+x_3}{x_0-x_3}\right)=\frac{1}{2}\ln \left(\frac{E+p_z}{E-p_z}\right).
\label{rapidity definition}
\end{equation}

\chapter{The {\it Gram-Schmidt} Procedure}
\label{Apendice B: Gram-Schmidt Procedure}

The {\it Gram-Schmidt} process is a simple algorithm used to produce an orthogonal or orthonormal basis for any non-null subspace. Being $\bra{\phi_i^A}\otimes\bra{\phi_j^B}=\bra{\phi_i^A\phi_j^B}$, consider the inner product:
\begin{equation}
\bra{\phi_i^A\phi_j^B}\ket{\psi_{AB}}=\alpha_n,
\label{B.0.1}
\end{equation}
with $|\alpha_n|^2\neq 0$.

Now, let $\ket{\phi_l^A}\in\mathscr{H}_A$, $\ni \ket{\phi_l^A}\perp\bra{\phi_i^A} $, and $\epsilon$ an arbitrary complex number, then:
\begin{equation}
||\ket{\phi_i^A}+\epsilon \ket{\phi_l^A}||^2=||\ket{\phi_i^A}||^2+|\epsilon|^2||\ket{\phi_l^A}||^2=1+\Theta(\epsilon^2).
\end{equation}

Disregarding second-order terms in $\epsilon$, the linear combination $\ket{\phi_i^A}+\epsilon \ket{\phi_l^A}$ becomes a unit vector. Now:
\begin{equation}
[\bra{\phi_i^A}+\epsilon \bra{\phi_l^A}]\otimes\bra{\phi_j^B}\ket{\psi_{AB}}=\bra{\phi_i^A\phi_j^B}\ket{\psi_{AB}}+\epsilon \bra{\phi_l^A}\otimes\bra{\phi_j^B}\ket{\psi_{AB}}=\alpha_n+\epsilon \bra{\phi_l^A}\otimes\bra{\phi_j^B}\ket{\psi_{AB}}.
\end{equation}

And:
\begin{equation}
\begin{split}
||[\bra{\phi_i^A}+\epsilon \bra{\phi_l^A}]\otimes\bra{\phi_j^B}\ket{\psi_{AB}}||^2&=||\alpha_n+\epsilon \bra{\phi_l^A}\otimes\bra{\phi_j^B}\ket{\psi_{AB}}||^2\\
&=||\alpha_n||^2+2\Re(\epsilon\alpha_n\bra{\phi_l^A}\otimes\bra{\phi_j^B}\ket{\psi_{AB}})+\Theta(\epsilon^2).
\end{split}
\label{B.0.4}
\end{equation}

The left-hand side of Eq.~[\ref{B.0.4}] is stationary with respect to any variation of $\ket{\phi_i^A}$, so:
\begin{equation}
\bra{\phi_l^A}\otimes\bra{\phi_j^B}\ket{\psi_{AB}}=0, \quad i\neq l \quad \forall \ket{\phi_l^A} \in \mathscr{H}_{A'},
\end{equation}
where $\mathscr{H}_{A'}$ is the set of all states belonging to $\mathscr{H}_{A}$ orthogonal to $\ket{\phi_i^A}$. Performing a similar procedure in the vector space $\mathscr{H}_{B}$, one can obtain:
\begin{equation}
\bra{\phi_i^A}\otimes\bra{\phi_k^B}\ket{\psi_{AB}}=0, \quad k\neq j \quad \forall \ket{\phi_k^B} \in \mathscr{H}_{B'}.
\end{equation}

Having the vector $\ket{\psi_{AB}'}$ given by:
\begin{equation}
\ket{\psi_{AB}'}=\ket{\psi_{AB}}-\alpha_n \ket{\phi_j^A}\otimes\ket{\phi_j^B} \quad \therefore \quad  \ket{\psi_{AB}} = \ket{\psi_{AB}'} + \alpha_n \ket{\phi_j^A}\otimes\ket{\phi_j^B},
\end{equation}
the definition of $\alpha_n$ results in:
\begin{equation}
\bra{\phi_i^A}\otimes\bra{\phi_j^B}\ket{\psi_{AB}'}=0.
\label{B.0.8}
\end{equation}

Thus, $\ket{\psi_{AB}'}\in \mathscr{H}_{A'}\otimes \mathscr{H}_{B'}$. The procedure outlined by equations [\ref{B.0.1}]-[\ref{B.0.8}] can be repeated to eliminate the $k$-th and $l$-th states and then the following ones, until the form is obtained:
\begin{equation}
\ket{\psi}=\sum_i\alpha_i\ket{a_i}\otimes\ket{b_i}.
\end{equation}

\chapter{Color Dipoles and the Balitsky-Kovchegov Equation} 
\label{apendice C: bk equation}

\begin{figure}[ht]
\centering
\includegraphics[width=0.5\linewidth]{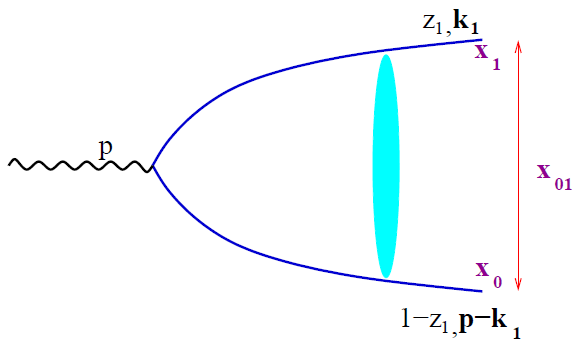}
\caption{Color dipole, in blue, generated from a virtual photon with 4-momentum $p$. The quark has a longitudinal momentum fraction $z_1$ and transverse momentum $\vec{k}_1$, while for the anti-quark, it has $1-z_1$ and $\vec{p}_t-\vec{k}_1$, respectively. The red line is the modulus of the vector $\vec{x}_{01}$. Reproduced from \cite{Anna}.}
\label{bkone}
\end{figure}

This section aims to briefly present the derivation and properties of the BK equation in the color dipole formalism. Initially, consider a quark-antiquark pair (Fig.~[\ref{bkone}]), a color dipole, with a wave function in the momentum representation denoted by $\psi_{\alpha\beta}^{(0)}(k_1, z_1)$, where $\vec{k}_1$ is the transverse momentum of the quark, $z_1=k^+_1/p^+$ is the fraction of the longitudinal momentum of the photon carried by the quark in light-cone variables, and $\alpha$ and $\beta$ are the color indices. This wave function is obtained from the Fock state expansion of the virtual photon state that generated the dipole.

The wave function in the transverse coordinate space $\psi_{\alpha\beta}^{(0)}(\vec{x}_0, \vec{x}_1, z_1)$ is obtained from a two-dimensional Fourier transform:
\begin{equation}
\psi_{\alpha\beta}^{(0)}(\vec{x}_0, \vec{x}_1, z_1)=\int\frac{d^2k_1}{(2\pi)^2}e^{\mathbbm{i}\vec{x}_{01}\cdot \vec{k}_1}\psi_{\alpha\beta}^{(0)}(k_1, z_1),  
\end{equation}
where in this equation, $\vec{x}_{01}$ comes from the definition:
\begin{equation}
    \vec{x}_{nm}=\vec{x}_n-\vec{x}_m,
\end{equation}
where $\vec{x}_0$ and $\vec{x}_1$ are the positions, in the nucleon rest frame, of the quark and the antiquark, respectively, configuring the endpoints of the dipole. Thus, the square of the wave function $\phi^{(0)}(\vec{x}_0, \vec{x}_1, z_1)$, the {\it probability of measuring a single dipole}, is given by:
\begin{equation}
\phi^{(0)}(\vec{x}_0, \vec{x}_1, z_1)=\sum_{\alpha,\beta}|\psi_{\alpha\beta}^{(0)}(\vec{x}_0, \vec{x}_1, z_1)|^2.    
\end{equation}

\begin{figure}[ht]
\centering
\includegraphics[width=0.7\linewidth]{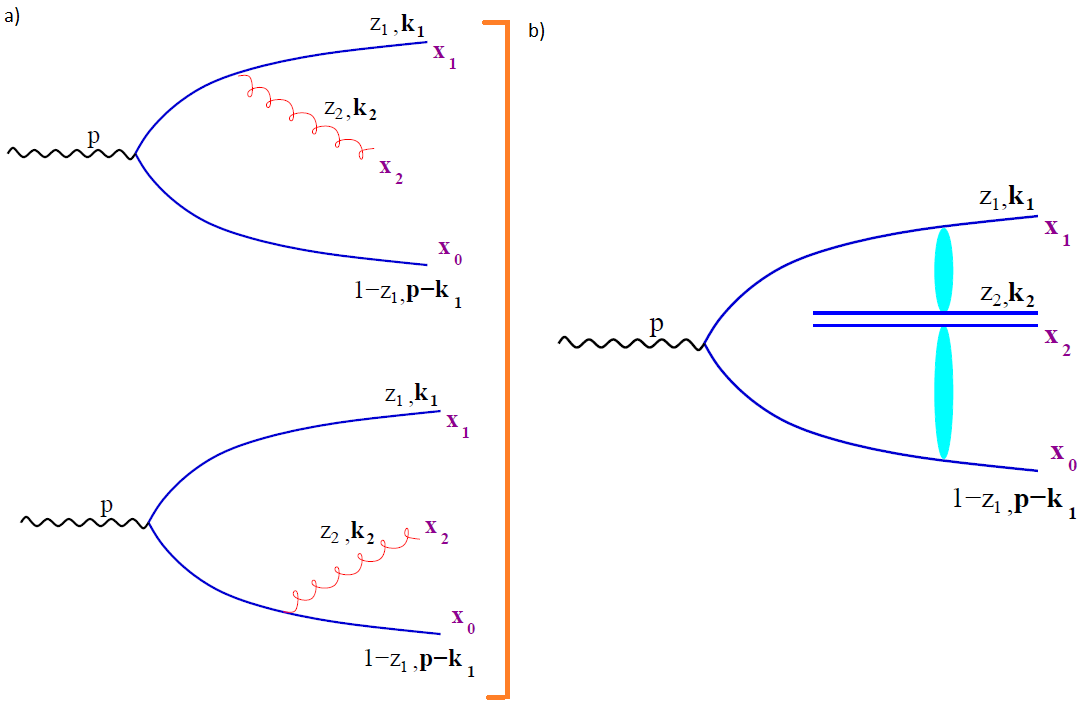}
\caption{Emission of a gluon from the quark or anti-quark (a), which equivalently represents a state with two dipoles in the high-color-number limit (b). Reproduced from \cite{Anna}.}
\label{bktwothree}
\end{figure}

Now, consider the emission of a soft gluon ($z_2/z_1 \ll 1$) from the original quark or anti-quark (Fig. \ref{bktwothree}), with longitudinal momentum fraction $z_2$ and transverse momentum $\vec{k}_2$. The probability of this emission $\phi^{(1)}(\vec{x}_0, \vec{x_1},z_1)$ can be obtained from $\phi^{(0)}$ with the relation:
\begin{equation}
\phi^{(1)}(\vec{x}_0, \vec{x_1},z_1)=\frac{\alpha_s C_F}{\pi^2}\int_{z_0}^{z_1}\frac{dz_2}{z_2}\int d^2 x_2 \frac{x_{01}^2}{x_{20}^2x_{12}^2} \phi^{(0)}(\vec{x}_0, \vec{x_1},z_1),
\end{equation}
where $C_F$ is the Casimir constant, given by:
\begin{equation}
C_F=\frac{N_c^2-1}{2N_c},
\end{equation}
and $\alpha_s$ is the strong coupling constant. Thus, the emission of an additional gluon is equivalent to the breakup of the original dipole $(0,1)$ into two dipoles $(0,2)$ and $(2,1)$ with a probability of measurement given by:
\begin{equation}
d^2x_2\frac{x_{01}^2}{x_{20}^2x_{12}^2}    
\end{equation}

\begin{figure}[ht]
\centering
\includegraphics[width=0.25\linewidth]{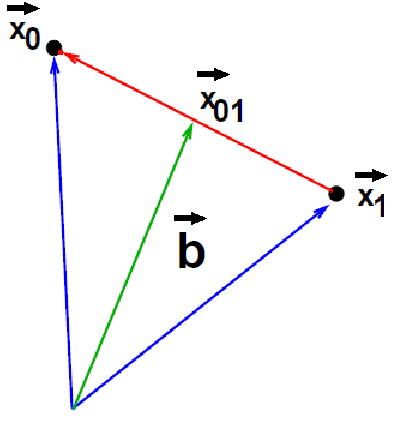}
\caption{Geometric representation of the impact parameter vector. Adapted from \cite{Anna}.}
\label{b}
\end{figure}

Thus, the process of subsequent soft gluon emission can be performed analogously, allowing the determination of the square of the wave function of a state with an arbitrary number of gluons. To describe this process, {\it Mueller}\cite{mumumu} introduced the {\it dipole generating function} $Z(\vec{x}_{01}, z_1,u)$. It must satisfy the normalization condition, i.e., $Z(\vec{x}_{01}, z_1,u=1)=1$. The use of this mathematical tool allows the square of the wave function with $n$ gluons, $\phi^{(n)}(\{\vec{x}_{n+1}\}, z_1)$, to be obtained using the expression:
\begin{equation}
\phi^{(n)}(\{\vec{x}_{n+1}\}, z_1)=\phi^{(0)}\prod_{j=2}^{n+1}\frac{\delta}{\delta u(\vec{x}_j)}Z(\vec{x}_0,\vec{x}_1, z_1,u)|_{u=0}.
\end{equation}
This equation relates the probability of finding $n$ dipoles created from the original quark-antiquark pair $(0,1)$, which will be produced at positions $\vec{x}_n$. The relation between the wave functions of $n$ and $n+1$ dipoles is given by the differential equation for the generating function:
\begin{equation}
\frac{d}{dY}Z(\vec{b}, \vec{x}_{01}, Y,u)=\int d^2x_2\frac{x_{01}^2}{x_{20}^2x_{12}^2}\left[Z(\vec{b}+\frac{\vec{x}_{12}}{2},\vec{x}_{20},Y,u)+Z(\vec{b}-\frac{\vec{x}_{20}}{2},\vec{x}_{12},Y,u)-Z(\vec{b}, \vec{x}_{01}, z_1,u)\right].
\label{muemue}
\end{equation}
In this equation, a dependence of the generating function on rapidity $Y=\ln 1/z_+$ and the impact parameter $\vec{b}$ (Fig.~[\ref{b}]) is introduced, given by:
\begin{equation}
\vec{b}=\frac{\vec{x}_0+\vec{x}_1}{2}.
\end{equation}

Using Eq.~[\ref{muemue}], it is possible to obtain the evolution equation for the dipole scattering amplitude on the target. To do this, we first define the dipole number density $n_k$:
\begin{equation}
n_k=\prod_{i=1}^k\frac{\delta}{\delta u(\vec{b},\vec{x}_i)}Z|_{u=1}.
\end{equation}
The scattering amplitude of a single dipole $N_1$ on the target is obtained from the convolution of the dipole number density with the propagator from the dipole to the nucleus:
\begin{equation}
N_1(\vec{x}_{01}, \vec{b}_{01},Y)=\int d[\mathscr{P}_1]n_1\vartheta_1,
\label{convul1}
\end{equation}
where $d[\mathscr{P}_1]=\frac{d^2x_1}{2\pi x_i^2}d^2b$ is the measure in phase space and $\vartheta\equiv\vartheta(\vec{x},\vec{b})$ is the propagator of a single dipole in the nucleon. Differentiating the equation for the generating function and using the relation [\ref{convul1}], we can obtain the equation for the dipole-target amplitude:
\begin{equation}
\frac{d}{dY}N_1(\vec{x}_{01}, \vec{b},Y)=\frac{\alpha_sN_c}{\pi}\int d^2x_2\frac{x_{01}^2}{x_{20}^2x_{12}^2}\left[N_1(\vec{b}+\frac{\vec{x}_{12}}{2},\vec{x}_{20},Y)+N_1(\vec{b}-\frac{\vec{x}_{20}}{2},\vec{x}_{12},Y)-N_1(\vec{b}, \vec{x}_{01}, Y)\right],
\end{equation}

In this derivation, only the contribution of a single dipole has been included. This equation is the dipole formalism version of the BFKL equation in transverse coordinate space. It is possible to generalize this equation by considering multiple scatterings of dipoles with the target. To do this, we consider the density of a number $k$ of dipoles and its convolution with $k$ propagators. In this process, the amplitude will be given by the expression:
\begin{equation}
N(\vec{x}_{01}, \vec{b}, Y)=\sum_{k=1}^{\infty}\int d[\mathscr{P}_k]n_k\prod_{j=1}^k\vartheta_j,
\end{equation}
so that the phase space average is now defined as:
\begin{equation}
[\mathscr{P}_k]=\prod_{i=1}^k\frac{d^2x_i}{2\pi x_i^2}d^2b.
\end{equation}
Analogously to the development of the equation for the amplitude of a single dipole, for multiple dipoles we obtain:
\begin{equation}
\begin{split}
\frac{dN}{dY} &= \frac{\alpha_sN_c}{\pi}\int \frac{d^2x_2 x_{01}^2}{x_{20}^2x_{12}^2}\\
&\times\left[N(\vec{b}+\frac{\vec{x}_{12}}{2},\vec{x}_{20}, Y)+N(\vec{b}-\frac{\vec{x}_{20}}{2},\vec{x}_{12}, Y)-N(\vec{b}, \vec{x}_{01}, Y)-N(\vec{b}+\frac{\vec{x}_{12}}{2},\vec{x}_{20}, Y)N(\vec{b}-\frac{\vec{x}_{20}}{2},\vec{x}_{12}, Y)\right].
\end{split}
\label{equationBK}
\end{equation}
This is the {\it Balitsky-Kovchegov} equation. It is a nonlinear expression that considers the interaction of $k$ dipoles with the target nucleon. It is an evolution equation with respect to rapidity $Y$, which requires initial conditions $N^{(0)}(\vec{b}, \vec{x}_{01}, Y=0)$, and is valid in the dominant logarithmic approximation. It also assumes a constant value for the strong coupling. The problem involves $(4+1)$ variables, that is, four degrees of freedom per dipole and one evolution variable, $Y$.

\chapter{Model-Independent Lévy Femtoscopic Imaging for Elastic Scattering}
\label{apendice D: model-independent femtoscopic lévy imaging for elastic scattering}

Lévy series are a generalization of the Lévy expansion methods proposed to analyze stable {\it Levy} source distributions in particle field femtoscopy\cite{levy1,levy2,levy3}. In this work, the focus is on momentum transfer with $t$-distributions in hadron-hadron elastic collisions. This model provides a systematic, model-independent method to characterize the variations of the approximate size of these distributions using a dimensionless variable, $z\equiv R^2|t|\geq 0$, and a completely orthonormal set of polynomials that are orthogonal to the weight function $\omega(z)=e^{-z^{\alpha}}$. The quantity $R$ denotes the scale parameter of {\it Levy}. In this appendix, we strictly follow the analysis of elastic cross-sections for $pp$ and $p\bar{p}$ processes performed in \cite{levy1}. A clear advantage of the Lévy method for proton imaging is that it provides the inelasticity profile of the proton as a function of energy and impact parameter.

In the $t$-momentum representation, the elastic differential cross-section is related to the modulus of the complex value of an elastic amplitude $T_{el}$. The sequence is expressed as the expansion of orthonormal series in terms of Lévy polynomials:
\begin{equation}
    \frac{d\sigma_{\text{el}}}{dt}=\frac{1}{4\pi}|T_{el}(s,t)|^2,
    \label{diferencial elastic cross section Levy I}
\end{equation}
with,
\begin{equation}
T_{el}(s,t)=\mathbbm{i}\sqrt{4\pi A}e^{-\frac{z^{\alpha}}{2}}\left(1+\sum_{i=1}^{\infty}c_il_i(z|\alpha)\right),
\label{levy elastic amplitude}
\end{equation}
where, in this equation, $c_i=a_i+\mathbbm{i}b_i$ are the coefficients of the complex expansion. The dimensionless variable $z$ is introduced as a measure of the magnitude of the square of the 4-momentum transfer $|t|$ multiplied by the square of the Lévy scale parameter, $R$, in natural units. The parameters for the expansion, $A$, $R$, $\alpha$, and the complex coefficients $c_i$ are available in Appendices A and B of reference \cite{apendix}. The quantities $l_i(z|\alpha)$ are the normalized Lévy polynomials of order $i$ and are given by:
\begin{equation}
    l_i(z|\alpha)=\frac{L_i(z|\alpha)}{\sqrt{D_i(\alpha)}\sqrt{D_{i+1}(\alpha)}},
    \label{normalized levy polinomials}
\end{equation}
for $j\geq 0$. These polynomials are built in terms of the unnormalized Lévy polynomials, which are generally:
\begin{equation}
    L_1(z|\alpha)=\det  \begin{pmatrix}
    \mu_0^{\alpha} & \mu_1^{\alpha} \\ 
    1              & z
    \end{pmatrix}
\label{unnormalized 1 levy polynomials}
\end{equation}

\begin{equation}
    L_2(z|\alpha)=\det  \begin{pmatrix}
    \mu_0^{\alpha} & \mu_1^{\alpha} & \mu_2^{\alpha} \\ 
    \mu_1^{\alpha} & \mu_2^{\alpha} & \mu_3^{\alpha} \\
    1              & z              & z^2   
    \end{pmatrix}
\label{unnormalized 2 levy polynomials}
\end{equation}

\begin{equation}
    L_m(z|\alpha)=\det  \begin{pmatrix}
    \mu_0^{\alpha} & ... & \mu_m^{\alpha} \\ 
    \vdots         & ... & \vdots         \\
    1              & ... & z^{m}   
    \end{pmatrix},
\label{unnormalized m levy polynomials}
\end{equation}
with $L_0(z|\alpha)=1$. In Eq.~[\ref{normalized levy polinomials}], $D_j(\alpha)$ are the Gram-determinants, defined as:
\begin{equation}
    D_1(\alpha)=\mu_0^{\alpha},
    \label{gram-determinant 1}
\end{equation}

\begin{equation}
    D_2(z|\alpha)=\det  \begin{pmatrix}
    \mu_0^{\alpha} & \mu_1^{\alpha} \\ 
    \mu_1^{\alpha} & \mu_2^{\alpha}
    \end{pmatrix},
\label{gram-determinant 2}
\end{equation}

\begin{equation}
    D_m(\alpha)=\det  \begin{pmatrix}
    \mu_0^{\alpha}      & ... & \mu_{m-1}^{\alpha} \\ 
    \vdots              & ... & \vdots         \\
    \mu_{m-1}^{\alpha}  & ... & \mu_{2m-2}^{\alpha}   
    \end{pmatrix},
\label{gram-determinant m}
\end{equation}
with,
\begin{equation}
\mu_{n}^{\alpha}=\frac{1}{\alpha}\Gamma\left(\frac{n+1}{\alpha}\right),
\label{mu levy coeficients}
\end{equation}
and $D_0(\alpha)\equiv 1$.

The total cross-section $\sigma_T\equiv \Im T_{el}(s,0)$ and the elastic cross-section are expressed in terms of the quantities below:
\begin{equation}
    \sigma_T=2\sqrt{4\pi A}\left(1+\sum_{i=1}^{\infty}a_il_i(0|\alpha)\right),
    \label{total cross section levy model}
\end{equation}

\begin{equation}
    \sigma_{\text{el}}=\frac{A}{R^2}\left[\frac{1}{\alpha}\Gamma\left(\frac{1}{\alpha}\right)+\sum_{i=1}^{\infty}(a_i^2+b_i^2)\right].
  \label{elastic cross section levy model}    
\end{equation}

It is demonstrated in reference \cite{levy1} that the expansion for $T_{el}(s,t)$ converges rapidly, and a third-order Lévy series is already sufficient to reproduce the measured data with $\sqrt{s}\leq 1$ TeV with high confidence levels for an appropriate statistical description.
 
\chapter{Wehrl Entropy}
\label{apendice E: Wehrl Entropy}

Entropy in classical phase space $f(q,p)$ is given by the expression:
\begin{equation}
    S=-k_B\int \frac{dpdq}{h'}f(q,p)\ln{f(q,p)},
    \label{classical entropy}
\end{equation}
where $h'$ is an elementary cell in this space. In the quantum case, there is no possibility of defining a phase space due to the uncertainty principle; however, entropy will be given by the {\it von Neumann} expression written in terms of the density matrix $\hat \rho$, here expressed in natural units:
\begin{equation}
    S_{vN}=-\Tr [\hat \rho \ln \hat \rho].
    \label{von neumann quantum entropy}
\end{equation}

The two entropies are not simply connected, {\it i.e.}, $S_{vN}$ does not reduce to the expression $S$ in the limit as $\hbar \rightarrow 0$. However, this conversion can be made from an intermediate definition of entropy given by {\it Alfred Wehrl}\cite{wehrl1979}. The classical expression for entropy given by Eq.~[\ref{classical entropy}] can take infinitely negative values due to the arbitrariness of the unit cell volume, which could violate the uncertainty principle. To adjust the model, we consider the basis of coherent states $\ket{c}$ with Gaussian packets of minimum uncertainty ($\sigma_p \sigma_q=\hbar/2$). Taking the trace of Eq.~[\ref{von neumann quantum entropy}] in the basis of coherent states, we get:
\begin{equation}
    S_{vN}=-\int \frac{dqdp}{2\pi\hbar}\bra{c}\hat\rho \ln{\hat\rho}\ket{c}.
\label{semi classical entropy}
\end{equation}

The Wehrl entropy $S_{W}$ is obtained by performing the {\it classical substitution}, which consists of replacing $\bra{c}\hat\rho \ln{\hat\rho}\ket{c}$ with $\bra{c}\hat\rho\ket{c} \ln{ \bra{c}\hat \rho \ket{c}}$. Thus:
\begin{equation}
S_{W}=-\int \frac{dqdp}{2\pi\hbar}\bra{c}\hat\rho\ket{c} \ln{\bra{c}\hat \rho \ket{c}}.
\label{wehrl entropy}
\end{equation}

\begin{figure}
\centering
\includegraphics[width=0.4\linewidth]{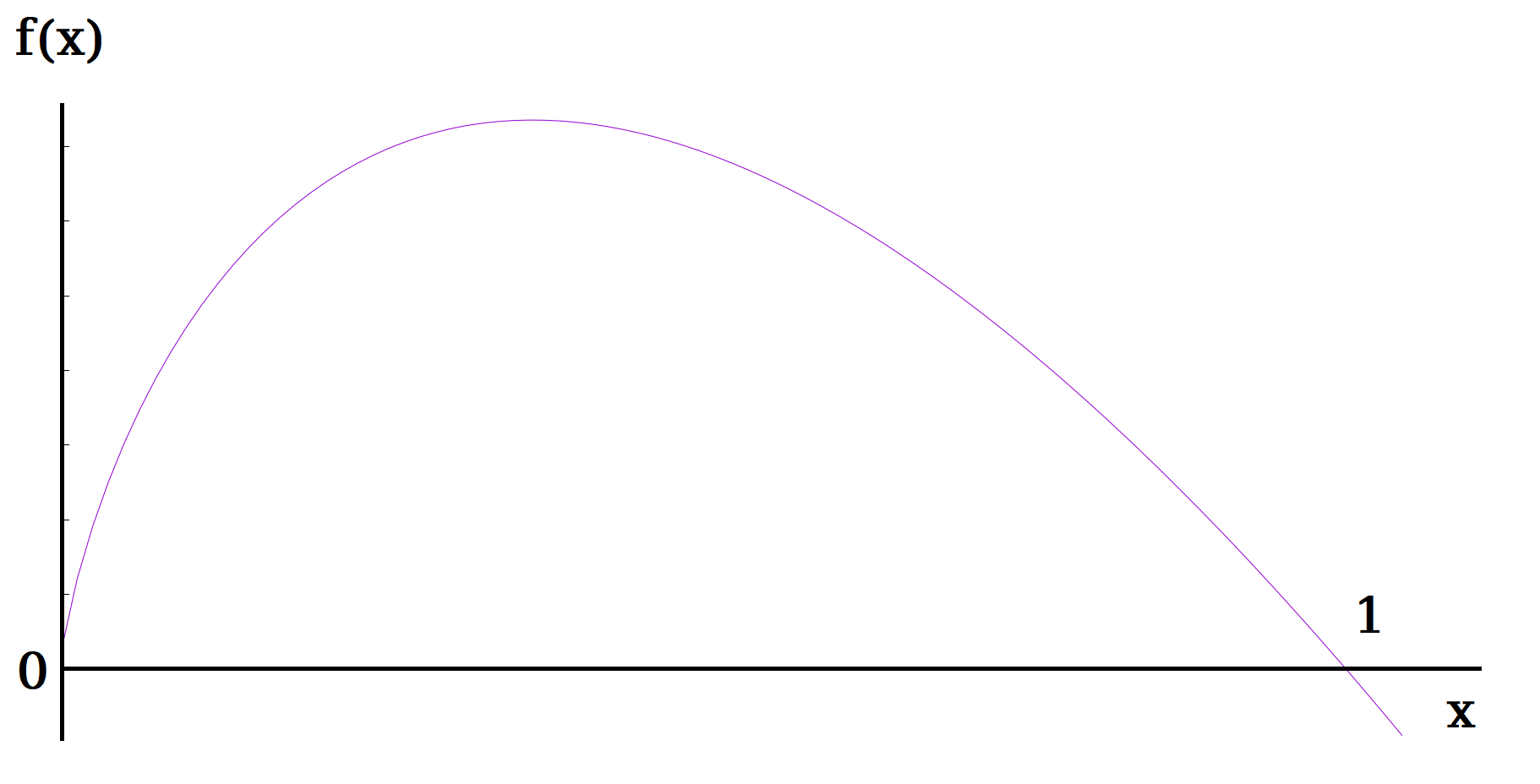}
\caption{Graph of $f(x)=-x\ln{x}$.}
\label{xlnx}
\end{figure}

Since $-x\ln{x}$ is a concave function (Fig.~[\ref{xlnx}]),
\begin{equation}
S_W>S_{vN}\geqslant 0.
\end{equation}
The equality $S_W=S_{vN}$ is impossible. This means that $S_W$ is always non-zero even for a pure state. For several physical systems subject to the classical substitution, a negligible error is obtained for smooth functions in phase space with a volume much larger than $\hbar$. However, if there are fluctuations concentrated in very small regions, this approximation is not a good model.

Furthermore, it would be useful to introduce a phase space to visualize what is happening in the quantum system of interest. However, this idealization contradicts the uncertainty principle. Now, by other means, some approximations can be made: Considering a one-dimensional quantum system with a generic pure state $\ket{\psi(t)}$, the so-called {\it Wigner distribution}\cite{wigner1932quantum} is defined as:
\begin{equation}
\begin{split}
W(q,p,t)&=\int_{-\infty}^{\infty}dxe^{-\mathbbm{i}px/\hbar}\bra{\psi(t)}\ket{q-x/2}\bra{q+x/2}\ket{\psi(t)} \\
&=\int_{-\infty}^{\infty}dxe^{-\mathbbm{i}px/\hbar}\bra{q+x/2}\hat{\rho}(t)\ket{q-x/2},   
\end{split}
\label{Wignereq}
\end{equation}
where $\hat{\rho}(t)$ is the density matrix of a pure state. The Wigner distribution is a function of both position $q$ and momentum $p$, satisfying the conditions:
\begin{equation}
\begin{cases}
\int\frac{dq}{2\pi\hbar}W(q,p,t)=|\bra{\psi(t)}\ket{p}|^2 \\
\int\frac{dp}{2\pi\hbar}W(q,p,t)=|\bra{\psi(t)}\ket{q}|^2 \\
\int\frac{dqdp}{2\pi\hbar}W(q,p,t)=1,  
\end{cases} 
\label{conditions}
\end{equation}
where the last property is the normalization. The set of properties [\ref{conditions}] makes it tempting to interpret $W$ as a probability distribution in phase space $(q,p)$. However, the Wigner distribution is strongly oscillatory and not positive definite, and thus it is a {\it quasi-distribution}. Nevertheless, it can still be used to investigate the properties of the system.

For an approximation of the phase space in the computation of Wehrl entropy, it is possible to use the Wigner distribution [\ref{Wignereq}] for cases where the analyzed physical system is compatible with a positive-definite distribution. However, the best we can do is describe the system in terms of the probabilities of finding the particle in a position given by the band $(q\pm\sigma_q/2,p\pm \sigma_p/2)$ with minimal uncertainty $\sigma_q\sigma_p=\hbar/2$.

Now, consider the Husimi distribution, which can be obtained from the Gaussian convolution of the Wigner distribution:
\begin{equation}
    H(q,p,t)=\frac{1}{\pi\hbar}\int dq'dp'e^{-m\omega(q-q')^2/\hbar-(p-p')^2/m\omega\hbar}W(q',p',t).
\label{husimi distribution}
\end{equation}
This expression is also known as the {\it Weistrass transform}\cite{zayed2019}. Here, $m$ is the mass of the particle and $\omega$ is an arbitrary parameter. The lengths of the Gaussian factors indicate that the distribution fills the configuration space with $\sigma_q =\sqrt{\hbar/2m\omega}$ and reciprocally in momentum space with $\sigma_p =\sqrt{\hbar m\omega/2}$. The different values taken by $\omega$ correspond to different resolution scales probed by the system. For oscillating systems, including radiation fields, $\omega$ is identified as the frequency.

Now, an important property of the Husimi distribution is that it is positive semi-definite:
\begin{equation}
H(q,p,t)\equiv\bra{c}\hat \rho \ket{c}=|\bra{\psi}\ket{c}|^2\geq 0,
\label{Husimi}
\end{equation}
This is essentially the trace of the density matrix in the coherent states basis. It is constructed in such a way that observables written in anti-normal order follow the optical equivalence theorem\cite{cahill1969}. This means that it is essentially the density matrix placed in normal order, that is, creation operators are placed to the left of annihilation operators:
\begin{equation}
    :\hat a \hat a^{\dagger}:=\hat a^{\dagger}\hat a.
    \label{normal}
\end{equation}
This procedure is also called Wick ordering and is essential in quantum field theory to avoid the appearance of infinities. Anti-normal ordering consists of reversing the logic built in [\ref{normal}].
Considering the definition [\ref{Husimi}], it is possible to write Wehrl entropy in the form:
\begin{equation}
    S_W=-\int\frac{dqdp}{2\pi\hbar}H(q,p)\ln{H(q,p)}.
    \label{Wehrltru}
\end{equation}
From [\ref{Wignereq}], it is also possible to define an alternative entropic definition:
\begin{equation}
    \bar{S}_W=-\int \frac{dqdp}{2\pi\hbar}W(q,p)\ln{W(q,p)}.
    \label{tilde}
\end{equation}

\begin{figure}[ht]
\centering
\includegraphics[width=0.95\linewidth]{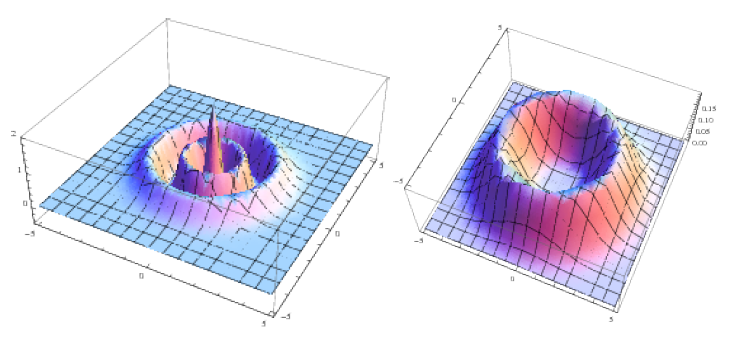}
\caption{The Wigner distributions (left) and Husimi distributions (right) for the fourth excited state of the harmonic oscillator in the $(q,p)$ plane. It can be noted that the Husimi distribution smoothes the abrupt oscillations that occur in the Wigner distribution, in addition to always being positive or null. This figure was taken from reference\cite{hatta2016}.}
\label{simulationwh}
\end{figure}

Now, for illustrative purposes, consider an example that can be solved analytically, the one-dimensional harmonic oscillator. This system has the classical Hamiltonian:
\begin{equation}
    \mathscr{H}=\frac{p^2}{2m}+\frac{m\omega^2q^2}{2}.
\end{equation}
For the $n$-th excited state, the Husimi distribution is given by:
\begin{equation}
H(q,p)=\frac{1}{n!}e^{-\mathscr{H}/\hbar\omega}\left(\frac{\mathscr{H}}{\hbar\omega}\right)^n. 
\end{equation}
Substituting this expression into [\ref{Wehrltru}], it is possible to show that:
\begin{equation}
    S_W=n+1+\ln{n!}-n\xi(n+1),
\end{equation}
where $\xi$ is the digamma function. Asymptotically, $S_W\approx \ln{\sqrt{n}}$. On the other hand, with the exception of the ground state, the expression for the conjugate entropy [\ref{tilde}], in this case, oscillates and becomes negative, which makes no sense for this specific problem; however, it is always possible to analyze both entropy notions and relate them through the Gaussian convolution. Fig.~[\ref{simulationwh}] shows the Husimi and Wigner distributions for the case of the harmonic oscillator in the fourth excited state, making good use of each of these distributions.

\section{Wehrl Entropy in QCD}
\label{apendice E: qcd wehrl entropy}

In the high-energy regime, partons are characterized by the fraction of longitudinal momentum $x$, transverse momentum $\vec{k}_T$, and transverse position or impact parameter $\vec{b}$. Thus, the system characterization can be obtained from the transverse momentum distribution TMD (from English {\it transverse momentum distribution}), $T(x, \vec{k}_T)$, and the Fourier transform of the generalized parton distribution GPD (from English, {\it generalized parton distribution}), $G(x, \vec{b})$.

Now, in the entropy calculation, both the information contained in $\vec{k}_T$ and $\vec{b}$ is necessary. Another example would be the decomposition of the nucleon spin obtained from the orbital angular momentum $\vec{b}\times\vec{k}_T$. One of the methods to obtain information from both variables, as they are conjugated, is the Wigner distribution already discussed in previous sections; however, now it is a function $W=W(x,\vec{b},\vec{k}_T)$.

In the quantum case, the Wigner distribution is given by [\ref{Wignereq}],
\begin{equation}
W(\vec{q},\vec{p},t)=\int_{-\infty}^{\infty}d^3xe^{-\mathbbm{i}\vec{p}\cdot \vec{x}}\bra{\vec{q}+\vec{x}/2}\hat{\rho}(t)\ket{\vec{q}-\vec{x}/2}.
\label{3dwigner}
\end{equation}

We will now discuss point by point the modifications that need to be made in [\ref{3dwigner}] for a construction consistent with QCD.

First, the change of variables:
\begin{equation}
t\rightarrow x, 
\end{equation}

\begin{equation}
\vec{q}\rightarrow \vec{b},   
\end{equation}

\begin{equation}
\vec{p}\rightarrow \vec{k}_T.    
\end{equation}
Thus, the uncertainty principle is ensured by:
\begin{equation}
\sigma_b\sigma_{k}\geq \frac{1}{2}.
\end{equation}

Eq.~[\ref{3dwigner}] is written in the position representation. In TQCs, it is customary to describe the operators in the momentum representation to apply the {\it Feynman} rules, for which a trivial two-dimensional Fourier transformation is needed. Furthermore, the fluctuation that respects the uncertainty principle in the Wigner distribution in quantum theory has its vectors labeled from $\pm \vec{x}/2$, symbolizing state vectors with minimum uncertainty as stipulated by the Heisenberg principle. In the context of QCD, $\pm \Delta/2$ is used, with $\Delta^{\mu}=(0,0,\vec{\Delta}_T)$, thus:
\begin{equation}
\int d^3xe^{-\mathbbm{i}\vec{p}\cdot \vec{x}} \rightarrow \int\frac{d^2\Delta_T}{(2\pi)^2}e^{-\mathbbm{i}\vec{\Delta}_T\cdot\vec{b}},
\end{equation}
remembering that $\vec{b}$ is the impact parameter bivector. Now, the pure state is represented by a hadron with 4-momentum $P^{\mu}$ respecting the minimum uncertainty limitation:
\begin{equation}
\ket{\vec{q}-\vec{x}/2}\rightarrow\ket{P-\Delta/2},
\end{equation}

\begin{equation}
\bra{\vec{q}+\vec{x}/2} \rightarrow \bra{P+\Delta/2}.
\end{equation}

Finally, the assimilation of the density matrix remains. It is given by:
\begin{equation}
\hat{\rho}(t)\rightarrow \int \frac{dz^-d^2z_T}{(2\pi)^3}e^{xP^+z^--\vec{k}_T\cdot \vec{z}_T}\Tr[F^{+\alpha}(z/2)U^{[+]}F_{\alpha}^+(-z/2)U^{[-]}],
\end{equation}

\begin{figure}[ht]
\centering
\includegraphics[width=0.9\linewidth]{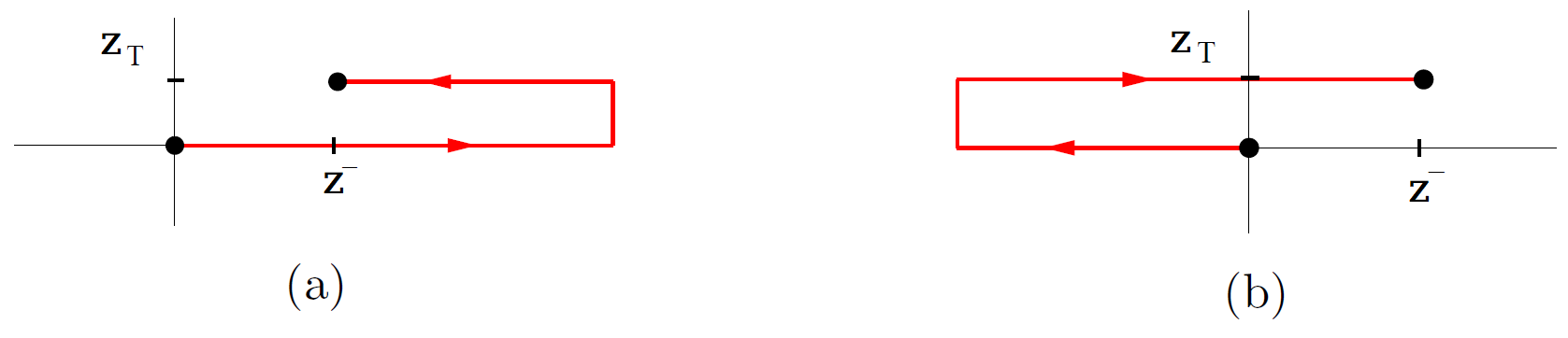}
\caption{Wilson lines in light-cone variables. In $(a)$ we have $U^{[+]}$ and in $(b)$ $U^{[-]}$.}
\label{linewilson}
\end{figure}

The Fourier transformation occurs for the same reasons as the substitution of the integral in $\vec{x}$, but now it is performed in light-cone variables. $F^{+\alpha}$ is the color field tensor and $U^{[\pm]}$ are the {\it Wilson} lines of form $U$ that keep the operators invariant under gauge transformations (for a discussion, see reference \cite{bomhof2006}). The {\it Wilson} lines here are shown in Fig.~[\ref{linewilson}].

Thus, the Wigner distribution in QCD can be written in the form:
\begin{equation}
\begin{split}
W(x,\vec{b},\vec{k}_T) &= \int \frac{dz^-d^2z_T}{(2\pi)^3}\frac{d^2\Delta_T}{(2\pi)^2}e^{-\mathbbm{i}(xP^+z^-+\vec{k}_T\cdot \vec{z}_T+\vec{\Delta}_T\cdot\vec{b})}\\
&\times\bra{P+\frac{\Delta}{2}}\Tr\left[F^{+\alpha}\left(\frac{z}{2}\right)U^{[+]}F_{\alpha}^+\left(-\frac{z}{2}\right)U^{[-]}\right]\ket{P-\frac{\Delta}{2}}.
\end{split}
\label{wignerdipolo}
\end{equation}
It is also possible to write the equation in a form more suitable for entropy per unit rapidity, with $xW(x,\vec{b},\vec{k}_T)$ given by:
\begin{equation}
xW= \int \frac{dz^-d^2z_T}{P^+(2\pi)^3}\frac{d^2\Delta_T}{(2\pi)^2}e^{-\mathbbm{i}(xP^+z^-+\vec{k}_T\cdot \vec{z}_T)}\bra{P+\frac{\Delta}{2}}\Tr\left[F^{+\alpha}\left(\vec{b}+\frac{z}{2}\right)U^{[+]}F_{\alpha}^+\left(\vec{b}-\frac{z}{2}\right)U^{[-]}\right]\ket{P-\frac{\Delta}{2}}.
\label{xwigner}
\end{equation}

Expression [\ref{wignerdipolo}] describes a distribution in transverse phase space characterized by the impact parameter $\vec{b}$ and transverse momentum $\vec{k}_T$ of the gluons that carry a fraction $x$ of the longitudinal momentum. If integration is performed over $\vec{b}$, the TMD $T(x,\vec{k}_T)$ is obtained; integrating over $\vec{k}_T$ gives the GPD $G(x,\vec{b})$. From it, the distribution related to the canonical orbital angular momentum of the longitudinally polarized nucleon can also be obtained:
\begin{equation}
L_W=\int dxd^2bd^2k(\vec{b}\times\vec{k}_T)W(x,\vec{b},\vec{k}_T).
\end{equation}

The Husimi distribution in QCD is given by:
\begin{equation}
xH(x,\vec{b},\vec{k}_T)=\frac{1}{\pi^2}\int d^2b'd^2k'e^{-(\vec{b}-\vec{b}')^2/\ell^2-\ell^2(\vec{k}_T-\vec{k}_T')^2}xW(x,\vec{b}',\vec{k}_T'),
\label{aquivai}
\end{equation}
where $\ell$ is an arbitrary parameter with dimensions of length\footnote{In some cases, $\ell$ is used as $R_h$, which is the hadronic radius. Another choice is $\ell^2=1/\expval{\vec{k}_T^2}$. Reference \cite{hatta2016} establishes a possible connection with high-energy physics where $\ell=1/Q_s$.}. Eq.~[\ref{aquivai}] seems to be an accurate extension of the Husimi distribution in the TQC's language, but its positivity is not guaranteed {\it a priori}, due to the recoil of the momentum $\Delta_T\neq0$, making there always be a difference between the initial and final states. Reference \cite{hatta2016} discusses some points of using expression [\ref{aquivai}]. Additionally, it is worth clarifying the positivity of the Husimi distribution as a working hypothesis. Thus, it is possible to write the Wehrl entropy in QCD:
\begin{equation}
 S_W(x)\equiv-\int d^2bd^2kxH(x,\vec{b},\vec{k}_T)\ln [xH(x,\vec{b},\vec{k}_T)],
 \label{entropiaqcdwehrl}
\end{equation}
and the definition of $\bar{S}_W(x)$ is also maintained, using the Wigner distribution:
\begin{equation}
\bar{S}_W(x)\equiv-\int d^2bd^2kxW(x,\vec{b},\vec{k}_T)\ln [xW(x,\vec{b},\vec{k}_T)],
\label{entropiaqcdwehrl2}
\end{equation}

A trivial example is a free quark or electron moving in the positive $z$ direction. The Husimi and Wigner distributions with $x=1$ are:
\begin{equation}
xH(\vec{b},\vec{k}_T)=\frac{e^{-b^2/\ell^2-\ell^2k^2}}{\pi^2},  
\end{equation}

\begin{equation}
xW(\vec{b},\vec{k}_T)=\delta^{(2)}(\vec{b})\delta^{(2)}(\vec{k}_T).    
\end{equation}

Although the Wigner distribution is positive-definite, its logarithm does not make sense, so the Wehrl entropy is obtained from Eq.~[\ref{entropiaqcdwehrl}],
\begin{equation}
S_W=\frac{1}{\pi^2}\int d^2bd^2ke^{-b^2/\ell^2-\ell^2k^2}\left(\frac{b^2}{\ell^2}+\ell^2k^2\right)=2.
\end{equation}

The fact that this entropy does not vanish reflects the inability to define position and momentum simultaneously due to the uncertainty principle.

\section{Wehrl Entropy for Partons}
\label{chapter 5 - section 1 - subsection 2: wehrl entropy for partons}

There are two approaches to dealing with the Wehrl entropy generated by the Wigner distributions of gluons in the small-$x$ region: the dipole formalism with the distribution $xW_{dip}$\cite{hatta2016b} and the Weiszacker-Williams (WW) gluons\cite{kovchegov1998} with the distribution $xW_{WW}$,

\begin{equation}
xW_{dip}(x, \vec{b}, \vec{k}_T)=\frac{2N_c}{\alpha_s(2\pi)^2}\int \frac{d^2r_T}{(2\pi)^2}e^{\mathbbm{i}\vec{k}_T\cdot\vec{r}_T}\left(\frac{\partial}{\partial b^2}b^2\frac{\partial}{\partial b^2}+k^2\right)\hat S(x,\vec{b},\vec{r}_T). 
\label{wigner dipolos}
\end{equation}

\begin{equation}
xW_{WW}(x, \vec{b}, \vec{k}_T)= \frac{C_F}{2\pi^4\alpha_s}\int d^2r\frac{e^{\mathbbm{i}\vec{r}_T\cdot \vec{k}_T}}{r_T^2}[1-\tilde{S}(x,r_T,b)]
\label{wigner ww}
\end{equation}

In these equations, $\hat S$ and $\tilde{S}$ are the S matrices for a dipole of size $r_T$ with an impact parameter $b$ in a scattering with a hadron, and the adjoint formulation of this operator, respectively. Evaluating these objects in the GBW model:
\begin{equation}
    \hat S = e^{-\frac{1}{4}r_T^2Q_{s}^2(x,b)}, \quad \quad  \tilde{S} = e^{-\frac{1}{4}r_T^2\tilde{Q}_{s}^2(x,b)},
    \label{matrizes S}
\end{equation}
where:
\begin{equation}
\tilde{Q}_{s}(x,b)=\frac{N_c}{C_F}Q_s^2(x,b)=\frac{N_c}{C_F}\left(\frac{x_0}{x}\right)^{\lambda}e^{-b^2/2\gamma_sB_{CGC}}.
\label{saturation scale in b-CGC}
\end{equation}
In this expression, the saturation scale depends on the impact parameter and is given by the b-CGC model, where $\gamma_s$ is the anomalous dimension and $B_{CGC}$ is a parameter. The constants were fitted with data obtained via DIS for small values of $x$\cite{rezaeian2013}
\begin{equation}
    \begin{cases}
    x_0=0.00105;\\
    \lambda=0.2063;\\
    \gamma_s=0.6599;\\
    B_{CGC}=5.5\quad GeV^{-2}.\\
    \end{cases}
    \label{bCGC values}
\end{equation}

The evaluation of the Wigner distribution case for dipoles will not be positive definite, requiring the calculation of the Husimi distribution, whose development was carried out in the work \cite{dissertacaogabriel}. In this thesis, the case of Wehrl entropy for WW gluons will be developed. Thus, it is possible to write expression [\ref{wigner ww}] as:
\begin{equation}
xW=\beta\int d^2r e^{\mathbbm{i}\vec{r}_T\cdot \vec{k}_T}f(x,r_T,b)=\beta \mathscr{F}^{-1}\{f(x,r_T,b)\},
\end{equation}
with the definition $\beta\equiv C_F/(2\alpha_s\pi^4)$, and the operation $\mathscr{F}^{-1}$ being the computation of the inverse two-dimensional Fourier transform for the function $f(x,r_T,b)$ given by:
\begin{equation}
f(x,r_T,b)=\frac{1-e^{-\frac{1}{4}r_T^2\tilde{Q}_{s}(x,b)}}{r_T^2},
\label{auxiliar f}
\end{equation}
such that:
\begin{equation}
\mathscr{F}^{-1}\{f(x,r_T,b)\}=\pi \Gamma\left(0,\frac{k^2_T}{\tilde{Q}_s^2(x,b)}\right), 
\label{fourier transform: incomplete gamma}
\end{equation}
where $\Gamma(0,x)$ is the positive-definite incomplete gamma function. Then, the Wigner distribution is given by:
\begin{equation}
xW_{WW}(x,k,b)=\frac{C_F}{2\pi^3\alpha_s}\Gamma\left(0,\frac{k^2_T}{\tilde{Q}_s^2(x,b)}\right).
\label{wigner WW em kT}
\end{equation}

Substituting Eq.~[\ref{wigner WW em kT}] into the definition of Wehrl entropy [\ref{wehrl entropy}] and neglecting the constant $\pi\beta$ in the factor involving the logarithm, we obtain:
\begin{equation}
    S_W=-\frac{C_F}{2\pi\alpha_s}\int_0^\infty db^2F(\tau)\tilde{Q}_s^2(x,b),
    \label{wehrl entropy with integral}
\end{equation}
where:
\begin{equation}
    F(\tau)=\int_0^{\tau}\Gamma(0,\tilde{\tau}) \ln \Gamma(0,\tilde{\tau}).
    \label{F definition}
\end{equation}
The integrand in expression $F$ was changed by a variable substitution in the form of $\tilde{\tau}=k^2/\tilde{Q}_s^2$, and to introduce a dependence on the resolution scale, $Q^2$ was inserted instead of $\infty$ in the integration. In expression [\ref{F definition}], $\tau=Q^2/\tilde{Q}_s^2(x,b)$. In the limit where $Q^2$ (and consequently $\tau$) tends to infinity, $F$ is just a number:
\begin{equation}
    \lim_{\tau \rightarrow \infty}F(\tau) = -0.248.
    \label{F limit}
\end{equation}
To simplify calculations, no numerical integration is performed on the impact parameter, taking advantage of the fact that the saturation scale has a maximum at $b=0$, with $\tilde{Q}_{s,max}^2(x)=\tilde{Q}_s^2(x,b=0)=(N_c/C_F)Q_s^2(x)$. In any case, in the small-$x$ region, the typical saturation scale is of the order of $1$ GeV. Thus, using $\tau=Q^2/\expval{\tilde{Q}_s^2}$ with $\expval{\tilde{Q}_s^2}=1$ GeV$^2$, we obtain $F\approx -0.095377$ for $Q^2=2$ GeV$^2$ and $F\approx -0.247802$ for $Q^2=10$ GeV$^2$. For any $Q^2$, after integrating over the impact parameter, we have:
\begin{equation}
    S_W(x)\approx-\frac{2F\gamma_sB_{CGC}N_c}{2\pi \alpha_s}Q_s^2(x),
    \label{final wehrl entropy}
\end{equation}
that is, $S_W \sim Q_s^2$.

\bibliographystyle{aip} 


\end{document}